\documentclass[twocolumn,times,astrosymb,tighten]{aastex701}

\newcommand{\epsindA}{$\epsilon$~Ind~A}
\newcommand{\epsindAb}{$\epsilon$~Ind~Ab}

\begin{document}

\title{Worlds Next Door. IV. Mapping the Late Stages of Giant Planet Evolution with a Precise Dynamical Mass and Luminosity for \epsindAb}

\correspondingauthor{Aniket Sanghi}

\author[0000-0002-1838-4757]{Aniket Sanghi}
\altaffiliation{NSF Graduate Research Fellow}
\affiliation{Cahill Center for Astronomy and Astrophysics, California Institute of Technology, 1200 E. California Boulevard, MC 249-17, Pasadena, CA 91125, USA}
\email[show]{asanghi@caltech.edu}

\author[0000-0001-5684-4593]{William Thompson}
\affiliation{NRC Herzberg Astronomy and Astrophysics, 5071 West Saanich Road, Victoria, BC, V9E 2E7, Canada}
\email{william.thompson@nrc-cnrc.gc.ca}

\author[0000-0001-5864-9599]{James Mang}
\altaffiliation{NSF Graduate Research Fellow}
\affiliation{Department of Astronomy, University of Texas at Austin, Austin, TX 78712, USA}
\email{j_mang@utexas.edu}

\author[0000-0002-6618-1137]{Jerry W. Xuan}
\altaffiliation{51 Pegasi b Fellow}
\affiliation{Department of Earth, Planetary, and Space Sciences, University of California, Los Angeles, CA 90095, USA}
\email{jerryxuan@g.ucla.edu}

\author[0000-0002-8895-4735]{Dimitri Mawet}
\affiliation{Cahill Center for Astronomy and Astrophysics, California Institute of Technology, 1200 E. California Boulevard, MC 249-17, Pasadena, CA 91125, USA}
\affiliation{Jet Propulsion Laboratory, California Institute of Technology, Pasadena, CA 91109, USA}
\email{dmawet@astro.caltech.edu}

\author[0000-0003-2233-4821]{Jean-Baptiste Ruffio}
\affiliation{Department of Astronomy \& Astrophysics, University of California, San Diego, La Jolla, CA 92093, USA}
\email{jruffio@ucsd.edu}

\author[0000-0003-0097-4414]{Yapeng Zhang}
\altaffiliation{51 Pegasi b Fellow}
\affiliation{Cahill Center for Astronomy and Astrophysics, California Institute of Technology, 1200 E. California Boulevard, MC 249-17, Pasadena, CA 91125, USA}
\email{yapzhang@caltech.edu}

\author[0000-0003-0774-6502]{Jason J. Wang}
\affiliation{Center for Interdisciplinary Exploration and Research in Astrophysics (CIERA), Northwestern University, 1800 Sherman Avenue, Evanston, IL, 60201, USA}
\affiliation{Department of Physics and Astronomy, Northwestern University, 2145 Sheridan Road, Evanston, IL 60208, USA}
\email{jason.wang@northwestern.edu}

\author[0000-0002-4404-0456]{Caroline V. Morley}
\affiliation{Department of Astronomy, University of Texas at Austin, Austin, TX 78712, USA}
\email{cmorley@utexas.edu}

\author[0000-0001-6975-9056]{Eric Nielsen}
\affiliation{Department of Astronomy, New Mexico State University, 1320 Frenger Mall, Las Cruces, NM 88003, USA}
\email{nielsen@nmsu.edu}

\author[0009-0008-9687-1877]{William Roberson}
\affiliation{Department of Astronomy, New Mexico State University, 1320 Frenger Mall, Las Cruces, NM 88003, USA}
\email{wcr@nmsu.edu}

\author[0000-0003-0593-1560]{Elisabeth Matthews}
\affiliation{Max-Planck-Institut f\"ur Astronomie, K\"onigstuhl 17, D-69117 Heidelberg, Germany}
\email{matthews@mpia.de}

\author[0000-0001-5365-4815]{Aarynn L. Carter}
\affiliation{Space Telescope Science Institute, 3700 San Martin Drive, Baltimore, MD 21218, USA}
\email{aacarter@stsci.edu}

\author[]{Ian J. M. Crossfield}
\affiliation{Department of Physics and Astronomy, University of Kansas, Lawrence, KS, USA}
\email{ianc@ku.edu}

\author[0000-0002-2918-8479]{Mathilde M\^alin}
\affiliation{Space Telescope Science Institute, 3700 San Martin Drive, Baltimore, MD 21218, USA}
\affiliation{Department of Physics \& Astronomy, Johns Hopkins University, 3400 N. Charles Street, Baltimore, MD 21218, USA}
\email{mmalin@stsci.edu}

\author[0000-0001-5578-1498]{Bj\"orn Benneke}
\affiliation{Department of Earth, Planetary, and Space Sciences, University of California, Los Angeles, CA 90095, USA}
\email{bbenneke@ucla.edu}

\author[0000-0002-9799-2303]{Alexis Bidot}
\affiliation{Space Telescope Science Institute, 3700 San Martin Drive, Baltimore, MD 21218, USA}
\email{abidot@stsci.edu}

\author[0000-0001-8612-3236]{Andr\'as G\'asp\'ar}
\affiliation{Steward Observatory, University of Arizona, Tucson, AZ 85721, USA}
\email{agaspar@arizona.edu}

\author[]{Carrie He}
\affiliation{Department of Earth, Planetary, and Space Sciences, University of California, Los Angeles, CA 90095, USA}
\email{carriehe@g.ucla.edu}

\author[0000-0001-9708-8667]{Katelyn Horstman}
\affiliation{Cahill Center for Astronomy and Astrophysics, California Institute of Technology, 1200 E. California Boulevard, MC 249-17, Pasadena, CA 91125, USA}
\email{khorstma@astro.caltech.edu}

\author[0000-0001-7443-6550]{Alexander Madurowicz}
\affiliation{Space Telescope Science Institute, 3700 San Martin Drive, Baltimore, MD 21218, USA}
\email{amadurowicz@stsci.edu}

\author[0000-0002-4164-4182]{Christian Marois}
\affiliation{NRC Herzberg Astronomy and Astrophysics, 5071 West Saanich Road, Victoria, BC, V9E 2E7, Canada}
\affiliation{Department of Physics and Astronomy, University of Victoria, 3800 Finnerty Road, Elliot Building, Victoria, BC V8P 5C2, Canada}
\email{christian.marois@nrc-cnrc.gc.ca}

\author[0000-0001-7130-7681]{Rebecca Oppenheimer}
\affiliation{American Museum of Natural History, New York, NY, USA}
\email{bro@amnh.org}

\author[0000-0002-3191-8151]{Marshall Perrin}
\affiliation{Space Telescope Science Institute, 3700 San Martin Drive, Baltimore, MD 21218, USA}
\email{mperrin@stsci.edu}

\shorttitle{JWST NIRCam and MIRI Imaging of \epsindA}
\shortauthors{Sanghi et al.}

\begin{abstract}
We present new JWST/NIRCam 4--5~$\mu$m (F410M, F430M) and JWST/MIRI 18--25~$\mu$m (F1800W, F2100W, F2550W) imaging detections of the nearby (3.6~pc) cold (275~K) gas giant exoplanet \epsindAb. The F2550W detection of \epsindAb\ constitutes the longest wavelength image of an exoplanet acquired to date. Combining three decades of radial velocity (RV) monitoring, Gaia-Hipparcos absolute astrometry, and relative astrometry from direct imaging (including the new NIRCam astrometry), we conduct a comprehensive re-analysis of \epsindAb's orbit and obtain a dynamical mass $M_{\rm Ab} = 6.5^{+0.7}_{-0.6}\;M_{\rm Jup}$. \added{Additionally, we demonstrate that the combination of unmodeled stellar activity, partial orbital coverage, and stitching together RVs from different instruments biased the eccentricity in previous orbit fits and led to an incorrect prediction for the planet's position at the JWST imaging discovery epoch.} Using \epsindAb's NIRCam and MIRI photometry, we assemble the first 4--25~$\mu$m spectral energy distribution (SED) of a cold gas giant outside the Solar System. The NIRCam photometry supports a metal-enriched atmosphere for \epsindAb\ based on analysis with atmospheric model grids, consistent with predictions from the giant planet mass-metallicity relation. While the current data do not provide definitive evidence for or against the presence of water ice clouds, we tentatively find that the H$_2$O vapor absorption-dominated F2550W photometry is systematically brighter ($>1\sigma$, but $<2\sigma$) than predictions from cloud-free/rainout chemistry models \added{and better explained by a cloudy model}. We calculate a bolometric luminosity of $\log L_{\rm bol}/L_\odot = -7.23 \pm 0.03$~dex by directly integrating \epsindAb's SED. Combining this with the planet's dynamical mass and age ($3.5 \pm 1.0$~Gyr), we demonstrate excellent agreement with evolutionary model predictions in a new regime of low luminosities, low masses, and old ages. Our results establish \epsindAb\ as a benchmark system for planetary evolution studies and set the stage for the detailed atmospheric characterization of this temperate extrasolar world.
\end{abstract}


\section{Introduction} 
The direct detection and characterization of extrasolar Jupiter-analog gas giants is central to developing our understanding of their formation and evolution \citep[e.g.,][]{madhusudhan_exoplanetary_2019, ikoma_formation_2025, snellen_exoplanet_2025}, placing the Solar System in the broader exoplanetary context through comparative planetology studies \citep[e.g.,][]{gelino_variability_2000, ge_rotational_2019, coulter_jupiter_2022}, and investigating their influence on the formation and long-term habitability of Earth-like worlds \citep[e.g.,][]{noble_orbital_2002, raymond_search_2006, georgakarakos_giant_2018, antoniadou_puzzling_2018, kong_how_2024}. Indeed, over the past two decades, the technique of high-contrast imaging has successfully combined high-order adaptive optics systems \citep{guyon_extreme_2018}, starlight suppression technologies \citep[][]{kenworthy_high-contrast_2025}, and innovative post-processing strategies \citep{cantalloube_exoplanet_2020} to assemble a remarkable sample of young ($<1$~Gyr) gas giants for detailed atmospheric characterization studies \citep{bowler_imaging_2016, 2023ASPC..534..799C}. 


Prior to the launch of JWST, nearly all direct imaging detections were made at near-infrared wavelengths ($JHK$ and $L$ bands, $\approx$1--4~$\mu$m), where a significant fraction of the bolometric flux of young, hot ($\sim1000$~K) gas giants is emitted. However, mature gas giants ($\sim$Gyr ages) are significantly colder ($\lesssim300$~K) and exhibit orders of magnitude more favorable contrast in the mid-infrared wavelengths ($>10$~$\mu$m), from a combination of being near the peak of the planet's thermal emission but in the Rayleigh-Jeans tail of the host star \citep[e.g.,][]{spiegel_spectral_2012}. These wavelengths are extremely challenging to access from the ground due to the rapidly increasing thermal background beyond 4~$\mu$m, though notable efforts were made with the VLT/NEAR experiment, which demonstrated the potential for detection of low mass planets in the $N$~band \citep[10--12.5~$\mu$m, ][]{kasper_near_2017,kasper_near_2019, wagner_imaging_2021, pathak_high-contrast_2021, viswanath_constraints_2021}. Now, JWST's unparalleled thermal infrared sensitivity \citep[4--25~$\mu$m,][]{rigby_science_2023} has opened the doors to the direct detection of gas giants more similar in age and temperature to Jupiter and Saturn in our own Solar System \citep[e.g.,][]{matthews_temperate_2024, beichman_worlds_2025, sanghi_worlds_2025, bowens-rubin_nircam_2025, gagliuffi_jwst_2025, sanghi_worlds_2026-1}. The nearest stars are the most favorable targets in this endeavor, since planet flux $f_p \propto d^{-2}$ and the planet's on-sky angular separation (for fixed physical separation) $\rho \propto d^{-1}$, where $d$ is distance to the system.

At a distance of $\sim$3.6~pc, \epsindA\ (K5V, 0.78~$M_\odot$, Table~\ref{tab:prop}) is one of the nearest Sun-like stars to Earth and \added{a high priority}, Tier A target in the search for habitable Exo-Earths with NASA's planned Habitable Worlds Observatory \citep[\added{no known dust disks, no known close binary companions, and high planet search completeness};][]{mamajek_nasa_2024}. The star's age is estimated to be $3.5\pm1.0$~Gyr, similar to our Solar System's age of $\sim$4.6~Gyr, based on activity indicators \citep{chen_precise_2022}. Early imaging discovered a co-moving brown dwarf companion $\epsilon$~Ind~B at a separation of $\sim$1460~au \citep{scholz_varepsilon_2003}. Follow-up high angular resolution near-infrared imaging with VLT NAOS/CONICA revealed that the companion was in fact a binary, consisting of the T1 and T6 dwarfs $\epsilon$~Ind~Ba and $\epsilon$~Ind~Bb, respectively \citep{mccaughrean_e_2004}. Long term radial velocity (RV) monitoring of the system identified a trend that could not be explained by the brown dwarf binary or secular acceleration of the system \citep{endl_planet_2002, zechmeister_planet_2013}, thereby pointing to the presence of an additional companion in the system. The combination of RV and Hipparcos-Gaia absolute astrometry suggested that the additional companion was a cold Jupiter, \epsindAb\ \citep[\added{initially estimated to be} $\sim$3~$M_{\rm Jup}$,][]{feng_detection_2019, feng_revised_2023}. Subsequent JWST/MIRI coronagraphic observations at 10.65~$\mu$m and 15.5~$\mu$m (and reprocessing of archival VLT/NEAR observations) provided the first direct detection of \epsindAb, but found it at a location inconsistent with the RV and astrometry orbit prediction \citep{matthews_temperate_2024}. The inconsistency was attributed to biases from modeling RVs covering a limited time span and the dynamical mass was revised to $\sim7\;M_{\rm Jup}$ \citep{feng_lessons_2025}. The mid-infrared photometry indicate that \epsindAb\ has an effective temperature $\sim$275~K, making it one of the coldest exoplanets imaged to date. Additionally, \citet{matthews_temperate_2024} found that the non-detection of the planet between 3.5--5.0~$\mu$m in previous ground-based imaging suggested an unknown source of flux suppression, possibly from high metallicity and/or carbon-to-oxygen ratio. \epsindAb's proximity, wide angular separation (3\arcsec--4\arcsec), and measured dynamical mass make it a prime benchmark target for detailed atmospheric characterization with JWST.

In this paper, we present new JWST/NIRCam 4--5~$\mu$m (F410M, F430M) and JWST/MIRI 18--25~$\mu$m (F1800W, F2100W, F2550W) imaging detections of \epsindAb, which, together, enable us to assemble the first 4--25~$\mu$m spectral energy distribution of a cold gas giant exoplanet. The paper is organized as follows. Section~\ref{sec:obs} summarizes the observational sequences executed with JWST, the data reduction, and procedures to estimate companion astrometry and photometry. Section~\ref{sec:orbit} re-considers the orbit of \epsindAb\ with the addition of the new high-precision NIRCam astrometry, revises the planet's dynamical mass, \added{and provides a solution to the issue of the incorrect original orbit prediction from joint modeling of the RVs and Hipparcos-Gaia absolute astrometry.} Section~\ref{sec:atmo-evo} compares \epsindAb's SED to that of Y dwarf WISE 0855, presents an initial reconnaissance of \epsindAb's atmospheric properties (metallicity, water ice clouds, disequilibrium chemistry), and calculates the planet's bolometric luminosity. Section~\ref{sec:evomodel} conducts comprehensive consistency tests between \epsindAb's measured fundamental properties and predictions from evolutionary models. Finally, Section~\ref{sec:concl} lists our conclusions. Appendices~\ref{sec:app-ref} and \ref{app:orbit-post} present the MIRI reference library and the posteriors for all orbit models considered in this work, respectively.

\begin{deluxetable}{lcc}
\tabletypesize{\footnotesize}
\tablecaption{\label{tab:prop}Stellar Properties of \epsindA}
\tablehead{\colhead{Property} & \colhead{Value} & \colhead{Ref.}}
\startdata
$\alpha_{2016.0}$ ($^\circ$) & 330.8724 $\pm$ 0.0687 & 1\\
$\delta_{2016.0}$ ($^\circ$) & --56.7973 $\pm$ 0.0655 & 1\\
$\mu_{\alpha^*}$\tablenotemark{a} (mas\,$\mathrm{yr^{-1}}$) & 3966.661 $\pm$ 0.086 & 1 \\
$\mu_{\delta}$ (mas\,$\mathrm{yr^{-1}}$) & --2536.192 $\pm$ 0.092 & 1 \\
$\varpi$ (mas) & $274.8431 \pm 0.0956$ & 1\\
Distance (pc) & $3.648^{+0.022}_{-0.009}$ & 2\\
SpT & K5V & 3\\
Mass ($M_\odot$) & $0.782 \pm 0.023$ & 4 \\
Radius ($R_{\odot}$) & 0.713 $\pm$ 0.006 & 2, 4, 5 \\
Age (Gyr) & $3.48^{+0.78}_{-1.03}$ & 6 \\
$T_{\mathrm{eff}}$ (K) & 4700 $\pm$ 65 & 4 \\
$\mathrm{[Fe/H]}$ (dex) & --0.17 $\pm$ 0.03 & 4 \\
$v \sin i$ ($\mathrm{km/s}$) & 2.0 & 5 \\
$\log(R'_{HK})$ (dex) & --4.72 & 7\\
$P_\mathrm{rot}$ (d) & --35.732$^{+0.006}_{-0.003}$& 8\\
$\mathrm{RUWE_{DR3}}$ & 1.148 & 1\\
\enddata
\tablenotetext{a}{Proper motion in R.A. includes a factor of $\cos \delta$.}
\tablerefs{(1)~\citet{gaia_collaboration_gaia_2023}; (2)~\citet{creevey_gaia_2023}; (3)~\citet{torres_search_2006}; (4)~\citet{lundkvist_low-amplitude_2024}; (5)~\citet{rains_precision_2020}; (6)~\citet{chen_precise_2022}; (7)~\citet{pace_chromospheric_2013}; (8)~\citet{feng_detection_2019}. 
}
\end{deluxetable}

\section{Observations and Data Reduction} 
\label{sec:obs}
\subsection{JWST/NIRCam Imaging}

\begin{figure*}
    \centering
    \includegraphics[width=\linewidth]{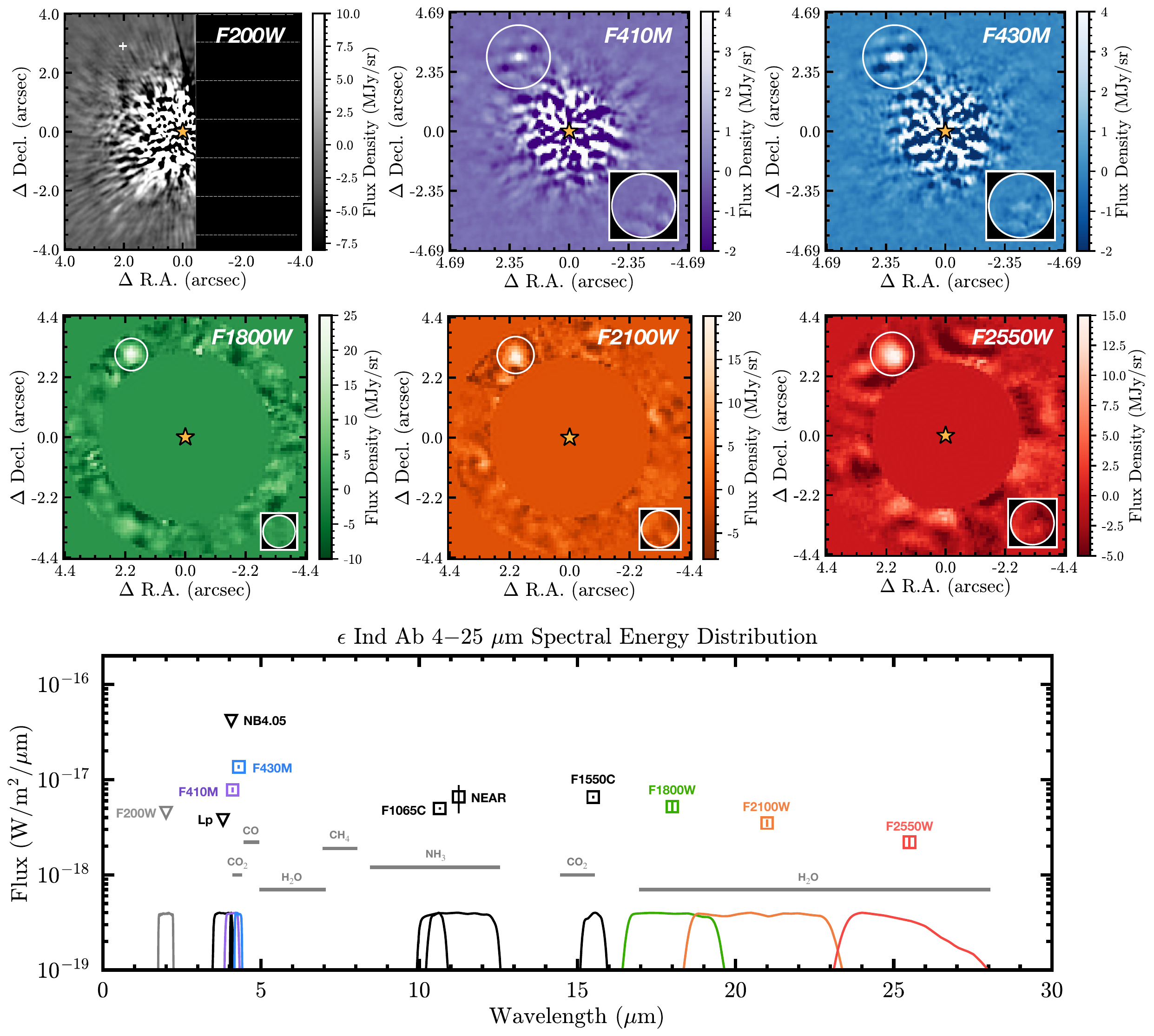}
    \caption{Gallery of PSF-subtracted images centered on \epsindA\ in NIRCam coronagraphic (top row) and MIRI non-coronagraphic (center row) imaging filters. All images are oriented North up and East left. Exoplanet \epsindAb\ is detected in all filters except F200W (expected location marked with a `$+$'). An inset image (bottom right corner) shows the residuals in the aperture centered on \epsindAb\ after the best-fit planet forward model is subtracted. The aperture encompasses the six-lobed structure of the planet PSF in the NIRCam images and has a radius = 1 FWHM in the MIRI images. The F2550W image of \epsindAb\ is the longest wavelength image of an exoplanet taken to date. The bottom panel shows the complete 4--25~$\mu$m spectral energy distribution of \epsindAb. Boxes with error bars represent the measured photometry and inverted triangles represent upper limits, with the filter profiles plotted below. Solid gray horizontal lines mark the dominant opacity source in the corresponding wavelength region.}
    \label{fig:detection}
\end{figure*}

\subsubsection{Executed Sequences}
We obtained F200W (1.755--2.227~$\mu$m), F410M (3.865--4.301~$\mu$m), and F430M (4.167--4.398~$\mu$m) JWST/NIRCam imaging of \epsindA\ on UT 2025~August~28 as part of Cycle 4 program GO 8714 (PI: J.~Xuan, Co-PI: A.~Sanghi, J.~B.~Ruffio, Y.~Zhang). We chose two-roll angular differential imaging \citep[ADI:][]{liu_substructure_2004, marois_angular_2006} to detect the giant planet \epsindAb\ given its wide separation ($>$3\arcsec). No reference stars were observed. The observations were carried out between UT 2025 August 28 03:56:22 and 06:10:59 in a non-interruptible sequence for a total of $\approx$2.2 hours (including observatory overheads) simultaneously with the F200W/F410M filter combination followed by the F200W/F430M filter combination, and using the MASK430R coronagraph at telescope V3 position angles 2.54$^\circ$ and 13.49$^\circ$ ($\approx11^\circ$ roll). We used the \texttt{SHALLOW4} readout pattern with 10 groups per integration and 23 integrations per exposure for all observations. The total exposure time on \epsindA\ is 80.4 minutes in F200W and 40.2 minutes in F410M and F430M across the two-roll sequence.

\subsubsection{Data Processing}
The uncalibrated Stage 0 (*uncal.fits) data products were acquired from the Barbara A. Mikulski Archive for Space Telescopes (MAST\footnote{The data described here may be obtained from the MAST archive at\dataset[10.17909/nqx2-yw28]{https://doi.org/10.17909/nqx2-yw28}.}) and processed with \texttt{spaceKLIP} \citep{girard_jwstnircam_2022, kammerer_performance_2022, carter_jwst_2023, carter_spaceklip_2025}, a community developed pipeline for high contrast imaging with JWST. \texttt{spaceKLIP} wraps the \texttt{jwst} pipeline \citep{bushouse_jwst_2025} for basic data processing steps with modifications for coronagraphic imaging reduction. Data reduction via \texttt{spaceKLIP} for our observations implements best-practices discussed in various works \citep{girard_jwstnircam_2022, kammerer_performance_2022, carter_jwst_2023, franson_jwstnircam_2024, gagliuffi_jwst_2025} and, more specifically, closely follows \citet{gagliuffi_jwst_2025}. We used \texttt{spaceKLIP} v2.1, via github commit \#11df3a1, and \texttt{jwst} v1.19.1; the calibration files were from CRDS v12.1.11 and the \texttt{jwst\_1413.pmap} CRDS context\footnote{\url{https://jwst-crds.stsci.edu/}}. The data were fit ``up the ramp" using the ``Likely" algorithm described in \citet{brandt_optimal_2024, brandt_likelihood-based_2024} and transformed from Stage 0 images into Stage 2 (*calints.fits) images, using a jump threshold of 4 and 4 pseudo-reference pixels on all sides of the subarrays. The $1/f$ noise was mitigated using a median filter computed along each column. We skipped dark current subtraction \citep{carter_jwst_2023}. Pixels flagged by the \texttt{jwst} pipeline as well as 5$\sigma$ outliers detected using sigma clipping were replaced with a 2-dimensional interpolation based on a 9-pixel kernel. We also identified additional pixels affected by cosmic rays by flagging pixels with significant temporal flux variations across integrations and replaced them by their temporal median. Subsequently, the images were blurred above the Nyquist sampling criterion, using a Gaussian with a FWHM of 2.55 for the F410M/LW detector, 2.30 for the F430M/LW detector, and 2.78 for the F200W/SW detector. These values ensure that sharp features in the data do not create artifacts when Fourier shifts are applied to the undersampled images. The position of the star behind the coronagraph was estimated by fitting a model coronagraphic point spread function (PSF) from \texttt{webbpsf\_ext} \citep{leisenring_webbpsf_2025} to the first science integration. All subsequent images were shifted by this initial offset. Images were cross-correlated to the first science integration, and these small shifts were applied to each integration to center the entire observing sequence to the first science integration. The position of \epsindA\ behind the mask differed by $\approx$18~mas between the two rolls.

We carry out PSF subtraction with \texttt{spaceKLIP}, which uses the Karhunen-Lo\'eve Image Projection \citep[KLIP:][]{soummer_detection_2012} algorithm implemented in \texttt{pyKLIP} \citep{wang_pyklip_2015} to model and subtract the host star PSF. \added{The reduction is parameterized by the number of annular subtraction zones (\texttt{annulus}), the number of azimuthal subsections each annulus is divided into (\texttt{subsections}), and the number of principal components used (\texttt{numbasis}).} Using \texttt{annulus}=1, \texttt{subsections}=1, and \texttt{numbasis}=1, \epsindAb\ is detected at signal-to-noise ratios (S/Ns) of 17.9 and 25.4 in F410M and F430M (Figure~\ref{fig:detection}), respectively. The S/N is estimated by comparing the companion contrast to the annular contrast. The planet is not recovered in F200W (Figure~\ref{fig:detection}). \added{Note that the star appears at the right edge of the SW subarray due to the projected proximity of MASK430R to the vertical gap between the SW detectors\footnote{\url{https://jwst-docs.stsci.edu/jwst-near-infrared-camera/nircam-observing-modes/nircam-coronagraphic-imaging}}}. Besides a known background source in the South-East direction at $\approx$11\arcsec separation \citep[also seen in Spitzer images,][]{matthews_temperate_2024}, no other significant sources are detected in the full-frame images.

\subsubsection{Astrometry and Photometry}
To measure the astrometry and photometry of \epsindAb, we use the KLIP forward modeling \citep[KLIP-FM:][]{pueyo_detection_2016} framework implemented in \texttt{spaceKLIP} and \texttt{pyKLIP}. This approach accounts for measurement biases introduced in the primary subtraction procedures. A flexible forward model is generated and then fit to a given source in a post-processed image within a Bayesian framework \citep{wang_orbit_2016}. We employ the \texttt{emcee} affine-invariant Markov Chain Monte Carlo (MCMC) ensemble sampler \citep{foreman-mackey_emcee_2013} to simultaneously fit the planet separation, position angle, and flux. The PSF model is produced by \texttt{stpsf} \citep{perrin_simulating_2012, perrin_updated_2014}. A total of 50 walkers over $5\times10^5$ total steps (10000 per walker) are used to sample the posteriors of the three model parameters. We discard the first 5000 steps of each chain as burn-in. To account for correlated speckle noise, we model the noise distribution via a Gaussian process with a Mat\'ern ($\nu$ = 3/2) kernel \citep{wang_orbit_2016}. The measurements in detector units are transformed into physical units accounting for coronagraph transmission, filter zero points, and system distance. 

The astrometric and photometric uncertainties are computed as the standard deviation of the MCMC chains. Additionally, the adopted uncertainties incorporate a systematic uncertainty of 0.05 pixel on the star-behind-mask position along each axis \citep[for astrometry,][]{girard_optimizing_2024} and 0.01 mag on the stellar magnitude (for photometry). The stellar magnitude in all filters is estimated from stellar spectra drawn from the posteriors obtained by \citet[][see their Extended Data Fig.~7 for the best-fit spectrum]{matthews_temperate_2024} after fitting ground- and space-based photometry of \epsindA. The resulting planet astrometry and photometry are shown in Table~\ref{tab:prop-planet}.

We measure the 5$\sigma$ upper limit in F200W through PSF injection-recovery. Synthetic sources generated with \texttt{stpsf} are injected at the same separation as \epsindAb\ across seven position angles (75$^\circ$ to 165$^\circ$ in steps of 15$^\circ$). The S/N of each injected source is then measured by comparing contrast at the injected location to annular contrast, masking the position of \epsindAb\ and the region outside the detector field-of-view (FOV; the star is offset in the short-wavelength channel, see Figure~\ref{fig:detection}). The injected flux is adjusted to produce a S/N of 5$\sigma$. The mean injected flux across all position angles is taken as the 5$\sigma$ upper limit (Table~\ref{tab:prop-planet}).

\subsection{JWST/MIRI Imaging}

\subsubsection{Executed Sequences and Archival Data}
We obtained F1800W (16.519--19.502~$\mu$m), F2100W (18.477--23.159~$\mu$m), and F2550W (23.301--26.733~$\mu$m) JWST/MIRI imaging of \epsindA\ on UT 2025~September~4 as part of Cycle 4 program GO 8714 (PI: J.~Xuan, Co-PI(s): A.~Sanghi, J.~B.~Ruffio, Y.~Zhang). The observations were conducted in the MIRI MRS simultaneous imaging mode\footnote{\url{https://jwst-docs.stsci.edu/jwst-mid-infrared-instrument/miri-operations/miri-mrs-simultaneous-imaging}} during dedicated background observations for spectroscopy of \epsindAb\ (Obs~\#5). This mode enables simultaneous data collection by both the MIRI imager and medium resolution spectrometer. The observations were carried out between UT 2025~September~4 04:00:53 and 07:26:05 for a total of $\approx$3.5 hours (including observatory overheads) sequentially in the F1800W, F2100W, and F2550W filters. A 4-point cycling dither pattern was employed for each filter to enable accurate background subtraction (the choice of the dither pattern was optimized for the MRS dedicated backgrounds). We used the \texttt{FULL} subarray (74\arcsec$\times$113\arcsec) and the \texttt{FASTR1} readout pattern with 5 groups per integration (the minimum recommended) and 53 integrations per exposure for all observations. The choice of subarray is not optimal as it saturates the PSF core in each filter and the first Airy ring in F1800W. However, the smaller subarrays (\texttt{SUB64}, \texttt{SUB128}, \texttt{SUB256}, \texttt{BRIGHTSKY}) all result in the star and planet outside their FOV in at least one of the four dither positions. Given \epsindAb's wide angular separation ($\approx3\farcs5$), the saturation was deemed acceptable. The total exposure time on \epsindA\ is 58.9 minutes in each filter combining the four dither sequence.

\begin{deluxetable*}{lcclllllrrc}
    \label{tab:prop-planet}
    \centering
    \tablecolumns{11}
    \tablecaption{Astrometry and Photometry of \epsindAb}
    \tablehead{\colhead{Instrument} & \colhead{Filter} & \colhead{Date} & \colhead{$\rho$} & \colhead{$\theta$} & \colhead{$m_\mathrm{*}$} & \colhead{$\Delta$(mag)}  & \colhead{$m_\mathrm{b}$} & \multicolumn{2}{c}{Flux} & \colhead{Ref.} \\ \colhead{} & \colhead{} & \colhead{} & \colhead{(\arcsec)} & \colhead{($^\circ$)} & \colhead{(Vega mag)} & \colhead{(Vega mag)} & \colhead{(Vega mag)} & \colhead{(Wm$^{-2}\mu$m$^{-1}$)} & \colhead{(Jy)} & \colhead{}} 
        \startdata
        JWST/NIRCam & F200W & 2025-08-28 & \nodata & \nodata & 2.187$\pm$0.008 & $>$18.06  & $>$20.25 & $<$4.5$\times10^{-18}$ & $<$6.0$\times10^{-6}$& 1 \\
        \hline
        VLT/NaCo & \emph{L'} & 2018-10-12 & \nodata & \nodata & 2.142$\pm$0.007 & $>$15.66 & $>$17.80 & $<$3.78$\times10^{-18}$ & $<$1.83$\times10^{-5}$& 2, 3 \\
        &  & 2018-10-26 &  &  &  &  &  &  & &  \\
        &  & 2018-11-03 &  &  &  &  &  &  & &  \\
        &  & 2018-11-04 &  &  &  &  &  &  & & \\
        \hline
        VLT/NaCo & NB4.05 & 2008-10-31 & \nodata & \nodata & 2.115$\pm$0.008 & $>$12.78 & $>$14.89 & $<$4.16$\times10^{-17}$ & $<$2.28$\times10^{-4}$& 2, 4 \\
         & & 2008-09-02 &  & &  &  &  &  &  & \\
         \hline
        JWST/NIRCam & F410M & 2025-08-28 & 3.551$\pm$0.003 & 34.97$\pm$0.05 & 2.131$\pm$0.007 &  14.58$\pm$0.06& 16.71$\pm$0.06 & (7.8$\pm$0.5)$\times10^{-18}$ & (4.37$\pm$0.28)$\times10^{-5}$& 1 \\
        \hline
        JWST/NIRCam &  F430M & 2025-08-28 & 3.551$\pm$0.003 & 34.97$\pm$0.05 & 2.154$\pm$0.008 & 13.75$\pm$0.05 & 15.90$\pm$0.05 & (1.36$\pm$0.06)$\times10^{-17}$ & (8.4$\pm$0.4)$\times10^{-5}$& 1 \\
        \hline
        VLT/VISIR & NEAR & 2019-09-14 & 4.820$\pm$0.164 & 42.5$\pm$2.5 & 2.237$\pm$0.008 & 10.5$\pm$0.4 & 12.7$\pm$0.4 & (6.6$\pm$2.2)$\times10^{-18}$ & (2.8$\pm$0.9)$\times10^{-4}$& 2 \\
         &  & 2019-09-15 &  &  &  &  &  &  & &  \\
         &  & 2019-09-17 &  &  &  &  &  &  & & \\
         \hline
        JWST/MIRI & F1065C & 2023-07-03 & 4.095$\pm$0.010 & 38.17$\pm$0.44 & 2.209$\pm$0.008 & 10.93$\pm$0.03 & 13.14$\pm$0.03 &(4.98$\pm$0.15)$\times10^{-18}$ & (1.88$\pm$0.06)$\times10^{-4}$& 2 \\
        \hline
        JWST/MIRI & F1550C & 2023-07-03 & 4.114$\pm$0.010 & 37.39$\pm$0.43 & 2.173$\pm$0.007 & 9.04$\pm$0.03 & 11.21$\pm$0.03 &(6.55$\pm$0.20)$\times10^{-18}$ & (5.25$\pm$0.16)$\times10^{-4}$& 2 \\
        \hline
        JWST/MIRI & F1800W & 2025-09-04 & 3.64$\pm$0.09 & 34.4$\pm$1.5 & 2.166$\pm$0.008 & 8.64$\pm$0.14 & 10.81$\pm$0.14  & (5.2$\pm$0.7)$\times10^{-18}$ & (5.6$\pm$0.8)$\times10^{-4}$& 1 \\
        \hline
        JWST/MIRI & F2100W & 2025-09-04 & 3.54$\pm$0.07 & 33.6$\pm$1.2 & 2.160$\pm$0.008 & 8.47$\pm$0.08 & 10.63$\pm$0.08 & (3.5$\pm$0.3)$\times10^{-18}$ & (5.2$\pm$0.4)$\times10^{-4}$& 1 \\
        \hline
        JWST/MIRI & F2550W & 2025-09-04 & 3.61$\pm$0.12 & 32.6$\pm$1.9 & 2.198$\pm$0.008 & 8.05$\pm$0.15 & 10.25$\pm$0.15 & (2.2$\pm$0.3)$\times10^{-18}$ & (4.8$\pm$0.7)$\times10^{-4}$& 1 \\
        \enddata
    \tablerefs{(1)~This Work; (2)~\citet{matthews_temperate_2024}; (3)~\citet{viswanath_constraints_2021}; (4)~\citet{janson_imaging_2009}.}
    \tablecomments{$\rho$ is the planet's separation in arcseconds, $\theta$ is the planet's position angle measured in degrees East of North, $m_\mathrm{*}$ is the stellar apparent magnitude, $\Delta$(mag) is the planet contrast with respect to the star, $m_\mathrm{b}$ is the planet's apparent magnitude.}
\end{deluxetable*}

Since the MIRI images of \epsindA\ were acquired in MIRI MRS simultaneous imaging mode as a bonus at no additional cost, it was not possible to conduct dedicated reference star observations or obtain two roll angles. To enable PSF subtraction, we constructed a PSF library for each MIRI filter using publicly available archival MIRI imaging observations on MAST. This is motivated by past success in both ground- and space-based high-contrast imaging applications \citep{choquet_first_2016, xuan_characterizing_2018, sanghi_efficiently_2022, xie_reference-star_2022, sanghi_efficiently_2024}. We selected observations where a bright PSF (not saturated to a greater extent than \epsindA) could be extracted with a sufficient FOV (accounting for the dither pattern) to perform reference star differential imaging (RDI) out to $\approx$4\farcs5. No further selection cuts were applied due to the limited number of observations available. The full list of datasets used for the archival PSF reference library are presented in Appendix~\ref{sec:app-ref}. \added{For added PSF diversity, we also included synthetic \texttt{stpsf} models in our reference library. We generated a set of 81 synthetic PSFs, per filter, dithered across a grid of [$-$2, 2] pixels in steps of 0.5 pixels.} 

\subsubsection{Data Processing}
\label{sec:miri-proc}
The calibrated Stage 1 (*cal.fits) data products (both \epsindA\ and selected archival reference observations) were acquired from MAST\footnote{The data described here may be obtained from the MAST archive at\dataset[10.17909/nqx2-yw28]{https://doi.org/10.17909/nqx2-yw28}.} and processed to Stage~3 (*stage3\_asn\_skysub\_i2d.fits) with \texttt{MAGIC} \citep[\url{https://github.com/kevin218/Magic}, e.g.,][]{bowens-rubin_nircam_2025}, a custom pipeline developed to address the non-uniform background structure in the MIRI flat-field. The Stage~2 background subtraction procedure in \texttt{MAGIC} follows an approach adapted from an STScI JWebbinar demo\footnote{\url{ https://github.com/spacetelescope/jwebbinar_prep/blob/jwebbinar31/jwebbinar31/miri/Pipeline_demo_subtract_imager_background-platform.ipynb}}, which applied a custom method utilizing multiple dithers per source. The Stage 3 reduction is identical to the \texttt{jwst} pipeline. We used \texttt{MAGIC} v1.0 and \texttt{jwst} v1.19.1; the calibration files were from CRDS v12.1.11 and the \texttt{jwst\_1413.pmap} CRDS context. 

We carry out PSF subtraction with \texttt{vip\_hci} \citep{gomez_gonzalez_vip_2017, christiaens_vip_2023}, which uses the KLIP algorithm to model and subtract the host star PSF. First, the \epsindA\ PSF is centered on the frame by registering it to a \texttt{stpsf} model using the \texttt{chi2\_shift} function in \texttt{image\_registration}\footnote{\url{https://image-registration.readthedocs.io/}} (masking the saturated core). Second, all PSFs are cropped about their center to a frame size of 81~$\times$~81 pixels (8\farcs91~$\times$~8\farcs91 FOV) in F1800W and F2100W, and 101~$\times$~101 pixels (11\farcs11~$\times$~11\farcs11 FOV) in F2550W. The centers of the archival MIRI PSFs are determined visually in \texttt{SAO DS9}. Third, we performed annular KLIP reductions (\texttt{pca\_annular} \added{in \texttt{vip\_hci}}) at the radial separation of \epsindAb\ (from NIRCam astrometry) for various annular subtraction geometries parameterized by annulus size (\texttt{asize}), number of sections (\texttt{n\_segments}), position angle of the first section (\texttt{theta\_init}), and radial distance to the center of the annulus (\texttt{rad}). \epsindAb\ is robustly detected in all three MIRI filters at its expected separation and position angle (Figure~\ref{fig:detection}). The reductions presented use \texttt{asize}=\{2, 2, 2.5\}~FWHM \added{(where the FWHM is set according to the observation wavelength\footnote{\url{https://jwst-docs.stsci.edu/jwst-mid-infrared-instrument/miri-performance/miri-point-spread-functions}})}, \texttt{n\_segments}=\{6, 8, 1\} \texttt{theta\_init}=\{0$^\circ$, 40$^\circ$, 0$^\circ$\}, \texttt{rad}=\{35, 34, 35\} pixels, and \{20, 17, 13\} principal components for \{F1800W, F2100W, F2550W\}, detecting the planet at S/N = \{7.7, 14.3, 7.5\}, respectively. The F2550W detection of \epsindAb\ constitutes the longest wavelength image of an exoplanet taken to date.

\subsubsection{Astrometry and Photometry}
Astrometric and photometric measurements of \epsindAb\ are extracted using the negative fake companion injection (NEGFC) method \citep[][]{lagrange_giant_2010, marois_exoplanet_2010}. We apply NEGFC to our observations using \texttt{vip\_hci}'s \texttt{firstguess} function. First, we generate a \texttt{stpsf} model with the appropriate MIRI filter, using the closest-in-time on-sky optical path difference (OPD) map to the observations, and including detector effects. \added{The OPD map includes measurements of the wavefront error across the telescope primary mirror segments.} The \texttt{stpsf} model in each filter is normalized to the exit pupil of the optical system so that the sum of the PSF is 1.0 in an infinite aperture. We inject the model PSF at the NIRCam measured position for varying \emph{negative} flux values, perform PSF subtraction, and minimize a figure of merit in a \{0.5, 0.5, 1\}~FWHM radius aperture at the location of injection for \{F1800W, F2100W, F2550W\} to obtain a first estimate of the planet's flux. The initial parameter estimates are input to a simplex Nelder–Mead minimization algorithm \citep[][implemented with argument \texttt{simplex = True}]{nelder_simplex_1965}, which provides a more accurate estimate of the planet's separation, position angle, and
flux. We minimize two figures of merit: (1) the reduced $\chi^2$ and (2) the residual sum of squares (RSS). The two solutions (found to be consistent with each other) are averaged to obtain the final astrometry and photometry (Table~\ref{tab:prop-planet}). 

\added{To obtain a first estimate of the astrometric uncertainty, we assume the radial and tangential directions are independent and calculate their uncertainties as FWHM/(S/N) \citep[see e.g.,][]{zhou_hubble_2021}. As discussed in \S\ref{sec:miri-proc}, because of core saturation, \epsindA\ was centered on the frame using the unsaturated PSF wings and by comparison with an \texttt{stpsf} model. This model does not perfectly represent the observed PSF. The estimate of the PSF center also changes depending on the specific unsaturated region chosen to perform the alignment procedure. Based on the observed variance in center coordinates, we adopt a centering uncertainty of 0.5~pixels ($\approx$50~mas) and add it in quadrature to the above. 

The photometric uncertainty is dominated by speckle noise. A first estimate can be made using the standard deviation of the integrated flux in non-overlapping 1~FWHM radius apertures at the separation of \epsindAb, accounting for the effect of small sample statistics \citep{mawet_fundamental_2014}. We note that the number of independent samples for standard deviation estimation exceeds ten at each wavelength and the correction is small.

We also implemented \texttt{vip\_hci}'s nested sampling-based injection-recovery approach to directly estimate the uncertainty from the posterior distributions for separation, position angle, and flux. This procedure yielded similar or slightly smaller values compared to our initial estimates. We adopt the uncertainties computed with the first method as they are larger and thus the more conservative choice. While the photometric estimates are expected to be robust, we would advise caution when using the astrometry given that the exact PSF center is unknown.}

\begin{deluxetable}{llll}
    \label{tab:priors}
    \centering
    \tablecaption{\epsindAb\ Orbit Modeling Parameters and Adopted Priors}
    \tablehead{\colhead{Parameter} & \colhead{Description} & \colhead{Unit} & \colhead{Prior}} 
        \startdata
        $M_A$ & Primary Mass & $M_\odot$ & $\mathcal{N}$(0.782, 0.023) \\
        $M_b$ & Secondary Mass & $M_{\rm Jup}$ & $\mathcal{U}$(0.01, 20) \\
        $\varpi$ & Parallax & mas & $\mathcal{N}$(274.8431, 0.0956) \\
        $\mu_{\alpha^*}$ & Proper Motion (R.A.) & mas/yr & $\mathcal{U}$(3965.661, 3967.661) \\
        $\mu_{\delta}$ & Proper Motion (Decl.) & mas/yr & $\mathcal{U}$($-$2537.192, $-$2535.192) \\
        $\gamma_*$ & Systemic RV & km/s & $\mathcal{N}$($-$40.035, 1) \\
        \hline
        $a$ & Semi-major Axis & au & $\mathcal{U}$(1, 100), $\mathcal{U}$($\log{1}$, $\log{100}$)\tablenotemark{a} \\
        $e$ & Eccentricity & \nodata & $\mathcal{U}$(0, 0.999) \\
        $i$ & Sky Inclination & \nodata & Sine \\
        $\omega$ & Argument of Periastron & rad & $\mathcal{U}$(0, 2$\pi$) \\
        $\Omega$ & Longitude of Ascending Node & rad & $\mathcal{U}$(0, 2$\pi$) \\
        $\theta$ & Position Angle at Reference Epoch & rad & $\mathcal{U}$(0, 2$\pi$) \\
        \hline
        $B$ & GP Amplitude & m/s & $\mathcal{U}$(0.00001, 2000000) \\
        $C$ & GP Amplitude Ratio & \nodata & $\mathcal{U}$(0.00001, 200) \\
        $L$ & GP Decay Timescale & days & $\mathcal{U}$(2, 2000) \\
        $P_{\rm rot}$ & GP Rotation Period & days & $\mathcal{U}$(1, 100) \\
        \hline
        $m$ & RV Slope & m/s/yr & $\mathcal{N}$(0, 1) \\
        $\gamma_{\rm LC}$ & CES (LC) Zero Point & m/s & $\mathcal{U}$($-$100, 100) \\
        $\gamma_{\rm VLC}$ & CES (VLC) Zero Point & m/s & $\mathcal{U}$($-$100, 100) \\
        $\gamma_{\rm HARPS03}$ & HARPS03 Zero Point & m/s & $\mathcal{U}$($-$100, 100) \\
        $\gamma_{\rm HARPS15}$ & HARPS15 Zero Point & m/s & $\mathcal{U}$($-$100, 100) \\
        $\gamma_{\rm HARPS20}$ & HARPS20 Zero Point & m/s & $\mathcal{U}$($-$100, 100) \\
        $\gamma_{\rm ESPR18}$ & ESPR18 Zero Point & m/s & $\mathcal{U}$($-$100, 100) \\
        $\gamma_{\rm ESPR19}$ & ESPR19 Zero Point & m/s & $\mathcal{U}$($-$100, 100) \\
        $\gamma_{\rm UVES}$ & UVES Zero Point & m/s & $\mathcal{U}$($-$100, 100) \\
        $\sigma_{\rm LC}$ & CES (LC) Jitter & m/s & $\mathcal{U}$($\log{0.01}$, $\log{100}$) \\
        $\sigma_{\rm VLC}$ & CES (VLC) Jitter & m/s & $\mathcal{U}$($\log{0.01}$, $\log{100}$) \\
        $\sigma_{\rm HARPS03}$ & HARPS03 Jitter & m/s & $\mathcal{U}$($\log{0.01}$, $\log{100}$) \\
        $\sigma_{\rm HARPS15}$ & HARPS15 Jitter & m/s & $\mathcal{U}$($\log{0.01}$, $\log{100}$) \\
        $\sigma_{\rm HARPS20}$ & HARPS20 Jitter & m/s & $\mathcal{U}$($\log{0.01}$, $\log{100}$) \\
        $\sigma_{\rm ESPR18}$ & ESPR18 Jitter & m/s & $\mathcal{U}$($\log{0.01}$, $\log{100}$) \\
        $\sigma_{\rm ESPR19}$ & ESPR19 Jitter & m/s & $\mathcal{U}$($\log{0.01}$, $\log{100}$) \\
        $\sigma_{\rm UVES}$ & UVES Jitter & m/s & $\mathcal{U}$($\log{0.01}$, $\log{100}$) \\
        \enddata  
    \tablenotetext{a}{The flat prior is used for ``observable-based" prior model fits and the log-flat prior is used for the ``uniform'' prior model fit.}
\end{deluxetable}

\section{Orbital Properties}
\label{sec:orbit}
The JWST imaging search for \epsindAb\ \citep{matthews_temperate_2024} was guided by evidence of a planet from a combined analysis of RV and absolute astrometry \citep{feng_detection_2019, feng_revised_2023}. However, the location of the imaged planet was discrepant with the original orbit prediction. This highlighted the challenge of modeling the orbits of wide separation companions, particularly when the RV time series consists of observations by several different instruments with independent offsets. \added{A solution to this issue is discussed later in this section.} First, we combine three decades of radial velocity monitoring, Gaia-Hipparcos absolute astrometry, and relative astrometry from direct imaging (including our new NIRCam detection) to perform a comprehensive re-analysis of \epsindAb's orbit. Though the orbit was recently revised by \citet{feng_lessons_2025}, we find significant differences in our solution and discuss the reasons for this.

\subsection{Radial Velocity Data}
We use publicly available radial velocity (RV) data on \epsindA\ from the Data \& Analysis Center for Exoplanets (DACE\footnote{\url{https://dace.unige.ch/daceTeam/}}) platform. This includes 73 RVs from the Long Camera (LC) and 55 RVs from the Very Long Camera (VLC) of the Coud\'e Echelle Spectrometer at the European Southern Observatory \citep[ESO, ][]{zechmeister_planet_2013}. The High Accuracy Radial-velocity Planet Searcher (HARPS) spectrograph has collected 4500+ RVs of \epsindA\ \citep{trifonov_public_2020}. We binned data points collected by HARPS within a single night in DACE before using them in our analysis. This gives us 106 RVs from the pre-2015 upgrade (HARPS03), 38 RVs from the post-2015 upgrade (HARPS15), and 50 RVs from the post-2020 shutdown (HARPS20) period. The data above are treated as three separate instruments to account for any baseline shifts. We also use measurements from the Very Large Telescope (VLT) Echelle SPectrograph for Rocky Exoplanets and Stable Spectroscopic Observations \citep[ESPRESSO,][]{pepe_espresso_2021}. A technical intervention was conducted in 2019 \citep{pepe_espresso_2021} and we accordingly model the available data as two separate instruments. There are 48 ESPRESSO RVs available pre-2019 intervention and 70 ESPRESSO RVs available post-2019 intervention. Finally, we incorporate 164 RVs from the Ultraviolet and Visual Echelle Spectrograph \citep[UVES,][]{feng_detection_2019}. This gave us a total of 604 data points, from 8 independent instruments, covering a period of 29 years from November 3, 1992 to September 22, 2021.

For all of the above, we use the RVs on DACE ignoring the $\approx$1.8~m/s/yr acceleration caused by the proper motion (secular acceleration). The RVs are not corrected for the changing light travel-time between \epsindA\ and the Earth due to Earth's motion around the Sun and \epsindA's changing distance from Earth. We handle both of these terms, \added{the secular acceleration and effect of changing light travel-time}, directly in our orbit model. 


\subsection{Relative Astrometry Data}
We incorporate the relative astrometry of \epsindAb\ from previously published 2019 VLT/NEAR and 2023 JWST/MIRI F1550C observations \citep[the F1065C astrometry is unreliable,][]{matthews_temperate_2024} in addition to our new 2025 JWST/NIRCam measurement (Table~\ref{tab:prop-planet}). Together, the imaging observations (with detections) have a baseline of $\approx$~6~years.

\subsection{Absolute Astrometry Data}
We include absolute astrometry of \epsindA\ from Hipparcos and Gaia's Third Data Release (DR3) in our orbit modeling. We use the Hipparcos proper motion from the DR3 Hipparcos-Gaia Catalog of Accelerations \citep[HGCA:][]{brandt_hipparcos-gaia_2021}, calculated from a mix of the \citet{esa_hipparcos_1997} and \citet{van_leeuwen_hipparcos_2007} Hipparcos data releases, and calibrated against the Gaia DR3 velocity reference frame. We also include the Gaia DR3 proper motion and Gaia-Hipparcos long-term proper motion and uncertainties from the HGCA. We undo the non-linear corrections \citep[$\gamma$ term in Eq.~21 and 22 in][]{brandt_hipparcos-gaia_2018} incorporated by the HGCA to account for secular acceleration and curvature in the R.A.\ and Decl.\ coordinate system, in favor of handling this motion inside our model. Note that the above corrections ensure accurate cross-calibration between the Hipparcos and Gaia reference frames \citep[see \S5.1 in][]{brandt_hipparcos-gaia_2021}; the Hipparcos and Gaia DR3 proper motions reported \emph{at their respective epochs} are already corrected for secular acceleration using the systemic RV during the data reduction procedure (see Table 1.2.3 in \citet{esa_hipparcos_1997}\footnote{Note that \citet{van_leeuwen_hipparcos_2007} did not make the correction to the proper motion derived in their re-reduction. However, the magnitude of the correction is negligibly small ($<$~0.05~$\rm{mas/yr^{-2}}$ in each coordinate direction) over the Hipparcos baseline and does not impact orbit fits.} and \S3.1 in \citet{lindegren_gaia_2021}).

\subsection{Model Implementation}
\label{sec:model-implement}
We implemented our orbit model in the \texttt{Julia}-based \citep{bezanson_julia_2012} \texttt{Octofitter} framework \citep{thompson_deep_2023} and sample using stabilized variational nonreversible parallel tempering \citep{surjanovic_parallel_2022, syed_non-reversible_2022} implemented in \texttt{Pigeons.jl} \citep{surjanovic_pigeonsjl_2023}. We jointly model the RVs, Hipparcos and Gaia absolute astrometry, and relative astrometry. The complete model incorporating every listed data source has a total of 33 different parameters. The majority of these are instrument-specific nuisance parameters, which we marginalize over when presenting results. The full list of parameters and our adopted priors are listed in Table~\ref{tab:priors}. 

\begin{figure*}
    \centering
    \includegraphics[width=\linewidth]{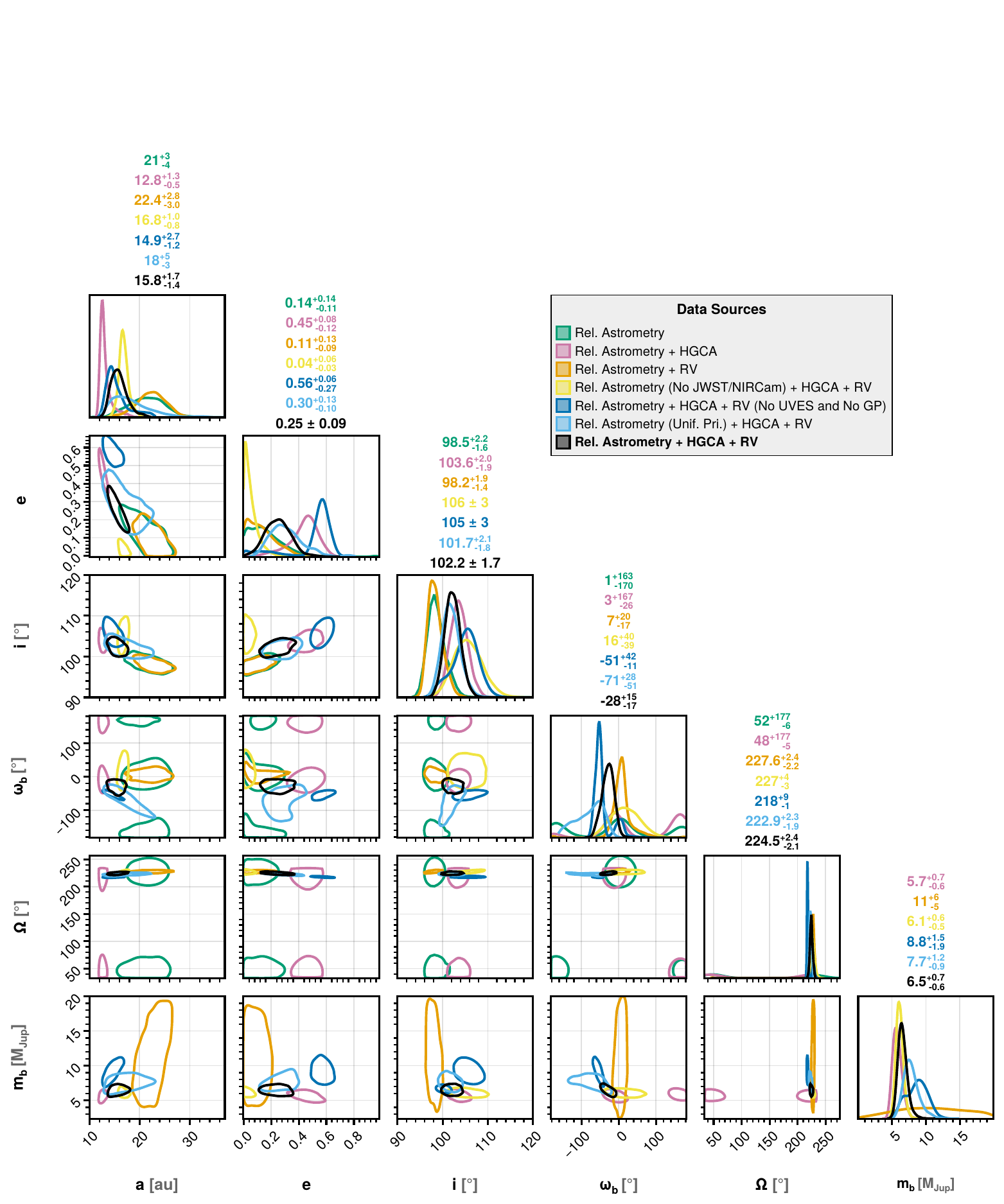}
    \caption{A comparison between posteriors from the seven converged (out of eight) orbit models considered for \epsindAb, listed in the legend. All contours in the corner plot encompass the volume contained within the 1$\sigma$ 2D Gaussian equivalent. The final adopted model is in black and yields a dynamical mass $6.5^{+0.7}_{-0.6}\;M_{\rm Jup}$. A key takeaway is that the new JWST/NIRCam astrometry provides evidence for a moderately eccentric orbit (yellow vs black posteriors).}
    \label{fig:orbit-post}
\end{figure*}

\begin{figure*}
    \centering
    \includegraphics[width=\linewidth]{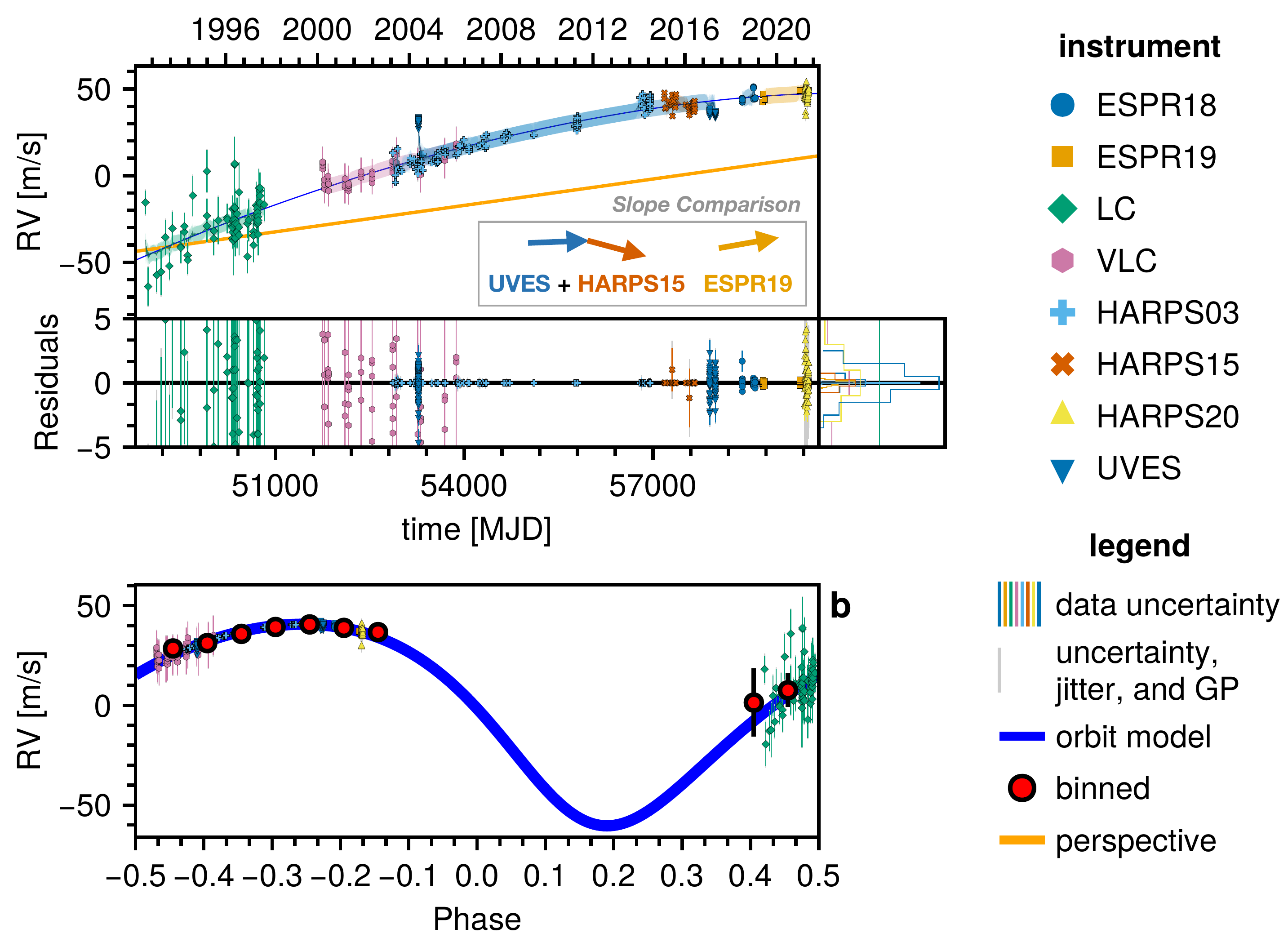}
    \caption{Summary of orbit fit for the complete relative astrometry + RV + HGCA model. \emph{Top:} The full RV timeseries with the model fit (dark blue, top sub-panel) and residuals (bottom sub-panel, in units of m/s, with an extension showing the distribution of residuals for each instrument). Data from distinct instruments are represented with different colors and symbols. \added{An orange line shows the contribution of secular (perspective) acceleration to the RV time series.} The colored line connecting the points is the GP model and the colored band shows the uncertainty from the GP model. An inset depicts the average slope vectors for the UVES, HARPS15, and ESPR19 time series (see the discussion in \S\ref{sec:UVES}). \emph{Bottom:} The phase-folded model RV curve in shown in blue. The binned data are shown as red points.}
    \label{fig:orbit-RV}
\end{figure*}

\begin{figure*}
    \centering
    \includegraphics[width=\linewidth]{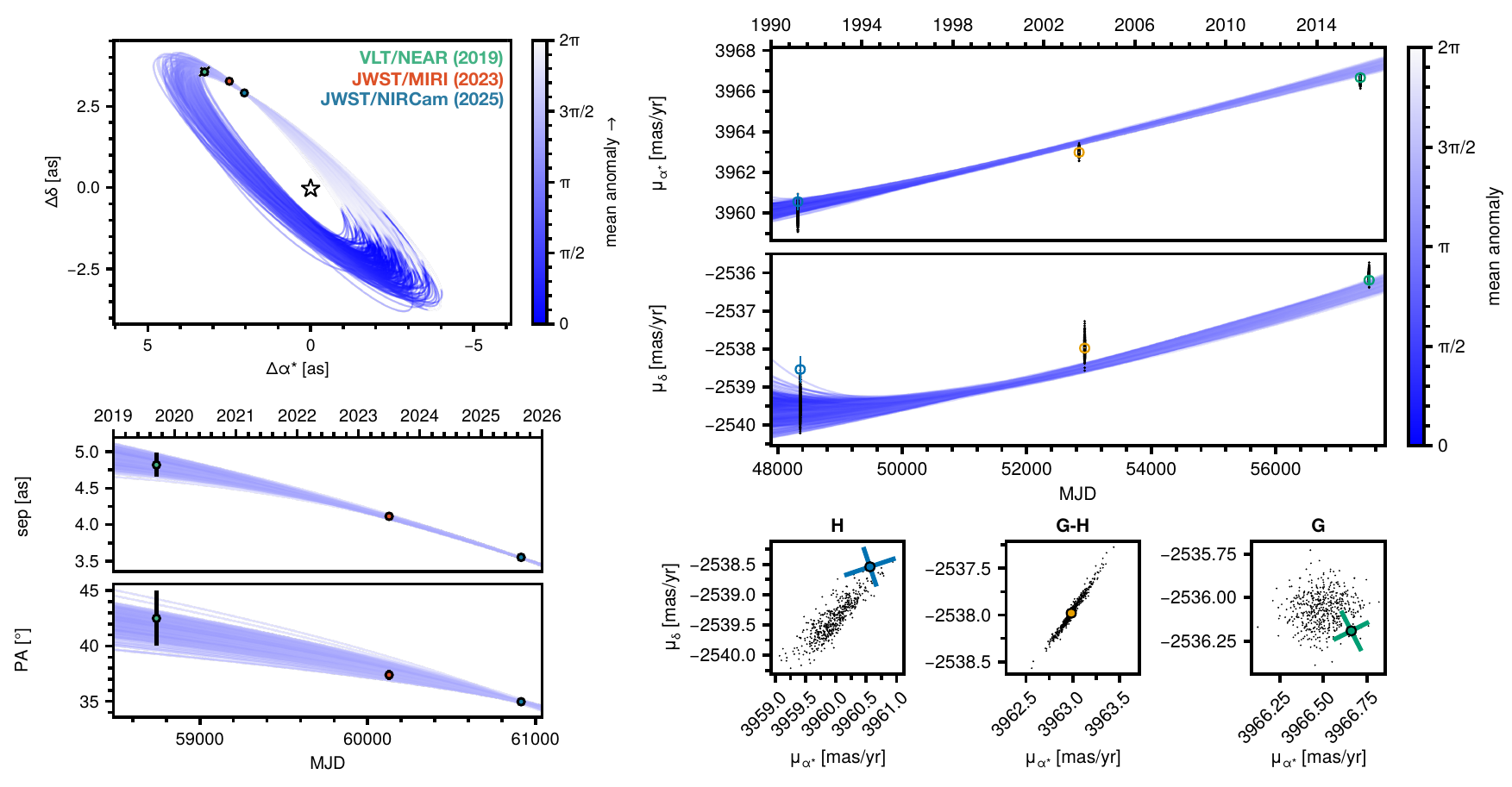}
    \caption{Summary of orbit fit for the complete relative astrometry + RV + HGCA model. \emph{Top left:} 250 randomly sampled sky-projected orbits with the measured relative astrometry overlaid. A mean anomaly of zero corresponds to periastron passage. \emph{Bottom left:} The projected separation and position angle vs time for the sampled orbits. \emph{Top right:} 500 random samples of the proper motion (PM) in R.A.\ and Decl.\ vs time with measurements from HGCA. \emph{Bottom right:} Covariances between R.A. and Decl. velocity at each PM epoch. The ``H" epoch is the net PM at the Hipparcos epoch calibrated to the DR3 velocity reference by the HGCA, ``G-H" is the long-term PM derived from the HGCA, and ``G" is the Gaia DR3 PM.}
    \label{fig:orbit-astrometry}
\end{figure*}


Non-linear effects in all of the above data such as secular acceleration and the changing light travel-time between \epsindA\ and Earth are accounted for within the \texttt{Octofitter} model, identical to \citet{thompson_revised_2025}. To be conservative, we nonetheless include a linear trend for the RV datasets in our model. A single slope parameter ($m$) is shared between all RV datasets (system-wide parameter). Additionally, we use a separate RV offset/zero point ($\gamma$) and jitter ($\sigma$) for each instrument. Stellar activity is known to induce RV variability over multiple time-scales in the case of \epsindA\ consistent with the star's rotation period of $\approx$36~days and magnetic cycle \citep{feng_detection_2019, lundkvist_low-amplitude_2024}. We modeled the activity signal from the star using a Gaussian process (GP) with a quasi-periodic kernel. The GP, with parameters $B$, $C$, $L$, and $P_{\rm rot}$ (similar to $\epsilon$~Eridani in \citet{roettenbacher_expres_2022, thompson_revised_2025}), is implemented in our model using the \texttt{Celerite} \citep{foreman-mackey_fast_2017} framework through the \texttt{Celerite.jl} package. We fit a single GP to the complete RV timeseries (system-wide parameters). We also conduct an orbit fit without a GP for comparison.

Given the low phase coverage of the relative astrometry, we use the framework of ``observable-based" priors introduced in \citet{oneil_improving_2019} and implemented in \texttt{Octofitter}\footnote{\url{https://sefffal.github.io/Octofitter.jl/stable/rel-astrom-obs/}}. The prior is based on uniformity in orbital observables rather than in orbital parameters, and has the impact of reducing bias towards high eccentricity orbits and those where periastron passage occurs near the observed planet position \citep{do_o_orbital_2023}. We also conduct an orbit fit with the ``uniform" (standard) priors for comparison. All orbit models assume a single planet.

\begin{figure*}
    \centering
    \includegraphics[width=\linewidth]{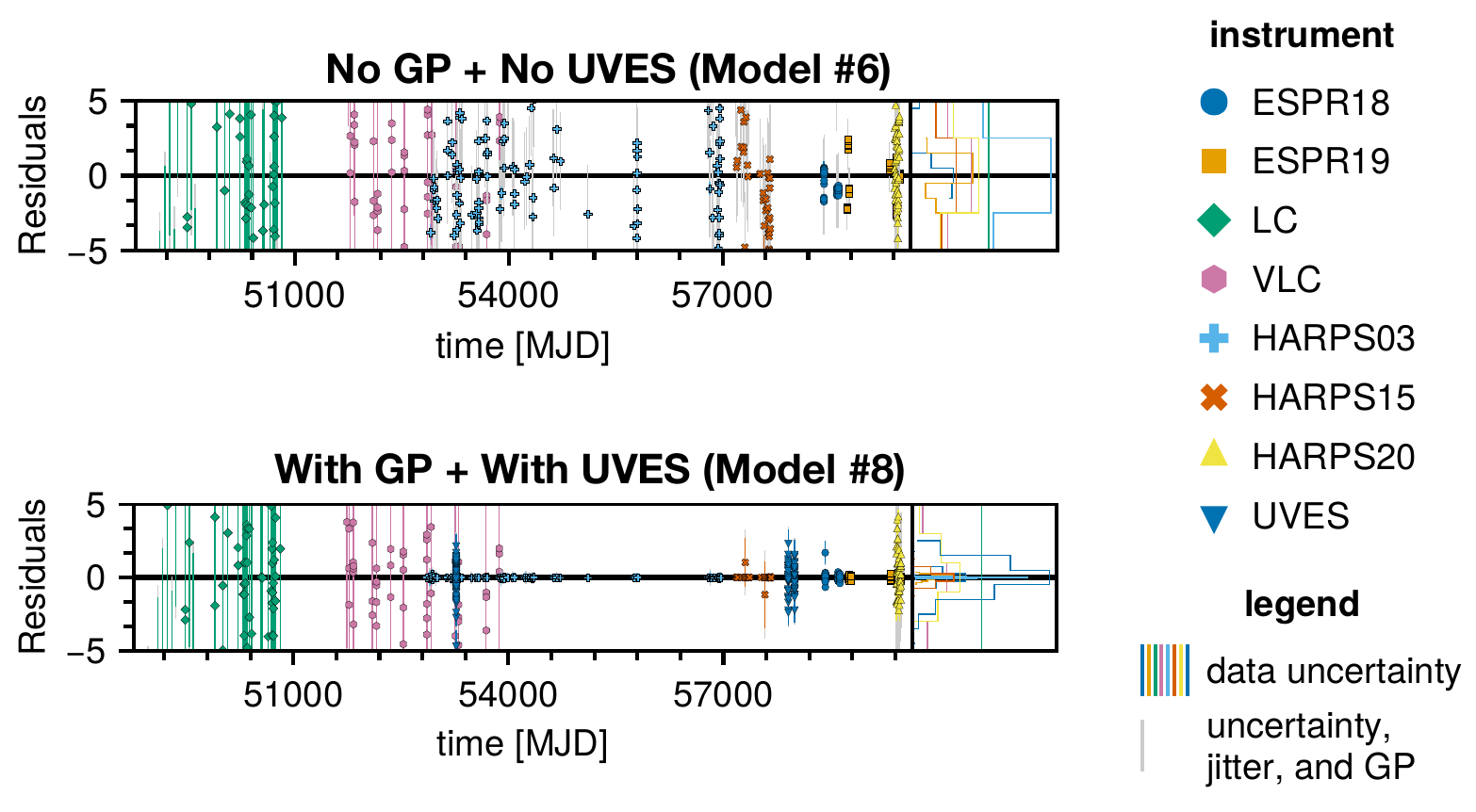}
    \caption{A comparison of the RV residuals (in units of m/s) obtained after fitting models without (top) and with (bottom) a GP, for the maximum posterior orbit. An extension to the right of each panel shows the distribution of residuals for each instrument. Without a GP model, additional instrumental jitter terms (gray bars) are necessary to explain variations in the RV residuals. With a GP model, motivated by evidence of stellar activity, the RV residuals are consistent with zero within the measurement uncertainties (colored bars).}
    \label{fig:GP}
\end{figure*}

\subsection{List of Models}
\label{sec:orb_model}
\added{The orbit model for \epsindAb\ is quite complex. It combines data acquired using different measurement techniques (radial velocity, imaging, absolute astrometry) and additionally, within each technique, data from multiple instruments. None of datasets cover a full orbital period of the planet. Thus, it is important to carefully consider how our inference of the orbital parameters depends on the data used and modeling choices made. We explore a range of models in our analysis:}
\begin{enumerate}
    \item Relative astrometry only
    \item Relative astrometry + HGCA
    \item Relative astrometry + RV
    \item Relative astrometry (No JWST/NIRCam) + HGCA + RV
    \item Relative astrometry + HGCA + RV (No GP)
    \item Relative astrometry + HGCA + RV (No UVES and No GP)
    \item Relative astrometry (Uniform Priors) + HGCA + RV 
    \item Relative astrometry + HGCA + RV
\end{enumerate}

Whenever we sample from a model that ignores a certain dataset, we remove the related instrument-specific nuisance parameters. By default, all fits use observable-based priors for the relative astrometry and include a GP in the RV model, unless specified otherwise. We consider model \#8 as our complete, fiducial model. We sample each model until the log-evidence converges between sampling rounds and the $\hat{R}$ diagnostic \citep{gelman_inference_1992, vehtari_rank-normalization_2021} is $\approx$ 1 to at least two or more decimal places. For most models, this required 15 sampling rounds with \texttt{Octofitter}.

\subsection{Results}
We report parameter constraints both using credible intervals (CI) and highest density intervals (HDI), with the $\pm$ notation used to indicate the median and 16\%/84\% quantiles, and the $[$lower, upper$]$ notation to indicate the 75\% HDI interval, similar to \citet{thompson_revised_2025}. \added{We use 75\% to convey the arbitrariness of the shape of the specific probability mass fraction (useful for non-Gaussian posteriors).} Tables for the above summarizing the posteriors of each model are presented in Appendix~\ref{app:orbit-post}. \added{The posteriors can be retrieved from Zenodo (\dataset[10.5281/zenodo.19931118]{10.5281/zenodo.19931118})}. Figure~\ref{fig:orbit-post} compares the different models tested against the fiducial model. Figures~\ref{fig:orbit-RV} and \ref{fig:orbit-astrometry} visualize the best-fit orbit obtained using the fiducial model. 

\subsubsection{Reliability of the UVES Data and Need for a GP Activity Model}
\label{sec:UVES}
Orbit fits for \epsindAb\ in the literature find a turnover in the RV trend with the inclusion of the HARPS20 data \citep[][]{matthews_temperate_2024, feng_lessons_2025}. However, our best-fit model shows that the RV turnover has not been observed yet. The reason for this is a slope difference between the (UVES, HARPS15) and ESPRESSO data. The orbit fits of \citet[][]{feng_detection_2019, feng_revised_2023, philipot_updated_2023, matthews_temperate_2024, feng_lessons_2025} do not include the two sets of new ESPRESSO RVs. In those cases, the (UVES, HARPS15) data anchors the offset of the HARPS20 time series. \added{Internally, the UVES time series has a nearly flat slope and the HARPS15 time series has an average negative (downward) slope (Figure~\ref{fig:orbit-RV}).} It is then not surprising that previous orbit models found a turnover. However, it is apparent in Figure~\ref{fig:orbit-RV} that the \added{flat and downward slope of the UVES and HARPS15 data, respectively, is incompatible with the ESPRESSO data.} The latter shows that \epsindA's RV is still trending upward (Figure~\ref{fig:orbit-RV}). Indeed, we find that the orbit fit with all the data without using a GP (model \#5) did not converge to a solution. 

The inconsistency in slope between the (UVES, HARPS15) and ESPRESSO datasets is likely due to stellar activity. In an analysis of the long-term HARPS time series and corresponding activity indicators, \citet{feng_detection_2019} find significant signals at periods of 2500~days and 278~days, related to the star's magnetic cycle, and at periods of 11~days, 18~days, and 36~days, related to the star's rotation period (alias, half period, and full period, respectively). Figure~\ref{fig:GP} shows that our fiducial model (\#8), including a GP to mitigate stellar activity signals, \added{results in residuals consistent with zero within the measurement uncertainties} (and can explain the UVES, HARPS15, and ESPRESSO RVs simultaneously) compared to a fit without the GP, \added{which requires additional instrumental jitter to fit the data} (excluding UVES, which has the strongest impact because of its long time baseline, to ensure convergence; model \#6). Further, the GP period parameter $P_{\rm rot}$ converges to $18 \pm 1$~days, equal to the half-rotation period, which \citet{feng_detection_2019} found to be more significant than the full rotation period in activity indicators, possibly due to spots. This consistency indicates that the GP is indeed modeling an activity signal in the RVs.

Together, this shows that using a GP in modeling the RVs of \epsindA\ is physically-motivated and improves the reliability of the fit to the full RV time series. In particular, we emphasize that inclusion of the full UVES time series without a GP model will yield incorrect orbit estimates. Comparing results between the fiducial model (with GP) and the model without UVES and without a GP (model \#6), we find consistent values for all parameters within 1$\sigma$--2$\sigma$. However, the median eccentricity obtained without using a GP ($e=0.56$) is $2\times$ higher than if a GP is used ($e=0.25$). The two results are formally consistent because the eccentricity posterior in the fit without the GP is not Gaussian and has a long tail towards low eccentricities. As such, given the previous discussion on the motivation for using a GP, we caution that the high eccentricity in the fit without a GP is likely biased.

\subsubsection{Observable-based Priors vs Uniform Priors}
Comparing results between the fiducial model (\#8) with observable-based priors \citep{oneil_improving_2019} and the model with uniform priors (\#7), we find that all the orbital parameters are consistent within 1$\sigma$--2$\sigma$. This can be visually seen as an overlap between the respective 1$\sigma$ contours in Figure~\ref{fig:orbit-post} and indicates that our conclusions are robust to the choice of priors for the relative astrometry. The parameter uncertainties for the uniform prior fit are similar or larger than those for the observable-based prior fit.

\begin{figure*}[!tb]
    \centering
    \includegraphics[width=0.75\linewidth]{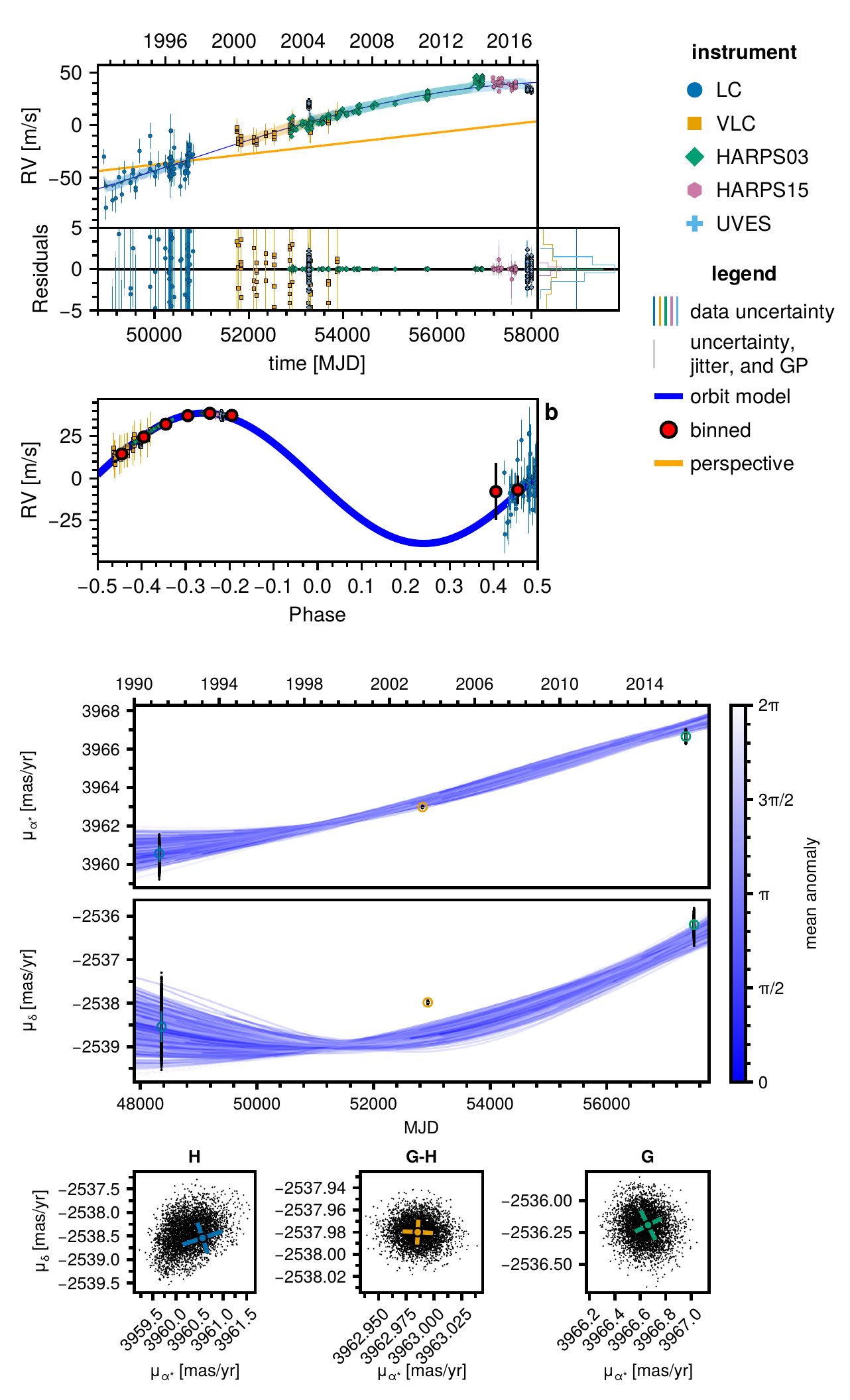}
    \caption{Best-fit circular orbit for RV + HGCA model using datasets prior to the JWST imaging detection \citep[i.e., using the exact datasets from][]{feng_revised_2023}. The individual panels are identical in format to those presented in Figures~\ref{fig:orbit-RV} and \ref{fig:orbit-astrometry}.}
    \label{fig:prediction}
\end{figure*}

\begin{figure}[!tb]
    \centering
    \includegraphics[width=0.95\linewidth]{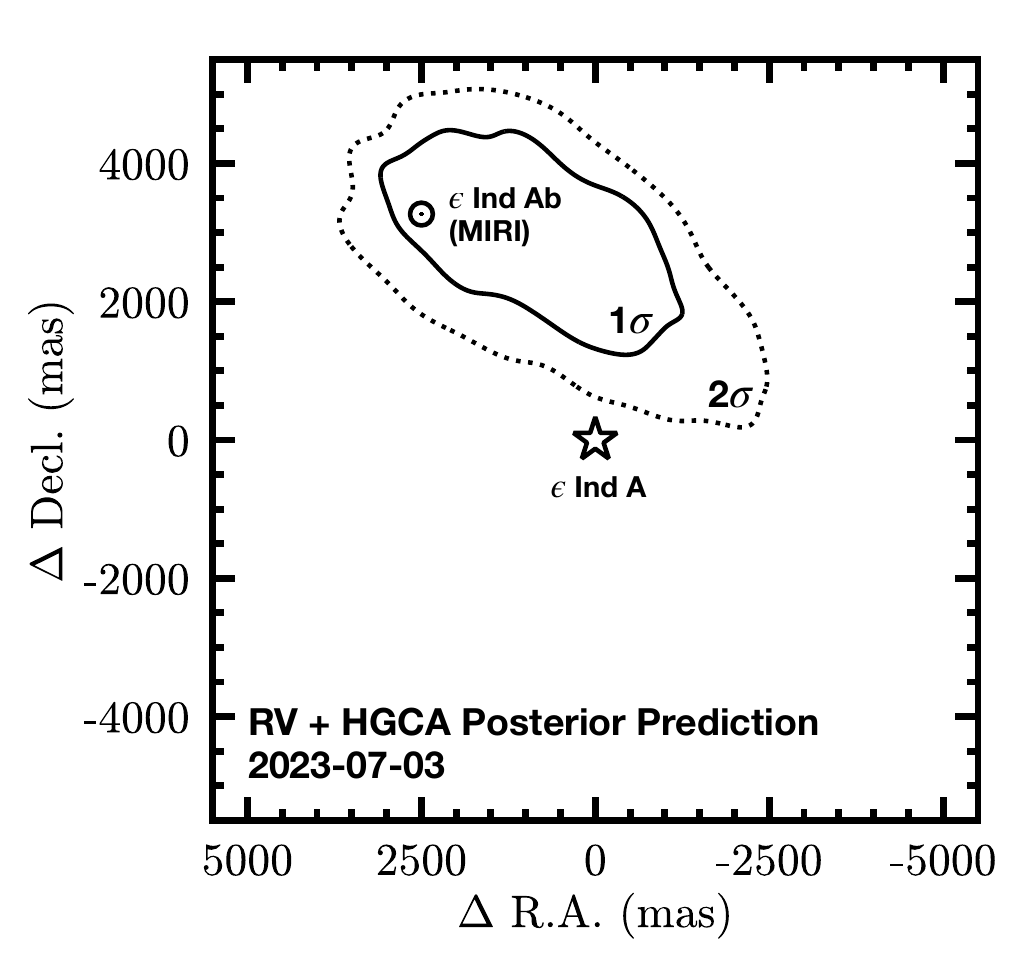}
    \caption{Predicted position of \epsindAb\ for the MIRI imaging discovery epoch (2023-07-03) using the best-fit RV + HGCA circular orbit model. The measured position of \epsindAb\ \citep[circle,][]{matthews_temperate_2024} is consistent with the prediction.}
    \label{fig:prediction-location}
\end{figure}

\begin{figure*}
    \centering
    \includegraphics[width=\linewidth]{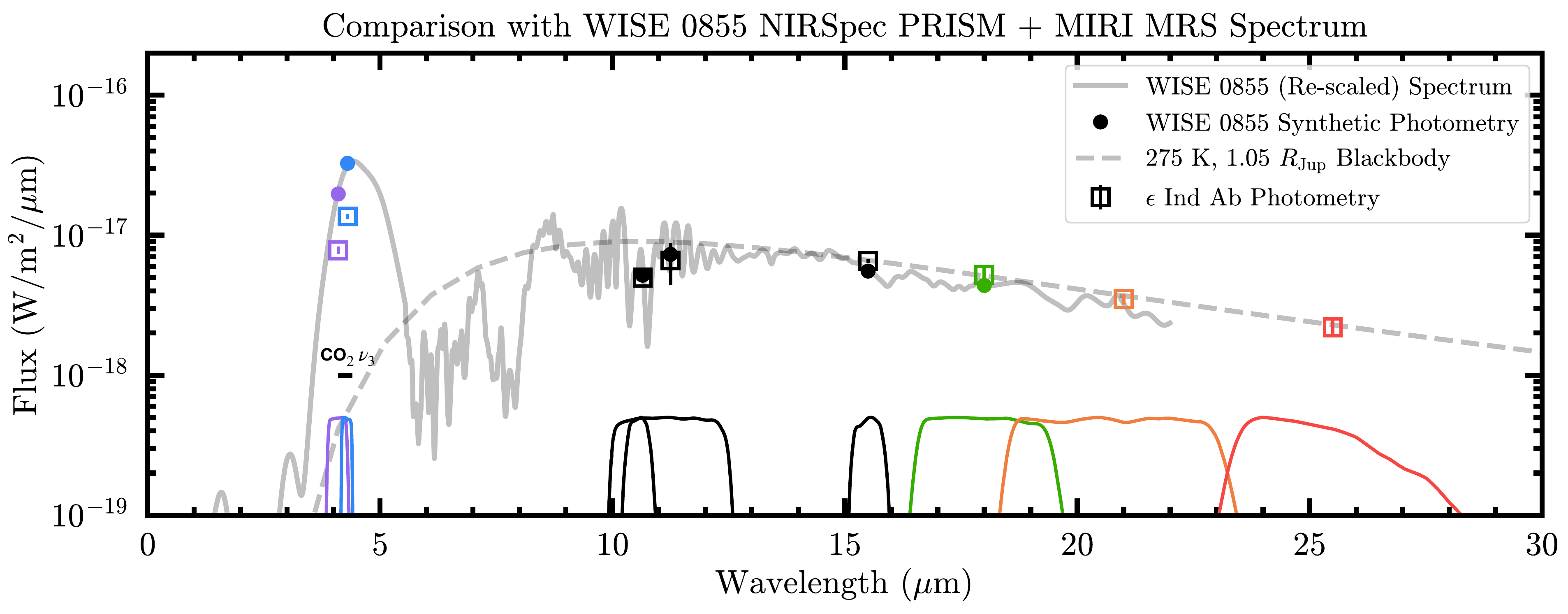}
    \caption{Comparison between \epsindAb's and $\sim$285~K brown dwarf WISE~0855's spectral energy distributions. The colored open boxes with error bars represent \epsindAb's measured photometry. WISE~0855's spectrum (flux re-scaled to match the distance to \epsindA\ and binned to $R\sim 50$) is shown in solid gray and the corresponding synthetic photometry is shown as colored filled circles. A reference blackbody spectrum is shown in dashed gray. The filter profiles are plotted below the photometry. The 4.1--4.3~$\mu$m CO$_2$ absorption region is marked with a solid horizontal black line.}
    \label{fig:SED}
\end{figure*}

\subsubsection{Evidence for Moderate Orbital Eccentricity}
The orbital eccentricity of \epsindAb\ is of particular interest as it is a key dynamical tracer of the planet's formation and evolutionary history \citep{bowler_population-level_2020, nagpal_impact_2023, do_o_orbital_2023}. While previous orbit fits \citep[][]{matthews_temperate_2024, feng_lessons_2025} suggested a higher, non-zero, eccentricity ($\sim$0.4), the estimate was biased because of the impact of stellar activity on the RVs (see \S\ref{sec:UVES}). In fact, an orbit fit to the relative astrometry (excluding the new JWST/NIRCam epoch) + HGCA + RV (model~\#4) converges to a circular orbit solution for the planet (75\% highest density interval is $e = [0.00, 0.08]$; also see Figure~\ref{fig:orbit-post}). Including the new JWST/NIRCam epoch in the orbit fit (model \#8) results in a moderate, non-zero, orbital eccentricity $e = 0.25 \pm 0.09$, though the data are consistent with a circular orbit within the $3\sigma$ uncertainty range. It is worth noting that the relative astrometry + HGCA only fit (model \#2) finds the highest eccentricity among all models considered (not including the case without a GP, see \S\ref{sec:UVES}), $e = 0.45^{+0.08}_{-0.12}$. This suggests that there may be some tension between the RV dataset and the absolute astrometry from the HGCA. Continued RV monitoring of the system, with the goal of capturing the turnover, can help better constrain the eccentricity of the planet.

\subsubsection{Solving the Mystery of the Incorrect Orbit Prediction}
\label{app:revisit}

The original prediction of \epsindAb's orbital location \citep{feng_revised_2023} was discrepant by a significant margin compared to the imaging discovery location \citep{matthews_temperate_2024}. \added{So far, the solution to this issue has not been presented in literature. It is an important problem to tackle as the strategy of imaging systems exhibiting RV + absolute astrometry trends will become routine with Gaia's fourth data release (DR4) and accurate orbit predictions will be needed for observation planning.} Here, we perform a new RV + HGCA orbit fit considering only the data used in \citet{feng_revised_2023} (i.e., no relative astrometry from imaging and only RVs captured before the JWST detection) and demonstrate that adopting a conservative circular orbit model would have yielded an accurate prediction of \epsindAb's location for the JWST direct imaging discovery epoch. 

For our orbit fit, we use the LC, VLC, UVES, HARPS03, and HARPS15 RV time series, similar to \citet{feng_revised_2023}. The implementation of the joint RV and HGCA orbit model is identical to that described in \S\ref{sec:model-implement}. The only change is that we fix the eccentricity to zero (circular orbit). Previous studies have shown that the eccentricity of a planet obtained from RV data can be significantly biased by incorrect modeling of the signal, non-Gaussian or correlated noise sources, and/or the presence of additional planets \citep[e.g.,][]{hara_bias_2019}. The \epsindA\ RV time series does not sample a full planet orbital period, is composed of data from multiple instruments fit by independent offsets, and is affected by stellar activity. This makes the data particularly susceptible to eccentricity biases and is the motivation for adopting a circular orbit. 

Indeed, our best-fit RV + HGCA circular orbit model (Figure~\ref{fig:prediction}) correctly pre-predicts the position of \epsindAb\ at the MIRI discovery imaging epoch (2023-07-23; Figure~\ref{fig:prediction-location}). This orbit fit yields $a = 15.2 \pm 1.9$~au, $i = 112^{+13^\circ}_{-9^\circ}$, and $M_b = 5.0^{+1.7}_{-1.4}\;M_{\rm Jup}$, consistent within 1$\sigma$--2$\sigma$ of the values inferred using our fiducial relative astrometry + HGCA + RV orbit model. The lesson with \epsindAb\ is to exercise caution when applying eccentric orbit models to RV and absolute astrometry data. Using the Bayes Factor to statistically evaluate the significance of eccentric solutions over a circular orbit and performing leave-one-out cross validation are example strategies to mitigate biases in data modeling \citep[see application to $\epsilon$~Eridani~b in][]{thompson_revised_2025}.

\section{Atmospheric Properties}
\label{sec:atmo-evo}

\subsection{Comparison to WISE~0855}
\label{sec:wise0855}

WISE J085510.83-071442.5 (also referred to as WISE~0855) is an isolated, late Y dwarf with a temperature of $\sim$285~K and is the coldest brown dwarf discovered to date \citep{luhman_discovery_2014, skemer_first_2016}. We compare \epsindAb's photometry with synthetic photometry derived from the NIRSpec PRISM \citep{luhman_jwstnirspec_2023} and MIRI MRS \citep{kuhnle_water_2025} \added{spectra} of WISE 0855 (Figure~\ref{fig:SED}). The F1800W$-$F1550C color serves as a good indicator of effective temperature as the spectral energy distribution is well-approximated by a blackbody at these wavelengths (Figure~\ref{fig:SED}). We find a similar F1800W$-$F1550C color for \epsindAb\  ($m_{\rm F1800W-F1550C} = 0.25$~mag) and WISE~0855 ($m_{\rm F1800W-F1550C} = 0.26$~mag) showing that the two substellar objects have similar effective temperatures. This is additionally supported by the evolutionary model analysis in \S\ref{sec:evomodel}. Interestingly, \epsindAb's NIRCam F410M/F430M fluxes are $\approx$2.5$\times$ fainter than those of WISE~0855 (Figure~\ref{fig:SED}). The 4.1--4.3~$\mu$m CO$_2$ absorption feature dominates in these NIRCam filters and its strength depends strongly on atmospheric metallicity \citep{lodders_atmospheric_2002, zahnle_atmospheric_2009, rustamkulov_early_2023, balmer_jwst-tst_2025, ruffio_jupiter-like_2026, xuan_compositions_2026, sanghi_worlds_2026-1}. The retrieved volume mixing ratio for $\rm{CO_2}$ in WISE~0855's atmosphere is extremely small, $\sim$$10^{-9}$ \citep{kuhnle_water_2025}. A higher atmospheric metallicity compared to WISE~0855 and, consequently, an elevated CO$_2$ abundance in \epsindAb's atmosphere could explain the difference in NIRCam fluxes. The atmospheric metallicity of WISE 0855 is not well-determined and \citet{kuhnle_water_2025} find that different grid models yield different values. A second possibility is that water ice clouds in the atmosphere suppress the 4--5~$\mu$m flux \citep[e.g.,][]{lacy_self-consistent_2023}. Both scenarios are explored in more detail in the following section.

\begin{deluxetable*}{lc}
\centering
\tablecaption{\label{tab:models}Atmospheric Model Grids}
\tablehead{\colhead{Parameter} & \colhead{Values}}
\startdata
\multicolumn{2}{c}{Sonora Elf Owl} \\
\hline
$T_{\rm eff}$ (K) & $\{275, 300\}$ \\ 
$g$ (m/$\rm{s^2}$)& \{100, 178, 316\} \\ 
$[\rm{M/H}]$ (dex) & $\{-1.0, -0.5, 0, +0.5, +0.7, +1.0\}$ \\
C/O ($\times$ solar) & $\{0.5, 1.0, 1.5, 2.5\}$ \\
log $[$$K_{\rm{zz}}$ ($\rm cm^2\;s^{-1}$)$]$ & \{2, 4, 7, 8, 9\} \\
\hline
\multicolumn{2}{c}{Sonora Flame Skimmer} \\
\hline
$T_{\rm eff}$ (K) & $\{250, 275, 300\}$ \\ 
$g$ (m/$\rm{s^2}$)& \{100, 178, 316\} \\ 
$[\rm{M/H}]$ (dex) & $\{-1.0, -0.5, 0, +0.5, +1.0, +1.5, +2.0\}$ \\
C/O ($\times$ solar) & $\{0.5, 1.0, 1.5, 2.5\}$ \\
log $[$$K_{\rm{zz}}$ ($\rm cm^2\;s^{-1}$)$]$ & \{0, 2, 4, 7, 8, 9\} \\
\hline
\multicolumn{2}{c}{Custom PICASO Cloudy Models} \\
\hline
$T_{\rm eff}$ (K) & $\{250, 275, 300\}$ \\ 
$g$ (m/$\rm{s^2}$)& \{100, 178, 316\} \\ 
$[\rm{M/H}]$ (dex) & $\{0, +0.5, +1.0\}$ \\
C/O ($\times$ solar) & $\{1.0, 2.5\}$ \\
log $[$$K_{\rm{zz}}$ ($\rm cm^2\;s^{-1}$)$]$ & \{0, 2, 7, 9\} \\
$f_{\rm sed}$ & \{4, 6, 8\} \\
\hline
\multicolumn{2}{c}{\citet{lacy_self-consistent_2023}} \\
\hline
$T_{\rm eff}$ (K) & $\{250, 275, 300\}$ \\ 
$\log\;[g\;(\rm{cm/s^2})]$ & \{3.50, 3.75, 4.00, 4.25, 4.50, 4.75\} \\ 
$[\rm{M/H}]$ (dex) & $\{-0.5, 0, +0.5\}$ \\
log $[$$K_{\rm{zz}}$ ($\rm cm^2\;s^{-1}$)$]$ & \{0, 6\} \\
Cloud Type & {None, E10, AEE10} \\
\hline
\multicolumn{2}{c}{ATMO 2020} \\
\hline
$T_{\rm eff}$ (K) & $\{250, 300\}$ \\ 
$\log\;[g\;(\rm{cm/s^2})]$ & \{2.5, 3.0, 3.5, 4.0, 4.5\} \\ 
$[\rm{M/H}]$ (dex) & $\{-0.5, 0, +0.5\}$ \\
Chemistry Type & \{CEQ, NEQ Weak, NEQ Strong\} \\
\hline
\multicolumn{2}{c}{ATMO 2020++ (PH$_3$ Removed)} \\
\hline
$T_{\rm eff}$ (K) & $\{250, 275, 300\}$ \\ 
$\log\;[g\;(\rm{cm/s^2})]$ & \{2.5, 3.0, 3.5, 4.0, 4.5\} \\ 
$[\rm{M/H}]$ (dex) & $\{-0.5, 0, +0.3\}$ \\
\hline
\multicolumn{2}{c}{ATMO 2020++ (PH$_3$ Included)} \\
\hline
$T_{\rm eff}$ (K) & $\{250, 275, 300\}$ \\ 
$\log\;[g\;(\rm{cm/s^2})]$ & \{2.5, 3.0, 3.5, 4.0, 4.5\} \\ 
$[\rm{M/H}]$ (dex) & $\{-1.0, -0.5, 0, +0.3\}$ \\
\enddata
\tablecomments{Models with all possible combinations of listed parameters, when available (not all model grids are rectangular), are fit to the data. Chemical equilibrium models ($K_{\rm zz} = 0$) are noted above as log $[$$K_{\rm{zz}}$ ($\rm cm^2\;s^{-1}$)$] = 0$~dex for convenience only.}
\end{deluxetable*}

\subsection{Self-Consistent Grid Model Analysis} 

In this section, we compare the 4--25~$\mu$m photometric spectral energy distribution of \epsindAb\ to precomputed grids of self-consistent radiative-convective equilibrium models. We present initial results on the atmospheric properties of \epsindAb\ based on the best-fitting models and derive the planet's bolometric luminosity for comparisons with evolutionary model predictions.

\begin{figure*}
    \centering
    \includegraphics[width=\linewidth]{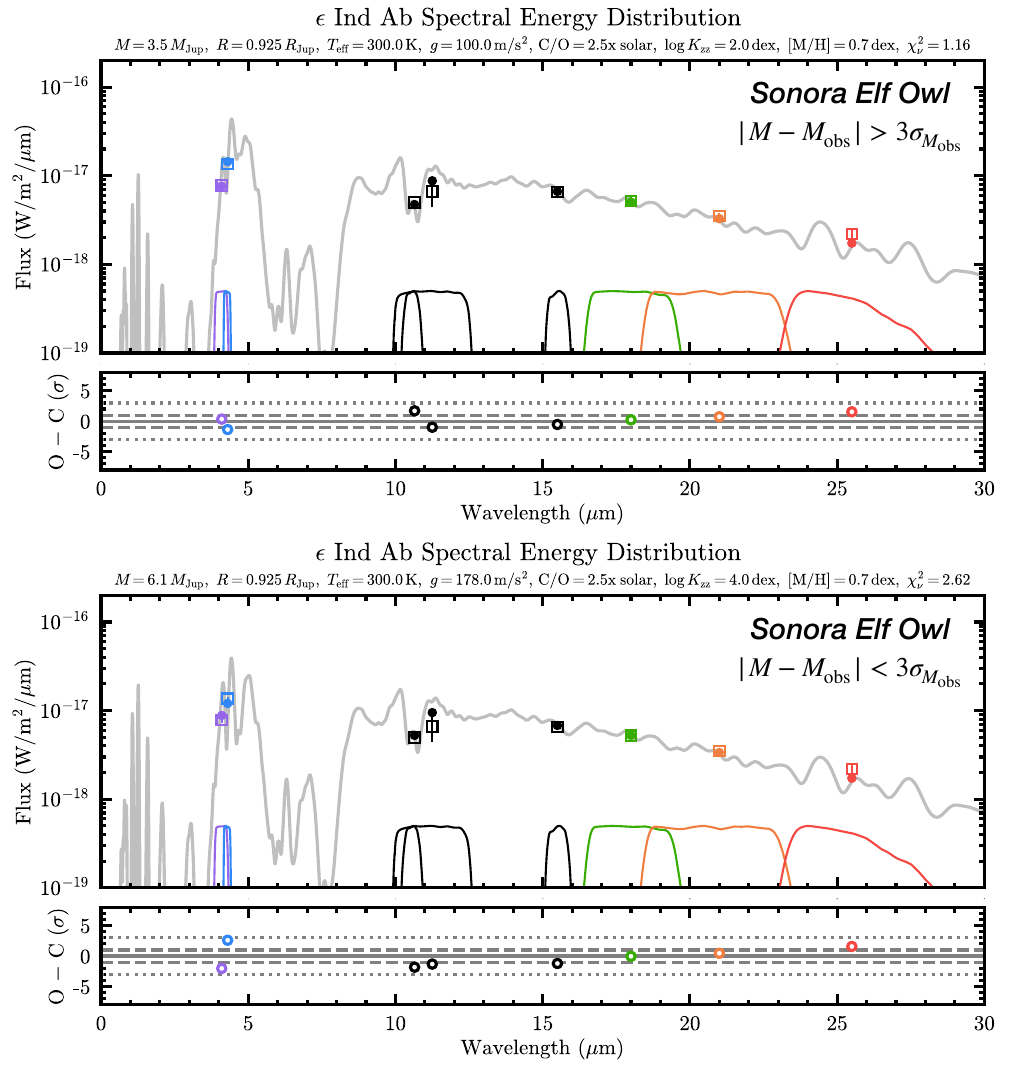}
    \caption{Atmospheric model fits with the (cloud-free, no rainout chemistry) Sonora Elf Owl grid. The top panel presents the lowest $\chi^2_\nu$ fit from the model grid. The bottom panel presents the lowest $\chi^2_\nu$ fit from the model grid that also predicts a mass consistent with the dynamical mass within $3\sigma$. In each panel, the colored open boxes with error bars represent \epsindAb's measured photometry. The model atmosphere is shown in gray (binned to $R\sim 50$) and the corresponding synthetic photometry are shown as colored filled circles. The filter profiles are plotted below the photometry. A sub-panel below the spectral energy distribution shows the residuals (in units of $\sigma$). Gray dashed and dotted lines mark 1$\sigma$ and 3$\sigma$ intervals.}
    \label{fig:sonora}
\end{figure*}

\begin{figure*}
    \centering
    \includegraphics[width=\linewidth]{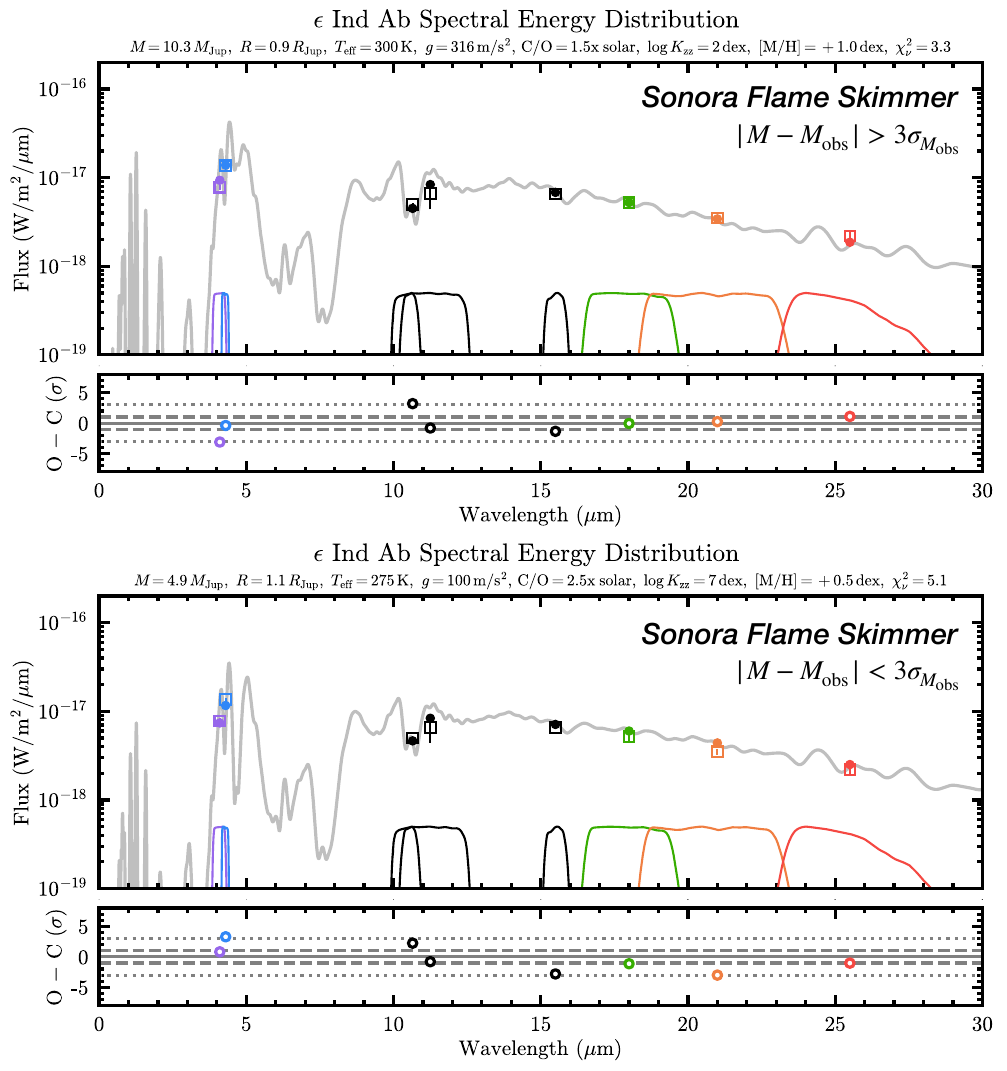}
    \caption{The overall lowest $\chi^2_\nu$ atmospheric model fit (top) and the lowest $\chi^2_\nu$ atmospheric model fit also consistent with the dynamical mass within 3$\sigma$ (bottom) from the (cloud-free, rainout chemistry) Sonora Flame Skimmer grid. The format of the individual panels is identical to that described in Figure~\ref{fig:sonora}.}
    \label{fig:sonora_alt}
\end{figure*}

\begin{figure*}
    \centering
    \includegraphics[width=\linewidth]{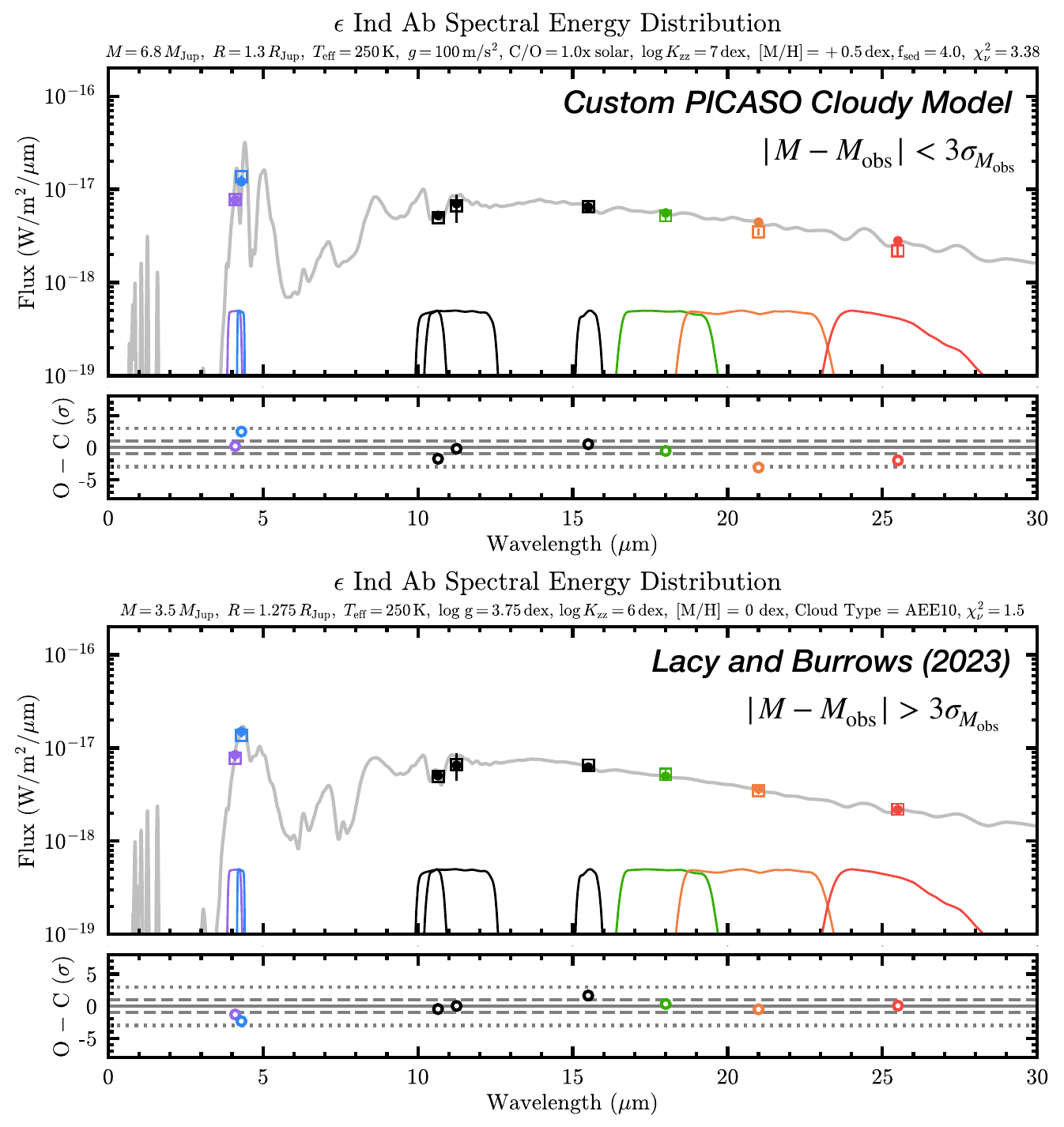}
    \caption{The lowest $\chi^2_\nu$ atmospheric model fit from the custom PICASO water ice cloud models (top, which is also consistent with the dynamical mass within $3\sigma$) and the \citet{lacy_self-consistent_2023} water ice cloud models (bottom). None of the \citet{lacy_self-consistent_2023} models are consistent with the dynamical mass within $3\sigma$. The format of the individual panels is identical to that described in Figure~\ref{fig:sonora}.}
    \label{fig:other-models-1}
\end{figure*}

\begin{figure*}
    \centering
    \includegraphics[width=0.8\linewidth]{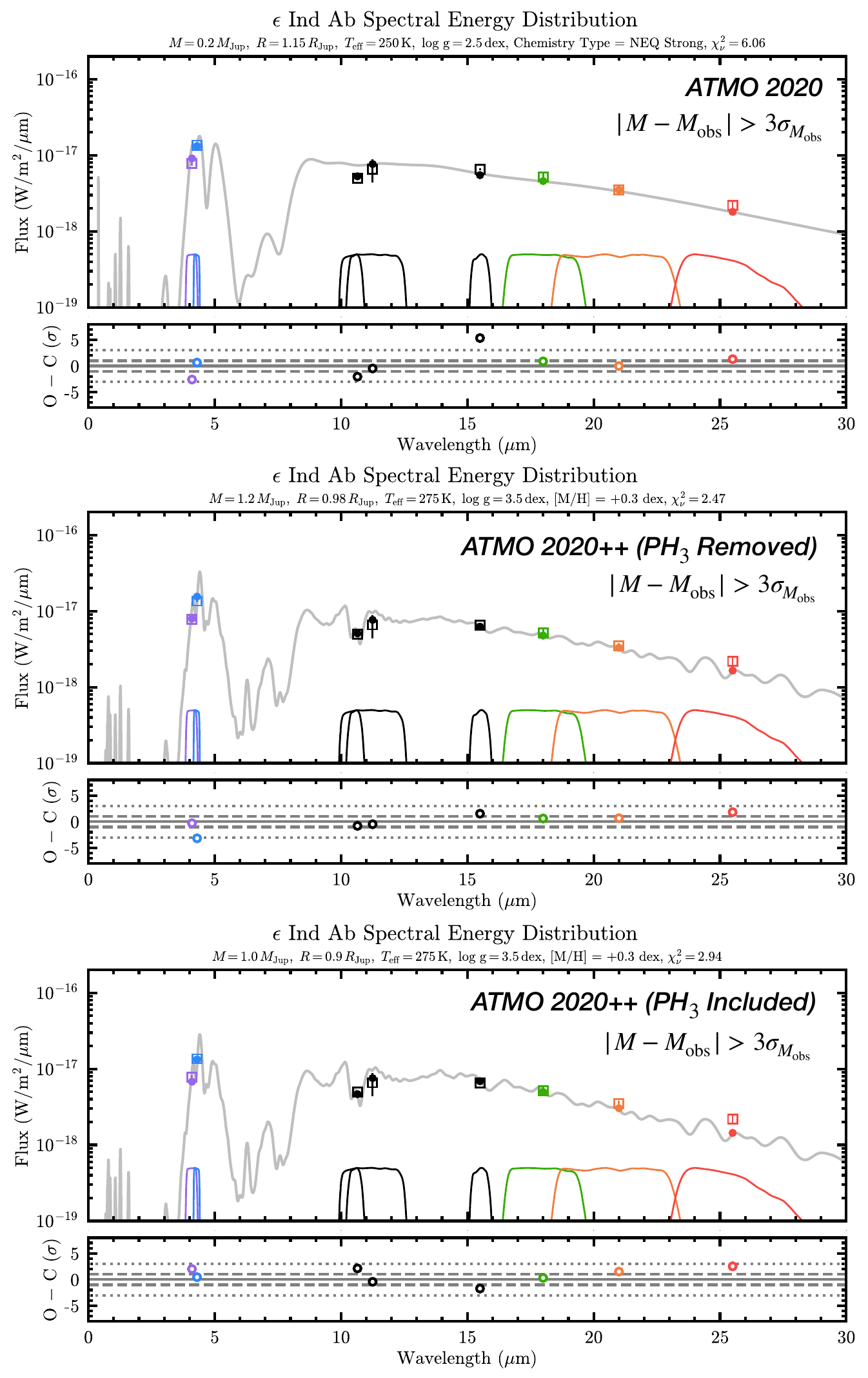}
    \caption{The lowest $\chi^2_\nu$ atmospheric model fit from the ATMO 2020 (top), ATMO 2020++ without PH$_3$ (center), and ATMO 2020++ with PH$_3$ (bottom). None of the models from these grids yield a planet mass consistent with the dynamical mass within $3\sigma$. The format of the individual panels is identical to that described in Figure~\ref{fig:sonora}.}
    \label{fig:other-models-2}
\end{figure*}

\subsubsection{Model Grids}

We consider the following model grids for our atmospheric analysis. The specific parameter values explored by each grid are summarized in Table~\ref{tab:models}. 

\emph{Sonora Elf Owl.}~The Sonora Elf Owl grid \citep{mukherjee_sonora_2024} consists of cloud-free atmospheric models computed using \texttt{PICASO 3.0} \citep{mukherjee_picaso_2023}, a 1D radiative–convective equilibrium climate model, that includes the effects of disequilibrium chemistry (parameterized by diffusion parameter $K_{\rm zz}$) for a wide range of substellar temperatures, surface gravities, atmospheric metallicities, and C/O ratios. We use the latest version of the models presented in \citet{wogan_sonora_2025} that correctly estimate the CO$_2$ quenching.

\emph{Sonora Flame Skimmer.}~The Sonora Flame Skimmer grid (Mang et al., in prep) extends the cloud-free Sonora Elf Owl grid \citep{mukherjee_sonora_2024, wogan_sonora_2025} to colder effective temperatures, lower surface gravities, and a broader range of metallicities. These models additionally incorporate rainout chemistry for H$_2$O, CH$_4$, and NH$_3$—even in cloud-free atmospheres—similar to the treatment in Sonora Bobcat. They address the underestimation of CO$_2$ found in the Sonora Elf Owl models \citep{mukherjee_sonora_2024}, which has since been revised in \citet{wogan_sonora_2025}.

\emph{Custom PICASO Water Ice Cloud Models.} We generate a custom grid of cloudy (H$_2$O) atmospheric models using \texttt{PICASO} \citep{batalha_exoplanet_2019,mukherjee_picaso_2023,mang_picaso_2026}. \texttt{PICASO} now supports fully self-consistent cloudy atmosphere calculations through integration with \texttt{Virga} \citep{batalha_natashabatalhavirga_2020, batalha_condensation_2025,moran_fractal_2025}, the Python implementation of the \citet{ackerman_precipitating_2001} cloud model. The clouds are parameterized with a sedimentation efficiency ($f_{\rm sed}$) parameter. Generally, the optical thickness and vertical extent of the cloud decreases and particle size increases with increasing $f_{\rm sed}$.

\emph{\citet{lacy_self-consistent_2023} Models.} This grid consists of a suite of 1D radiative-convective equilibrium models computed with coolTLUSTY \citep{hubeny_non-lte_1995, sudarsky_phase_2005, burrows_optical_2008} including updated opacities and the effects of both disequilibrium chemistry and water ice clouds in Y dwarf atmospheres. Specifically, two water ice cloud types, labeled ``E10" and ``AEE10", are presented. The modal cloud particle size is 10~$\mu$m in both cases. The key difference is vertical extent: the E10 clouds are vertically compact and the AEE10 clouds are vertically extended.
                
\emph{ATMO 2020.}~ATMO 2020 \citep{phillips_new_2020} is a set of solar metallicity atmospheric models that incorporate the effects of non-equilibrium chemistry due to vertical mixing in the atmosphere (``NEQ Weak" and ``NEQ Strong", corresponding to different mixing strengths). Key improvements in ATMO 2020 over previous model grids include a new H-He equation of state, updated molecular opacities, and better treatment of the collisionally broadened potassium resonance doublet. 

\emph{ATMO 2020++.}~Using the ATMO 2020 models with strong vertical mixing as a starting point, ATMO 2020++ \citep{leggett_measuring_2021, meisner_exploring_2023, leggett_james_2024} modifies the adiabatic ideal gas index $\gamma$ (and thus atmospheric temperature gradient) to account for the effect of processes responsible for producing a non-adiabatic cooling curve in giant planet and brown dwarf atmospheres. These processes include complex atmospheric dynamics (e.g., zones, spots, waves) due to rapid rotation, compositional changes due to condensation, upper atmosphere heating by cloud decks or breaking gravity waves, etc. Recent modeling with JWST data has shown that this grid provides an improved fit to Y dwarf spectra compared to the standard-adiabat models \citep{leggett_first_2023, leggett_redshifting_2024, luhman_jwstnirspec_2023}. We use both the model grids with and without PH$_3$.

\startlongtable
\begin{deluxetable*}{ccccccccccc}
    \label{tab:model-fitting}
    \tabletypesize{\scriptsize}
    \setlength{\tabcolsep}{4pt}
    \centering
    \tablecaption{Atmospheric Model Fits to the Measured Planet Photometry with $\chi^2_\nu < 10$}
    \tablehead{\colhead{$T_{\rm eff}$} & \colhead{$g\; \rm{or} \log g$} & \colhead{[M/H]} & \colhead{$\log K_{\rm zz}$ or Chemistry} & \colhead{C/O} & \colhead{Cloud Type or $f_{\rm sed}$} & \colhead{$R$} & \colhead{$M$} & \colhead{$\log L_{\rm bol}/L_\odot$} & \colhead{$\chi^2_\nu$} & \colhead{$3\sigma_{M_b}$}\\ \colhead{(K)} & \colhead{($\rm{m\:s^{-2}}$ or $\log\;[\rm{cm\:s^{-2}}]$)} & \colhead{(dex)} & \colhead{($\log\;[\rm{cm^2\:s^{-1}}]$)} & \colhead{($\times$ solar)} & \colhead{} & \colhead{($R_{\rm Jup}$)} & \colhead{($M_{\rm Jup}$)} & \colhead{(dex)} & \colhead{} & \colhead{}} 
        \startdata
        \multicolumn{10}{c}{Sonora Elf Owl} \\
        \hline
        \textbf{300} & \textbf{178} & \textbf{+0.7} & \textbf{4~dex} & \textbf{2.5} & \nodata & \textbf{0.925} & \textbf{6.1} & \textbf{$-$7.21} & \textbf{2.62} & $\checkmark$ \\
        275 & 178 & +0.5 & 7~dex & 2.5 & \nodata & 1.025 & 7.5 & $-7.22$ & 4.24 & $\checkmark$\\
        275 & 178 & +0.5 & 8~dex & 2.5 & \nodata & 1.025 & 7.5 & $-7.23$ & 4.39 & $\checkmark$\\
        275 & 100 & +0.7 & 4~dex & 1.5 & \nodata & 1.1 & 4.9 & $-7.20$ & 7.11 & $\checkmark$\\
        275 & 178 & +0.5 & 9~dex & 2.5 & \nodata & 1.05 & 7.9 & $-7.21$ & 9.64 & $\checkmark$\\
        \hline
        \textbf{300} & \textbf{100} & \textbf{+0.7} & \textbf{2~dex} & \textbf{2.5} & \nodata & \textbf{0.925} & \textbf{3.5} & \textbf{$-$7.21} & \textbf{1.16} & $\times$\\
        275 & 100 & +0.5 & 7~dex & 2.5 & \nodata & 1.025 & 4.2 & $-7.22$ & 2.17 & $\times$\\
        300 & 100 & +0.7 & 4~dex & 1.5 & \nodata & 0.925 & 3.5 & $-7.22$ & 3.43 & $\times$\\
        300 & 316 & +1.0 & 2~dex & 1.5 & \nodata & 0.925 & 10.9 & $-7.21$ & 4.55 & $\times$\\
        300 & 100 & +1.0 & 2~dex & 1.0 & \nodata & 0.95 & 3.6 & $-7.20$ & 5.6 & $\times$\\
        300 & 100 & +0.5 & 4~dex & 2.5 & \nodata & 0.875 & 3.1 & $-7.25$ & 7.15 & $\times$\\
        300 & 316 & +0.7 & 4~dex & 2.5 & \nodata & 0.875 & 9.8 & $-7.25$ & 7.31 & $\times$\\
        275 & 316 & +0.7 & 7~dex & 2.5 & \nodata & 1.025 & 13.4 & $-7.22$ & 8.7 & $\times$\\
        \hline
        \multicolumn{10}{c}{Sonora Flame Skimmer} \\
        \hline
        275 & 100 & +0.5 & 7~dex & 2.5 & \nodata & 1.1 & 4.9 & $-7.19$ & 5.10 & $\checkmark$\\
        275 & 178 & +0.5 & 7~dex & 2.5 & \nodata & 1.075 & 8.3 & $-7.22$ & 5.82 & $\checkmark$\\
        300 & 178 & +1.0 & 2~dex & 1.5 & \nodata & 0.9 & 5.8 & $-7.21$ & 6.28 & $\checkmark$\\
        275 & 100 & +0.5 & 7~dex & 1.5 & \nodata & 1.05 & 4.4 & $-7.23$ & 8.84 & $\checkmark$\\
        275 & 100 & +0.5 & 8~dex & 1.5 & \nodata & 1.1 & 4.9 & $-7.19$ & 9.03 & $\checkmark$\\
        300 & 178 & +0.5 & 7~dex & 2.5 & \nodata & 0.9 & 5.8 & $-7.22$ & 9.35 & $\checkmark$\\
        \hline
        300 & 316 & +1.0 & 2~dex & 1.5 & \nodata & 0.9 & 10.3 & $-7.22$ & 3.30 & $\times$\\
        300 & 100 & +1.0 & 2~dex & 1.0 & \nodata & 0.925 & 3.5 & $-7.20$ & 5.65 & $\times$\\
        275 & 178 & +0.5 & 8~dex & 2.5 & \nodata & 1.1 & 8.7 & $-7.20$ & 5.81 & $\times$\\
        300 & 100 & +0.5 & 4~dex & 2.5 & \nodata & 0.875 & 3.1 & $-7.24$ & 6.07 & $\times$\\
        300 & 100 & +0.5 & 7~dex & 2.5 & \nodata & 0.9 & 3.3 & $-7.22$ & 9.22 & $\times$\\
        \hline
        \multicolumn{10}{c}{Custom PICASO Cloudy Models} \\
        \hline
        250 & 100 & +0.5 & 7~dex & 1.0 & 4.0 & 1.3 & 6.8 & -7.21 & 3.38 & $\checkmark$\\
        250 & 100 & 0 & 0~dex & 1.0 & 8.0 & 1.25 & 6.3 & -7.23 & 4.38 & $\checkmark$\\
        250 & 100 & +0.5 & 0~dex & 1.0 & 6.0 & 1.275 & 6.6 & -7.22 & 4.77 & $\checkmark$\\
        275 & 178 & +0.5 & 7~dex & 2.5 & 6.0 & 1.05 & 7.9 & -7.23 & 5.23 & $\checkmark$\\
        250 & 100 & 0 & 9~dex & 2.5 & 4.0 & 1.225 & 6.1 & -7.26 & 5.24 & $\checkmark$\\
        250 & 100 & 0 & 0~dex & 2.5 & 8.0 & 1.275 & 6.6 & -7.22 & 5.48 & $\checkmark$\\
        275 & 178 & +0.5 & 7~dex & 2.5 & 8.0 & 1.075 & 8.3 & -7.22 & 5.71 & $\checkmark$\\
        250 & 100 & 0 & 0~dex & 1.0 & 6.0 & 1.275 & 6.6 & -7.22 & 7.64 & $\checkmark$\\
        250 & 100 & 0 & 2~dex & 1.0 & 6.0 & 1.275 & 6.6 & -7.22 & 7.71 & $\checkmark$\\
        275 & 100 & +1.0 & 0~dex & 2.5 & 6.0 & 1.05 & 4.4 & -7.22 & 7.81 & $\checkmark$\\
        250 & 100 & 0 & 2~dex & 2.5 & 8.0 & 1.275 & 6.6 & -7.22 & 8.08 & $\checkmark$\\
        250 & 100 & +0.5 & 2~dex & 2.5 & 4.0 & 1.275 & 6.6 & -7.22 & 8.48 & $\checkmark$\\
        300 & 178 & +0.5 & 7~dex & 2.5 & 6.0 & 0.875 & 5.5 & -7.24 & 9.90 & $\checkmark$\\
        \hline
        275 & 316 & +1.0 & 0~dex & 1.0 & 8.0 & 1.05 & 14.1 & -7.22 & 3.55 & $\times$\\
        250 & 316 & 0 & 0~dex & 1.0 & 8.0 & 1.275 & 20.7 & -7.23 & 4.49 & $\times$\\
        250 & 316 & 0 & 0~dex & 1.0 & 6.0 & 1.275 & 20.7 & -7.24 & 5.53 & $\times$\\
        275 & 316 & +0.5 & 9~dex & 2.5 & 8.0 & 1.05 & 14.1 & -7.23 & 5.70 & $\times$\\
        275 & 316 & +0.5 & 9~dex & 2.5 & 6.0 & 1.05 & 14.1 & -7.22 & 6.44 & $\times$\\
        250 & 316 & +0.5 & 9~dex & 1.0 & 8.0 & 1.275 & 20.7 & -7.22 & 6.65 & $\times$\\
        275 & 316 & +0.5 & 9~dex & 2.5 & 4.0 & 1.025 & 13.4 & -7.25 & 6.76 & $\times$\\
        250 & 316 & 0 & 7~dex & 2.5 & 8.0 & 1.2 & 18.4 & -7.27 & 7.03 & $\times$\\
        250 & 178 & 0 & 0~dex & 1.0 & 8.0 & 1.275 & 11.7 & -7.22 & 7.12 & $\times$\\
        250 & 316 & 0 & 0~dex & 2.5 & 6.0 & 1.275 & 20.7 & -7.23 & 7.33 & $\times$\\
        250 & 316 & +0.5 & 2~dex & 2.5 & 4.0 & 1.25 & 19.9 & -7.24 & 7.40 & $\times$\\
        250 & 316 & 0 & 0~dex & 2.5 & 8.0 & 1.275 & 20.7 & -7.23 & 7.43 & $\times$\\
        250 & 316 & 0 & 2~dex & 1.0 & 6.0 & 1.25 & 19.9 & -7.24 & 8.10 & $\times$\\
        275 & 316 & +0.5 & 0~dex & 1.0 & 6.0 & 1.0 & 12.7 & -7.27 & 8.54 & $\times$\\
        250 & 316 & +1.0 & 0~dex & 1.0 & 8.0 & 1.275 & 20.7 & -7.22 & 8.85 & $\times$\\
        300 & 100 & +0.5 & 7~dex & 2.5 & 8.0 & 0.9 & 3.3 & -7.22 & 9.15 & $\times$\\
        \hline
        \multicolumn{10}{c}{\citet{lacy_self-consistent_2023}} \\
        \hline
        250 & 3.75~dex & 0 & 6~dex & \nodata & AEE10 & 1.275 & 3.5 & $-7.25$ & 1.50 & $\times$\\
        275 & 3.75~dex & +0.5 & 0~dex & \nodata & AEE10 & 1.05 & 2.4 & $-7.22$ & 1.57 & $\times$\\
        275 & 4.00~dex & +0.5 & 0~dex & \nodata & AEE10 & 1.05 & 4.3 & $-7.22$ & 3.45 & $\times$\\
        275 & 3.50~dex & 0 & 6~dex & \nodata & E10 & 1.025 & 1.3 & $-7.24$ & 4.60 & $\times$\\
        \hline
        \multicolumn{10}{c}{ATMO 2020} \\
        \hline
        250 & 2.5~dex & 0 & NEQ Strong & \nodata & \nodata & 1.15 & 0.2 & $-7.30$ & 6.06 & $\times$ \\
        \hline
        \multicolumn{10}{c}{ATMO 2020++ (PH$_3$ Removed)} \\
        \hline
        275 & 3.5~dex & +0.3 & \nodata & \nodata & \nodata & 0.98 & 1.2 & $-7.26$ & 2.47 & $\times$\\
        250 & 2.5~dex & 0 & \nodata & \nodata & \nodata & 1.18 & 0.2 & $-7.24$ & 4.08 & $\times$\\
        \hline
        \multicolumn{10}{c}{ATMO 2020++ (PH$_3$ Included)} \\
        \hline
        275 & 3.5~dex & +0.3 & \nodata & \nodata & \nodata & 0.90 & 1.0 & $-7.28$ & 2.94 & $\times$\\
        275 & 3.0~dex & 0 & \nodata & \nodata & \nodata & 0.85 & 0.3 & $-7.30$ & 5.81 & $\times$\\
        300 & 3.5~dex & +0.3 & \nodata & \nodata & \nodata & 0.75 & 0.7 & $-7.30$ & 7.33 & $\times$\\
        \hline
        \enddata
\tablecomments{The atmospheric models are sorted in ascending order of their $\chi^2_{\nu}$ and divided into two categories (where applicable) based on consistency of the predicted mass with the dynamical mass ($6.5 \pm 0.7\; M_{\rm Jup}$) within $3\sigma$, as noted in the last column ($3\sigma_{M_b}$). The overall lowest $\chi^2_{\nu}$ model fit and the lowest $\chi^2_{\nu}$ model fit consistent with the dynamical mass within 3$\sigma$ are both marked as bolded entries.}
\end{deluxetable*}

\subsubsection{Model Fitting}
We jointly fit the F410M and F430M JWST/NIRCam flux and the F1065C, F1550C, F1800W, F2100W, and F2550W JWST/MIRI flux to all of the atmosphere models listed in Table~\ref{tab:models} (upper limits and the NEAR photometry are not considered). The fitting procedure synthesizes model photometry in the JWST bandpasses using the transmission curve from the Spanish Virtual Observatory (SVO) filter profile service\footnote{\url{http://svo2.cab.inta-csic.es/theory/fps/}} and finds the radius (within a range of 0.5--1.5~$R_{\rm Jup}$, though we note the extremes are unphysical according to evolutionary models, \added{see \S\ref{sec:teff-radius}}) that minimizes the $\chi^2_{\nu}$ with respect to the measured photometry. Models that fit the observed photometry with $\chi^2_\nu < 10$ are listed in Table~\ref{tab:model-fitting}. For each model, we also consider whether the predicted mass (from model surface gravity and fit radius) is consistent with the dynamical mass within $3\sigma$. The lowest $\chi^2_\nu$ model fits from each atmospheric grid are visualized in Figures~\ref{fig:sonora}, \ref{fig:sonora_alt}, \ref{fig:other-models-1}, and \ref{fig:other-models-2}.
\begin{figure*}
    \centering
    \includegraphics[width=\linewidth]{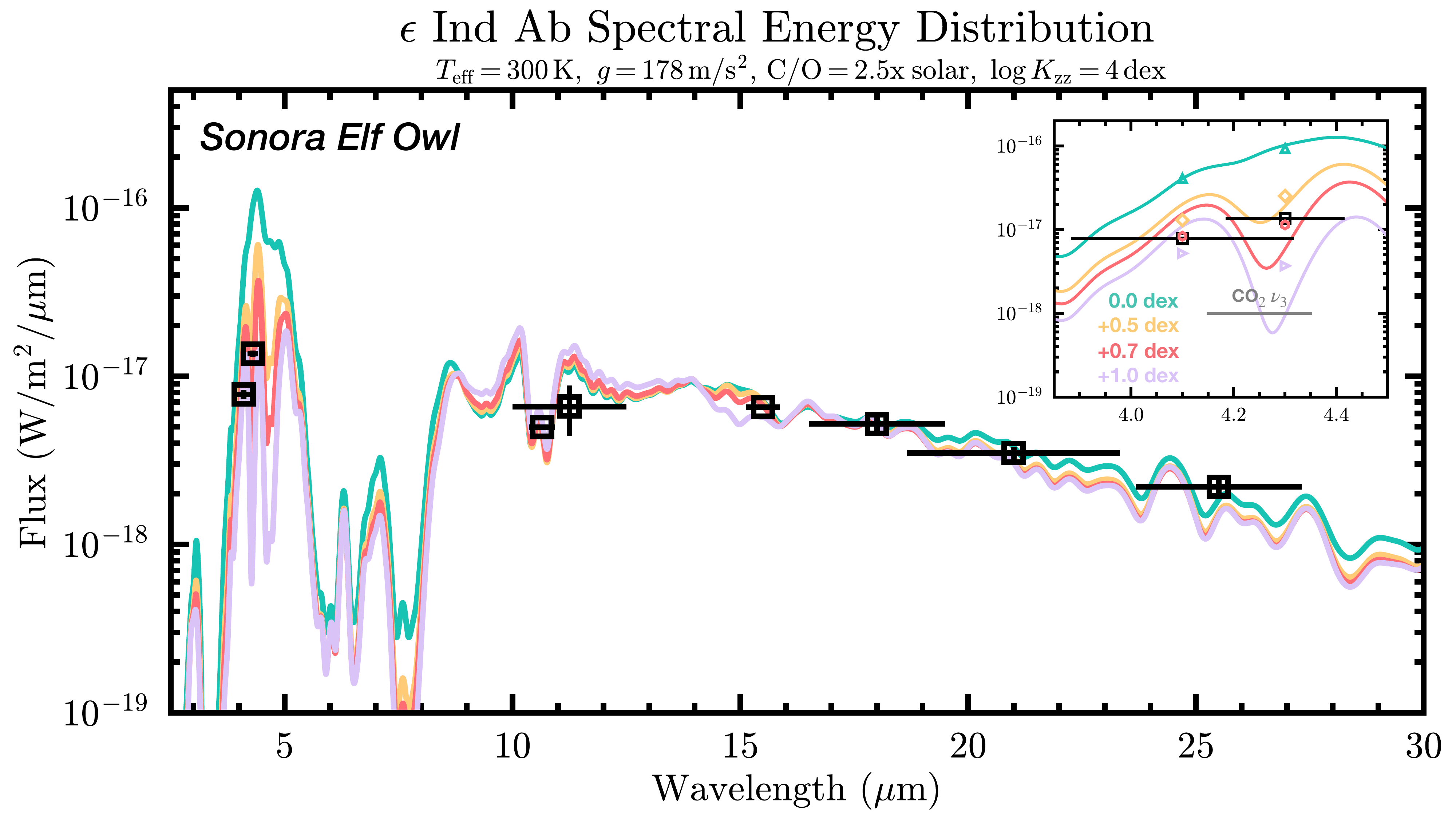}
    \caption{Example Sonora Elf Owl model spectra for varying atmospheric metallicities (colored lines, binned to $R\sim 50$), keeping parameters listed in the title fixed to the values corresponding to the lowest $\chi^2_\nu$ model consistent with the dynamical mass within $3\sigma$ (bolded entry in Table~\ref{tab:model-fitting}). Each model is scaled to minimize the $\chi^2_\nu$ with respect to the $>10$~$\mu$m photometry only. Black squares correspond to the measured \epsindAb\ photometry. The inset panel is a zoom-in view of the 4--5~$\mu$m region with synthetic F410M/F430M photometry from each model. A metal-enriched atmosphere increases the strength of the CO$_2$ absorption feature and can explain the observed flux suppression.}
    \label{fig:metal}
\end{figure*}

\subsubsection{Atmospheric Metallicity}
\added{The formation of a gas giant planet in a protoplanetary disk via the core accretion mechanism \citep[][]{mizuno_formation_1980, stevenson_formation_1982, pollack_formation_1996} involves the simultaneous accretion of gas and solids onto a massive core. The interplay between solid and gas accretion processes together with the formation location of the planet relative to the ice lines sets the overall composition and metallicity of the atmosphere that we observe today \citep[e.g.,][]{oberg_effects_2011, hasegawa_origin_2018, ikoma_formation_2025, yunerman_icy_2026}.} The transiting giant exoplanets follow a well-established empirical mass-metallicity relation \citep{guillot_correlation_2006,miller_heavy-element_2011,thorngren_mass-metallicity_2016} consistent with core accretion theory. Atmospheric studies of distant imaged planets ($\gtrsim$10~au) also find evidence of super-stellar atmospheric metallicities, suggesting that distant giants likely also accrete significant metal content \citep[e.g.,][]{nasedkin_four---kind_2024, balmer_jwst-tst_2025,balmer_vltigravity_2025,madurowicz_direct_2025, ruffio_jupiter-like_2026, xuan_compositions_2026}. 

Recently, \citet{chachan_revising_2025} revised the giant planet mass-metallicity relation based on estimates of the bulk heavy element content of 147 warm ($<1000$~K) giant planets with measured masses and radii. Using the best-fit relation, $Z_\mathrm{p}/Z_* = a/M_\mathrm{p}+b$, where $Z_\mathrm{p}$ is the bulk metallicity of the planet, $Z_*$ is the stellar metallicity, $M_\mathrm{p}$ is planet mass in Jupiter masses, $a = 2.6 \pm 0.3$, and $b = 3.3 \pm 0.5$, we find $Z_\mathrm{p}/Z_* = 3.7 \pm 0.5$ for \epsindAb\ (for the known dynamical mass of the planet). To convert the above ratio to atmospheric metallicity [M/H], we use the method of \citet{thorngren_connecting_2019} adapted by \citet[][based on Equations~18--20]{nasedkin_four---kind_2024}. \added{This is summarized below. First, we have,
\begin{equation}
    Z_p = (Z_\mathrm{p}/Z_*) \cdot 0.014 \cdot 10^{\rm [Fe/H]},
\end{equation}
where \epsindA's [Fe/H]~$ = -0.17 \pm 0.03$ \citep{lundkvist_low-amplitude_2024}. Then,
\begin{equation}
    [\mathrm{M/H}] = \log_{10}\left[ \frac{(1 + Y/X)}{(Z_p^{-1} - 1) \cdot (\mu_Z/\mu_H) \cdot (Z/H_\odot)}\right],
\end{equation}
where X, Y, and Z are the solar hydrogen, helium and metal mass fractions \citep[$Y/X = 0.3383$, $Z/H_\odot = 1.04 \times 10^{-3}$;][]{asplund_chemical_2009} and $\mu$ is the mean molecular weight of the metal content of the atmosphere \citep[$\mu_Z = 18$, $\mu_H = 1$ following][]{thorngren_connecting_2019}. Substituting, we find [M/H]~$=0.4 \pm 0.1$~dex.} The calculation assumes the interior is well-mixed such that the atmospheric metallicity is a good approximation of the planet's bulk metallicity. This is not an unrealistic assumption for massive planets ($\gtrsim5\;M_{\rm Jup}$) \added{as large-scale convection may quickly destroy compositional gradients \citep{muller_synthetic_2021, howard_giant_2025}}. \added{For example, a similar calculation by \citet{balmer_vltigravity_2025} for AF~Lep~b ($3.75 \pm 0.50\;M_{\rm Jup}$) also shows that their estimate is in agreement with the mass-metallicity relation.} 

From the atmospheric model fits, we find 45/63 of the models with $\chi^2_\nu < 10$ (Table~\ref{tab:model-fitting}) and 9/10 of the overall best-fitting (lowest $\chi^2_\nu$) models support an enhanced metallicity for \epsindAb. This is consistent with expectations from the giant planet mass-metallicity relation. The preference for a metal-enriched atmosphere can be attributed to the measured planet flux in F410M and F430M, \added{which is fainter than predictions from solar metallicity models due to stronger CO$_2$ absorption.} The dependence of the strength of CO$_2$ absorption in the NIRCam filters on atmospheric metallicity is visualized in Figure~\ref{fig:metal}. These results highlight that atmospheric metallicity is a key parameter to consider when evaluating the NIRCam detectability of cold gas giants \citep[in addition to potential water ice clouds, see e.g.,][]{bowens-rubin_nircam_2025}.

\begin{deluxetable*}{lllll|l|lll}
\tabletypesize{\scriptsize}
\tablecaption{\label{tab:consistency}Consistency of Best-fit Atmospheric Models with Observed \epsindAb\ Photometry}
\tablehead{\colhead{Model Grid} & \colhead{$\chi^2_\nu$} & \colhead{F410M} & \colhead{F430M} & \colhead{F1550C} & \colhead{F1065C} & \colhead{F1800W} & \colhead{F2100W} & \colhead{F2550W} \\ \colhead{} & \colhead{} & \colhead{(CO$_2$)} & \colhead{(CO$_2$)} & \colhead{(CO$_2$)} & \colhead{(NH$_3$)} & \colhead{(H$_2$O)} & \colhead{(H$_2$O)} & \colhead{(H$_2$O)}}
\startdata
\multicolumn{9}{c}{Clear Models} \\
\hline
Sonora Elf Owl & 1.16 & $\checkmark$ ($0.3\sigma$) & $\times$ ($-1.4\sigma$)  & $\checkmark$ ($-0.5\sigma$) & $\times$ ($1.7\sigma$) & $\checkmark$ ($0.2\sigma$) & $\checkmark$ ($0.7\sigma$) & $\times$ ($1.5\sigma$)\\
Sonora Flame Skimmer & 3.30 & $\times$ ($-3.1\sigma$) & $\checkmark$ ($-0.4\sigma$)  & $\times$ ($-1.3\sigma$) & $\times$ ($3.2\sigma$) & $\checkmark$ ($-0.1\sigma$) & $\checkmark$ ($0.3\sigma$) & $\times$ ($1.1\sigma$)\\
ATMO 2020 & 6.06 & $\times$ ($-2.6\sigma$) & $\checkmark$ ($0.6\sigma$)  & $\times$ ($5.3\sigma$) & $\times$ ($-2.1\sigma$) & $\checkmark$ ($0.9\sigma$) & $\checkmark$ ($-0.02\sigma$) & $\times$ ($1.3\sigma$)\\
ATMO 2020++ (PH$_3$ Removed)& 2.47 & $\checkmark$ ($-0.3\sigma$) & $\times$ ($-3.2\sigma$)  & $\times$ ($1.5\sigma$) & $\checkmark$ ($-0.8\sigma$) & $\checkmark$ ($0.6\sigma$) & $\checkmark$ ($0.7\sigma$) & $\times$ ($1.8\sigma$) \\
ATMO 2020++ (PH$_3$ Included)& 2.94 & $\times$ ($2.0\sigma$) & $\checkmark$ ($0.5\sigma$) & $\times$ ($-1.7\sigma$)  & $\times$ ($2.1\sigma$) & $\checkmark$ ($0.3\sigma$) & $\times$ ($1.5\sigma$) & $\times$ ($2.5\sigma$)\\
\hline
\multicolumn{9}{c}{Water Ice Cloud Models} \\
\hline
Custom PICASO Cloudy Model & 3.38 & $\checkmark$ ($0.2\sigma$) & $\times$ ($2.5\sigma$)  & $\checkmark$ ($0.5\sigma$) & $\times$ ($-1.8\sigma$) & $\checkmark$ ($-0.5\sigma$) & $\times$ ($-3.1\sigma$) & $\times$ ($-2.0\sigma$)\\
\citet{lacy_self-consistent_2023} & 1.50 & $\times$ ($-1.3\sigma$) & $\times$ ($-2.3\sigma$)  & $\times$ ($1.7\sigma$) & $\checkmark$ ($-0.5\sigma$) & $\checkmark$ ($0.3\sigma$) & $\checkmark$ ($-0.5\sigma$) & $\checkmark$ ($0.1\sigma$) \\
\enddata
\tablecomments{The dominant molecular absorber affecting each filter is noted in parentheses below the filter name. Models within 1$\sigma$ of the observed photometry in a given filter are marked with a `$\checkmark$'. The difference is reported as (observed $-$ model)/uncertainty in parentheses next to the checkmark or cross.}
\end{deluxetable*}

\subsubsection{Water Ice Clouds} 
Theoretical modeling suggests water vapor condenses into ice clouds in the atmospheres of the coolest extrasolar giant planets and brown dwarfs \citep[$\lesssim$400~K;][]{burrows_chemical_1999, marley_reflected_1999, burrows_beyond_2003, sudarsky_theoretical_2003, burrows_spectra_2004, sudarsky_phase_2005, morley_water_2014}. As an example, water ice clouds are observed in Jupiter's atmosphere \citep[e.g.,][]{sato_jupiters_1979, carlson_abundance_1992, banfield_jupiters_1998}. The presence of water ice clouds can suppress the emergent 4--5~$\mu$m planet flux due to their increasing optical thickness with decreasing planet temperature \citep[e.g.,][]{morley_water_2014, mang_microphysics_2022, lacy_self-consistent_2023, mang_microphysical_2024, bowens-rubin_nircam_2025, sanghi_worlds_2026-1}. Additionally, the depletion of water vapor in the upper atmosphere (through rainout or cloud formation) can modify the strength of mid-infrared water absorption bands \citep{kuhnle_water_2025}. Given \epsindAb's effective temperature is $\approx$275--300~K, its atmosphere is a candidate to host water ice clouds. We can frame our investigation around two possible questions. \\

\noindent
\emph{(1) Do the cloudy models fit the observed photometry better than the clear models?}  Considering $\chi^2_\nu$ for the various model fits, we find that the clear Sonora Elf Owl models better match the observed photometry compared to both the custom PICASO water ice cloud models and the \citet{lacy_self-consistent_2023} cloudy models (Table~\ref{tab:model-fitting}), though the difference in $\chi^2_\nu$ is small in the latter case. It is also worth noting that no single value of the $f_{\rm sed}$ parameter is strongly preferred in the model fits. \\

\noindent
\emph{(2) Are there systematic trends in the flux residuals for clear vs cloudy model fits?} The presence of clouds can affect the observed strength of molecular absorption features. In such a scenario, clear models may systematically under- or over-estimate the flux in certain photometric bandpasses. To investigate this, we evaluate the consistency (within 1$\sigma$) of both the best-fit clear and cloudy models with individual filter photometry. The results are presented in Table~\ref{tab:consistency}. 

\begin{figure*}
    \centering
    \includegraphics[width=\linewidth]{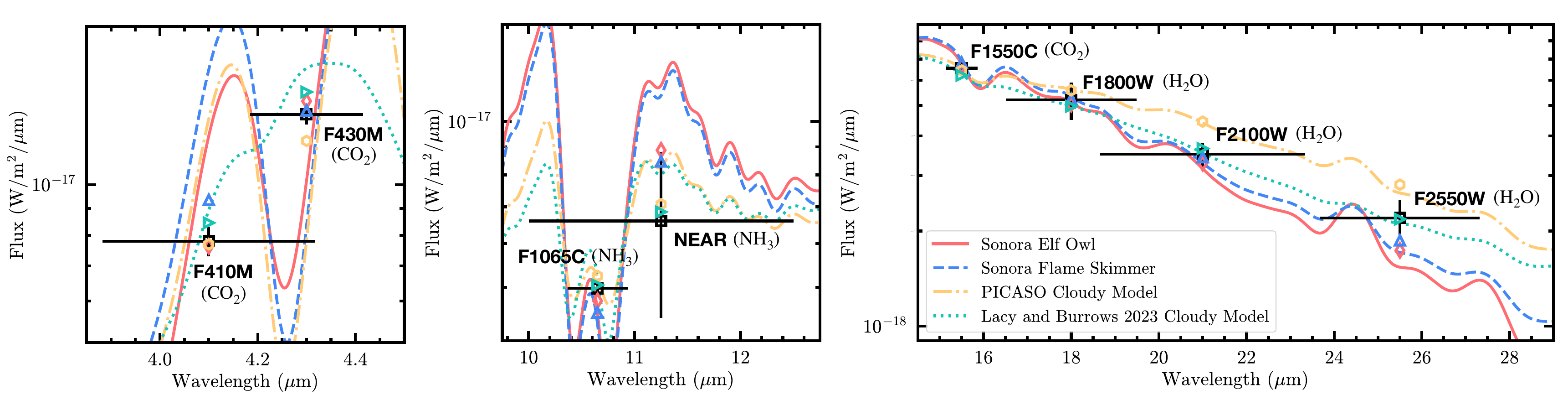}
    \caption{Comparison between the best-fit (lowest $\chi^2_\nu$) clear Sonora Elf Owl model (solid red), clear rainout chemistry Sonora Flame Skimmer model (dashed blue), cloudy PICASO model (dash-dotted yellow), and cloudy \citet{lacy_self-consistent_2023} (dotted green) model (binned to $R\sim 50$; see Table~\ref{tab:model-fitting} for model parameters). Black squares correspond to the observed photometry for \epsindAb\ and the dominant absorber in each filter is noted in parentheses. Synthetic model photometry are marked with diamond (Sonora Elf Owl), top-facing triangle (Sonora Flame Skimmer), hexagon (PICASO Cloudy), and right-facing triangle \citep{lacy_self-consistent_2023} symbols.}
    \label{fig:vary-clouds}
\end{figure*}

First, we find that no single model matches all three CO$_2$ dominated bandpasses \citep[F410M, F430M, and F1550C;][]{burrows_spectra_2014, mukherjee_sonora_2024} within $1\sigma$ (Table~\ref{tab:consistency}). There does not appear to be a systematic trend of flux under- or over-estimation between the clear and cloudy model fits that are discrepant with the observed photometry in these bandpasses. The inconsistency (at $1\sigma$ level) in at least one bandpass is statistically expected and also reflects the challenge of disentangling degeneracies between the effects of metallicity, disequilibrium chemistry, and clouds at these wavelengths \citep[e.g.,][]{morley_water_2014, mukherjee_sonora_2024, crotts_follow-up_2025, gagliuffi_jwst_2025, sanghi_worlds_2026-1}. 

Second, the F1065C photometry \citep[bandpass dominated by NH$_3$;][]{cushing_spitzer_2006, lacy_self-consistent_2023} is not well-fit by the majority of models. If we exclude the ATMO 2020 model (which has the worst overall fit among all grids), we observe that the inconsistent clear models \added{(Elf Owl, Flame Skimmer, and ATMO 2020++ with $\rm{PH_3}$; Table~\ref{tab:consistency})} systematically underestimate the F1065C flux, i.e., predict stronger NH$_3$ absorption than observed. The cloudy \citet{lacy_self-consistent_2023} model does improve the consistency to within 1$\sigma$ but the PICASO cloudy model overshoots and instead predicts weaker NH$_3$ absorption than observed. Overall, given that one of the ATMO2020++ clear models is still consistent with the observed photometry, it is unclear whether clouds are solely responsible for the discrepancy in F1065C photometry with the other clear models. 

Third, among the H$_2$O dominated bandpasses, we find that the F1800W and F2100W photometry are well-fit by the majority of models but the F2550W photometry is inconsistent at the 1$\sigma$ level with all clear models (which also include models incorporating water condensation through rainout chemistry). Specifically, all the clear models systematically underestimate the F2550W flux, i.e., predict stronger H$_2$O absorption than observed (Table~\ref{tab:consistency}). The custom PICASO water ice cloud model shows the opposite behavior, predicting weaker H$_2$O absorption than observed. The only model that is within 1$\sigma$ of the F2550W photometry is the cloudy \citet{lacy_self-consistent_2023} model (Figure~\ref{fig:vary-clouds}).

Fourth, we consider the fit of the custom PICASO cloudy model, the Sonora Flame Skimmer (clear, with rainout chemistry), and the Sonora Elf Owl (clear, no rainout chemistry) model in more detail (Figure~\ref{fig:vary-clouds}). The water ice clouds in the PICASO model lead to high suppression of F430M flux (stronger than observed), redistributing it to longer wavelengths and making the planet brighter than observed in F1065C, F2100W and F2550W. However, the clear Sonora Elf Owl model shows the opposite behavior, there is not enough flux suppression in the F430M filter and the planet is fainter in both F1065C and F2550W than observed. This indicates that there may be an intermediate solution, for example, one with patchy water ice clouds \citep[e.g.,][]{morley_spectral_2014, morley_water_2014}, that can consistently explain all the photometry points. Indeed, patchy clouds may be a more realistic scenario for cold giant planets, analogous to the spatially heterogeneous band structures observed on Jupiter \citep{carlson_abundance_1992,roos-serote_water_2004,arregi_phase_2006}. The current PICASO models all assume 100\% atmospheric cloud coverage fraction. We also find that the clear, rainout chemistry Sonora Flame Skimmer model, \added{in which water precipitates,} improves the fit to the F2550W photometry over the clear Sonora Elf Owl model, \added{in which water does not precipitate} (from $1.5\sigma$ to $1.1\sigma$, Figure~\ref{fig:vary-clouds} and Table~\ref{tab:consistency}). This indicates that water condensation may play a role in the chemistry of \epsindAb's upper atmosphere. 

Finally, we highlight that the \citet{lacy_self-consistent_2023} models and the PICASO models are two independent parameterized water ice cloud models and differences in quality-of-fit are likely due to varying model assumptions. While the \citet{lacy_self-consistent_2023} cloudy model provides an excellent fit to the flux in the NH$_3$ and H$_2$O dominated bandpasses, it does not explain the flux in the CO$_2$ dominated bandpasses well.  \\

To summarize, we consider the results from fitting self-consistent grid models to \epsindAb's photometry inconclusive as to the presence or absence of water ice clouds. While trends observed with the F2550W photometry and a comparison of the best-fit Sonora Elf Owl, Sonora Flame Skimmer, and cloudy PICASO models hint at the possibility of water ice clouds or rainout depletion in \epsindAb's atmosphere, we do not find any statistically incontrovertible signatures. Water ice clouds are not detected in a retrieval analysis of WISE 0855's NIRSpec+MIRI spectrum \citep{kuhnle_water_2025}, which has a similar effective temperature, and plausibly similar surface gravity \citep[$\log g =$3.5--4.5~dex in cgs units,][]{miles_observations_2020}, as \epsindAb. However, \citet{kuhnle_water_2025} do find evidence for water depletion in WISE 0855's upper atmosphere based on the strength of the water absorption bands in the MIRI MRS spectrum. The photometric results presented here motivate and pave the way for a similar analysis with the MIRI MRS spectrum of \epsindAb\ acquired as part of the same JWST observing program (\#8714).

\begin{figure*}
    \centering
    \includegraphics[width=\linewidth]{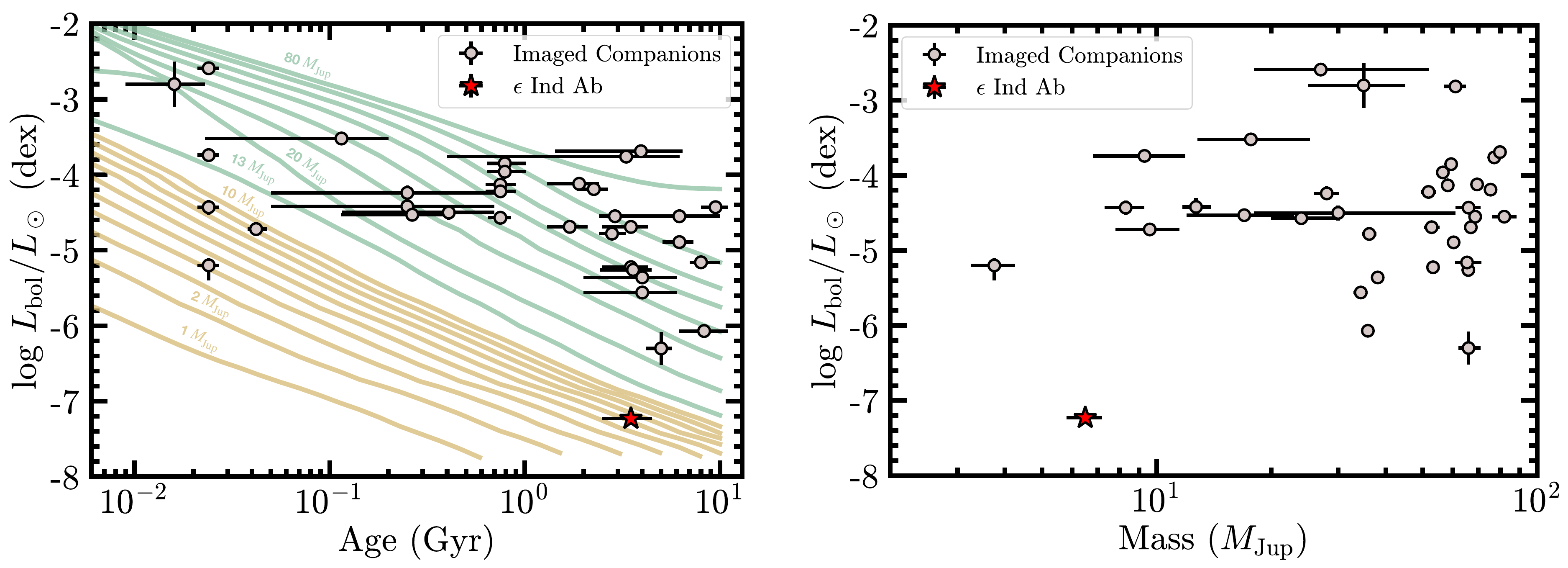}
    \caption{Imaged substellar companions with dynamical masses, bolometric luminosities, and age constraints. Based on data compiled by \citet[][see their Table~8]{li_test_2025}. We show the bolometric luminosity vs age (left) with isomass cooling tracks (yellow for planetary mass objects and green for brown dwarf mass objects) from the \citet{chabrier_impact_2023} evolutionary models and the bolometric luminosity vs mass (right). \epsindAb\ uniquely populates a previously unexplored parameter space of low mass, low luminosity, and old age.}
    \label{fig:fundprop-pop}
\end{figure*}

\subsubsection{Disequilibrium Chemistry} 
There is strong evidence that disequilibrium processes impact Solar System gas giant and cool substellar atmospheres \citep[e.g.,][]{fegley_equilibrium_1985, noll_detection_1997, saumon_ammonia_2006, leggett_36-79_2007, luhman_jwstnirspec_2023}. The increasing strength of CO absorption in the 4--5~$\mu$m region is a signature of vertical mixing in the atmosphere \citep{miles_observations_2020, mukherjee_sonora_2024, balmer_jwst-tst_2025}. In this case, the atmospheric model fits for \epsindAb\ do not find a clear preference for any single value of $\log K_{\rm zz}$ (Table~\ref{tab:model-fitting}), \added{where $K_{\rm zz}$ is the vertical eddy diffusion parameter. A high $K_{\rm zz}$ value represents strong vertical mixing and a low $K_{\rm zz}$ value represents slow, inefficient mixing in the atmosphere  \citep{mukherjee_sonora_2024}.} This is not surprising as our F410M/F430M photometry do not cover the 4.5--4.8~$\mu$m CO feature. The JWST/NIRSpec spectrum of \epsindAb, acquired as part of the same program (\#8714), and to be presented in a future paper, should provide better constraints on $\log K_{\rm zz}$.

\subsubsection{Bolometric Luminosity of \texorpdfstring{\epsindAb}{eps Ind Ab}}
The multi-wavelength photometry of \epsindAb\ enables a calculation of the planet's bolometric flux ($F_{\rm bol}$) by directly integrating the complete near- to mid-infrared spectral energy distribution (SED) \citep[e.g.,][]{filippazzo_fundamental_2015, sanghi_hawaii_2023}. Combining the bolometric flux with the known parallactic distance ($d$) yields the bolometric luminosity ($L_{\rm bol} = F_{\rm bol} \cdot 4\pi d^2$). We note that the uncertainty in the parallactic distance is at least an order of magnitude smaller than the uncertainty in $F_{\rm bol}$, which is discussed below.

The $F_{\rm bol}$ derived from a photometric SED relies on the use of atmosphere models to estimate flux contributed at wavelengths not covered by the photometry. The two sources of uncertainty in $F_{\rm bol}$ are the observational uncertainties in the photometry and the model-contributed uncertainty. The model-contributed uncertainty arises from variations in the flux (not covered by the photometry) calculated from distinct models that can explain the observed photometry (model spectra have no intrinsic uncertainty). In the case of \epsindAb, we can account for both uncertainty sources by estimating the bolometric flux using all atmosphere models that fit the observed photometry with $\chi^2_\nu < 10$ (Table~\ref{tab:model-fitting}). This choice of the $\chi^2_\nu$ threshold translates to a requirement that, on average, the synthetic model photometry is within $\approx$3$\sigma$ of the measured value. Thus, this method incorporates the observed photometric uncertainties. By calculating $F_{\rm bol}$ using models spanning the seven different grids considered (Table~\ref{tab:models}) that satisfy the above criterion, we are also accounting for the model-contributed uncertainty.

Given a model spectrum fit to the data with $\chi^2_\nu < 10$, the bolometric flux is determined by directly integrating the model spectrum across its full wavelength range. The model coverage on the short wavelength end is sufficient ($\gtrsim0.1\;\mu$m) to capture the flux contribution accurately (the flux of the planet rapidly drops shortward of 4~$\mu$m). However, on the long wavelength end, several models cut-off at $\lambda \sim 30\;\mu$m. To estimate the flux in the tail, we integrate a blackbody spectrum with the same effective temperature and radius as the fit model spectrum from the cut-off wavelength to an arbitrarily large wavelength ($10^6\;\mu$m). This is a good approximation as we are in the Rayleigh-Jeans limit at these wavelengths. The two contributions are added to obtain $F_{\rm bol}$ and thus, $L_{\rm bol}$ (noted in Table~\ref{tab:model-fitting}). We adopt the mean and standard deviation of the individual $L_{\rm bol}$ estimates as our final measurement, $\log L_{\rm bol}/L_\odot = -7.23 \pm 0.03$~dex (corresponds to $\approx$7\% uncertainty in $L_{\rm bol}$), where $L_\odot = 3.828 \times 10^{26}$~W \citep{mamajek_iau_2015}.

While the estimated $L_{\rm bol}$ is not independent of the atmosphere models, we highlight that the observed JWST photometry captures a significant fraction, $\approx$~40\%, of the total flux. The model contribution is $\approx$~30\% between 4.4--10.3~$\mu$m, $\approx$~20\% between 12--15~$\mu$m, and $\approx$~10\% in the tail beyond 28~$\mu$m. The overall consistency of $L_{\rm bol}$ calculated using different atmosphere models, as indicated by the small $L_{\rm bol}$ uncertainty, shows that the estimate is robust for comparisons to predictions from evolutionary models in the following section.

\section{Evolutionary Model Analysis}
\label{sec:evomodel}
Evolutionary models are a critical and widely-used tool to infer substellar fundamental properties from observables. Our new dynamical mass ($M_{\rm obs} = 6.5 \pm 0.7\;M_{\rm Jup}$) and bolometric luminosity ($L_{\rm obs} = -7.23 \pm 0.03$, expressed as $\log L_{\rm bol}/L_\odot$), combined with existing estimates of the stellar age ($t_{\rm obs} = 3.5 \pm 1.0$~Gyr), make \epsindAb\ the first planet to serve as a model benchmark in a new regime of low luminosities (cold temperatures) and old ages (Figure~\ref{fig:fundprop-pop}). Two evolutionary model grids covering the (mass, age, luminosity) of \epsindAb\ are considered for the following analysis.

\emph{Sonora Bobcat.} Sonora Bobcat \citep{marley_sonora_2021} builds upon the heritage \citet{saumon_evolution_2008} ``hybrid" evolutionary models and includes a number of advancements such as accounting for metals in the equation of state (EOS) to enable calculation of models for [M/H] = $\{-0.5, 0, +0.5\}$~dex and using improved nuclear-reaction screening factors \citep{potekhin_thermonuclear_2012}. \added{One difference is that the \citet{saumon_evolution_2008} models include the effects of clouds for L-type objects whereas the Sonora Bobcat models are cloud-free (given \epsindAb's cold temperature, $\sim$275~K, silicate clouds do not play a role in its atmosphere).} The self-consistently computed Sonora Bobcat model atmospheres serve as a boundary condition for the evolution models and include updated opacities (see their Table~2) and atmospheric chemistry.

\begin{figure*}
    \centering
    \includegraphics[width=\linewidth]{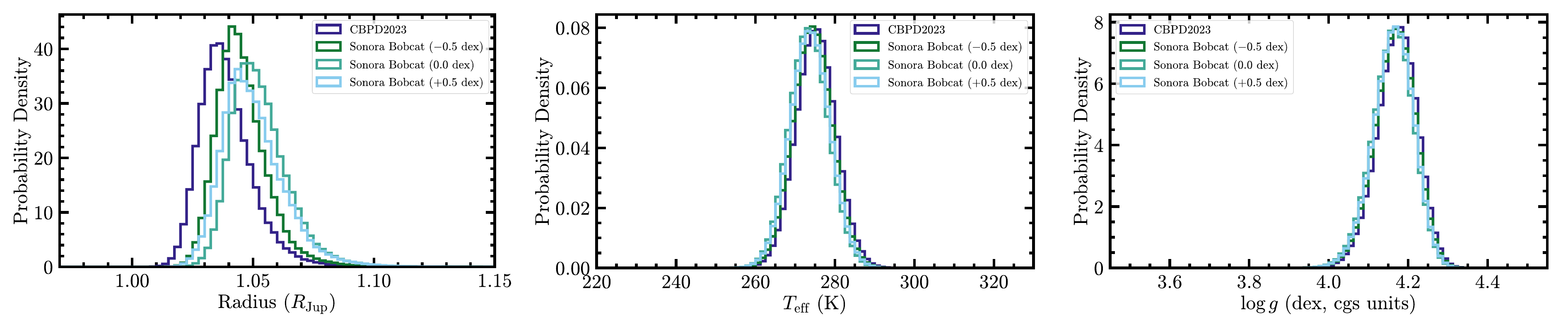}
    \caption{Posterior distributions for \epsindAb's radius (left), effective temperature ($T_{\rm eff}$, center) and surface gravity (right) using the ATMO2020 and Sonora Bobcat evolutionary models given the planet's dynamical mass, age, and bolometric luminosity.}
    \label{fig:teff-radius-logg}
\end{figure*}

\begin{figure*}
    \centering
    \includegraphics[width=\linewidth]{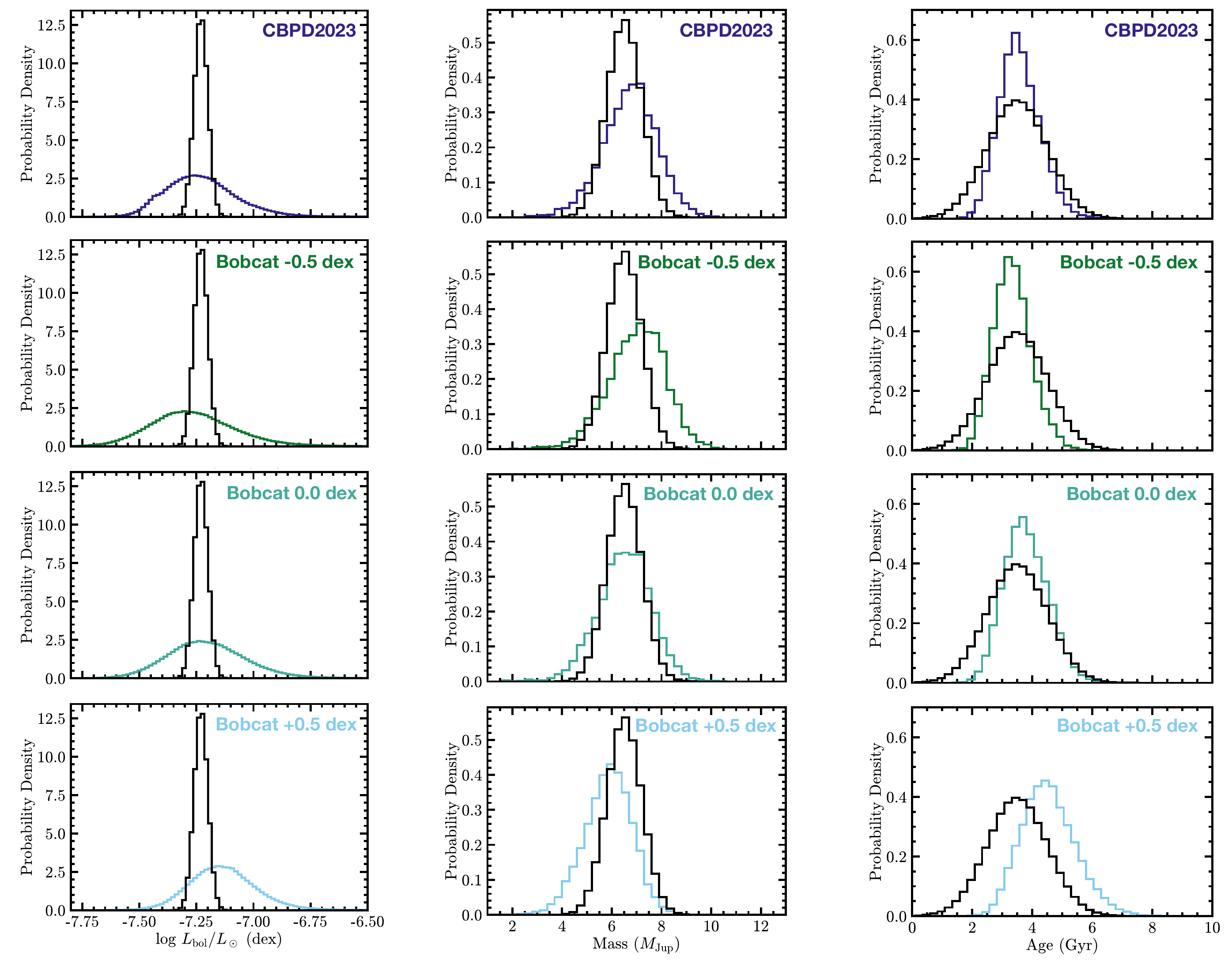}
    \caption{Comparison between \epsindAb's observed (black) and model-predicted (colored) bolometric luminosity (left), mass (center), and age (right). The observed and model-predicted parameters for \epsindAb\ are consistent with each other.}
    \label{fig:evocompare}
\end{figure*}

\emph{\citet[][CBPD2023]{chabrier_impact_2023} Models.} The CBPD2023 evolution models are based on the Lyon stellar evolution code \citep{chabrier_structure_1997, baraffe_evolutionary_1998, baraffe_evolutionary_2003} and incorporate the updated EOS for H-He mixtures from \citet{chabrier_new_2021}\footnote{Note that the CBPD2023 models are distinct from the ATMO~2020 evolution models \citep{phillips_new_2020}, which use the \citet{chabrier_new_2019} EOS. The CBPD2023 models are labeled as ``ATMO\_2020\_neweos\_update" on \url{https://noctis.erc-atmo.eu/fsdownload/zyU96xA6o/phillips2020}.}. The solar metallicity ATMO 2020 atmosphere models \citep{phillips_new_2020} serve as a boundary condition for the CBPD2023 evolution models and include both expanded opacity \citep[e.g., CH$_4$ and NH$_3$ from][]{tennyson_exomol_2018} and chemistry.

\subsection{Effective Temperature and Radius}
\label{sec:teff-radius}
Evolutionary models predict substellar radii as a function of mass, age, and metallicity. We can combine our empirical measurements of bolometric luminosity ($L_{\rm bol}$) and mass ($M$) with evolutionary model-predicted radii to obtain effective temperature $T_{\rm eff} = \left(L_{\rm bol}/4\pi R^2 \sigma_{\rm SB}\right)^{1/4}$ and surface gravity $\log g = \log (GM/R^2)$. First, to derive the posterior distribution for \epsindAb's radius, we draw $10^6$ normally distributed random values for mass and age following their observational uncertainties. The evolutionary model grids are bilinearly interpolated at each (mass, age) point to obtain radius. Next, $T_{\rm eff}$ and $\log g$ are calculated using the posterior samples with the above equations. The resulting distributions are shown in Figure~\ref{fig:teff-radius-logg} and the 68\% confidence intervals are reported in Table~\ref{tab:evo}. We find \epsindAb's radius is $1.05 \pm 0.01\;R_{\rm Jup}$, $T_{\rm eff} = 275 \pm 5$~K, and $\log g = 4.17 \pm 0.05$~dex (cgs units). The $T_{\rm eff}$ is indeed similar to that of WISE0855 as was determined from the mid-infrared color comparison (\S\ref{sec:wise0855}). \added{Note that the above estimates only present the statistical uncertainty in each parameter. However, any systematic uncertainties from the models are likely to be small given that the model predictions agree with the empirical mass, luminosity, and age within measurement uncertainties (see the next section).}

\begin{deluxetable*}{lccccccc}
\tabletypesize{\scriptsize}
\setlength{\tabcolsep}{3pt}
\tablecaption{\label{tab:evo}Evolutionary Model Analysis for \epsindAb}
\tablehead{\colhead{Evolutionary Model} & \colhead{Radius} & \colhead{$T_{\rm eff}$} & \colhead{$\log g$} & \colhead{$L_{\rm obs} - L_{\rm mod}(M_{\rm obs}, t_{\rm obs})$} & \colhead{$M_{\rm obs} - M_{\rm mod}(L_{\rm obs}, t_{\rm obs})$} & \colhead{$t_{\rm obs} - t_{\rm mod}(L_{\rm obs}, M_{\rm obs})$} & \colhead{$D^2_{\sigma}$} \\ \colhead{} & \colhead{($R_{\rm Jup}$)} & \colhead{(K)} & \colhead{(dex, cgs)} & \colhead{(dex)} & \colhead{($M_{\rm Jup}$)} & \colhead{(Gyr)} & \colhead{}}
\startdata
CBPD2023 & $1.038^{+0.012}_{-0.009}$ & $275 \pm 5$ & $4.17 \pm 0.05$& $0.01^{+0.14}_{-0.16}$ & $-0.27^{+1.31}_{-1.29}$ & $-0.07^{+1.19}_{-1.21}$ & $0.08\sigma$ \\
Sonora Bobcat ($-0.5$ dex) & $1.044^{+0.011}_{-0.008}$ & $274 \pm 5$& $4.17 \pm 0.05$ &  $0.04^{+0.17}_{-0.19}$ & $-0.5^{+1.4}_{-1.2}$ & $0.15\pm1.18$ & $0.26\sigma$\\
Sonora Bobcat ($0.0$ dex) & $1.051^{+0.013}_{-0.009}$  & $273 \pm 5$ & $4.16 \pm 0.05 $& $-0.02^{+0.16}_{-0.18}$ & $-0.006^{+1.283}_{-1.246}$ & $-0.27^{+1.21}_{-1.24}$ & $0.10\sigma$\\
Sonora Bobcat ($+0.5$ dex) & $1.048^{+0.014}_{-0.010}$ & $274 \pm 5$ & $4.16 \pm 0.05$ & $-0.09^{+0.14}_{-0.15}$& $0.6^{+1.2}_{-1.1}$ & $-1.07^{+1.33}_{-1.35}$ & $0.62\sigma$\\
\enddata
\tablecomments{The reported parameter estimates correspond to the 68\% confidence interval. For the evolutionary model comparisons, we use $L_{\rm obs} = -7.23 \pm 0.03$~dex (expressed as $\log L_{\rm bol}/L_\odot$), $M_{\rm obs} = 6.5 \pm 0.7\;M_{\rm Jup}$, $t_{\rm obs} = 3.5 \pm 1.0$~Gyr. Note that the above estimates do not account for potential systematic uncertainties arising from the treatment of heavy metals, their distribution in the interior, the adopted atmospheric boundary conditions, and the influence of fuzzy cores \citep[e.g.,][]{sur_next-generation_2025}.}
\end{deluxetable*}

\subsection{Luminosity, Mass, Age Consistency Tests and Goodness-of-Fit}
\label{sec:consistency}

Given independent estimates of a planet's luminosity, mass, and age, evolutionary models can infer any one of the three properties using the other two. This enables a direct consistency check between the observed and model-derived parameter estimate, testing substellar evolutionary models. 

To obtain the posterior distribution for luminosity ($L_{\rm mod}$, expressed as log~$L_{\rm bol}/L_\odot$), we first draw $10^6$ normally distributed random values for mass and age following their observational uncertainties. The model grid is bilinearly interpolated at each (mass, age) point to obtain $L_{\rm mod}$. 

To infer the posterior distribution for mass $M_{\rm mod}$ given observational measurements of age and luminosity, we use the Bayesian rejection sampling technique\footnote{This approach is preferred over Markov Chain Monte Carlo (MCMC) since the inference problem is one-dimensional and rejection sampling has the advantage of producing independent and identically distributed samples unlike MCMC, where the samples are correlated.} \citep[e.g.,][]{dupuy_individual_2017, sanghi_hawaii_2023, dupuy_masses_2023, hurt_uniform_2024, li_test_2025}. We begin by drawing $10^6$ uniformly distributed $M_{\rm mod}$ samples across the full range available in the model grid and $10^6$ normally distributed age samples following the observational uncertainty. The model grid is bilinearly interpolated at each ($M_{\rm mod}$, age) point to derive $L_{\rm mod}$. We compute $\chi^2$ for each sample,
\begin{equation}
    \chi^2 = \frac{(L_{\rm obs} - L_{\rm mod})^2}{\sigma^2_{L_{\rm obs}}}.
\end{equation}
The $\chi^2$ for each sample is converted to a probability $p$, normalizing by the sample with minimum $\chi^2$ as follows,
\begin{equation}
    p = \mathrm{exp}\left[-\frac{1}{2}(\chi^2 - \chi^2_{\rm min})\right].
\end{equation}
Next, we randomly draw $10^6$ uniformly distributed variates ($u$) in range from 0 to 1 and reject any $M_{\rm mod}$ samples where $p < u$. The remaining samples determine the distribution of $M_{\rm mod}$. \added{The posterior distribution for age $t_{\rm mod}$ is inferred using an identical rejection sampling procedure where mass and luminosity are the observed parameters.} Figure~\ref{fig:evocompare} presents a comparison of the observed and model-predicted posterior distributions for each of the three fundamental parameters. Additionally, Table~\ref{tab:evo} provides the 68\% confidence interval for the parameter bias, \added{which is calculated as the difference between the $x_{\rm mod}$ samples and samples from the observed distribution of parameter $x$.} Overall, we find excellent agreement between the observed and model-derived luminosity, mass, and age of \epsindAb. The median bias in each parameter ($M$, $L$, $t$) is within the observed value's 1$\sigma$ uncertainty (with the exception of the $+0.5$~dex Sonora Bobcat model, which has a larger bias in luminosity). 

A generalized goodness-of-fit metric that can be calculated between the models and the observation is the Mahalanobis distance $D^2$ \citep{mahalanobis1930tests}. The evolutionary models are two-dimensional manifolds in the three dimensional parameter space of mass, age, and luminosity. $D^2$ evaluates how close the observed (mass, age, luminosity) value for \epsindAb\ lies to the model manifold and reduces to the $\chi^2$ metric in the absence of correlations between the three quantities. This is a reasonable assumption in our case \citep[e.g.,][]{li_test_2025} since the parallax uncertainty is negligible. This analysis is distinct from the consistency tests previously discussed as it does not fix the value of any of the three observables. First, we bilinearly interpolate the evolutionary model over a fine grid of $M \in [1, 20]\;M_{\rm Jup}$ and $t \in [0.1, 10]$~Gyr to determine luminosity. We compute $D^2$ between ($M_{\rm obs}$, $t_{\rm obs}$, $L_{\rm obs}$) and each interpolated (mass, age, luminosity) point on the model manifold to determine the minimum distance $D^2_{\rm min}$. Given $D^2 \sim \chi^2$ \citep[1 degree of freedom;][]{li_test_2025}, we can calculate the two-sided tail probability $p$ that a random draw from this distribution exceeds $D^2_{\rm min}$ using the corresponding survival function \added{(complement of the cumulative distribution function)}. Finally, the probability $p$ is converted to an equivalent Gaussian $\sigma$ \added{(denoted as $D^2_\sigma$)} using the inverse survival function of a standard normal distribution. The $D^2_\sigma$ value for each model is $<1\sigma$ (Table~\ref{tab:evo}) and supports the conclusion that the evolutionary model predictions are consistent with the \epsindAb\ observations.

As the next generation of evolutionary models including a wider array of physical processes are developed \citep[e.g.,][]{sur_next-generation_2025}, \epsindAb\ will serve as a critical benchmark at cold temperatures and old ages.

\begin{figure*}[htb]
    \centering
    \includegraphics[width=\linewidth]{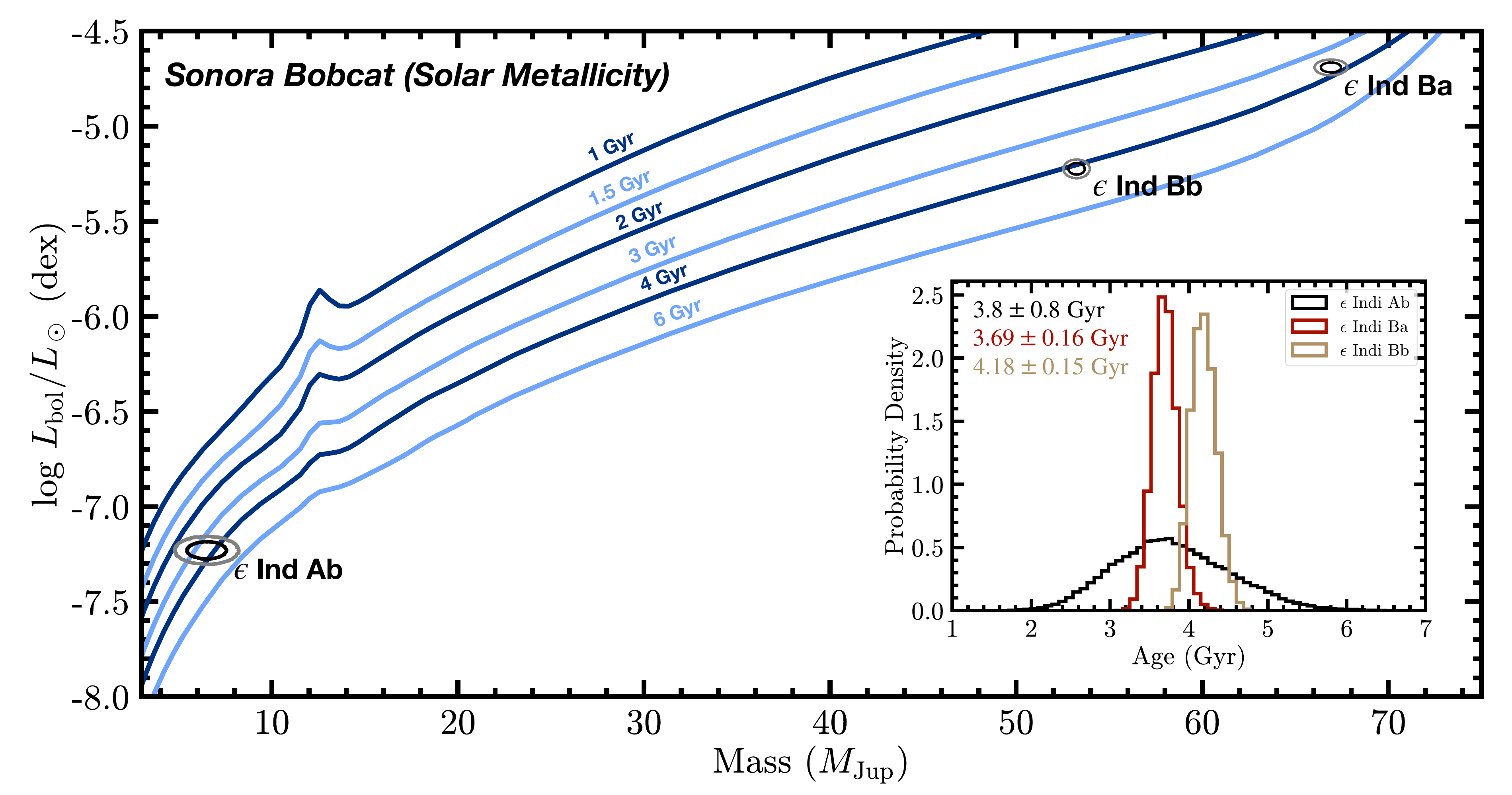}
    \caption{Isochrones (light and dark blue) from the solar metallicity Sonora Bobcat evolutionary models ranging from 1~to~6~Gyr. Black and gray contours show the joint 1$\sigma$ and 2$\sigma$ confidence intervals of the masses and luminosities of \epsindAb, $\epsilon$~Ind~Ba, and $\epsilon$~Ind~Bb. All three objects are expected to be coeval and, indeed, remarkably, lie along a single isochrone (4~Gyr). The inset figure shows the posterior distribution for system age derived with the solar metallicity Sonora Bobcat model using the dynamical mass and observed luminosity of each of the three objects.}
    \label{fig:coevality}
\end{figure*}

\subsection{Coevality of the \texorpdfstring{$\epsilon$~Ind}{eps~Ind} Substellar Objects}
The $\epsilon$~Ind system additionally hosts the binary brown dwarfs $\epsilon$~Ind~Ba and $\epsilon$~Ind~Bb. \citet{chen_precise_2022} noted that the $\epsilon$~Ind~B binary represents a rare case where one can test coevality of the two objects using model isochrones, given their precise dynamical mass and bolometric luminosity measurements. Now, with the addition of the above measurements for \epsindAb\ in this work, $\epsilon$~Ind is truly unique among benchmark systems, amenable to a coevality test over two orders of magnitude in luminosity and one order of magnitude in mass.

\citet{chen_precise_2022} demonstrated that only the \citet{saumon_evolution_2008} ``hybrid" evolutionary models yielded consistent cooling ages for the two brown dwarfs (but predict an older age, $\sim$5~Gyr, at odds with observational determinations). Analysis of the brown dwarf binary with the CBPD2023 models is presented in \citet{chabrier_impact_2023}. The CBPD2023 models do not satisfy coevality but the cooling ages inferred for the two brown dwarfs are consistent with the observational age estimate within uncertainties. Here, we consider the Sonora Bobcat models for the coevality test. Overplotting the observed masses and luminosities of the $\epsilon$~Ind substellar objects on the Sonora Bobcat solar metallicity model isochrones, we find that all three companions, remarkably, lie approximately along the 4~Gyr isochrone (Figure~\ref{fig:coevality}). 

For a quantitative comparison, we derive the age posterior from the observed mass and luminosity measurements of each objects following the rejection sampling procedure in \S\ref{sec:consistency}. \epsindAb's cooling age is consistent within $1\sigma$ with those of the binary brown dwarfs (Figure~\ref{fig:coevality}). The uncertainty in \epsindAb's dynamical mass dominates the uncertainty in the cooling age. The median cooling ages for the two brown dwarfs are more broadly consistent at the $3\sigma$ level (Figure~\ref{fig:coevality}). The cooling ages of all three objects are consistent with the observational estimate of $3.5 \pm 1.0$~Gyr for the host star \citep{chen_precise_2022}. Investigating the larger difference between the cooling ages of the two brown dwarfs is beyond the scope of this work (and is the goal of JWST program \#5765, PI: E.~Matthews), however, \added{we note that improved consistency requires a particularly steep slope for the mass-luminosity relation, $\Delta\log L/\Delta \log M = 5.37 \pm 0.08$ as calculated by \citet[][]{chen_precise_2022}. Indeed, the need for a steeper slope, with Sonora Bobcat, can be visually seen in Figure~\ref{fig:coevality}.} The metal poor and metal rich Sonora Bobcat models result in a larger discrepancy in the cooling ages between the two brown dwarfs (but not \epsindAb, as shown in the previous section) and are thus not presented here.

\section{Conclusions}
\label{sec:concl}
In this work, we assembled the first 4--25~$\mu$m spectral energy distribution for a cold ($< 300$~K) exoplanet using new JWST/NIRCam 4--5~$\mu$m and JWST/MIRI 18--25~$\mu$m imaging of the super-Jupiter \epsindAb. Our main conclusions are summarized below:
\begin{enumerate}

\item \epsindAb\ is detected at a separation of $\approx3.5$\arcsec, position angle of $\approx35^\circ$ East of North, and S/N = \{17.9, 25.4, 7.7, 14.3, 7.5\} in \{F410M, F430M, F1800W, F2100W, F2550W\}, respectively. \added{The MIRI detections, in particular, were obtained without a coronagraph, with the PSF core saturated, and using archival observations to construct a reference library for PSF subtraction. This highlights the power of MIRI direct imaging for cold planet detection.}

\item We performed a comprehensive re-analysis of \epsindAb's orbit with the new high-precision NIRCam astrometry. Our joint RV, absolute astrometry, and relative astrometry model used the framework of observable-based priors to reduce biases in fitting the relative astrometry data, mitigated the impact of stellar activity in the RV time series with a Gaussian process, and accounted for non-linear effects such as secular acceleration. We estimated $M_{\rm Ab} = 6.5^{+0.7}_{-0.6}\;M_{\rm Jup}$, $a = 15.8^{+1.7}_{-1.4}$~au, $e = 0.25 \pm 0.09$, and $i = 102.2^\circ \pm 1.7^\circ$. \added{Additionally, we provide a solution to why the initial RV + absolute astrometry only orbit prediction was incorrect. The combination of unmodeled stellar activity, partial orbital coverage, and stitching together RVs from different instruments with independent offsets biased the orbit fit towards an (incorrectly high) eccentric solution. A circular orbit (zero eccentricity) fit to the same data correctly pre-predicted \epsindAb's imaging discovery epoch position.}

\item An initial comparison between the SED of the coldest Y-dwarf ($\sim$285~K) WISE 0855 and \epsindAb\ yielded similar F1800W$-$F1550C color, and thus, effective temperature, for the two objects. We confirmed suppression of the 4--5~$\mu$m flux, compared to WISE 0855, in \epsindAb's atmosphere. This was hinted at by a non-detection of the planet in ground-based imaging.

\item Fitting self-consistent atmospheric models (both clear and cloudy, across multiple grids) to \epsindAb's SED, we found a preference for enhanced atmospheric metallicity, consistent with expectations from the giant planet mass-metallicity relation (which predicted [M/H]~$=0.4\pm0.1$~dex, assuming a well-mixed interior). This preference is driven by the measured F410M and F430M photometry, \added{which is fainter than predictions from solar metallicity atmosphere models. An enhanced metallicity increases the abundance of CO$_2$ and thus the strength of its 4.1--4.3~$\mu$m absorption feature.} No constraint is obtained on the strength of disequilibrium chemistry in the atmosphere since our photometry do not cover the 4.5--4.8~$\mu$m CO feature.

\item We did not find definitive evidence for the presence or absence of water ice clouds in \epsindAb's atmosphere. The overall best-fit atmospheric model was a clear Sonora Elf Owl model ($\chi^2_\nu = 1.16$), though a cloudy \citet{lacy_self-consistent_2023} model also agreed well with the photometry ($\chi^2_\nu = 1.50$). Notably, the F2550W flux (affected by H$_2$O vapor absorption) was systematically underestimated ($>$1$\sigma$, but $<2\sigma$) by all clear and rainout chemistry models. Only a cloudy \citet{lacy_self-consistent_2023} model could explain the F2550W flux within 1$\sigma$. This should be investigated further using \epsindAb's MIRI/MRS spectrum.

\item By directly integrating \epsindAb's SED, we estimated a bolometric luminosity $\log L_{\rm bol}/L_\odot = -7.23 \pm 0.03$~dex for the planet. The observed photometry accounts for 40\% of the bolometric luminosity and the quoted error bar incorporates the model-contributed uncertainty.

\item Using the Sonora Bobcat and \citet{chabrier_impact_2023} evolutionary models, we estimated \epsindAb's $R\approx1.05\;R_{\rm Jup}$, $T_{\rm eff} \approx 275$~K, and $\log g \approx 4.17$~dex. Combining the planet's dynamical mass, bolometric luminosity, and age ($3.5 \pm 1.0$~Gyr), we conducted the first test of substellar evolution models in a new regime of low masses, low luminosities, and old ages. All three measurements were in excellent agreement ($<1\sigma$) with model predictions. We also showed that the Sonora Bobcat models agree with \epsindAb, $\epsilon$~Ind~Ba, and $\epsilon$~Ind~Bb being co-eval objects.
\end{enumerate}

The addition of a precise dynamical mass and bolometric luminosity for \epsindAb\ establishes the $\epsilon$~Ind system as a unique benchmark for substellar evolution studies at temperatures and ages more similar to that of Jupiter in our own Solar System. Upcoming medium-resolution JWST/NIRSpec and MIRI MRS spectroscopy of \epsindAb\ will measure elemental and isotopic abundances in its atmosphere and investigate the relative importance of different accretion processes operating during gas giant formation. More broadly, this work sets the stage for the detailed characterization of a rapidly expanding sample of imaged extremely cold worlds ($\lesssim$300~K): the $\alpha$~Cen~Ab candidate \citep[][]{wagner_imaging_2021, beichman_worlds_2025, sanghi_worlds_2025}, TWA 7b \citep[][]{lagrange_evidence_2025, crotts_follow-up_2025}, 14 Her c \citep[][]{gagliuffi_jwst_2025}, and the candidate/confirmed planets orbiting white dwarfs \citep{limbach_miri_2024, limbach_thermal_2025}.

\begin{acknowledgments}
\added{We thank the anonymous referee for many helpful comments that improved this manuscript.} The author thanks William Balmer, Jorge Llop-Sayson, and Rodrigo Ferrer-Chavez for discussions on JWST/NIRCam data processing, Rachel Bowens-Rubin and Kevin Stevenson for their guidance on using the \texttt{MAGIC} pipeline for JWST/MIRI data reduction, Helena Kühnle for sharing their reduced MIRI MRS spectrum of WISE~0855, Floor van Leeuwen, Clarissa R.\ Do \'O, Quang Tran, and Kareem El-Badry for helpful discussions on orbit modeling, Michael Liu, Trent Dupuy, and Yaguang Li for productive conversations on the evolutionary model analysis, and Gabriel-Dominique Marleau and Markus Kasper for providing the filter profile for NEAR. We are grateful for the support of Dean Hines, Julien Girard, and the STScI planning and operations team in scheduling and executing this program. A.S acknowledges support from the National Science Foundation Graduate Research Fellowship under Grant No.~2139433. J.W.X is grateful for support from the Heising-Simons Foundation 51~Pegasi~b Fellowship (grant \#2025-5887). J.M. acknowledges support from the National Science Foundation Graduate Research Fellowship Program under Grant No. DGE~2137420.

This publication makes use of The Data \& Analysis Center for Exoplanets (DACE), which is a facility based at the University of Geneva (CH) dedicated to extrasolar planets data visualisation, exchange and analysis. DACE is a platform of the Swiss National Centre of Competence in Research (NCCR) PlanetS, federating the Swiss expertise in Exoplanet research. The DACE platform is available at \url{https://dace.unige.ch}. This publication makes use of VOSA, developed under the Spanish Virtual Observatory (\url{https://svo.cab.inta-csic.es}) project funded by MCIN/AEI/10.13039/501100011033/ through grant PID2020-112949GB-I00. VOSA has been partially updated by using funding from the European Union's Horizon 2020 Research and Innovation Programme, under grant Agreement \#776403 (EXOPLANETS-A). The computations presented here were conducted in the Resnick High Performance Computing Center, a facility supported by Resnick Sustainability Institute at the California Institute of Technology.
\end{acknowledgments}

\begin{contribution}
A.~Sanghi led the overall analysis and writing and submission of this manuscript. A.~Sanghi and J.~Xuan designed the NIRCam and MIRI imaging observations. J.~Xuan led the JWST proposal and facilitated interpretation of the results. W.~Thompson conducted an initial investigation of the planet's orbit and advised A.~Sanghi on the orbit fitting. J.~Mang generated the Sonora Flame Skimmer and custom PICASO atmospheric models used in this work and contributed to interpreting the model fits. All authors assisted with the preparation of the original JWST proposal and/or provided feedback on the manuscript.
\end{contribution}

\facilities{JWST(NIRCam), JWST(MIRI), Hipparcos, Gaia}

\software{\texttt{astropy
} \citep{astropy_collaboration_astropy_2013, astropy_collaboration_astropy_2018, astropy_collaboration_astropy_2022}, \texttt{matplotlib
} \citep{hunter_matplotlib_2007}, \texttt{numpy
} \citep{harris_array_2020}, \texttt{pandas
} \citep{mckinney_data_2010, team_pandas-devpandas_2025}, \texttt{python
} \citep{van_rossum_python_2009}, \texttt{scipy
} \citep{virtanen_scipy_2020, gommers_scipyscipy_2023}, \texttt{astroquery
} \citep{ginsburg_astroquery_2019, ginsburg_astropyastroquery_2024}, \texttt{scikit-image
} \citep{van_der_walt_scikit-image_2014}, \texttt{STPSF
} \citep{perrin_simulating_2012, perrin_updated_2014}, \texttt{jwst} \citep{bushouse_jwst_2025}, \texttt{pyKLIP} \citep{wang_pyklip_2015}, \texttt{vip} \citep{gomez_gonzalez_vip_2017, christiaens_vip_2023}, \texttt{spaceKLIP} \citep{kammerer_performance_2022, carter_jwst_2023, carter_spaceklip_2025}, \texttt{emcee} \citep{foreman-mackey_emcee_2013}, \texttt{webbpsf\_ext} \citep{leisenring_webbpsf_2025}, \texttt{PICASO} \citep{batalha_exoplanet_2019, mukherjee_picaso_2023, mang_picaso_2026}, and
\texttt{Virga} \citep{virga_code, virga_paper1, virga_paper2}.}

\appendix
\restartappendixnumbering

\section{MIRI Imaging Reference Library}
\label{sec:app-ref}
Table~\ref{tab:ref} presents the archival observations used to construct a PSF library for MIRI PSF subtraction.

\startlongtable
\begin{deluxetable*}{lccccl}
    \label{tab:ref}
    \centering
    \tablecaption{MIRI PSF Library}
    \tablehead{\colhead{Program} & \colhead{Target Name} & \colhead{Filters} & \colhead{Detector (x, y)\tablenotemark{a}} & \colhead{Dithers} & \colhead{Reference}} 
        \startdata
        COM 1024 & 2MASS J05220207-6930388\tablenotemark{b} & F1800W & (734, 827) & 4 & PI: Glasse, Alistair \\
        COM 1024 & 2MASS J05220207-6930388\tablenotemark{b} & F1800W & (700, 928) & 4 & PI: Glasse, Alistair \\
        COM 1024 & 2MASS J05220207-6930388\tablenotemark{b} & F1800W & (574, 456) & 4 & PI: Glasse, Alistair \\
        COM 1027 & BD+60 1753 & F1800W & (806, 580) & 4 & PI: Garcia Marin, Macarena \\
        GTO 1193 & * del Eri & F1800W & (324, 346) & 4 & PI: Beichman, Charles A. \\
        CAL 1523 & HD 2811 & F1800W & (70, 62) & 4 & PI: Law, David R. \\
        CAL 1536 & HD 180609 & F1800W & (157, 115) & 4 & PI: Gordon, Karl D. \\
        CAL 1536 & HD 163466 & F1800W & (60, 52) & 4 & PI: Gordon, Karl D. \\
        CAL 1536 & del UMi & F1800W & (49, 48) & 4 & PI: Gordon, Karl D. \\
        CAL 1536 & HD 163466 & F1800W & (60, 52) & 4 & PI: Gordon, Karl D. \\
        CAL 1538 & HD 106252 & F1800W & (50, 49) & 4 & PI: Gordon, Karl D. \\
        CAL 1538 & 16 Cyg B & F1800W & (60, 52) & 4 & PI: Gordon, Karl D. \\
        CAL 1538 & HD 37962 & F1800W & (49, 48) & 4 & PI: Gordon, Karl D. \\
        \hline
        COM 1024 & 2MASS J05220207-6930388\tablenotemark{b} & F2100W & (748, 830) & 4 & PI: Glasse, Alistair \\
        COM 1024 & 2MASS J05220207-6930388\tablenotemark{b} & F2100W & (588, 459) & 4 & PI: Glasse, Alistair \\
        COM 1024 & 2MASS J05220207-6930388\tablenotemark{b} & F2100W & (714, 931) & 4 & PI: Glasse, Alistair \\
        COM 1027 & BD+60 1753 & F2100W & (807, 579) & 4 & PI: Garcia Marin, Macarena \\
        GTO 1193 & * del Eri & F2100W & (325, 346) & 4 & PI: Beichman, Charles A. \\
        GTO 1194 & HD 218261 & F2100W & (170, 171) & 4 & PI: Beichman, Charles A. \\
        GTO 1194 & HR8799 & F2100W & (197, 221) & 4 & PI: Beichman, Charles A. \\
        GTO 1237 & DIGELC1-A & F2100W & (644, 1005) & 4 & PI: Ressler, Michael E. \\
        CAL 1523 & HD 2811 & F2100W & (78, 60) & 4 & PI: Law, David R. \\
        GO 6122 & Ross 154 & F2100W & (200, 210) & 4 & PI: Bowens-Rubin, Rachel \\
        GO 6122 & Wolf 359 & F2100W & (198, 209) & 4 & PI: Bowens-Rubin, Rachel \\
        \hline
        GTO 1193 & * alf PsA\tablenotemark{c} & F2550W & (324, 348) & 4 & PI: Beichman, Charles A. \\
        GTO 1193 & * alf PsA\tablenotemark{c} & F2550W & (324, 346) & 4 & PI: Beichman, Charles A. \\
        GTO 1193 & * alf Lyr\tablenotemark{c} & F2550W & (324, 346) & 4 & PI: Beichman, Charles A. \\
        GTO 1193 & * alf Lyr\tablenotemark{c} & F2550W & (324, 348) & 4 & PI: Beichman, Charles A. \\
        GTO 1193 & * del Eri & F2550W & (325, 346) & 4 & PI: Beichman, Charles A. \\
        GTO 1193 & * 19 PsA & F2550W & (321, 349) & 4 & PI: Beichman, Charles A. \\
        GTO 1193 & HD 169305 & F2550W & (324, 348) & 4 & PI: Beichman, Charles A. \\
        \enddata
        \tablenotetext{a}{Zero-indexed detector coordinates for the PSF in the stacked image after background subtraction with \texttt{MAGIC}.}
        \tablenotetext{b}{Three distinct source PSFs are extracted from the full detector image.}
        \tablenotetext{c}{Observations of the same target conducted at different times.}
\end{deluxetable*}

\section{Orbit Posteriors}
\label{app:orbit-post}
Table~\ref{tab:post-CI} presents the 68\% confidence interval and Table~\ref{tab:post-HDI} presents the 75\% highest density interval for all orbital parameters, computed from their respective posterior distributions, for all (converged) orbit models in \S\ref{sec:orb_model}. \added{The posteriors can be retrieved from Zenodo (\dataset[10.5281/zenodo.19931118]{10.5281/zenodo.19931118})}. 

\begin{rotatetable*}
\begin{deluxetable*}{l|l|l|l|l|l|l|l}
    \label{tab:post-CI}
    \centerwidetable
    \tabletypesize{\scriptsize}
    \centering
    \tablecaption{Posterior 68\% Credible Intervals}
    \tablehead{\colhead{Parameter (Unit)} & \colhead{RA\tablenotemark{$*$}} & \colhead{RA + HGCA} & \colhead{RA + RV} & \colhead{RA + HGCA + RV} & \colhead{RA + HGCA + RV} & \colhead{A + HGCA + RV} & \colhead{\textbf{RA + HGCA + RV}} \\ \colhead{} & \colhead{} & \colhead{} & \colhead{} & \colhead{(No NIRCam)} & \colhead{(No GP and No UVES)} & \colhead{(Unif.~Pri.)} & \colhead{\textbf{(Fiducial Model)}}} 
        \startdata
        $M_A$ ($M_{\odot}$)& $ 0.80^{+ 0.02 } _{- 0.02 } $& $ 0.80 ^{+ 0.02 } _{- 0.02 } $& $ 0.80^{+ 0.02 } _{- 0.02 } $& $ 0.78 ^{+ 0.02 } _{- 0.02 } $& $ 0.79 ^{+ 0.02 } _{- 0.02 } $ & $ 0.79 ^{+ 0.02 } _{- 0.02 } $& \boldmath{$ 0.80^{+ 0.02 } _{- 0.02 } $}\\
        $M_b$ ($M_{\rm Jup}$)& \nodata & $ 5.73 ^{+ 0.74 } _{- 0.60 } $& $ 10.76 ^{+ 5.63 } _{- 5.21 } $& $ 6.10^{+ 0.55 } _{- 0.49 } $ & $ 8.76 ^{+ 1.47 } _{- 1.87 } $& $ 7.71 ^{+ 1.22 } _{- 0.94 } $ & \boldmath{$ 6.50^{+ 0.72 } _{- 0.59 } $}\\
        $\varpi$ (mas)& $ 274.85 ^{+ 0.10} _{- 0.10} $ & $ 274.85 ^{+ 0.09 } _{- 0.09 } $ & $ 274.84 ^{+ 0.10} _{- 0.10} $& $ 274.84 ^{+ 0.10} _{- 0.10} $& $ 274.84 ^{+ 0.10} _{- 0.09 } $& $ 274.85 ^{+ 0.10} _{- 0.09 } $& \boldmath{$ 274.85 ^{+ 0.09 } _{- 0.10} $} \\
        $\mu_{\alpha^*}$ (mas/yr)& $ 3966.66 ^{+ 0.68 } _{- 0.68 } $ &$ 3966.67 ^{+ 0.11 } _{- 0.13 } $ & $ 3966.66 ^{+ 0.68 } _{- 0.69 } $& $ 3966.77 ^{+ 0.23 } _{- 0.27 } $& $ 3966.27 ^{+ 0.15 } _{- 0.35 } $& $ 3966.40^{+ 0.19 } _{- 0.26 } $& \boldmath{$ 3966.46 ^{+ 0.16 } _{- 0.24 } $} \\
        $\mu_{\delta}$ (mas/yr)& $ -2536.20^{+ 0.68 } _{- 0.68 } $& $ -2536.32 ^{+ 0.15 } _{- 0.14 } $& $ -2536.18 ^{+ 0.68 } _{- 0.68 } $& $ -2536.07 ^{+ 0.22 } _{- 0.21 } $& $ -2536.14 ^{+ 0.15 } _{- 0.69 } $ & $ -2536.20^{+ 0.19 } _{- 0.23 } $& \boldmath{$ -2536.34 ^{+ 0.19 } _{- 0.22 } $} \\
        $\gamma_*$ (m/s)& $ -40022.29 ^{+ 931.80} _{- 925.07 } $ & $ -40230.62 ^{+ 929.20 } _{- 908.51 } $& $ -40033.47 ^{+ 939.97 } _{- 934.21 } $& $ -40002.85 ^{+ 931.11 } _{- 913.51 } $& $ -40478.56 ^{+ 929.74 } _{- 851.92 } $ &$ -40447.97 ^{+ 927.88 } _{- 855.10} $ & \boldmath{$ -40430.15 ^{+ 934.66 } _{- 849.73 } $} \\
        \hline
        $a$ (au)& $ 21.46 ^{+ 3.38 } _{- 3.66 } $ & $ 12.79 ^{+ 1.35 } _{- 0.52 } $& $ 22.42 ^{+ 2.82 } _{- 2.95 } $& $ 16.78 ^{+ 1.00} _{- 0.79 } $& $ 14.87 ^{+ 2.70} _{- 1.23 } $& $ 17.78 ^{+ 5.41 } _{- 3.02 } $& \boldmath{$ 15.76 ^{+ 1.73 } _{- 1.39 } $} \\
        $P$\tablenotemark{$!$} (yrs)& $ 111.22 ^{+ 26.94 } _{- 26.83 } $ & $ 50.88 ^{+ 8.12 } _{- 2.66 } $& $ 118.08 ^{+ 22.42 } _{- 21.98 } $& $ 77.29 ^{+ 7.08 } _{- 5.50} $& $ 64.48 ^{+ 18.16 } _{- 7.87 } $ & $ 83.85 ^{+ 41.12 } _{- 20.50} $& \boldmath{$ 69.67 ^{+ 12.01 } _{- 9.04 } $} \\
        $e$ & $ 0.14 ^{+ 0.14 } _{- 0.11 } $& $ 0.45 ^{+ 0.08 } _{- 0.12 } $& $ 0.11 ^{+ 0.13 } _{- 0.09 } $& $ 0.04 ^{+ 0.06 } _{- 0.03 } $& $ 0.56 ^{+ 0.06 } _{- 0.27 } $ & $ 0.30^{+ 0.13 } _{- 0.10} $& \boldmath{$ 0.25 ^{+ 0.09 } _{- 0.09 } $} \\
        $i$ (deg)& $ 98.46 ^{+ 2.17 } _{- 1.62 } $ & $ 103.56 ^{+ 1.97 } _{- 1.86 } $
        & $ 98.23 ^{+ 1.85 } _{- 1.38 } $& $ 105.50^{+ 3.27 } _{- 3.00} $& $ 105.20^{+ 2.51 } _{- 3.30} $& $ 101.75 ^{+ 2.06 } _{- 1.83 } $& \boldmath{$ 102.18 ^{+ 1.72 } _{- 1.67 } $}\\
        $\omega$ (deg)& $1.06^{+163.01}_{-169.83}$ & $2.62_{-25.67}^{+166.54}$& $7.32_{-16.80}^ {+19.81}$& $16.42_{-39.17}^{+40.3}$ & $-51.08_{-10.73}^{+42.49}$& $-71.12_{-50.68}^{ +28.45}$& \boldmath{$-28.13_{-17.02}^{+15.32}$} \\
        $\Omega$ (deg)& $ 51.61 ^{+ 176.69 } _{- 5.67 } $ & $ 48.05 ^{+ 176.68 } _{- 5.34 } $& $ 227.62 ^{+ 2.39 } _{- 2.22 } $& $ 226.58 ^{+ 3.86 } _{- 2.67 } $&  $ 217.98 ^{+ 8.83 } _{- 1.02 } $ & $ 222.87 ^{+ 2.33 } _{- 1.93 } $& \boldmath{$ 224.46 ^{+ 2.36 } _{- 2.12 } $} \\
        $t_{\rm p}$\tablenotemark{$!$} (MJD)& $ 34483.94 ^{+ 5338.17 } _{- 7817.37 } $ & $ 47028.14 ^{+ 694.17 } _{- 2352.93 } $& $ 33095.84 ^{+ 4928.88 } _{- 6448.36 } $& $ 44695.08 ^{+ 3057.46 } _{- 3184.01 } $& $ 40785.64 ^{+ 2525.93 } _{- 3832.55 } $ & $ 36772.43 ^{+ 7009.51 } _{- 13211.06 } $& \boldmath{$ 41420.26 ^{+ 2675.44 } _{- 3216.73 } $} \\
        \hline
        $B$ (m/s) & \nodata & \nodata & $ 26.09 ^{+ 5.99 } _{- 4.34 } $& $ 25.20^{+ 5.69 } _{- 4.22 } $& \nodata & $ 25.38 ^{+ 5.72 } _{- 4.20} $& \boldmath{$ 25.83 ^{+ 5.93 } _{- 4.36 } $} \\
        $C$ & \nodata & \nodata & $ 1.37 ^{+ 3.55 } _{- 1.03 } $& $ 1.35 ^{+ 3.60} _{- 1.02 } $& \nodata & $ 1.29 ^{+ 3.53 } _{- 0.98 } $& \boldmath{$ 1.37 ^{+ 3.69 } _{- 1.02 } $}\\
        $L$ (days)& \nodata & \nodata & $ 35.17 ^{+ 12.42 } _{- 10.33 } $& $ 33.46 ^{+ 12.09 } _{- 10.34 } $& \nodata & $ 33.70^{+ 12.08 } _{- 10.24 } $& \boldmath{$ 34.52 ^{+ 12.65 } _{- 10.45 } $} \\
        $P_{\rm rot}$ (days)& \nodata & \nodata & $ 18.33 ^{+ 0.86 } _{- 1.15 } $& $ 18.30^{+ 0.89 } _{- 1.22 } $& \nodata & $ 18.33 ^{+ 0.89 } _{- 1.19 } $& \boldmath{$ 18.33 ^{+ 0.88 } _{- 1.19 } $} \\
        \hline
        $m$ (m/s/yr) & \nodata & \nodata & $ 0.01 ^{+ 0.00} _{- 0.00} $& $ 0.00^{+ 0.00} _{- 0.00} $& $ 0.00^{+ 0.01 } _{- 0.00} $ & $ 0.00^{+ 0.00} _{- 0.00} $& \boldmath{$ 0.00^{+ 0.00} _{- 0.00} $} \\
        $\gamma_{\rm LC}$ (m/s)& \nodata & \nodata & $ -21.06 ^{+ 30.41 } _{- 31.19 } $& $ 37.83 ^{+ 9.50} _{- 11.04 } $
        & $ 28.34 ^{+ 8.94 } _{- 18.38 } $ & $ 19.02 ^{+ 14.73 } _{- 15.72 } $& \boldmath{$ 27.69 ^{+ 9.57 } _{- 10.98 } $} \\
        $\gamma_{\rm VLC}$ (m/s) & \nodata & \nodata & $ -14.10^{+ 30.02 } _{- 29.09 } $& $ 25.89 ^{+ 5.63 } _{- 6.53 } $& $ 22.05 ^{+ 7.72 } _{- 11.43 } $ & $ 14.31 ^{+ 10.86 } _{- 12.99 } $& \boldmath{$ 21.16 ^{+ 6.28 } _{- 7.29 } $} \\
        $\gamma_{\rm HARPS03}$ (m/s) & \nodata & \nodata & $ -25.33 ^{+ 27.90} _{- 26.51 } $& $ -0.40^{+ 3.07 } _{- 3.32 } $& $ -1.94 ^{+ 7.20} _{- 7.87 } $ & $ -9.30^{+ 9.71 } _{- 12.18 } $& \boldmath{$ -1.07 ^{+ 5.14 } _{- 5.27 } $} \\
        $\gamma_{\rm HARPS15}$ (m/s) & \nodata & \nodata & $ 1.00^{+ 23.29 } _{- 22.03 } $& $ 7.84 ^{+ 4.90} _{- 5.45 } $& $ 2.89 ^{+ 11.01 } _{- 11.79 } $ & $ -1.46 ^{+ 9.44 } _{- 13.81 } $& \boldmath{$ 10.68 ^{+ 5.15 } _{- 5.64 } $}\\
        $\gamma_{\rm HARPS20}$ (m/s) & \nodata & \nodata & $ 19.42 ^{+ 18.53 } _{- 17.65 } $& $ 19.07 ^{+ 6.33 } _{- 7.03 } $& $ 3.46 ^{+ 22.88 } _{- 13.50} $ & $ 6.98 ^{+ 10.36 } _{- 14.47 } $& \boldmath{$ 21.70^{+ 6.19 } _{- 7.86 } $} \\
        $\gamma_{\rm ESPR18}$ (m/s) & \nodata & \nodata & $ 2.74 ^{+ 21.27 } _{- 19.66 } $& $ 5.54 ^{+ 5.98 } _{- 6.71 } $& $ -4.62 ^{+ 16.87 } _{- 12.75 } $ & $ -5.52 ^{+ 10.25 } _{- 14.32 } $& \boldmath{$ 8.32 ^{+ 5.84 } _{- 6.99 } $} \\
        $\gamma_{\rm ESPR19}$ (m/s) & \nodata & \nodata & $ 8.70^{+ 19.05 } _{- 18.18 } $& $ 8.82 ^{+ 5.97 } _{- 6.74 } $& $ -4.63 ^{+ 21.41 } _{- 13.31 } $ & $ -2.86 ^{+ 10.05 } _{- 14.52 } $& \boldmath{$ 11.67 ^{+ 5.63 } _{- 7.37 } $} \\
        $\gamma_{\rm UVES}$ (m/s) & \nodata & \nodata & $ -48.84 ^{+ 26.26 } _{- 24.88 } $& $ -26.46 ^{+ 4.25 } _{- 4.24 } $& \nodata & $ -36.87 ^{+ 10.02 } _{- 12.94 } $& \boldmath{$ -27.38 ^{+ 5.88 } _{- 5.91 } $} \\
        $\sigma_{\rm LC}$ (m/s) & \nodata & \nodata & $ 0.17 ^{+ 1.08 } _{- 0.15 } $& $ 0.17 ^{+ 1.10} _{- 0.15 } $& $ 2.28 ^{+ 3.25 } _{- 2.21 } $ & $ 0.17 ^{+ 1.03 } _{- 0.15 } $& \boldmath{$ 0.18 ^{+ 1.11 } _{- 0.15 } $} \\
        $\sigma_{\rm VLC}$ (m/s) & \nodata & \nodata & $ 0.12 ^{+ 0.55 } _{- 0.09 } $& $ 0.12 ^{+ 0.54 } _{- 0.09 } $& $ 0.12 ^{+ 0.61 } _{- 0.10} $ & $ 0.12 ^{+ 0.55 } _{- 0.10} $& \boldmath{$ 0.12 ^{+ 0.55 } _{- 0.10} $} \\
        $\sigma_{\rm HARPS03}$ (m/s) & \nodata & \nodata & $ 0.05 ^{+ 0.14 } _{- 0.04 } $& $ 0.05 ^{+ 0.14 } _{- 0.04 } $& $ 3.29 ^{+ 0.24 } _{- 0.21 } $ & $ 0.05 ^{+ 0.13 } _{- 0.04 } $& \boldmath{$ 0.05 ^{+ 0.14 } _{- 0.04 } $} \\
        $\sigma_{\rm HARPS15}$ (m/s) & \nodata & \nodata & $ 0.08 ^{+ 0.30} _{- 0.06 } $& $ 0.08 ^{+ 0.30} _{- 0.06 } $& $ 3.89 ^{+ 0.52 } _{- 0.43 } $ & $ 0.08 ^{+ 0.30} _{- 0.06 } $& \boldmath{$ 0.08 ^{+ 0.30} _{- 0.06 } $} \\
        $\sigma_{\rm HARPS20}$ (m/s) & \nodata & \nodata & $ 2.41 ^{+ 0.37 } _{- 0.32 } $& $ 2.41 ^{+ 0.38 } _{- 0.32 } $& $ 3.33 ^{+ 0.37 } _{- 0.32 } $& $ 2.41 ^{+ 0.37 } _{- 0.32 } $& \boldmath{$ 2.41 ^{+ 0.37 } _{- 0.32 } $} \\
        $\sigma_{\rm ESPR18}$ (m/s) & \nodata & \nodata & $ 0.15 ^{+ 0.04 } _{- 0.04 } $& $ 0.14 ^{+ 0.04 } _{- 0.04 } $& $ 2.16 ^{+ 0.24 } _{- 0.20} $ & $ 0.15 ^{+ 0.04 } _{- 0.04 } $& \boldmath{$ 0.15 ^{+ 0.04 } _{- 0.04 } $} \\
        $\sigma_{\rm ESPR19}$ (m/s) & \nodata & \nodata & $ 0.08 ^{+ 0.03 } _{- 0.04 } $& $ 0.08 ^{+ 0.03 } _{- 0.04 } $& $ 1.36 ^{+ 0.13 } _{- 0.11 } $ & $ 0.08 ^{+ 0.03 } _{- 0.04 } $& \boldmath{$ 0.08 ^{+ 0.03 } _{- 0.04 } $} \\
        $\sigma_{\rm UVES}$ (m/s) & \nodata & \nodata & $ 0.57 ^{+ 0.09 } _{- 0.09 } $& $ 0.57 ^{+ 0.09 } _{- 0.09 } $& \nodata & $ 0.57 ^{+ 0.09 } _{- 0.09 } $& \boldmath{$ 0.57 ^{+ 0.09 } _{- 0.09 } $} \\
        \enddata
\tablenotetext{$*$}{Relative astrometry only constrains the total system mass.}
\tablenotetext{$!$}{Derived parameter.}
\end{deluxetable*}
\end{rotatetable*}

\movetabledown=7mm
\begin{rotatetable*}
    \begin{deluxetable*}{l|l|l|l|l|l|l|l}
        \label{tab:post-HDI}
        \centerwidetable
        \setlength{\tabcolsep}{2pt}
        \tabletypesize{\scriptsize}
        \centering
        \tablecaption{Posterior 75\% Highest Density Intervals}
        \tablehead{\colhead{Parameter (Unit)} & \colhead{RA\tablenotemark{$*$}} & \colhead{RA + HGCA} & \colhead{RA + RV} & \colhead{RA + HGCA + RV} & \colhead{RA + HGCA + RV} & \colhead{RA + HGCA + RV} & \colhead{\textbf{RA + HGCA + RV}} \\ \colhead{} & \colhead{} & \colhead{} & \colhead{} & \colhead{(No NIRCam)} & \colhead{(No GP and No UVES)} & \colhead{(Unif.~Pri.)} & \colhead{\textbf{(Fiducial Model)}}} 
            \startdata
            $M_A$ ($M_{\odot}$)& $[ 0.77 , 0.82 ]$& $[ 0.77 , 0.82 ]$& $[ 0.77 , 0.83 ]$& $[ 0.76 , 0.81 ]$& $[ 0.76 , 0.81 ]$ & $[ 0.76 , 0.82 ]$& \boldmath{$[ 0.77 , 0.83 ]$}\\
            $M_b$ ($M_{\rm Jup}$)& \nodata & $[ 4.91 , 6.42 ]$& $[ 5.13 , 17.37 ]$& $[ 5.47 , 6.67 ]$ & $[ 6.74 , 10.52 ]$ & $[ 6.44 , 8.89 ]$ & \boldmath{$[ 5.73 , 7.22 ]$}\\
            $\varpi$ (mas)& $[ 274.73 , 274.95 ]$ & $[ 274.73 , 274.95 ]$ & $[ 274.74 , 274.95 ]$& $ 274.84 ^{+ 0.10} _{- 0.10} $& $[ 274.73 , 274.95 ]$ & $[ 274.74 , 274.95 ]$& \boldmath{$[ 274.74 , 274.96 ]$} \\
            $\mu_{\alpha^*}$ (mas/yr)& $[ 3965.76 , 3967.25 ]$ &$[ 3966.55 , 3966.82 ]$& $[ 3966.12 , 3967.62 ]$& $[ 274.74 , 274.96 ]$& $[ 3966.06 , 3966.52 ]$ & $[ 3966.16 , 3966.66 ]$& \boldmath{$[ 3966.24 , 3966.69 ]$} \\
            $\mu_{\delta}$ (mas/yr)& $[ -2537.06 , -2535.56 ]$& $[ -2536.48 , -2536.15 ]$& $[ -2536.70, -2535.21 ]$& $[ -2536.32 , -2535.83 ]$& $[ -2536.33 , -2535.87 ]$ & $[ -2536.43 , -2535.95 ]$& \boldmath{$[ -2536.56 , -2536.09 ]$} \\
            $\gamma_*$ (m/s)& $[ -41085.58 , -38958.21 ]$ & $[ -41323.37 , -39218.36 ]$& $[ -41104.65 , -38953.95 ]$& $[ -41029.87 , -38919.29 ]$& $[ -41592.59 , -39579.20]$ &$[ -41557.32 , -39543.90]$ & \boldmath{$[ -41557.45 , -39532.64 ]$} \\
            \hline
            $a$ (au)& $[ 17.27 , 25.40]$ & $[ 11.96 , 13.74 ]$& $[ 19.41 , 26.12 ]$& $[ 15.71 , 17.81 ]$& $[ 12.86 , 16.65 ]$ & $[ 13.16 , 21.24 ]$& \boldmath{$[ 13.71 , 17.18 ]$} \\
            $P$\tablenotemark{$!$} (yrs)& $[ 80.29 , 142.42 ]$ & $[ 46.20, 55.58 ]$& $[ 93.08 , 144.78 ]$& $[ 69.46 , 84.15 ]$& $[ 51.83 , 76.17 ]$ & $[ 52.78 , 109.20]$& \boldmath{$[ 56.52 , 79.40]$} \\
            $e$ & $[ 0.00, 0.23 ]$& $[ 0.35 , 0.57 ]$& $[ 0.00, 0.19 ]$& $[ 0.00, 0.08 ]$& $[ 0.48 , 0.66 ]$ & $[ 0.15 , 0.40]$& \boldmath{$[ 0.14 , 0.35 ]$} \\
            $i$ (deg)& $[ 96.25 , 100.54 ]$ & $[ 101.28 , 105.67 ]$ & $[ 96.21 , 99.86 ]$& $[ 101.99 , 109.18 ]$& $[ 102.10, 108.78 ]$ & $[ 99.48 , 103.94 ]$& \boldmath{$[ 100.21 , 104.10]$}\\
            $\omega$ (deg)& $[-179.99, 44.28]$& $[-16.56, 178.14]$& $[-17.7, 25.05]$& $[-28.03, 63.57]$ & $[-70.74, -39.68]$& $[-109.29, -23.93]$& \boldmath{$[-45.98, -8.56]$} \\
            $\Omega$ (deg)& $[ 44.89 , 228.20]$ & $[ 42.30, 225.17 ]$& $[ 224.61 , 229.88 ]$& $[ 222.85 , 230.05 ]$& $[ 216.01 , 219.68 ]$ & $[ 220.49 , 225.45 ]$& \boldmath{$[ 221.75 , 227.01 ]$} \\
            $t_{\rm p}$\tablenotemark{$!$} (MJD)& $[ 27509.27 , 42345.80]$ & $[ 45718.32 , 48183.52 ]$& $[ 26672.87 , 39636.71 ]$& $[ 41419.73 , 48638.75 ]$& $[ 37372.02 , 44393.46 ]$ & $[ 23437.34 , 46002.84 ]$& \boldmath{$[ 38525.03 , 45029.21 ]$} \\
            \hline
            $B$ (m/s) & \nodata & \nodata & $[ 19.87 , 31.38 ]$& $[ 19.27 , 30.32 ]$& \nodata & $[ 19.40, 30.44 ]$& \boldmath{$[ 19.50, 30.95 ]$} \\
            $C$ & \nodata & \nodata & $[ 0.00, 3.11 ]$&$[ 0.00, 3.06 ]$& \nodata & $[ 0.00, 3.04 ]$& \boldmath{$[ 0.00, 3.15 ]$}\\
            $L$ (days)& \nodata & \nodata & $[ 22.08 , 48.33 ]$& $[ 19.71 , 45.64 ]$& \nodata & $[ 20.18 , 46.02 ]$& \boldmath{$[ 20.59 , 47.25 ]$} \\
            $P_{\rm rot}$ (days)& \nodata & \nodata & $[ 17.15 , 19.57 ]$& $[ 17.03 , 19.57 ]$& \nodata & $[ 17.10, 19.61 ]$& \boldmath{$[ 17.13 , 19.62 ]$} \\
            \hline
            $m$ (m/s/yr) & \nodata & \nodata & $[ 0.00, 0.01 ]$& $[ 0.00, 0.00]$& $[ 0.00, 0.00]$ & $[ 0.00, 0.00]$& \boldmath{$[ 0.00, 0.00]$} \\
            $\gamma_{\rm LC}$ (m/s)& \nodata & \nodata & $[ -56.59 , 12.99 ]$& $[ 26.79 , 50.31 ]$ & $[ 15.91 , 43.37 ]$ & $[ 3.48 , 37.95 ]$& \boldmath{$[ 16.61 , 40.10]$} \\
            $\gamma_{\rm VLC}$ (m/s) & \nodata & \nodata & $[ -49.02 , 18.10]$& $[ 19.39 , 33.29 ]$& $[ 11.65 , 32.84 ]$ & $[ 2.27 , 28.88 ]$& \boldmath{$[ 14.02 , 29.59 ]$} \\
            $\gamma_{\rm HARPS03}$ (m/s) & \nodata & \nodata & $[ -57.14 , 4.57 ]$& $[ -4.09 , 3.30]$& $[ -10.15 , 7.22 ]$ & $[ -20.71 , 3.86 ]$& \boldmath{$[ -6.95 , 5.04 ]$} \\
            $\gamma_{\rm HARPS15}$ (m/s) & \nodata & \nodata & $[ -26.10, 25.41 ]$& $[ 1.81 , 13.71 ]$& $[ -8.44 , 16.98 ]$ & $[ -14.69 , 11.59 ]$& \boldmath{$[ 4.73 , 17.29 ]$}\\
            $\gamma_{\rm HARPS20}$ (m/s) & \nodata & \nodata & $[ -1.52 , 39.78 ]$& $[ 11.75 , 27.09 ]$& $[ -20.75 , 18.82 ]$ & $[ -6.19 , 21.95 ]$& \boldmath{$[ 14.30, 30.27 ]$} \\
            $\gamma_{\rm ESPR18}$ (m/s) & \nodata & \nodata & $[ -22.06 , 24.62 ]$& $[ -1.76 , 12.86 ]$& $[ -16.56 , 16.25 ]$ & $[ -18.95 , 8.79 ]$& \boldmath{$[ 1.26 , 16.06 ]$} \\
            $\gamma_{\rm ESPR19}$ (m/s) & \nodata & \nodata & $[ -13.49 , 29.06 ]$& $[ 1.83 , 16.38 ]$& $[ -17.26 , 20.75 ]$ & $[ -15.62 , 12.02 ]$& \boldmath{$[ 4.17 , 19.01 ]$} \\
            $\gamma_{\rm UVES}$ (m/s) & \nodata & \nodata & $[ -80.66 , -22.46 ]$& $[ -31.47 , -21.69 ]$& \nodata & $[ -49.02 , -23.02 ]$& \boldmath{$[ -33.88 , -20.31 ]$} \\
            $\sigma_{\rm LC}$ (m/s) & \nodata & \nodata & $[ 0.01 , 0.73 ]$& $[ 0.01 , 0.75 ]$& $[ 0.01 , 4.72 ]$ & $[ 0.01 , 0.71 ]$& \boldmath{$[ 0.01 , 0.75 ]$} \\
            $\sigma_{\rm VLC}$ (m/s) & \nodata & \nodata & $[ 0.01 , 0.41 ]$& $[ 0.01 , 0.40]$&$[ 0.01 , 0.45 ]$  & $[ 0.01 , 0.41 ]$& \boldmath{$[ 0.01 , 0.41 ]$} \\
            $\sigma_{\rm HARPS03}$ (m/s) & \nodata & \nodata & $[ 0.01 , 0.13 ]$& $[ 0.01 , 0.13 ]$& $[ 3.03 , 3.55 ]$ & $[ 0.01 , 0.13 ]$& \boldmath{$[ 0.01 , 0.13 ]$} \\
            $\sigma_{\rm HARPS15}$ (m/s) & \nodata & \nodata & $[ 0.01 , 0.25 ]$& $[ 0.01 , 0.25 ]$& $[ 3.33 , 4.41 ]$ & $[ 0.01 , 0.25 ]$& \boldmath{$[ 0.01 , 0.25 ]$} \\
            $\sigma_{\rm HARPS20}$ (m/s) & \nodata & \nodata & $[ 2.00, 2.79 ]$& $[ 2.01 , 2.81 ]$& $[ 2.91 , 3.70]$ & $[ 2.00, 2.79 ]$& \boldmath{$[ 1.98 , 2.77 ]$} \\
            $\sigma_{\rm ESPR18}$ (m/s) & \nodata & \nodata & $[ 0.10, 0.19 ]$& $[ 0.10, 0.19 ]$& $[ 1.89 , 2.40]$ & $[ 0.10, 0.19 ]$& \boldmath{$[ 0.10, 0.19 ]$} \\
            $\sigma_{\rm ESPR19}$ (m/s) & \nodata & \nodata & $[ 0.04 , 0.13 ]$& $[ 0.04 , 0.13 ]$& $[ 1.22 , 1.50]$ & $[ 0.04 , 0.13 ]$& \boldmath{$[ 0.04 , 0.13 ]$} \\
            $\sigma_{\rm UVES}$ (m/s) & \nodata & \nodata & $[ 0.46 , 0.67 ]$& $[ 0.46 , 0.67 ]$& \nodata & $[ 0.46 , 0.68 ]$& \boldmath{$[ 0.46 , 0.67 ]$} \\
            \enddata
    \tablenotetext{$*$}{Relative astrometry only constrains the total system mass.}
    \tablenotetext{$!$}{Derived parameter.}
    \end{deluxetable*}
\end{rotatetable*}

\bibliography{references.bib, additional-references.bib}

@article{mahalanobis1930tests,
  title={On tests and measures of group divergence},
  author={Mahalanobis, Prasanta Chandra},
  journal={J. Asiat. Soc. Bengal},
  volume={26},
  pages={541--588},
  year={1930}
}

@INPROCEEDINGS{2023ASPC..534..799C,
       author = {{Currie}, T. and {Biller}, B. and {Lagrange}, A. and {Marois}, C. and {Guyon}, O. and {Nielsen}, E.~L. and {Bonnefoy}, M. and {De Rosa}, R.~J.},
        title = "{Direct Imaging and Spectroscopy of Extrasolar Planets}",
    booktitle = {Protostars and Planets VII},
         year = 2023,
       editor = {{Inutsuka}, S. and {Aikawa}, Y. and {Muto}, T. and {Tomida}, K. and {Tamura}, M.},
       series = {Astronomical Society of the Pacific Conference Series},
       volume = {534},
        month = jul,
        pages = {799},
          doi = {10.48550/arXiv.2205.05696},
archivePrefix = {arXiv},
       eprint = {2205.05696},
 primaryClass = {astro-ph.EP},
       adsurl = {https://ui.adsabs.harvard.edu/abs/2023ASPC..534..799C}
}

@software{virga_code,
  author       = {Natasha Batalha and
                  caoimherooney11 and
                  sagnickm},
  title        = {natashabatalha/virga: Initial Release},
  month        = apr,
  year         = 2020,
  publisher    = {Zenodo},
  version      = {v0.0},
  doi          = {10.5281/zenodo.3759888},
  url          = {https://doi.org/10.5281/zenodo.3759888}
}

@ARTICLE{virga_paper1,
       author = {{Batalha}, Natasha E. and {Rooney}, Caoimhe M. and {Visscher}, Channon and {Moran}, Sarah E. and {Marley}, Mark S. and {Sengupta}, Aditya R. and {Kiefer}, Sven and {Lodge}, Matt G. and {Mang}, James and {Morley}, Caroline V. and {Mukherjee}, Sagnick and {Fortney}, Jonathan J. and {Gao}, Peter and {Lewis}, Nikole K. and {Mayorga}, L.~C. and {Pearce}, Logan A. and {Wakeford}, Hannah R.},
        title = "{Condensation Clouds in Substellar Atmospheres with Virga}",
      journal = {arXiv e-prints},
         year = 2025,
        month = aug,
          eid = {arXiv:2508.15102},
        pages = {arXiv:2508.15102},
          doi = {10.48550/arXiv.2508.15102},
archivePrefix = {arXiv},
       eprint = {2508.15102},
 primaryClass = {astro-ph.EP},
       adsurl = {https://ui.adsabs.harvard.edu/abs/2025arXiv250815102B}
}

@ARTICLE{virga_paper2,
       author = {{Moran}, Sarah E. and {Lodge}, Matt G. and {Batalha}, Natasha E. and {Ohno}, Kazumasa and {Vahidinia}, Sanaz and {Marley}, Mark S. and {Wakeford}, Hannah R. and {Leinhardt}, Zo{\"e} M.},
        title = "{Fractal Aggregate Aerosols in the Virga Cloud Code. I. Model Description and Application to a Benchmark Cloudy Exoplanet}",
      journal = {\apj},
         year = 2025,
        month = nov,
       volume = {994},
       number = {1},
          eid = {116},
        pages = {116},
          doi = {10.3847/1538-4357/ae0583},
archivePrefix = {arXiv},
       eprint = {2509.06708},
 primaryClass = {astro-ph.EP},
       adsurl = {https://ui.adsabs.harvard.edu/abs/2025ApJ...994..116M}
}

@article{marois_exoplanet_2010,
	title = {Exoplanet imaging with {LOCI} processing: photometry and astrometry with the new {SOSIE} pipeline},
	volume = {7736},
	shorttitle = {Exoplanet imaging with {LOCI} processing},
	url = {https://ui.adsabs.harvard.edu/abs/2010SPIE.7736E..1JM},
	doi = {10.1117/12.857225},
	urldate = {2025-01-23},
	journal = {SPIE},
	publisher = {SPIE},
	author = {Marois, Christian and Macintosh, Bruce and Véran, Jean-Pierre},
	month = jul,
	year = {2010},
	note = {Conference Name: Adaptive Optics Systems II
ADS Bibcode: 2010SPIE.7736E..1JM},
	pages = {77361J},
}

@book{van_leeuwen_hipparcos_2007,
	title = {Hipparcos, the {New} {Reduction} of the {Raw} {Data}},
	volume = {350},
	url = {https://ui.adsabs.harvard.edu/abs/2007ASSL..350.....V},
	doi = {10.1007/978-1-4020-6342-8},
	urldate = {2025-11-07},
	publisher = {Astrophysics and Space Science Library},
	author = {van Leeuwen, Floor},
	month = jan,
	year = {2007},
	note = {Publication Title: Astrophysics and Space Science Library
ADS Bibcode: 2007ASSL..350.....V},
}

@article{howard_giant_2025,
	title = {Giant exoplanet composition: {The} impact of the hydrogen─helium equation of state and interior structure},
	volume = {693},
	issn = {0004-6361},
	shorttitle = {Giant exoplanet composition},
	url = {https://ui.adsabs.harvard.edu/abs/2025A&A...693L...7H},
	doi = {10.1051/0004-6361/202452783},
	urldate = {2026-04-29},
	journal = {Astronomy and Astrophysics},
	publisher = {EDP},
	author = {Howard, S. and Helled, R. and Müller, S.},
	month = jan,
	year = {2025},
	note = {ADS Bibcode: 2025A\&A...693L...7H},
	pages = {L7},
}

@article{muller_synthetic_2021,
	title = {Synthetic evolution tracks of giant planets},
	volume = {507},
	issn = {0035-8711},
	url = {https://ui.adsabs.harvard.edu/abs/2021MNRAS.507.2094M},
	doi = {10.1093/mnras/stab2250},
	urldate = {2026-04-29},
	journal = {Monthly Notices of the Royal Astronomical Society},
	publisher = {OUP},
	author = {Müller, Simon and Helled, Ravit},
	month = oct,
	year = {2021},
	note = {ADS Bibcode: 2021MNRAS.507.2094M},
	pages = {2094--2102},
}

@misc{yunerman_icy_2026,
	title = {Icy {Volatile} {Enhancements} in {Evolving} {Protoplanetary} {Disks}},
	url = {https://ui.adsabs.harvard.edu/abs/2026arXiv260414124Y},
	doi = {10.48550/arXiv.2604.14124},
	urldate = {2026-04-29},
	publisher = {arXiv},
	author = {Yunerman, Elizabeth and Price, Ellen and Öberg, Karin},
	month = apr,
	year = {2026},
	note = {ADS Bibcode: 2026arXiv260414124Y},
}

@article{oberg_effects_2011,
	title = {The {Effects} of {Snowlines} on {C}/{O} in {Planetary} {Atmospheres}},
	volume = {743},
	issn = {0004-637X},
	url = {https://ui.adsabs.harvard.edu/abs/2011ApJ...743L..16O},
	doi = {10.1088/2041-8205/743/1/L16},
	urldate = {2026-04-29},
	journal = {The Astrophysical Journal},
	publisher = {IOP},
	author = {Öberg, Karin I. and Murray-Clay, Ruth and Bergin, Edwin A.},
	month = dec,
	year = {2011},
	note = {ADS Bibcode: 2011ApJ...743L..16O},
	pages = {L16},
}

@article{pollack_formation_1996,
	title = {Formation of the {Giant} {Planets} by {Concurrent} {Accretion} of {Solids} and {Gas}},
	volume = {124},
	issn = {0019-1035},
	url = {https://ui.adsabs.harvard.edu/abs/1996Icar..124...62P},
	doi = {10.1006/icar.1996.0190},
	urldate = {2026-04-29},
	journal = {Icarus},
	publisher = {Elsevier},
	author = {Pollack, James B. and Hubickyj, Olenka and Bodenheimer, Peter and Lissauer, Jack J. and Podolak, Morris and Greenzweig, Yuval},
	month = nov,
	year = {1996},
	note = {ADS Bibcode: 1996Icar..124...62P},
	pages = {62--85},
}

@article{stevenson_formation_1982,
	title = {Formation of the giant planets},
	volume = {30},
	issn = {0032-0633},
	url = {https://ui.adsabs.harvard.edu/abs/1982P&SS...30..755S},
	doi = {10.1016/0032-0633(82)90108-8},
	urldate = {2026-04-29},
	journal = {Planetary and Space Science},
	publisher = {Elsevier},
	author = {Stevenson, D. J.},
	month = aug,
	year = {1982},
	note = {ADS Bibcode: 1982P\&SS...30..755S},
	pages = {755--764},
}

@article{mizuno_formation_1980,
	title = {Formation of the {Giant} {Planets}},
	volume = {64},
	issn = {0033-06840033-068X},
	url = {https://ui.adsabs.harvard.edu/abs/1980PThPh..64..544M},
	doi = {10.1143/PTP.64.544},
	urldate = {2026-04-29},
	journal = {Progress of Theoretical Physics},
	author = {Mizuno, H.},
	month = aug,
	year = {1980},
	note = {ADS Bibcode: 1980PThPh..64..544M},
	pages = {544--557},
}

@article{burrows_spectra_2014,
	title = {Spectra as windows into exoplanet atmospheres},
	volume = {111},
	issn = {0027-8424},
	url = {https://ui.adsabs.harvard.edu/abs/2014PNAS..11112601B},
	doi = {10.1073/pnas.1304208111},
	urldate = {2026-04-29},
	journal = {Proceedings of the National Academy of Science},
	author = {Burrows, Adam S.},
	month = sep,
	year = {2014},
	note = {ADS Bibcode: 2014PNAS..11112601B},
	pages = {12601--12609},
}

@article{cushing_spitzer_2006,
	title = {A {Spitzer} {Infrared} {Spectrograph} {Spectral} {Sequence} of {M}, {L}, and {T} {Dwarfs}},
	volume = {648},
	issn = {0004-637X},
	url = {https://iopscience.iop.org/article/10.1086/505637},
	doi = {10.1086/505637},
	language = {en},
	number = {1},
	urldate = {2026-04-29},
	journal = {The Astrophysical Journal},
	publisher = {IOP Publishing},
	author = {Cushing, Michael C. and Roellig, Thomas L. and Marley, Mark S. and Saumon, D. and Leggett, S. K. and Kirkpatrick, J. Davy and Wilson, John C. and Sloan, G. C. and Mainzer, Amy K. and Cleve, Jeff E. Van and Houck, James R.},
	month = sep,
	year = {2006},
	pages = {614},
}

@inproceedings{cantalloube_exoplanet_2020,
	title = {Exoplanet imaging data challenge: benchmarking the various image processing methods for exoplanet detection},
	volume = {11448},
	shorttitle = {Exoplanet imaging data challenge},
	url = {https://www.spiedigitallibrary.org/conference-proceedings-of-spie/11448/114485A/Exoplanet-imaging-data-challenge--benchmarking-the-various-image-processing/10.1117/12.2574803.full},
	doi = {10.1117/12.2574803},
	urldate = {2026-04-28},
	booktitle = {Adaptive {Optics} {Systems} {VII}},
	publisher = {SPIE},
	author = {Cantalloube, F. and Gomez-Gonzalez, C. and Absil, O. and Cantero, C. and Bacher, R. and Bonse, M. J. and Bottom, M. and Dahlqvist, C.-H. and Desgrange, C. and Flasseur, O. and Fuhrmann, T. and Henning, Th and Jensen-Clem, R. and Kenworthy, M. and Mawet, D. and Mesa, D. and Meshkat, T. and Mouillet, D. and Müller, A. and Nasedkin, E. and Pairet, B. and Piérard, S. and Ruffio, J.-B. and Samland, M. and Stone, J. and Droogenbroeck, M. Van},
	month = dec,
	year = {2020},
	pages = {1027--1062},
}

@article{kenworthy_high-contrast_2025,
	title = {High-{Contrast} {Coronagraphy}},
	volume = {63},
	issn = {0066-4146},
	url = {https://ui.adsabs.harvard.edu/abs/2025ARA&A..63..179K},
	doi = {10.1146/annurev-astro-021225-022840},
	urldate = {2026-04-28},
	journal = {Annual Review of Astronomy and Astrophysics},
	author = {Kenworthy, Matthew A. and Haffert, Sebastiaan Y.},
	month = aug,
	year = {2025},
	note = {ADS Bibcode: 2025ARA\&A..63..179K},
	pages = {179--216},
}

@article{guyon_extreme_2018,
	title = {Extreme {Adaptive} {Optics}},
	volume = {56},
	issn = {0066-4146},
	url = {https://ui.adsabs.harvard.edu/abs/2018ARA&A..56..315G},
	doi = {10.1146/annurev-astro-081817-052000},
	urldate = {2026-04-28},
	journal = {Annual Review of Astronomy and Astrophysics},
	author = {Guyon, Olivier},
	month = sep,
	year = {2018},
	note = {ADS Bibcode: 2018ARA\&A..56..315G},
	pages = {315--355},
}

@inproceedings{girard_jwstnircam_2022,
	address = {eprint: arXiv:2208.00998},
	title = {{JWST}/{NIRCam} coronagraphy: commissioning and first on-sky results},
	volume = {12180},
	shorttitle = {{JWST}/{NIRCam} coronagraphy},
	url = {https://ui.adsabs.harvard.edu/abs/2022SPIE12180E..3QG},
	doi = {10.1117/12.2629636},
	urldate = {2026-04-28},
	booktitle = {{SPIE}},
	author = {Girard, Julien H. and Leisenring, Jarron and Kammerer, Jens and Gennaro, Mario and Rieke, Marcia and Stansberry, John and Rest, Armin and Egami, Eiichi and Sunnquist, Ben and Boyer, Martha and Canipe, Alicia and Correnti, Matteo and Hilbert, Bryan and Perrin, Marshall D. and Pueyo, Laurent and Soummer, Remi and Allen, Marsha and Bushouse, Howard and Aguilar, Jonathan and Brooks, Brian and Coe, Dan and DiFelice, Audrey and Golimowski, David and Hartig, George and Hines, Dean C. and Koekemoer, Anton and Nickson, Bryony and Nikolov, Nikolay and Kozhurina-Platais, Vera and Pirzkal, Nor and Robberto, Massimo and Sivaramakrishnan, Anand and Sohn, Sangmo Tony and Telfer, Randal and Wu, Chi Rai and Beatty, Thomas and Florian, Michael and Hainline, Kevin and Kelly, Doug and Misselt, Karl and Schlawin, Everett and Sun, Fengwu and Williams, Christina and Willmer, Christopher and Stark, Christopher and Ygouf, Marie and Carter, Aarynn and Beichman, Charles and Greene, Thomas P. and Roellig, Thomas and Krist, John and Adams Redai, Jéa. and Wang, Jason and Clark, Charles R. and Lewis, Dan and Ferry, Malcolm},
	month = aug,
	year = {2022},
	note = {ADS Bibcode: 2022SPIE12180E..3QG},
	pages = {121803Q},
}

@article{sanghi_worlds_2026-1,
	title = {Worlds {Next} {Door}. {III}. {Indirect} {Evidence} for {Enhanced} {Atmospheric} {Metallicity} and/or the {Presence} of {Water} {Clouds} in the {Nearest} {Jupiter}-analog \${\textbackslash}epsilon\$ {Eri} b},
	volume = {171},
	issn = {1538-3881},
	url = {https://doi.org/10.3847/1538-3881/ae4909},
	doi = {10.3847/1538-3881/ae4909},
	language = {en},
	number = {4},
	urldate = {2026-03-13},
	journal = {The Astronomical Journal},
	publisher = {The American Astronomical Society},
	author = {Sanghi, Aniket and Mang, James and Llop-Sayson, Jorge and Mamajek, Eric E. and Thompson, William and Sur, Ankan and Beichman, Charles and Bryden, Geoffrey and Gáspár, András and Leisenring, Jarron and Mawet, Dimitri and Morley, Caroline V. and Ruffio, Jean-Baptiste and Wolff, Schuyler G. and Ygouf, Marie},
	month = mar,
	year = {2026},
	pages = {225},
}

@article{thompson_revised_2025,
	title = {Revised {Mass} and {Orbit} of ∊ {Eridani} b: {A} 1 {MJup} {Planet} on a {Near}-circular {Orbit}},
	volume = {170},
	issn = {0004-6256},
	shorttitle = {Revised {Mass} and {Orbit} of ∊ {Eridani} b},
	url = {https://ui.adsabs.harvard.edu/abs/2025AJ....170..301T},
	doi = {10.3847/1538-3881/ae0cbd},
	urldate = {2026-03-03},
	journal = {The Astronomical Journal},
	publisher = {IOP},
	author = {Thompson, William and Nielsen, Eric L. and Ruffio, Jean-Baptiste and Blunt, Sarah and Marois, Christian},
	month = dec,
	year = {2025},
	note = {ADS Bibcode: 2025AJ....170..301T},
	pages = {301},
}

@article{choquet_first_2016,
	title = {{FIRST} {IMAGES} {OF} {DEBRIS} {DISKS} {AROUND} {TWA} 7, {TWA} 25, {HD} 35650, {AND} {HD} 377},
	volume = {817},
	issn = {2041-8205},
	url = {https://doi.org/10.3847/2041-8205/817/1/L2},
	doi = {10.3847/2041-8205/817/1/L2},
	language = {en},
	number = {1},
	urldate = {2026-03-02},
	journal = {The Astrophysical Journal Letters},
	publisher = {The American Astronomical Society},
	author = {Choquet, Elodie and Perrin, Marshall D. and Chen, Christine H. and Soummer, Remi and Pueyo, Laurent and Hagan, James B. and Gofas-Salas, Elena and Rajan, Abhijith and Golimowski, David A. and Hines, Dean C. and Schneider, Glenn and Mazoyer, Johan and Augereau, Jean-Charles and Debes, John and Stark, Christopher C. and Wolff, Schuyler and N’Diaye, Mamadou and Hsiao, Kevin},
	month = jan,
	year = {2016},
	pages = {L2},
}

@article{xie_reference-star_2022,
	title = {Reference-star differential imaging on {SPHERE}/{IRDIS}},
	volume = {666},
	issn = {0004-6361},
	url = {https://ui.adsabs.harvard.edu/abs/2022A&A...666A..32X},
	doi = {10.1051/0004-6361/202243379},
	urldate = {2026-03-02},
	journal = {Astronomy and Astrophysics},
	publisher = {EDP},
	author = {Xie, Chen and Choquet, Elodie and Vigan, Arthur and Cantalloube, Faustine and Benisty, Myriam and Boccaletti, Anthony and Bonnefoy, Mickael and Desgrange, Celia and Garufi, Antonio and Girard, Julien and Hagelberg, Janis and Janson, Markus and Kenworthy, Matthew and Lagrange, Anne-Marie and Langlois, Maud and Menard, François and Zurlo, Alice},
	month = oct,
	year = {2022},
	note = {ADS Bibcode: 2022A\&A...666A..32X},
	pages = {A32},
}

@article{xuan_characterizing_2018,
	title = {Characterizing the {Performance} of the {NIRC2} {Vortex} {Coronagraph} at {W}. {M}. {Keck} {Observatory}},
	volume = {156},
	issn = {0004-6256},
	url = {https://ui.adsabs.harvard.edu/abs/2018AJ....156..156X},
	doi = {10.3847/1538-3881/aadae6},
	urldate = {2026-03-02},
	journal = {The Astronomical Journal},
	publisher = {IOP},
	author = {Xuan, W. Jerry and Mawet, Dimitri and Ngo, Henry and Ruane, Garreth and Bailey, Vanessa P. and Choquet, Élodie and Absil, Olivier and Alvarez, Carlos and Bryan, Marta and Cook, Therese and Femenía Castellá, Bruno and Gomez Gonzalez, Carlos and Huby, Elsa and Knutson, Heather A. and Matthews, Keith and Ragland, Sam and Serabyn, Eugene and Zawol, Zoë},
	month = oct,
	year = {2018},
	note = {ADS Bibcode: 2018AJ....156..156X},
	pages = {156},
}

@misc{mang_picaso_2026,
	title = {{PICASO} 4.0: {Clouds} and {Photochemistry} in {Climate} {Models} of {Brown} {Dwarfs} and {Exoplanets}},
	shorttitle = {{PICASO} 4.0},
	url = {http://arxiv.org/abs/2602.22468},
	doi = {10.48550/arXiv.2602.22468},
	urldate = {2026-02-27},
	publisher = {arXiv},
	author = {Mang, James and Batalha, Natasha E. and Morley, Caroline V. and Wogan, Nicholas F. and Mukherjee, Sagnick and Visscher, Channon and Marley, Mark S. and Fortney, Jonathan J. and Chubb, Katy L. and Gao, Peter and Malsky, Isaac},
	month = feb,
	year = {2026},
	note = {arXiv:2602.22468 [astro-ph]},
}

@article{philipot_updated_2023,
	title = {Updated characterization of long-period single companion by combining radial velocity, relative astrometry, and absolute astrometry},
	volume = {670},
	issn = {0004-6361},
	url = {https://ui.adsabs.harvard.edu/abs/2023A&A...670A..65P},
	doi = {10.1051/0004-6361/202245396},
	urldate = {2026-02-20},
	journal = {Astronomy and Astrophysics},
	publisher = {EDP},
	author = {Philipot, F. and Lagrange, A.-M. and Rubini, P. and Kiefer, F. and Chomez, A.},
	month = feb,
	year = {2023},
	note = {ADS Bibcode: 2023A\&A...670A..65P},
	pages = {A65},
}

@article{trifonov_public_2020,
	title = {Public {HARPS} radial velocity database corrected for systematic errors},
	volume = {636},
	issn = {0004-6361},
	url = {https://ui.adsabs.harvard.edu/abs/2020A&A...636A..74T},
	doi = {10.1051/0004-6361/201936686},
	urldate = {2026-02-20},
	journal = {Astronomy and Astrophysics},
	publisher = {EDP},
	author = {Trifonov, Trifon and Tal-Or, Lev and Zechmeister, Mathias and Kaminski, Adrian and Zucker, Shay and Mazeh, Tsevi},
	month = apr,
	year = {2020},
	note = {ADS Bibcode: 2020A\&A...636A..74T},
	pages = {A74},
}

@article{ruffio_jupiter-like_2026,
	title = {Jupiter-like uniform metal enrichment in a system of multiple giant exoplanets},
	copyright = {2026 The Author(s), under exclusive licence to Springer Nature Limited},
	issn = {2397-3366},
	url = {https://www.nature.com/articles/s41550-026-02783-z},
	doi = {10.1038/s41550-026-02783-z},
	language = {en},
	urldate = {2026-02-19},
	journal = {Nature Astronomy},
	publisher = {Nature Publishing Group},
	author = {Ruffio, Jean-Baptiste and Xuan, Jerry W. and Chachan, Yayaati and Kesseli, Aurora and Lee, Eve J. and Beichman, Charles and Hodapp, Klaus and Balmer, William O. and Konopacky, Quinn and Perrin, Marshall D. and Mawet, Dimitri and Knutson, Heather A. and Bryden, Geoffrey and Greene, Thomas P. and Johnstone, Doug and Leisenring, Jarron and Meyer, Michael and Ygouf, Marie},
	month = feb,
	year = {2026},
	pages = {1--11},
}

@article{snellen_exoplanet_2025,
	title = {Exoplanet {Atmospheres} at {High} {Spectral} {Resolution}},
	volume = {63},
	issn = {0066-4146, 1545-4282},
	url = {https://www.annualreviews.org/content/journals/10.1146/annurev-astro-052622-031342},
	doi = {10.1146/annurev-astro-052622-031342},
	language = {en},
	number = {Volume 63, 2025},
	urldate = {2026-02-19},
	journal = {Annual Review of Astronomy and Astrophysics},
	publisher = {Annual Reviews},
	author = {Snellen, Ignas A. G.},
	month = aug,
	year = {2025},
	pages = {83--125},
}

@article{sanghi_efficiently_2024,
	title = {Efficiently {Searching} for {Close}-in {Companions} {Around} {Young} {M} {Dwarfs} {Using} a {Multiyear} {PSF} {Library}},
	volume = {168},
	issn = {0004-6256},
	url = {https://ui.adsabs.harvard.edu/abs/2024AJ....168..215S},
	doi = {10.3847/1538-3881/ad769f},
	urldate = {2026-02-13},
	journal = {The Astronomical Journal},
	publisher = {IOP},
	author = {Sanghi, Aniket and Xuan, Jerry W. and Wang, Jason J. and Mawet, Dimitri and Bowler, Brendan P. and Ngo, Henry and Bryan, Marta L. and Ruane, Garreth and Absil, Olivier and Huby, Elsa},
	month = nov,
	year = {2024},
	note = {ADS Bibcode: 2024AJ....168..215S},
	pages = {215},
}

@misc{xuan_compositions_2026,
	title = {The compositions of the {HR} 8799 planets reflect accretion of both solids and metal-enriched gas},
	url = {http://arxiv.org/abs/2602.09422},
	doi = {10.48550/arXiv.2602.09422},
	urldate = {2026-02-12},
	publisher = {arXiv},
	author = {Xuan, Jerry W. and Ruffio, Jean-Baptiste and Chachan, Yayaati and Ohno, Kazumasa and Kesseli, Aurora Y. and Murray-Clay, Ruth A. and Lee, Eve J. and Moses, Julianne I. and Balmer, William O. and Baburaj, Aneesh and Blake, Geoffrey A. and Johnstone, Doug and Zhang, Yapeng and Knutson, Heather A. and Mawet, Dimitri and Beichman, Charles and Hodapp, Klaus W. and Perrin, Marshall D. and Konopacky, Quinn M. and Meyer, Michael R. and Bryden, Geoffrey and Greene, Thomas P. and Leisenring, Jarron and Ygouf, Marie and Benneke, Björn and Inglis, Julie and Wallack, Nicole L.},
	month = feb,
	year = {2026},
	note = {arXiv:2602.09422 [astro-ph]},
}

@article{hara_bias_2019,
	title = {Bias and robustness of eccentricity estimates from radial velocity data},
	volume = {489},
	issn = {0035-8711},
	url = {https://ui.adsabs.harvard.edu/abs/2019MNRAS.489..738H},
	doi = {10.1093/mnras/stz1849},
	urldate = {2026-02-10},
	journal = {Monthly Notices of the Royal Astronomical Society},
	publisher = {OUP},
	author = {Hara, Nathan C. and Boué, G. and Laskar, J. and Delisle, J.-B. and Unger, N.},
	month = oct,
	year = {2019},
	note = {ADS Bibcode: 2019MNRAS.489..738H},
	pages = {738--762},
}

@article{spiegel_spectral_2012,
	title = {Spectral and {Photometric} {Diagnostics} of {Giant} {Planet} {Formation} {Scenarios}},
	volume = {745},
	issn = {0004-637X},
	url = {https://ui.adsabs.harvard.edu/abs/2012ApJ...745..174S},
	doi = {10.1088/0004-637X/745/2/174},
	urldate = {2026-01-21},
	journal = {The Astrophysical Journal},
	publisher = {IOP},
	author = {Spiegel, David S. and Burrows, Adam},
	month = feb,
	year = {2012},
	note = {ADS Bibcode: 2012ApJ...745..174S},
	pages = {174},
}

@article{kasper_near_2019,
	title = {{NEAR}: {First} {Results} from the {Search} for {Low}-{Mass} {Planets} in α {Cen}},
	volume = {178},
	issn = {0722-6691},
	shorttitle = {{NEAR}},
	url = {https://ui.adsabs.harvard.edu/abs/2019Msngr.178....5K},
	doi = {10.18727/0722-6691/5163},
	urldate = {2026-01-21},
	journal = {The Messenger},
	author = {Kasper, M. and Arsenault, R. and Käufl, U. and Jakob, G. and Leveratto, S. and Zins, G. and Pantin, E. and Duhoux, P. and Riquelme, M. and Kirchbauer, J.-P. and Kolb, J. and Pathak, P. and Siebenmorgen, R. and Soenke, C. and Fuenteseca, E. and Sterzik, M. and Ageorges, N. and Gutruf, S. and Kampf, D. and Reutlinger, A. and Absil, O. and Delacroix, C. and Maire, A.-L. and Huby, E. and Guyon, O. and Klupar, P. and Mawet, D. and Ruane, G. and Karlsson, M. and Dohlen, K. and Vigan, A. and N'Diaye, M. and Quanz, S. and Carlotti, A.},
	month = dec,
	year = {2019},
	note = {ADS Bibcode: 2019Msngr.178....5K},
	pages = {5--9},
}

@article{kasper_near_2017,
	title = {{NEAR}: {Low}-mass {Planets} in α {Cen} with {VISIR}},
	volume = {169},
	issn = {0722-6691},
	shorttitle = {{NEAR}},
	url = {https://ui.adsabs.harvard.edu/abs/2017Msngr.169...16K},
	doi = {10.18727/0722-6691/5033},
	urldate = {2026-01-21},
	journal = {The Messenger},
	author = {Kasper, M. and Arsenault, R. and Käufl, H.-U. and Jakob, G. and Fuenteseca, E. and Riquelme, M. and Siebenmorgen, R. and Sterzik, M. and Zins, G. and Ageorges, N. and Gutruf, S. and Reutlinger, A. and Kampf, D. and Absil, O. and Carlomagno, B. and Guyon, O. and Klupar, P. and Mawet, D. and Ruane, G. and Karlsson, M. and Pantin, E. and Dohlen, K.},
	month = sep,
	year = {2017},
	note = {ADS Bibcode: 2017Msngr.169...16K},
	pages = {16--20},
}

@inproceedings{gelino_variability_2000,
	title = {Variability in an {Unresolved} {Jupiter}},
	volume = {212},
	url = {https://ui.adsabs.harvard.edu/abs/2000ASPC..212..322G},
	urldate = {2026-01-21},
	booktitle = {From {Giant} {Planets} to {Cool} {Stars}},
	author = {Gelino, C. and Marley, M.},
	month = jan,
	year = {2000},
	note = {ADS Bibcode: 2000ASPC..212..322G},
	pages = {322},
}

@article{ge_rotational_2019,
	title = {Rotational {Light} {Curves} of {Jupiter} from {Ultraviolet} to {Mid}-infrared and {Implications} for {Brown} {Dwarfs} and {Exoplanets}},
	volume = {157},
	issn = {1538-3881},
	url = {https://doi.org/10.3847/1538-3881/aafba7},
	doi = {10.3847/1538-3881/aafba7},
	language = {en},
	number = {2},
	urldate = {2026-01-21},
	journal = {The Astronomical Journal},
	publisher = {The American Astronomical Society},
	author = {Ge, Huazhi and Zhang, Xi and Fletcher, Leigh N. and Orton, Glenn S. and Sinclair, James and Fernandes, Josh and Momary, Tom and Kasaba, Yasumasa and Sato, Takao M. and Fujiyoshi, Takuya},
	month = feb,
	year = {2019},
	pages = {89},
}

@article{coulter_jupiter_2022,
	title = {Jupiter and {Saturn} as {Spectral} {Analogs} for {Extrasolar} {Gas} {Giants} and {Brown} {Dwarfs}},
	volume = {263},
	issn = {0067-0049},
	url = {https://ui.adsabs.harvard.edu/abs/2022ApJS..263...15C},
	doi = {10.3847/1538-4365/ac886a},
	urldate = {2026-01-21},
	journal = {The Astrophysical Journal Supplement Series},
	publisher = {IOP},
	author = {Coulter, Daniel J. and Barnes, Jason W. and Fortney, Jonathan J.},
	month = nov,
	year = {2022},
	note = {ADS Bibcode: 2022ApJS..263...15C},
	pages = {15},
}

@misc{mamajek_nasa_2024,
	title = {{NASA} {Exoplanet} {Exploration} {Program} ({ExEP}) {Mission} {Star} {List} for the {Habitable} {Worlds} {Observatory} (2023)},
	url = {https://ui.adsabs.harvard.edu/abs/2024arXiv240212414M},
	doi = {10.48550/arXiv.2402.12414},
	urldate = {2026-01-21},
	publisher = {arXiv},
	author = {Mamajek, Eric and Stapelfeldt, Karl},
	month = feb,
	year = {2024},
	note = {ADS Bibcode: 2024arXiv240212414M},
}

@article{kong_how_2024,
	title = {How the presence of a giant planet affects the outcome of terrestrial planet formation simulations},
	volume = {687},
	copyright = {© The Authors 2024},
	issn = {0004-6361, 1432-0746},
	url = {https://www.aanda.org/articles/aa/abs/2024/07/aa49043-23/aa49043-23.html},
	doi = {10.1051/0004-6361/202349043},
	language = {en},
	urldate = {2026-01-21},
	journal = {Astronomy \& Astrophysics},
	publisher = {EDP Sciences},
	author = {Kong, Zhihui and Johansen, Anders and Lambrechts, Michiel and Jiang, Jonathan H. and Zhu, Zong-Hong},
	month = jul,
	year = {2024},
	pages = {A121},
}

@article{georgakarakos_giant_2018,
	title = {Giant {Planets}: {Good} {Neighbors} for {Habitable} {Worlds}?},
	volume = {856},
	issn = {0004-637X},
	shorttitle = {Giant {Planets}},
	url = {https://doi.org/10.3847/1538-4357/aaaf72},
	doi = {10.3847/1538-4357/aaaf72},
	language = {en},
	number = {2},
	urldate = {2026-01-21},
	journal = {The Astrophysical Journal},
	publisher = {The American Astronomical Society},
	author = {Georgakarakos, Nikolaos and Eggl, Siegfried and Dobbs-Dixon, Ian},
	month = apr,
	year = {2018},
	pages = {155},
}

@article{antoniadou_puzzling_2018,
	title = {Puzzling out the coexistence of terrestrial planets and giant exoplanets. {The} 2/1 resonant periodic orbits},
	volume = {615},
	issn = {0004-6361},
	url = {https://ui.adsabs.harvard.edu/abs/2018A&A...615A..60A},
	doi = {10.1051/0004-6361/201732058},
	urldate = {2026-01-21},
	journal = {Astronomy and Astrophysics},
	publisher = {EDP},
	author = {Antoniadou, Kyriaki I. and Libert, Anne-Sophie},
	month = jul,
	year = {2018},
	note = {ADS Bibcode: 2018A\&A...615A..60A},
	pages = {A60},
}

@article{noble_orbital_2002,
	title = {Orbital {Stability} of {Terrestrial} {Planets} inside the {Habitable} {Zones} of {Extrasolar} {Planetary} {Systems}},
	volume = {572},
	issn = {0004-637X},
	url = {https://ui.adsabs.harvard.edu/abs/2002ApJ...572.1024N},
	doi = {10.1086/340430},
	urldate = {2026-01-21},
	journal = {The Astrophysical Journal},
	publisher = {IOP},
	author = {Noble, M. and Musielak, Z. E. and Cuntz, M.},
	month = jun,
	year = {2002},
	note = {ADS Bibcode: 2002ApJ...572.1024N},
	pages = {1024--1030},
}

@article{raymond_search_2006,
	title = {The {Search} for {Other} {Earths}: {Limits} on the {Giant} {Planet} {Orbits} {That} {Allow} {Habitable} {Terrestrial} {Planets} to {Form}},
	volume = {643},
	issn = {0004-637X},
	shorttitle = {The {Search} for {Other} {Earths}},
	url = {https://iopscience.iop.org/article/10.1086/505596},
	doi = {10.1086/505596},
	language = {en},
	number = {2},
	urldate = {2026-01-21},
	journal = {The Astrophysical Journal},
	publisher = {IOP Publishing},
	author = {Raymond, Sean N.},
	month = may,
	year = {2006},
	pages = {L131},
}

@article{endl_planet_2002,
	title = {The planet search program at the {ESO} {Coudé} {Echelle} spectrometer. {III}. {The} complete {Long} {Camera} survey results},
	volume = {392},
	issn = {0004-6361},
	url = {https://ui.adsabs.harvard.edu/abs/2002A&A...392..671E},
	doi = {10.1051/0004-6361:20020937},
	urldate = {2026-01-20},
	journal = {Astronomy and Astrophysics},
	publisher = {EDP},
	author = {Endl, M. and Kürster, M. and Els, S. and Hatzes, A. P. and Cochran, W. D. and Dennerl, K. and Döbereiner, S.},
	month = sep,
	year = {2002},
	note = {ADS Bibcode: 2002A\&A...392..671E},
	pages = {671--690},
}

@article{zechmeister_planet_2013,
	title = {The planet search programme at the {ESO} {CES} and {HARPS}. {IV}. {The} search for {Jupiter} analogues around solar-like stars},
	volume = {552},
	issn = {0004-6361},
	url = {https://ui.adsabs.harvard.edu/abs/2013A&A...552A..78Z},
	doi = {10.1051/0004-6361/201116551},
	urldate = {2026-01-20},
	journal = {Astronomy and Astrophysics},
	publisher = {EDP},
	author = {Zechmeister, M. and Kürster, M. and Endl, M. and Lo Curto, G. and Hartman, H. and Nilsson, H. and Henning, T. and Hatzes, A. P. and Cochran, W. D.},
	month = apr,
	year = {2013},
	note = {ADS Bibcode: 2013A\&A...552A..78Z},
	pages = {A78},
}

@article{mccaughrean_e_2004,
	title = {ɛ {Indi} {Ba},{Bb}: {The} nearest binary brown dwarf},
	volume = {413},
	issn = {0004-6361},
	shorttitle = {ɛ {Indi} {Ba},{Bb}},
	url = {https://ui.adsabs.harvard.edu/abs/2004A&A...413.1029M},
	doi = {10.1051/0004-6361:20034292},
	urldate = {2026-01-20},
	journal = {Astronomy and Astrophysics},
	publisher = {EDP},
	author = {McCaughrean, M. J. and Close, L. M. and Scholz, R.-D. and Lenzen, R. and Biller, B. and Brandner, W. and Hartung, M. and Lodieu, N.},
	month = jan,
	year = {2004},
	note = {ADS Bibcode: 2004A\&A...413.1029M},
	pages = {1029--1036},
}

@article{scholz_varepsilon_2003,
	title = {varepsilon  {Indi} {B}: {A} new benchmark {T} dwarf},
	volume = {398},
	issn = {0004-6361},
	shorttitle = {varepsilon  {Indi} {B}},
	url = {https://ui.adsabs.harvard.edu/abs/2003A&A...398L..29S},
	doi = {10.1051/0004-6361:20021847},
	urldate = {2026-01-20},
	journal = {Astronomy and Astrophysics},
	publisher = {EDP},
	author = {Scholz, R.-D. and McCaughrean, M. J. and Lodieu, N. and Kuhlbrodt, B.},
	month = feb,
	year = {2003},
	note = {ADS Bibcode: 2003A\&A...398L..29S},
	pages = {L29--L33},
}

@article{miller_heavy-element_2011,
	title = {The {Heavy}-element {Masses} of {Extrasolar} {Giant} {Planets}, {Revealed}},
	volume = {736},
	issn = {0004-637X},
	url = {https://ui.adsabs.harvard.edu/abs/2011ApJ...736L..29M},
	doi = {10.1088/2041-8205/736/2/L29},
	urldate = {2026-01-13},
	journal = {The Astrophysical Journal},
	publisher = {IOP},
	author = {Miller, Neil and Fortney, Jonathan J.},
	month = aug,
	year = {2011},
	note = {ADS Bibcode: 2011ApJ...736L..29M},
	pages = {L29},
}

@article{guillot_correlation_2006,
	title = {A correlation between the heavy element content of transiting extrasolar planets and the metallicity of their parent stars},
	volume = {453},
	issn = {0004-6361},
	url = {https://ui.adsabs.harvard.edu/abs/2006A&A...453L..21G},
	doi = {10.1051/0004-6361:20065476},
	urldate = {2026-01-13},
	journal = {Astronomy and Astrophysics},
	publisher = {EDP},
	author = {Guillot, T. and Santos, N. C. and Pont, F. and Iro, N. and Melo, C. and Ribas, I.},
	month = jul,
	year = {2006},
	note = {ADS Bibcode: 2006A\&A...453L..21G},
	pages = {L21--L24},
}

@article{thorngren_mass-metallicity_2016,
	title = {The {Mass}-{Metallicity} {Relation} for {Giant} {Planets}},
	volume = {831},
	issn = {0004-637X},
	url = {https://ui.adsabs.harvard.edu/abs/2016ApJ...831...64T},
	doi = {10.3847/0004-637X/831/1/64},
	urldate = {2026-01-13},
	journal = {The Astrophysical Journal},
	publisher = {IOP},
	author = {Thorngren, Daniel P. and Fortney, Jonathan J. and Murray-Clay, Ruth A. and Lopez, Eric D.},
	month = nov,
	year = {2016},
	note = {ADS Bibcode: 2016ApJ...831...64T},
	pages = {64},
}

@article{ikoma_formation_2025,
	title = {Formation of {Giant} {Planets}},
	volume = {63},
	issn = {0066-4146},
	url = {https://ui.adsabs.harvard.edu/abs/2025ARA&A..63..217I},
	doi = {10.1146/annurev-astro-052722-094843},
	urldate = {2026-01-13},
	journal = {Annual Review of Astronomy and Astrophysics},
	author = {Ikoma, Masahiro and Kobayashi, Hiroshi},
	month = aug,
	year = {2025},
	note = {ADS Bibcode: 2025ARA\&A..63..217I},
	pages = {217--258},
}

@article{hasegawa_origin_2018,
	title = {The {Origin} of the {Heavy}-element {Content} {Trend} in {Giant} {Planets} via {Core} {Accretion}},
	volume = {865},
	issn = {0004-637X},
	url = {https://ui.adsabs.harvard.edu/abs/2018ApJ...865...32H},
	doi = {10.3847/1538-4357/aad912},
	urldate = {2026-01-13},
	journal = {The Astrophysical Journal},
	publisher = {IOP},
	author = {Hasegawa, Yasuhiro and Bryden, Geoffrey and Ikoma, Masahiro and Vasisht, Gautam and Swain, Mark},
	month = sep,
	year = {2018},
	note = {ADS Bibcode: 2018ApJ...865...32H},
	pages = {32},
}

@article{arregi_phase_2006,
	title = {Phase dispersion relation of the 5-micron hot spot wave from a long-term study of {Jupiter} in the visible},
	volume = {111},
	copyright = {Copyright 2006 by the American Geophysical Union.},
	issn = {2156-2202},
	url = {https://onlinelibrary.wiley.com/doi/abs/10.1029/2005JE002653},
	doi = {10.1029/2005JE002653},
	language = {en},
	number = {E9},
	urldate = {2026-01-13},
	journal = {Journal of Geophysical Research: Planets},
	author = {Arregi, J. and Rojas, J. F. and Sánchez-Lavega, A. and Morgado, A.},
	year = {2006},
	note = {\_eprint: https://agupubs.onlinelibrary.wiley.com/doi/pdf/10.1029/2005JE002653},
}

@article{roos-serote_water_2004,
	series = {Jupiter after {Galileo} and {Cassini}},
	title = {On the water abundance in the atmosphere of {Jupiter}},
	volume = {52},
	issn = {0032-0633},
	url = {https://www.sciencedirect.com/science/article/pii/S0032063303002265},
	doi = {10.1016/j.pss.2003.06.007},
	number = {5},
	urldate = {2026-01-13},
	journal = {Planetary and Space Science},
	author = {Roos-Serote, M. and Atreya, S. K. and Wong, M. K. and Drossart, P.},
	month = apr,
	year = {2004},
	pages = {397--414},
}

@article{lodders_atmospheric_2002,
	title = {Atmospheric {Chemistry} in {Giant} {Planets}, {Brown} {Dwarfs}, and {Low}-{Mass} {Dwarf} {Stars}. {I}. {Carbon}, {Nitrogen}, and {Oxygen}},
	volume = {155},
	issn = {0019-1035},
	url = {https://ui.adsabs.harvard.edu/abs/2002Icar..155..393L},
	doi = {10.1006/icar.2001.6740},
	urldate = {2026-01-12},
	journal = {Icarus},
	publisher = {Elsevier},
	author = {Lodders, Katharina and Fegley, Bruce},
	month = feb,
	year = {2002},
	note = {ADS Bibcode: 2002Icar..155..393L},
	pages = {393--424},
}

@article{zahnle_atmospheric_2009,
	title = {Atmospheric {Sulfur} {Photochemistry} on {Hot} {Jupiters}},
	volume = {701},
	issn = {0004-637X},
	url = {https://ui.adsabs.harvard.edu/abs/2009ApJ...701L..20Z},
	doi = {10.1088/0004-637X/701/1/L20},
	urldate = {2026-01-12},
	journal = {The Astrophysical Journal},
	publisher = {IOP},
	author = {Zahnle, K. and Marley, M. S. and Freedman, R. S. and Lodders, K. and Fortney, J. J.},
	month = aug,
	year = {2009},
	note = {ADS Bibcode: 2009ApJ...701L..20Z},
	pages = {L20--L24},
}

@article{miles_observations_2020,
	title = {Observations of {Disequilibrium} {CO} {Chemistry} in the {Coldest} {Brown} {Dwarfs}},
	volume = {160},
	issn = {0004-6256},
	url = {https://ui.adsabs.harvard.edu/abs/2020AJ....160...63M},
	doi = {10.3847/1538-3881/ab9114},
	urldate = {2026-01-10},
	journal = {The Astronomical Journal},
	publisher = {IOP},
	author = {Miles, Brittany E. and Skemer, Andrew J. I. and Morley, Caroline V. and Marley, Mark S. and Fortney, Jonathan J. and Allers, Katelyn N. and Faherty, Jacqueline K. and Geballe, Thomas R. and Visscher, Channon and Schneider, Adam C. and Lupu, Roxana and Freedman, Richard S. and Bjoraker, Gordon L.},
	month = aug,
	year = {2020},
	note = {ADS Bibcode: 2020AJ....160...63M},
	pages = {63},
}

@article{leggett_36-79_2007,
	title = {3.6-7.9 μm {Photometry} of {L} and {T} {Dwarfs} and the {Prevalence} of {Vertical} {Mixing} in their {Atmospheres}},
	volume = {655},
	issn = {0004-637X},
	url = {https://ui.adsabs.harvard.edu/abs/2007ApJ...655.1079L},
	doi = {10.1086/510014},
	urldate = {2026-01-09},
	journal = {The Astrophysical Journal},
	publisher = {IOP},
	author = {Leggett, S. K. and Saumon, D. and Marley, M. S. and Geballe, T. R. and Golimowski, D. A. and Stephens, D. and Fan, X.},
	month = feb,
	year = {2007},
	note = {ADS Bibcode: 2007ApJ...655.1079L},
	pages = {1079--1094},
}

@article{saumon_ammonia_2006,
	title = {Ammonia as a {Tracer} of {Chemical} {Equilibrium} in the {T7}.5 {Dwarf} {Gliese} {570D}},
	volume = {647},
	issn = {0004-637X},
	url = {https://ui.adsabs.harvard.edu/abs/2006ApJ...647..552S},
	doi = {10.1086/505419},
	urldate = {2026-01-09},
	journal = {The Astrophysical Journal},
	publisher = {IOP},
	author = {Saumon, D. and Marley, M. S. and Cushing, M. C. and Leggett, S. K. and Roellig, T. L. and Lodders, K. and Freedman, R. S.},
	month = aug,
	year = {2006},
	note = {ADS Bibcode: 2006ApJ...647..552S},
	pages = {552--557},
}

@article{noll_detection_1997,
	title = {Detection of {Abundant} {Carbon} {Monoxide} in the {Brown} {Dwarf} {Gliese} {229B}},
	volume = {489},
	issn = {0004-637X},
	url = {https://ui.adsabs.harvard.edu/abs/1997ApJ...489L..87N},
	doi = {10.1086/310954},
	urldate = {2026-01-09},
	journal = {The Astrophysical Journal},
	publisher = {IOP},
	author = {Noll, Keith S. and Geballe, T. R. and Marley, Mark S.},
	month = nov,
	year = {1997},
	note = {ADS Bibcode: 1997ApJ...489L..87N},
	pages = {L87--L90},
}

@article{fegley_equilibrium_1985,
	title = {Equilibrium and nonequilibrium chemistry of {Saturn}'s atmosphere - {Implications} for the observability of {PH3}, {N2}, {CO}, and  {GeH4}.},
	volume = {299},
	issn = {0004-637X},
	url = {https://ui.adsabs.harvard.edu/abs/1985ApJ...299.1067F},
	doi = {10.1086/163775},
	urldate = {2026-01-09},
	journal = {The Astrophysical Journal},
	publisher = {IOP},
	author = {Fegley, Jr., B. and Prinn, R. G.},
	month = dec,
	year = {1985},
	note = {ADS Bibcode: 1985ApJ...299.1067F},
	pages = {1067--1078},
}

@article{moran_fractal_2025,
	title = {Fractal {Aggregate} {Aerosols} in the {Virga} {Cloud} {Code}. {I}. {Model} {Description} and {Application} to a {Benchmark} {Cloudy} {Exoplanet}},
	volume = {994},
	issn = {0004-637X},
	url = {https://ui.adsabs.harvard.edu/abs/2025ApJ...994..116M},
	doi = {10.3847/1538-4357/ae0583},
	urldate = {2026-01-09},
	journal = {The Astrophysical Journal},
	publisher = {IOP},
	author = {Moran, Sarah E. and Lodge, Matt G. and Batalha, Natasha E. and Ohno, Kazumasa and Vahidinia, Sanaz and Marley, Mark S. and Wakeford, Hannah R. and Leinhardt, Zoë M.},
	month = nov,
	year = {2025},
	note = {ADS Bibcode: 2025ApJ...994..116M},
	pages = {116},
}

@misc{batalha_condensation_2025,
	title = {Condensation {Clouds} in {Substellar} {Atmospheres} with {Virga}},
	url = {https://ui.adsabs.harvard.edu/abs/2025arXiv250815102B},
	doi = {10.48550/arXiv.2508.15102},
	urldate = {2026-01-09},
	publisher = {arXiv},
	author = {Batalha, Natasha E. and Rooney, Caoimhe M. and Visscher, Channon and Moran, Sarah E. and Marley, Mark S. and Sengupta, Aditya R. and Kiefer, Sven and Lodge, Matt G. and Mang, James and Morley, Caroline V. and Mukherjee, Sagnick and Fortney, Jonathan J. and Gao, Peter and Lewis, Nikole K. and Mayorga, L. C. and Pearce, Logan A. and Wakeford, Hannah R.},
	month = aug,
	year = {2025},
	note = {ADS Bibcode: 2025arXiv250815102B},
}

@misc{batalha_natashabatalhavirga_2020,
	title = {natashabatalha/virga: {Initial} {Release}},
	shorttitle = {natashabatalha/virga},
	url = {https://zenodo.org/records/3759888},
	doi = {10.5281/zenodo.3759888},
	urldate = {2026-01-09},
	publisher = {Zenodo},
	author = {Batalha, Natasha and Rooney, Caoimhe and Mukherjee, Sagnick},
	month = apr,
	year = {2020},
}

@article{ackerman_precipitating_2001,
	title = {Precipitating {Condensation} {Clouds} in {Substellar} {Atmospheres}},
	volume = {556},
	issn = {0004-637X},
	url = {https://ui.adsabs.harvard.edu/abs/2001ApJ...556..872A},
	doi = {10.1086/321540},
	urldate = {2026-01-09},
	journal = {The Astrophysical Journal},
	publisher = {IOP},
	author = {Ackerman, Andrew S. and Marley, Mark S.},
	month = aug,
	year = {2001},
	note = {ADS Bibcode: 2001ApJ...556..872A},
	pages = {872--884},
}

@article{hubeny_non-lte_1995,
	title = {Non-{LTE} {Line}-blanketed {Model} {Atmospheres} of {Hot} {Stars}. {I}. {Hybrid} {Complete} {Linearization}/{Accelerated} {Lambda} {Iteration} {Method}},
	volume = {439},
	issn = {0004-637X},
	url = {https://ui.adsabs.harvard.edu/abs/1995ApJ...439..875H},
	doi = {10.1086/175226},
	urldate = {2026-01-09},
	journal = {The Astrophysical Journal},
	publisher = {IOP},
	author = {Hubeny, I. and Lanz, T.},
	month = feb,
	year = {1995},
	note = {ADS Bibcode: 1995ApJ...439..875H},
	pages = {875},
}

@article{burrows_optical_2008,
	title = {Optical {Albedo} {Theory} of {Strongly} {Irradiated} {Giant} {Planets}: {The} {Case} of {HD} 209458b},
	volume = {682},
	issn = {0004-637X},
	shorttitle = {Optical {Albedo} {Theory} of {Strongly} {Irradiated} {Giant} {Planets}},
	url = {https://ui.adsabs.harvard.edu/abs/2008ApJ...682.1277B},
	doi = {10.1086/589824},
	urldate = {2026-01-09},
	journal = {The Astrophysical Journal},
	publisher = {IOP},
	author = {Burrows, A. and Ibgui, L. and Hubeny, I.},
	month = aug,
	year = {2008},
	note = {ADS Bibcode: 2008ApJ...682.1277B},
	pages = {1277--1282},
}

@article{sudarsky_phase_2005,
	title = {Phase {Functions} and {Light} {Curves} of {Wide}-{Separation} {Extrasolar} {Giant} {Planets}},
	volume = {627},
	issn = {0004-637X},
	url = {https://ui.adsabs.harvard.edu/abs/2005ApJ...627..520S},
	doi = {10.1086/430206},
	urldate = {2026-01-09},
	journal = {The Astrophysical Journal},
	publisher = {IOP},
	author = {Sudarsky, David and Burrows, Adam and Hubeny, Ivan and Li, Aigen},
	month = jul,
	year = {2005},
	note = {ADS Bibcode: 2005ApJ...627..520S},
	pages = {520--533},
}

@article{mukherjee_picaso_2023,
	title = {{PICASO} 3.0: {A} {One}-dimensional {Climate} {Model} for {Giant} {Planets} and {Brown} {Dwarfs}},
	volume = {942},
	issn = {0004-637X},
	shorttitle = {{PICASO} 3.0},
	url = {https://ui.adsabs.harvard.edu/abs/2023ApJ...942...71M},
	doi = {10.3847/1538-4357/ac9f48},
	urldate = {2026-01-09},
	journal = {The Astrophysical Journal},
	publisher = {IOP},
	author = {Mukherjee, Sagnick and Batalha, Natasha E. and Fortney, Jonathan J. and Marley, Mark S.},
	month = jan,
	year = {2023},
	note = {ADS Bibcode: 2023ApJ...942...71M},
	pages = {71},
}

@article{leggett_james_2024,
	title = {James {Webb} {Space} {Telescope} {Spectra} of {Cold} {Brown} {Dwarfs} are {Well}-reproduced by {Phosphine}-free, {Diabatic}, {ATMO2020}++ {Models}},
	volume = {8},
	issn = {2515-5172},
	url = {https://doi.org/10.3847/2515-5172/ad1b61},
	doi = {10.3847/2515-5172/ad1b61},
	language = {en},
	number = {1},
	urldate = {2026-01-08},
	journal = {Research Notes of the AAS},
	publisher = {The American Astronomical Society},
	author = {Leggett, S. K. and Tremblin, Pascal},
	month = jan,
	year = {2024},
	pages = {13},
}

@misc{mamajek_iau_2015,
	title = {{IAU} 2015 {Resolution} {B3} on {Recommended} {Nominal} {Conversion} {Constants} for {Selected} {Solar} and {Planetary} {Properties}},
	url = {http://arxiv.org/abs/1510.07674},
	doi = {10.48550/arXiv.1510.07674},
	urldate = {2026-01-07},
	publisher = {arXiv},
	author = {Mamajek, E. E. and Prsa, A. and Torres, G. and Harmanec, P. and Asplund, M. and Bennett, P. D. and Capitaine, N. and Christensen-Dalsgaard, J. and Depagne, E. and Folkner, W. M. and Haberreiter, M. and Hekker, S. and Hilton, J. L. and Kostov, V. and Kurtz, D. W. and Laskar, J. and Mason, B. D. and Milone, E. F. and Montgomery, M. M. and Richards, M. T. and Schou, J. and Stewart, S. G.},
	month = oct,
	year = {2015},
	note = {arXiv:1510.07674 [astro-ph]},
}

@article{filippazzo_fundamental_2015,
	title = {Fundamental {Parameters} and {Spectral} {Energy} {Distributions} of {Young} and {Field} {Age} {Objects} with {Masses} {Spanning} the {Stellar} to {Planetary} {Regime}},
	volume = {810},
	issn = {0004-637X},
	url = {https://ui.adsabs.harvard.edu/abs/2015ApJ...810..158F},
	doi = {10.1088/0004-637X/810/2/158},
	urldate = {2026-01-07},
	journal = {The Astrophysical Journal},
	publisher = {IOP},
	author = {Filippazzo, Joseph C. and Rice, Emily L. and Faherty, Jacqueline and Cruz, Kelle L. and Van Gordon, Mollie M. and Looper, Dagny L.},
	month = sep,
	year = {2015},
	note = {ADS Bibcode: 2015ApJ...810..158F},
	pages = {158},
}

@article{potekhin_thermonuclear_2012,
	title = {Thermonuclear fusion in dense stars. {Electron} screening, conductive cooling, and magnetic field effects},
	volume = {538},
	issn = {0004-6361},
	url = {https://ui.adsabs.harvard.edu/abs/2012A&A...538A.115P},
	doi = {10.1051/0004-6361/201117938},
	urldate = {2026-01-06},
	journal = {Astronomy and Astrophysics},
	publisher = {EDP},
	author = {Potekhin, A. Y. and Chabrier, G.},
	month = feb,
	year = {2012},
	note = {ADS Bibcode: 2012A\&A...538A.115P},
	pages = {A115},
}

@article{chabrier_new_2021,
	title = {A {New} {Equation} of {State} for {Dense} {Hydrogen}-{Helium} {Mixtures}. {II}. {Taking} into {Account} {Hydrogen}-{Helium} {Interactions}},
	volume = {917},
	issn = {0004-637X},
	url = {https://ui.adsabs.harvard.edu/abs/2021ApJ...917....4C},
	doi = {10.3847/1538-4357/abfc48},
	urldate = {2026-01-06},
	journal = {The Astrophysical Journal},
	publisher = {IOP},
	author = {Chabrier, Gilles and Debras, Florian},
	month = aug,
	year = {2021},
	note = {ADS Bibcode: 2021ApJ...917....4C},
	pages = {4},
}

@article{chabrier_impact_2023,
	title = {Impact of a new {H}/{He} equation of state on the evolution of massive brown dwarfs. {New} determination of the hydrogen burning limit},
	volume = {671},
	issn = {0004-6361},
	url = {https://ui.adsabs.harvard.edu/abs/2023A&A...671A.119C},
	doi = {10.1051/0004-6361/202243832},
	urldate = {2026-01-06},
	journal = {Astronomy and Astrophysics},
	publisher = {EDP},
	author = {Chabrier, Gilles and Baraffe, Isabelle and Phillips, Mark and Debras, Florian},
	month = mar,
	year = {2023},
	note = {ADS Bibcode: 2023A\&A...671A.119C},
	pages = {A119},
}

@article{tennyson_exomol_2018,
	title = {The {ExoMol} {Atlas} of {Molecular} {Opacities}},
	volume = {6},
	url = {https://ui.adsabs.harvard.edu/abs/2018Atoms...6...26T},
	doi = {10.3390/atoms6020026},
	urldate = {2026-01-06},
	journal = {Atoms},
	author = {Tennyson, Jonathan and Yurchenko, Sergei N.},
	month = may,
	year = {2018},
	note = {ADS Bibcode: 2018Atoms...6...26T},
	pages = {26},
}

@article{chabrier_new_2019,
	title = {A {New} {Equation} of {State} for {Dense} {Hydrogen}-{Helium} {Mixtures}},
	volume = {872},
	issn = {0004-637X},
	url = {https://ui.adsabs.harvard.edu/abs/2019ApJ...872...51C},
	doi = {10.3847/1538-4357/aaf99f},
	urldate = {2026-01-06},
	journal = {The Astrophysical Journal},
	publisher = {IOP},
	author = {Chabrier, G. and Mazevet, S. and Soubiran, F.},
	month = feb,
	year = {2019},
	note = {ADS Bibcode: 2019ApJ...872...51C},
	pages = {51},
}

@article{baraffe_evolutionary_2003,
	title = {Evolutionary models for cool brown dwarfs and extrasolar giant planets. {The} case of {HD} 209458},
	volume = {402},
	issn = {0004-6361},
	url = {https://ui.adsabs.harvard.edu/abs/2003A&A...402..701B},
	doi = {10.1051/0004-6361:20030252},
	urldate = {2026-01-06},
	journal = {Astronomy and Astrophysics},
	publisher = {EDP},
	author = {Baraffe, I. and Chabrier, G. and Barman, T. S. and Allard, F. and Hauschildt, P. H.},
	month = may,
	year = {2003},
	note = {ADS Bibcode: 2003A\&A...402..701B},
	pages = {701--712},
}

@misc{baraffe_evolutionary_1998,
	title = {Evolutionary models for solar metallicity low-mass stars: mass-magnitude relationships and color-magnitude diagrams},
	shorttitle = {Evolutionary models for solar metallicity low-mass stars},
	url = {https://ui.adsabs.harvard.edu/abs/1998A&A...337..403B},
	doi = {10.48550/arXiv.astro-ph/9805009},
	urldate = {2026-01-06},
	publisher = {arXiv},
	author = {Baraffe, I. and Chabrier, G. and Allard, F. and Hauschildt, P. H.},
	month = sep,
	year = {1998},
	note = {ISSN: 0004-6361
Volume: 337
ADS Bibcode: 1998A\&A...337..403B},
}

@misc{chabrier_structure_1997,
	title = {Structure and evolution of low-mass stars},
	url = {https://ui.adsabs.harvard.edu/abs/1997A&A...327.1039C},
	doi = {10.48550/arXiv.astro-ph/9704118},
	urldate = {2026-01-06},
	publisher = {arXiv},
	author = {Chabrier, Gilles and Baraffe, Isabelle},
	month = nov,
	year = {1997},
	note = {ISSN: 0004-6361
Volume: 327
ADS Bibcode: 1997A\&A...327.1039C},
}

@article{hurt_uniform_2024,
	title = {Uniform {Forward}-modeling {Analysis} of {Ultracool} {Dwarfs}. {III}. {Late}-{M} and {L} {Dwarfs} in {Young} {Moving} {Groups}, the {Pleiades}, and the {Hyades}},
	volume = {961},
	issn = {0004-637X},
	url = {https://ui.adsabs.harvard.edu/abs/2024ApJ...961..121H},
	doi = {10.3847/1538-4357/ad0b12},
	urldate = {2026-01-03},
	journal = {The Astrophysical Journal},
	publisher = {IOP},
	author = {Hurt, Spencer A. and Liu, Michael C. and Zhang, Zhoujian and Phillips, Mark and Allers, Katelyn N. and Deacon, Niall R. and Aller, Kimberly M. and Best, William M. J.},
	month = jan,
	year = {2024},
	note = {ADS Bibcode: 2024ApJ...961..121H},
	pages = {121},
}

@article{saumon_evolution_2008,
	title = {The {Evolution} of {L} and {T} {Dwarfs} in {Color}-{Magnitude} {Diagrams}},
	volume = {689},
	issn = {0004-637X},
	url = {https://ui.adsabs.harvard.edu/abs/2008ApJ...689.1327S},
	doi = {10.1086/592734},
	urldate = {2026-01-03},
	journal = {The Astrophysical Journal},
	publisher = {IOP},
	author = {Saumon, D. and Marley, Mark S.},
	month = dec,
	year = {2008},
	note = {ADS Bibcode: 2008ApJ...689.1327S},
	pages = {1327--1344},
}

@article{dupuy_masses_2023,
	title = {On the masses, age, and architecture of the {VHS} {J1256}-{1257AB} b system},
	volume = {519},
	issn = {0035-8711},
	url = {https://ui.adsabs.harvard.edu/abs/2023MNRAS.519.1688D},
	doi = {10.1093/mnras/stac3557},
	urldate = {2026-01-03},
	journal = {Monthly Notices of the Royal Astronomical Society},
	publisher = {OUP},
	author = {Dupuy, Trent J. and Liu, Michael C. and Evans, Elise L. and Best, William M. J. and Pearce, Logan A. and Sanghi, Aniket and Phillips, Mark W. and Bardalez Gagliuffi, Daniella C.},
	month = feb,
	year = {2023},
	note = {ADS Bibcode: 2023MNRAS.519.1688D},
	pages = {1688--1694},
}

@article{sanghi_hawaii_2023,
	title = {The {Hawaii} {Infrared} {Parallax} {Program}. {VI}. {The} {Fundamental} {Properties} of 1000+ {Ultracool} {Dwarfs} and {Planetary}-mass {Objects} {Using} {Optical} to {Mid}-infrared {Spectral} {Energy} {Distributions} and {Comparison} to {BT}-{Settl} and {ATMO} 2020 {Model} {Atmospheres}},
	volume = {959},
	issn = {0004-637X},
	url = {https://ui.adsabs.harvard.edu/abs/2023ApJ...959...63S},
	doi = {10.3847/1538-4357/acff66},
	urldate = {2026-01-03},
	journal = {The Astrophysical Journal},
	publisher = {IOP},
	author = {Sanghi, Aniket and Liu, Michael C. and Best, William M. J. and Dupuy, Trent J. and Siverd, Robert J. and Zhang, Zhoujian and Hurt, Spencer A. and Magnier, Eugene A. and Aller, Kimberly M. and Deacon, Niall R.},
	month = dec,
	year = {2023},
	note = {ADS Bibcode: 2023ApJ...959...63S},
	pages = {63},
}

@article{dupuy_individual_2017,
	title = {Individual {Dynamical} {Masses} of {Ultracool} {Dwarfs}},
	volume = {231},
	issn = {0067-0049},
	url = {https://ui.adsabs.harvard.edu/abs/2017ApJS..231...15D},
	doi = {10.3847/1538-4365/aa5e4c},
	urldate = {2026-01-03},
	journal = {The Astrophysical Journal Supplement Series},
	publisher = {IOP},
	author = {Dupuy, Trent J. and Liu, Michael C.},
	month = aug,
	year = {2017},
	note = {ADS Bibcode: 2017ApJS..231...15D},
	pages = {15},
}

@article{batalha_exoplanet_2019,
	title = {Exoplanet {Reflected}-light {Spectroscopy} with {PICASO}},
	volume = {878},
	issn = {0004-637X},
	url = {https://doi.org/10.3847/1538-4357/ab1b51},
	doi = {10.3847/1538-4357/ab1b51},
	language = {en},
	number = {1},
	urldate = {2025-12-25},
	journal = {The Astrophysical Journal},
	publisher = {The American Astronomical Society},
	author = {Batalha, Natasha E. and Marley, Mark S. and Lewis, Nikole K. and Fortney, Jonathan J.},
	month = jun,
	year = {2019},
	pages = {70},
}

@article{sato_jupiters_1979,
	title = {Jupiter's atmospheric composition and cloud structure deduced from absorption bands in reflected sunlight.},
	volume = {36},
	issn = {0022-4928},
	url = {https://ui.adsabs.harvard.edu/abs/1979JAtS...36.1133S},
	doi = {10.1175/1520-0469(1979)036<1133:JACACS>2.0.CO;2},
	urldate = {2025-12-21},
	journal = {Journal of the Atmospheric Sciences},
	publisher = {AMS},
	author = {Sato, M. and Hansen, J. E.},
	month = jul,
	year = {1979},
	note = {ADS Bibcode: 1979JAtS...36.1133S},
	pages = {1133--1167},
}

@article{morley_spectral_2014,
	title = {Spectral {Variability} from the {Patchy} {Atmospheres} of {T} and {Y} {Dwarfs}},
	volume = {789},
	issn = {0004-637X},
	url = {https://ui.adsabs.harvard.edu/abs/2014ApJ...789L..14M},
	doi = {10.1088/2041-8205/789/1/L14},
	urldate = {2025-12-21},
	journal = {The Astrophysical Journal},
	publisher = {IOP},
	author = {Morley, Caroline V. and Marley, Mark S. and Fortney, Jonathan J. and Lupu, Roxana},
	month = jul,
	year = {2014},
	note = {ADS Bibcode: 2014ApJ...789L..14M},
	pages = {L14},
}

@misc{li_test_2025,
	title = {A {Test} of {Substellar} {Evolutionary} {Models} with {High}-{Precision} {Ages} from {Asteroseismology} and {Gyrochronology} for the {Benchmark} {System} {HR} {7672AB}},
	url = {http://arxiv.org/abs/2512.06083},
	doi = {10.48550/arXiv.2512.06083},
	urldate = {2025-12-19},
	publisher = {arXiv},
	author = {Li, Yaguang and Liu, Michael C. and Dupuy, Trent J. and Huber, Daniel and Zhang, Jingwen and Hey, Daniel and Costa, R. R. and Larsen, Jens Reersted and Ong, J. M. Joel and Basu, Sarbani and Metcalfe, Travis S. and Zhou, Yixiao and Saders, Jennifer van and Bedding, Timothy R. and Hon, Marc and Kjeldsen, Hans and Campante, Tiago L. and Monteiro, Mário J. P. F. G. and Lundkvist, Mia Sloth and Winther, Mark Lykke and Chontos, Ashley and Saunders, Nicholas and Carmichael, Theron W. and Bouchez, Antonin and Alvarez, Carlos and Walker, Sam and Sepulveda, Aldo G. and Isaacson, Howard and Howard, Andrew W. and Gibson, Steven R. and Halverson, Samuel and Rider, Kodi and Roy, Arpita and Baker, Ashley D. and Edelstein, Jerry and Smith, Chris and Fulton, Benjamin J. and Walawender, Josh},
	month = dec,
	year = {2025},
	note = {arXiv:2512.06083 [astro-ph]},
}

@misc{sur_next-generation_2025,
	title = {Next-{Generation} {Improvements} in {Giant} {Exoplanet} {Evolutionary} and {Structural} {Models}},
	url = {http://arxiv.org/abs/2510.08681},
	doi = {10.48550/arXiv.2510.08681},
	urldate = {2025-12-16},
	publisher = {arXiv},
	author = {Sur, Ankan and Arevalo, Roberto Tejada and Burrows, Adam and Chen, Yi-Xian},
	month = oct,
	year = {2025},
	note = {arXiv:2510.08681 [astro-ph]},
}

@article{skemer_first_2016,
	title = {The {First} {Spectrum} of the {Coldest} {Brown} {Dwarf}},
	volume = {826},
	issn = {0004-637X},
	url = {https://ui.adsabs.harvard.edu/abs/2016ApJ...826L..17S},
	doi = {10.3847/2041-8205/826/2/L17},
	urldate = {2025-12-12},
	journal = {The Astrophysical Journal},
	publisher = {IOP},
	author = {Skemer, Andrew J. and Morley, Caroline V. and Allers, Katelyn N. and Geballe, Thomas R. and Marley, Mark S. and Fortney, Jonathan J. and Faherty, Jacqueline K. and Bjoraker, Gordon L. and Lupu, Roxana},
	month = aug,
	year = {2016},
	note = {ADS Bibcode: 2016ApJ...826L..17S},
	pages = {L17},
}

@article{luhman_discovery_2014,
	title = {Discovery of a {\textasciitilde}250 {K} {Brown} {Dwarf} at 2 pc from the {Sun}},
	volume = {786},
	issn = {0004-637X},
	url = {https://ui.adsabs.harvard.edu/abs/2014ApJ...786L..18L},
	doi = {10.1088/2041-8205/786/2/L18},
	urldate = {2025-12-12},
	journal = {The Astrophysical Journal},
	publisher = {IOP},
	author = {Luhman, K. L.},
	month = may,
	year = {2014},
	note = {ADS Bibcode: 2014ApJ...786L..18L},
	pages = {L18},
}

@article{luhman_jwstnirspec_2023,
	title = {{JWST}/{NIRSpec} {Observations} of the {Coldest} {Known} {Brown} {Dwarf}*},
	volume = {167},
	issn = {1538-3881},
	url = {https://doi.org/10.3847/1538-3881/ad0b72},
	doi = {10.3847/1538-3881/ad0b72},
	language = {en},
	number = {1},
	urldate = {2025-12-12},
	journal = {The Astronomical Journal},
	publisher = {The American Astronomical Society},
	author = {Luhman, K. L. and Tremblin, P. and Alves de Oliveira, C. and Birkmann, S. M. and Baraffe, I. and Chabrier, G. and Manjavacas, E. and Parker, R. J. and Valenti, J.},
	month = dec,
	year = {2023},
	pages = {5},
}

@article{carlson_abundance_1992,
	title = {The {Abundance} and {Distribution} of {Water} {Vapor} in the {Jovian} {Troposphere} as {Inferred} from {Voyager} {IRIS} {Observations}},
	volume = {388},
	issn = {0004-637X},
	url = {https://ui.adsabs.harvard.edu/abs/1992ApJ...388..648C},
	doi = {10.1086/171182},
	urldate = {2025-12-12},
	journal = {The Astrophysical Journal},
	publisher = {IOP},
	author = {Carlson, Barbara E. and Lacis, Andrew A. and Rossow, William B.},
	month = apr,
	year = {1992},
	note = {ADS Bibcode: 1992ApJ...388..648C},
	pages = {648},
}

@article{banfield_jupiters_1998,
	title = {Jupiter's {Cloud} {Structure} from {Galileo} {Imaging} {Data}},
	volume = {135},
	issn = {0019-1035},
	url = {https://ui.adsabs.harvard.edu/abs/1998Icar..135..230B},
	doi = {10.1006/icar.1998.5985},
	urldate = {2025-12-12},
	journal = {Icarus},
	publisher = {Elsevier},
	author = {Banfield, D. and Gierasch, P. J. and Bell, M. and Ustinov, E. and Ingersoll, A. P. and Vasavada, A. R. and West, Robert A. and Belton, M. J. S.},
	month = sep,
	year = {1998},
	note = {ADS Bibcode: 1998Icar..135..230B},
	pages = {230--250},
}

@article{burrows_spectra_2004,
	title = {Spectra and {Diagnostics} for the {Direct} {Detection} of {Wide}-{Separation} {Extrasolar} {Giant} {Planets}},
	volume = {609},
	issn = {0004-637X},
	url = {https://ui.adsabs.harvard.edu/abs/2004ApJ...609..407B},
	doi = {10.1086/420974},
	urldate = {2025-12-12},
	journal = {The Astrophysical Journal},
	publisher = {IOP},
	author = {Burrows, Adam and Sudarsky, David and Hubeny, Ivan},
	month = jul,
	year = {2004},
	note = {ADS Bibcode: 2004ApJ...609..407B},
	pages = {407--416},
}

@article{sudarsky_theoretical_2003,
	title = {Theoretical {Spectra} and {Atmospheres} of {Extrasolar} {Giant} {Planets}},
	volume = {588},
	issn = {0004-637X},
	url = {https://ui.adsabs.harvard.edu/abs/2003ApJ...588.1121S},
	doi = {10.1086/374331},
	urldate = {2025-12-12},
	journal = {The Astrophysical Journal},
	publisher = {IOP},
	author = {Sudarsky, David and Burrows, Adam and Hubeny, Ivan},
	month = may,
	year = {2003},
	note = {ADS Bibcode: 2003ApJ...588.1121S},
	pages = {1121--1148},
}

@article{marley_reflected_1999,
	title = {Reflected {Spectra} and {Albedos} of {Extrasolar} {Giant} {Planets}. {I}. {Clear} and {Cloudy} {Atmospheres}},
	volume = {513},
	issn = {0004-637X},
	url = {https://ui.adsabs.harvard.edu/abs/1999ApJ...513..879M},
	doi = {10.1086/306881},
	urldate = {2025-12-12},
	journal = {The Astrophysical Journal},
	publisher = {IOP},
	author = {Marley, Mark S. and Gelino, Christopher and Stephens, Denise and Lunine, Jonathan I. and Freedman, Richard},
	month = mar,
	year = {1999},
	note = {ADS Bibcode: 1999ApJ...513..879M},
	pages = {879--893},
}

@article{burrows_beyond_2003,
	title = {Beyond the {T} {Dwarfs}: {Theoretical} {Spectra}, {Colors}, and {Detectability} of the {Coolest} {Brown} {Dwarfs}},
	volume = {596},
	issn = {0004-637X},
	shorttitle = {Beyond the {T} {Dwarfs}},
	url = {https://ui.adsabs.harvard.edu/abs/2003ApJ...596..587B},
	doi = {10.1086/377709},
	urldate = {2025-12-12},
	journal = {The Astrophysical Journal},
	publisher = {IOP},
	author = {Burrows, Adam and Sudarsky, David and Lunine, Jonathan I.},
	month = oct,
	year = {2003},
	note = {ADS Bibcode: 2003ApJ...596..587B},
	pages = {587--596},
}

@article{burrows_chemical_1999,
	title = {Chemical {Equilibrium} {Abundances} in {Brown} {Dwarf} and {Extrasolar} {Giant} {Planet} {Atmospheres}},
	volume = {512},
	issn = {0004-637X},
	url = {https://ui.adsabs.harvard.edu/abs/1999ApJ...512..843B},
	doi = {10.1086/306811},
	urldate = {2025-12-12},
	journal = {The Astrophysical Journal},
	publisher = {IOP},
	author = {Burrows, Adam and Sharp, C. M.},
	month = feb,
	year = {1999},
	note = {ADS Bibcode: 1999ApJ...512..843B},
	pages = {843--863},
}

@article{nagpal_impact_2023,
	title = {The {Impact} of {Bayesian} {Hyperpriors} on the {Population}-level {Eccentricity} {Distribution} of {Imaged} {Planets}},
	volume = {165},
	issn = {0004-6256},
	url = {https://ui.adsabs.harvard.edu/abs/2023AJ....165...32N},
	doi = {10.3847/1538-3881/ac9fd2},
	urldate = {2025-12-11},
	journal = {The Astronomical Journal},
	publisher = {IOP},
	author = {Nagpal, Vighnesh and Blunt, Sarah and Bowler, Brendan P. and Dupuy, Trent J. and Nielsen, Eric L. and Wang, Jason J.},
	month = feb,
	year = {2023},
	note = {ADS Bibcode: 2023AJ....165...32N},
	pages = {32},
}

@article{bowler_population-level_2020,
	title = {Population-level {Eccentricity} {Distributions} of {Imaged} {Exoplanets} and {Brown} {Dwarf} {Companions}: {Dynamical} {Evidence} for {Distinct} {Formation} {Channels}},
	volume = {159},
	issn = {0004-6256},
	shorttitle = {Population-level {Eccentricity} {Distributions} of {Imaged} {Exoplanets} and {Brown} {Dwarf} {Companions}},
	url = {https://ui.adsabs.harvard.edu/abs/2020AJ....159...63B},
	doi = {10.3847/1538-3881/ab5b11},
	urldate = {2025-12-11},
	journal = {The Astronomical Journal},
	publisher = {IOP},
	author = {Bowler, Brendan P. and Blunt, Sarah C. and Nielsen, Eric L.},
	month = feb,
	year = {2020},
	note = {ADS Bibcode: 2020AJ....159...63B},
	pages = {63},
}

@article{pathak_high-contrast_2021,
	title = {High-contrast imaging at ten microns: {A} search for exoplanets around {Eps} {Indi} {A}, {Eps} {Eri}, {Tau} {Ceti}, {Sirius} {A}, and {Sirius} {B}},
	volume = {652},
	copyright = {© ESO 2021},
	issn = {0004-6361, 1432-0746},
	shorttitle = {High-contrast imaging at ten microns},
	url = {https://www.aanda.org/articles/aa/abs/2021/08/aa40529-21/aa40529-21.html},
	doi = {10.1051/0004-6361/202140529},
	language = {en},
	urldate = {2025-11-25},
	journal = {Astronomy \& Astrophysics},
	publisher = {EDP Sciences},
	author = {Pathak, P. and Roche, D. J. M. Petit dit de la and Kasper, M. and Sterzik, M. and Absil, O. and Boehle, A. and Feng, F. and Ivanov, V. D. and Janson, M. and Jones, H. R. A. and Kaufer, A. and Käufl, H.-U. and Maire, A.-L. and Meyer, M. and Pantin, E. and Siebenmorgen, R. and Ancker, M. E. van den and Viswanath, G.},
	month = aug,
	year = {2021},
	pages = {A121},
}

@article{kuhnle_water_2025,
	title = {Water depletion and {15NH3} in the atmosphere of the coldest brown dwarf observed with {JWST}/{MIRI}},
	volume = {695},
	copyright = {© The Authors 2025},
	issn = {0004-6361, 1432-0746},
	url = {https://www.aanda.org/articles/aa/abs/2025/03/aa52547-24/aa52547-24.html},
	doi = {10.1051/0004-6361/202452547},
	language = {en},
	urldate = {2025-11-23},
	journal = {Astronomy \& Astrophysics},
	publisher = {EDP Sciences},
	author = {Kühnle, H. and Patapis, P. and Mollière, P. and Tremblin, P. and Matthews, E. and Glauser, A. M. and Whiteford, N. and Vasist, M. and Absil, O. and Barrado, D. and Min, M. and Lagage, P.-O. and Waters, L. B. F. M. and Guedel, M. and Henning, Th and Vandenbussche, B. and Baudoz, P. and Decin, L. and Pye, J. P. and Royer, P. and Dishoeck, E. F. van and Östlin, G. and Ray, T. P. and Wright, G.},
	month = mar,
	year = {2025},
	pages = {A224},
}

@article{nasedkin_four---kind_2024,
	title = {Four-of-a-kind? {Comprehensive} atmospheric characterisation of the {HR} 8799 planets with {VLTI}/{GRAVITY}},
	volume = {687},
	issn = {0004-6361},
	shorttitle = {Four-of-a-kind?},
	url = {https://ui.adsabs.harvard.edu/abs/2024A&A...687A.298N},
	doi = {10.1051/0004-6361/202449328},
	urldate = {2025-11-21},
	journal = {Astronomy and Astrophysics},
	publisher = {EDP},
	author = {Nasedkin, E. and Mollière, P. and Lacour, S. and Nowak, M. and Kreidberg, L. and Stolker, T. and Wang, J. J. and Balmer, W. O. and Kammerer, J. and Shangguan, J. and Abuter, R. and Amorim, A. and Asensio-Torres, R. and Benisty, M. and Berger, J.-P. and Beust, H. and Blunt, S. and Boccaletti, A. and Bonnefoy, M. and Bonnet, H. and Bordoni, M. S. and Bourdarot, G. and Brandner, W. and Cantalloube, F. and Caselli, P. and Charnay, B. and Chauvin, G. and Chavez, A. and Choquet, E. and Christiaens, V. and Clénet, Y. and Coudé Du Foresto, V. and Cridland, A. and Davies, R. and Dembet, R. and Dexter, J. and Drescher, A. and Duvert, G. and Eckart, A. and Eisenhauer, F. and Förster Schreiber, N. M. and Garcia, P. and Garcia Lopez, R. and Gendron, E. and Genzel, R. and Gillessen, S. and Girard, J. H. and Grant, S. and Haubois, X. and Heißel, G. and Henning, Th. and Hinkley, S. and Hippler, S. and Houllé, M. and Hubert, Z. and Jocou, L. and Keppler, M. and Kervella, P. and Kurtovic, N. T. and Lagrange, A.-M. and Lapeyrère, V. and Le Bouquin, J.-B. and Lutz, D. and Maire, A.-L. and Mang, F. and Marleau, G.-D. and Mérand, A. and Monnier, J. D. and Mordasini, C. and Ott, T. and Otten, G. P. P. L. and Paladini, C. and Paumard, T. and Perraut, K. and Perrin, G. and Pfuhl, O. and Pourré, N. and Pueyo, L. and Ribeiro, D. C. and Rickman, E. and Ruffio, J. B. and Rustamkulov, Z. and Shimizu, T. and Sing, D. and Stadler, J. and Straub, O. and Straubmeier, C. and Sturm, E. and Tacconi, L. J. and van Dishoeck, E. F. and Vigan, A. and Vincent, F. and von Fellenberg, S. D. and Widmann, F. and Winterhalder, T. O. and Woillez, J. and Yazici, Ş. and {Gravity Collaboration}},
	month = jul,
	year = {2024},
	note = {ADS Bibcode: 2024A\&A...687A.298N},
	pages = {A298},
}

@article{thorngren_connecting_2019,
	title = {Connecting {Giant} {Planet} {Atmosphere} and {Interior} {Modeling}: {Constraints} on {Atmospheric} {Metal} {Enrichment}},
	volume = {874},
	issn = {0004-637X},
	shorttitle = {Connecting {Giant} {Planet} {Atmosphere} and {Interior} {Modeling}},
	url = {https://ui.adsabs.harvard.edu/abs/2019ApJ...874L..31T},
	doi = {10.3847/2041-8213/ab1137},
	urldate = {2025-11-21},
	journal = {The Astrophysical Journal},
	publisher = {IOP},
	author = {Thorngren, Daniel and Fortney, Jonathan J.},
	month = apr,
	year = {2019},
	note = {ADS Bibcode: 2019ApJ...874L..31T},
	pages = {L31},
}

@article{chachan_revising_2025,
	title = {Revising the {Giant} {Planet} {Mass}─{Metallicity} {Relation}: {Deciphering} the {Formation} {Sequence} of {Giant} {Planets}},
	volume = {994},
	issn = {0004-637X},
	shorttitle = {Revising the {Giant} {Planet} {Mass}─{Metallicity} {Relation}},
	url = {https://ui.adsabs.harvard.edu/abs/2025ApJ...994...43C},
	doi = {10.3847/1538-4357/ae0cbf},
	urldate = {2025-11-21},
	journal = {The Astrophysical Journal},
	publisher = {IOP},
	author = {Chachan, Yayaati and Fortney, Jonathan J. and Ohno, Kazumasa and Thorngren, Daniel and Murray-Clay, Ruth},
	month = nov,
	year = {2025},
	note = {ADS Bibcode: 2025ApJ...994...43C},
	pages = {43},
}

@article{balmer_vltigravity_2025,
	title = {{VLTI}/{GRAVITY} {Observations} of {AF} {Lep} b: {Preference} for {Circular} {Orbits}, {Cloudy} {Atmospheres}, and a {Moderately} {Enhanced} {Metallicity}},
	volume = {169},
	issn = {0004-6256},
	shorttitle = {{VLTI}/{GRAVITY} {Observations} of {AF} {Lep} b},
	url = {https://ui.adsabs.harvard.edu/abs/2025AJ....169...30B},
	doi = {10.3847/1538-3881/ad9265},
	urldate = {2025-11-21},
	journal = {The Astronomical Journal},
	publisher = {IOP},
	author = {Balmer, William O. and Franson, Kyle and Chomez, Antoine and Pueyo, Laurent and Stolker, Tomas and Lacour, Sylvestre and Nowak, Mathias and Nasedkin, Evert and Bonse, Markus J. and Thorngren, Daniel and Palma-Bifani, Paulina and Mollière, Paul and Wang, Jason J. and Zhang, Zhoujian and Chavez, Amanda and Kammerer, Jens and Blunt, Sarah and Bowler, Brendan P. and Bonnefoy, Mickael and Brandner, Wolfgang and Charnay, Benjamin and Chauvin, Gael and Henning, Th. and Lagrange, A.-M. and Pourré, Nicolas and Rickman, Emily and De Rosa, Robert and Vigan, Arthur and Winterhalder, Thomas},
	month = jan,
	year = {2025},
	note = {ADS Bibcode: 2025AJ....169...30B},
	pages = {30},
}

@article{rustamkulov_early_2023,
	title = {Early {Release} {Science} of the exoplanet {WASP}-39b with {JWST} {NIRSpec} {PRISM}},
	volume = {614},
	issn = {0028-0836},
	url = {https://ui.adsabs.harvard.edu/abs/2023Natur.614..659R},
	doi = {10.1038/s41586-022-05677-y},
	urldate = {2025-11-21},
	journal = {Nature},
	author = {Rustamkulov, Z. and Sing, D. K. and Mukherjee, S. and May, E. M. and Kirk, J. and Schlawin, E. and Line, M. R. and Piaulet, C. and Carter, A. L. and Batalha, N. E. and Goyal, J. M. and López-Morales, M. and Lothringer, J. D. and MacDonald, R. J. and Moran, S. E. and Stevenson, K. B. and Wakeford, H. R. and Espinoza, N. and Bean, J. L. and Batalha, N. M. and Benneke, B. and Berta-Thompson, Z. K. and Crossfield, I. J. M. and Gao, P. and Kreidberg, L. and Powell, D. K. and Cubillos, P. E. and Gibson, N. P. and Leconte, J. and Molaverdikhani, K. and Nikolov, N. K. and Parmentier, V. and Roy, P. and Taylor, J. and Turner, J. D. and Wheatley, P. J. and Aggarwal, K. and Ahrer, E. and Alam, M. K. and Alderson, L. and Allen, N. H. and Banerjee, A. and Barat, S. and Barrado, D. and Barstow, J. K. and Bell, T. J. and Blecic, J. and Brande, J. and Casewell, S. and Changeat, Q. and Chubb, K. L. and Crouzet, N. and Daylan, T. and Decin, L. and Désert, J. and Mikal-Evans, T. and Feinstein, A. D. and Flagg, L. and Fortney, J. J. and Harrington, J. and Heng, K. and Hong, Y. and Hu, R. and Iro, N. and Kataria, T. and Kempton, E. M.-R. and Krick, J. and Lendl, M. and Lillo-Box, J. and Louca, A. and Lustig-Yaeger, J. and Mancini, L. and Mansfield, M. and Mayne, N. J. and Miguel, Y. and Morello, G. and Ohno, K. and Palle, E. and Petit dit de la Roche, D. J. M. and Rackham, B. V. and Radica, M. and Ramos-Rosado, L. and Redfield, S. and Rogers, L. K. and Shkolnik, E. L. and Southworth, J. and Teske, J. and Tremblin, P. and Tucker, G. S. and Venot, O. and Waalkes, W. C. and Welbanks, L. and Zhang, X. and Zieba, S.},
	month = feb,
	year = {2023},
	note = {ADS Bibcode: 2023Natur.614..659R},
	pages = {659--663},
}

@article{asplund_chemical_2009,
	title = {The {Chemical} {Composition} of the {Sun}},
	volume = {47},
	issn = {0066-4146},
	url = {https://ui.adsabs.harvard.edu/abs/2009ARA&A..47..481A},
	doi = {10.1146/annurev.astro.46.060407.145222},
	urldate = {2025-11-18},
	journal = {Annual Review of Astronomy and Astrophysics},
	author = {Asplund, Martin and Grevesse, Nicolas and Sauval, A. Jacques and Scott, Pat},
	month = sep,
	year = {2009},
	note = {ADS Bibcode: 2009ARA\&A..47..481A},
	pages = {481--522},
}

@article{janson_imaging_2009,
	title = {Imaging search for the unseen companion to ɛ {Ind} {A} - improving the detection limits with 4 μm observations},
	volume = {399},
	issn = {0035-8711},
	url = {https://ui.adsabs.harvard.edu/abs/2009MNRAS.399..377J},
	doi = {10.1111/j.1365-2966.2009.15285.x},
	urldate = {2025-11-18},
	journal = {Monthly Notices of the Royal Astronomical Society},
	publisher = {OUP},
	author = {Janson, M. and Apai, D. and Zechmeister, M. and Brandner, W. and Kürster, M. and Kasper, M. and Reffert, S. and Endl, M. and Lafrenière, D. and Geißler, K. and Hippler, S. and Henning, Th.},
	month = oct,
	year = {2009},
	note = {ADS Bibcode: 2009MNRAS.399..377J},
	pages = {377--384},
}

@article{viswanath_constraints_2021,
	title = {Constraints on the nearby exoplanet ɛ {Indi} {Ab} from deep near- and mid-infrared imaging limits},
	volume = {651},
	issn = {0004-6361},
	url = {https://ui.adsabs.harvard.edu/abs/2021A&A...651A..89V},
	doi = {10.1051/0004-6361/202140730},
	urldate = {2025-11-18},
	journal = {Astronomy and Astrophysics},
	publisher = {EDP},
	author = {Viswanath, Gayathri and Janson, Markus and Dahlqvist, Carl-Henrik and Petit dit de la Roche, Dominique and Samland, Matthias and Girard, Julien and Pathak, Prashant and Kasper, Markus and Feng, Fabo and Meyer, Michael and Boehle, Anna and Quanz, Sascha P. and Jones, Hugh R. A. and Absil, Olivier and Brandner, Wolfgang and Maire, Anne-Lise and Siebenmorgen, Ralf and Sterzik, Michael and Pantin, Eric},
	month = jul,
	year = {2021},
	note = {ADS Bibcode: 2021A\&A...651A..89V},
	pages = {A89},
}

@inproceedings{girard_optimizing_2024,
	title = {Optimizing {NIRCam} coronagraphy: performance and science operations},
	volume = {13092},
	shorttitle = {Optimizing {NIRCam} coronagraphy},
	url = {https://www.spiedigitallibrary.org/conference-proceedings-of-spie/13092/1309251/Optimizing-NIRCam-coronagraphy-performance-and-science-operations/10.1117/12.3020243.full},
	doi = {10.1117/12.3020243},
	urldate = {2025-11-12},
	booktitle = {Space {Telescopes} and {Instrumentation} 2024: {Optical}, {Infrared}, and {Millimeter} {Wave}},
	publisher = {SPIE},
	author = {Girard, Julien H. and Gennaro, Mario and Rest, Armin and Leisenring, Jarron and Hilbert, Bryan and Canipe, Alicia and Boyer, Martha and Golimowski, David and Sohn, Tony and Sahoo, Ananya and Sunnquist, Ben and Brooks, Brian and Koekemoer, Anton M. and Kozhurina-Platais, Vera and Pierel, Justin R. and Carter, Aarynn and Stansberry, John and Perrin, Marshall and Pueyo, Laurent and Balmer, William and Kammerer, Jens and Lawson, Kellen and Bogat, Ell and Wang, Jason and Rieke, Marcia},
	month = aug,
	year = {2024},
	pages = {1309251},
}

@article{vehtari_rank-normalization_2021,
	title = {Rank-normalization, folding, and localization: {An} improved {R}-hat for assessing convergence of {MCMC} (with {Discussion})},
	volume = {16},
	shorttitle = {Rank-normalization, folding, and localization},
	url = {https://ui.adsabs.harvard.edu/abs/2021BayAn..16..667V},
	doi = {10.1214/20-BA1221},
	urldate = {2025-11-10},
	journal = {Bayesian Analysis},
	author = {Vehtari, Aki and Gelman, Andrew and Simpson, Daniel and Carpenter, Bob and Bürkner, Paul-Christian},
	month = jun,
	year = {2021},
	note = {ADS Bibcode: 2021BayAn..16..667V},
	pages = {667--718},
}

@article{gelman_inference_1992,
	title = {Inference from {Iterative} {Simulation} {Using} {Multiple} {Sequences}},
	volume = {7},
	url = {https://ui.adsabs.harvard.edu/abs/1992StaSc...7..457G},
	doi = {10.1214/ss/1177011136},
	urldate = {2025-11-10},
	journal = {Statistical Science},
	author = {Gelman, Andrew and Rubin, Donald B.},
	month = jan,
	year = {1992},
	note = {ADS Bibcode: 1992StaSc...7..457G},
	pages = {457--472},
}

@article{do_o_orbital_2023,
	title = {The {Orbital} {Eccentricities} of {Directly} {Imaged} {Companions} {Using} {Observable}-based {Priors}: {Implications} for {Population}-level {Distributions}},
	volume = {166},
	issn = {1538-3881},
	shorttitle = {The {Orbital} {Eccentricities} of {Directly} {Imaged} {Companions} {Using} {Observable}-based {Priors}},
	url = {https://doi.org/10.3847/1538-3881/acdc9a},
	doi = {10.3847/1538-3881/acdc9a},
	language = {en},
	number = {2},
	urldate = {2025-11-08},
	journal = {The Astronomical Journal},
	publisher = {The American Astronomical Society},
	author = {Do Ó, Clarissa R. and O’Neil, Kelly K. and Konopacky, Quinn M. and Do, Tuan and Martinez, Gregory D. and Ruffio, Jean-Baptiste and Ghez, Andrea M.},
	month = jul,
	year = {2023},
	pages = {48},
}

@article{lindegren_gaia_2021,
	title = {Gaia {Early} {Data} {Release} 3. {The} astrometric solution},
	volume = {649},
	issn = {0004-6361},
	url = {https://ui.adsabs.harvard.edu/abs/2021A&A...649A...2L},
	doi = {10.1051/0004-6361/202039709},
	urldate = {2025-11-08},
	journal = {Astronomy and Astrophysics},
	publisher = {EDP},
	author = {Lindegren, L. and Klioner, S. A. and Hernández, J. and Bombrun, A. and Ramos-Lerate, M. and Steidelmüller, H. and Bastian, U. and Biermann, M. and de Torres, A. and Gerlach, E. and Geyer, R. and Hilger, T. and Hobbs, D. and Lammers, U. and McMillan, P. J. and Stephenson, C. A. and Castañeda, J. and Davidson, M. and Fabricius, C. and Gracia-Abril, G. and Portell, J. and Rowell, N. and Teyssier, D. and Torra, F. and Bartolomé, S. and Clotet, M. and Garralda, N. and González-Vidal, J. J. and Torra, J. and Abbas, U. and Altmann, M. and Anglada Varela, E. and Balaguer-Núñez, L. and Balog, Z. and Barache, C. and Becciani, U. and Bernet, M. and Bertone, S. and Bianchi, L. and Bouquillon, S. and Brown, A. G. A. and Bucciarelli, B. and Busonero, D. and Butkevich, A. G. and Buzzi, R. and Cancelliere, R. and Carlucci, T. and Charlot, P. and Cioni, M.-R. L. and Crosta, M. and Crowley, C. and del Peloso, E. F. and del Pozo, E. and Drimmel, R. and Esquej, P. and Fienga, A. and Fraile, E. and Gai, M. and Garcia-Reinaldos, M. and Guerra, R. and Hambly, N. C. and Hauser, M. and Janßen, K. and Jordan, S. and Kostrzewa-Rutkowska, Z. and Lattanzi, M. G. and Liao, S. and Licata, E. and Lister, T. A. and Löffler, W. and Marchant, J. M. and Masip, A. and Mignard, F. and Mints, A. and Molina, D. and Mora, A. and Morbidelli, R. and Murphy, C. P. and Pagani, C. and Panuzzo, P. and Peñalosa Esteller, X. and Poggio, E. and Re Fiorentin, P. and Riva, A. and Sagristà Sellés, A. and Sanchez Gimenez, V. and Sarasso, M. and Sciacca, E. and Siddiqui, H. I. and Smart, R. L. and Souami, D. and Spagna, A. and Steele, I. A. and Taris, F. and Utrilla, E. and van Reeven, W. and Vecchiato, A.},
	month = may,
	year = {2021},
	note = {ADS Bibcode: 2021A\&A...649A...2L},
	pages = {A2},
}

@article{roettenbacher_expres_2022,
	title = {{EXPRES}. {III}. {Revealing} the {Stellar} {Activity} {Radial} {Velocity} {Signature} of ϵ {Eridani} with {Photometry} and {Interferometry}},
	volume = {163},
	issn = {0004-6256},
	url = {https://ui.adsabs.harvard.edu/abs/2022AJ....163...19R},
	doi = {10.3847/1538-3881/ac3235},
	urldate = {2025-11-07},
	journal = {The Astronomical Journal},
	publisher = {IOP},
	author = {Roettenbacher, Rachael M. and Cabot, Samuel H. C. and Fischer, Debra A. and Monnier, John D. and Henry, Gregory W. and Harmon, Robert O. and Korhonen, Heidi and Brewer, John M. and Llama, Joe and Petersburg, Ryan R. and Zhao, Lily L. and Kraus, Stefan and Le Bouquin, Jean-Baptiste and Anugu, Narsireddy and Davies, Claire L. and Gardner, Tyler and Lanthermann, Cyprien and Schaefer, Gail and Setterholm, Benjamin and Clark, Catherine A. and Jorstad, Svetlana G. and Kuehn, Kyler and Levine, Stephen},
	month = jan,
	year = {2022},
	note = {ADS Bibcode: 2022AJ....163...19R},
	pages = {19},
}

@article{foreman-mackey_fast_2017,
	title = {Fast and {Scalable} {Gaussian} {Process} {Modeling} with {Applications} to {Astronomical} {Time} {Series}},
	volume = {154},
	issn = {0004-6256},
	url = {https://ui.adsabs.harvard.edu/abs/2017AJ....154..220F},
	doi = {10.3847/1538-3881/aa9332},
	urldate = {2025-11-07},
	journal = {The Astronomical Journal},
	publisher = {IOP},
	author = {Foreman-Mackey, Daniel and Agol, Eric and Ambikasaran, Sivaram and Angus, Ruth},
	month = dec,
	year = {2017},
	note = {ADS Bibcode: 2017AJ....154..220F},
	pages = {220},
}

@article{syed_non-reversible_2022,
	title = {Non-{Reversible} {Parallel} {Tempering}: {A} {Scalable} {Highly} {Parallel} {MCMC} {Scheme}},
	volume = {84},
	issn = {1369-7412},
	shorttitle = {Non-{Reversible} {Parallel} {Tempering}},
	url = {https://doi.org/10.1111/rssb.12464},
	doi = {10.1111/rssb.12464},
	number = {2},
	urldate = {2025-11-07},
	journal = {Journal of the Royal Statistical Society Series B: Statistical Methodology},
	author = {Syed, Saifuddin and Bouchard-Côté, Alexandre and Deligiannidis, George and Doucet, Arnaud},
	month = apr,
	year = {2022},
	pages = {321--350},
}

@misc{surjanovic_pigeonsjl_2023,
	title = {Pigeons.jl: {Distributed} {Sampling} {From} {Intractable} {Distributions}},
	shorttitle = {Pigeons.jl},
	url = {https://ui.adsabs.harvard.edu/abs/2023arXiv230809769S},
	doi = {10.48550/arXiv.2308.09769},
	urldate = {2025-11-07},
	publisher = {arXiv},
	author = {Surjanovic, Nikola and Biron-Lattes, Miguel and Tiede, Paul and Syed, Saifuddin and Campbell, Trevor and Bouchard-Côté, Alexandre},
	month = aug,
	year = {2023},
	note = {ADS Bibcode: 2023arXiv230809769S},
}

@misc{surjanovic_parallel_2022,
	title = {Parallel {Tempering} {With} a {Variational} {Reference}},
	url = {https://ui.adsabs.harvard.edu/abs/2022arXiv220600080S},
	doi = {10.48550/arXiv.2206.00080},
	urldate = {2025-11-07},
	publisher = {arXiv},
	author = {Surjanovic, Nikola and Syed, Saifuddin and Bouchard-Côté, Alexandre and Campbell, Trevor},
	month = may,
	year = {2022},
	note = {ADS Bibcode: 2022arXiv220600080S},
}

@article{thompson_deep_2023,
	title = {Deep {Orbital} {Search} for {Additional} {Planets} in the {HR} 8799 {System}},
	volume = {165},
	issn = {0004-6256},
	url = {https://ui.adsabs.harvard.edu/abs/2023AJ....165...29T},
	doi = {10.3847/1538-3881/aca1af},
	urldate = {2025-11-07},
	journal = {The Astronomical Journal},
	publisher = {IOP},
	author = {Thompson, William and Marois, Christian and Do Ó, Clarissa R. and Konopacky, Quinn and Ruffio, Jean-Baptiste and Wang, Jason and Skemer, Andy J. and De Rosa, Robert J. and Macintosh, Bruce},
	month = jan,
	year = {2023},
	note = {ADS Bibcode: 2023AJ....165...29T},
	pages = {29},
}

@misc{bezanson_julia_2012,
	title = {Julia: {A} {Fast} {Dynamic} {Language} for {Technical} {Computing}},
	shorttitle = {Julia},
	url = {https://ui.adsabs.harvard.edu/abs/2012arXiv1209.5145B},
	doi = {10.48550/arXiv.1209.5145},
	urldate = {2025-11-07},
	publisher = {arXiv},
	author = {Bezanson, Jeff and Karpinski, Stefan and Shah, Viral B. and Edelman, Alan},
	month = sep,
	year = {2012},
	note = {ADS Bibcode: 2012arXiv1209.5145B},
}

@article{esa_hipparcos_1997,
	title = {The {HIPPARCOS} and {TYCHO} catalogues. {Astrometric} and photometric star catalogues derived from the {ESA} {HIPPARCOS} {Space} {Astrometry} {Mission}},
	volume = {1200},
	issn = {1609-042X},
	url = {https://ui.adsabs.harvard.edu/abs/1997ESASP1200.....E},
	urldate = {2025-11-07},
	journal = {ESA Special Publication},
	author = {{ESA}},
	month = jan,
	year = {1997},
	note = {ADS Bibcode: 1997ESASP1200.....E},
}

@article{brandt_hipparcos-gaia_2021,
	title = {The {Hipparcos}-{Gaia} {Catalog} of {Accelerations}: {Gaia} {EDR3} {Edition}},
	volume = {254},
	issn = {0067-0049},
	shorttitle = {The {Hipparcos}-{Gaia} {Catalog} of {Accelerations}},
	url = {https://ui.adsabs.harvard.edu/abs/2021ApJS..254...42B},
	doi = {10.3847/1538-4365/abf93c},
	urldate = {2025-11-07},
	journal = {The Astrophysical Journal Supplement Series},
	publisher = {IOP},
	author = {Brandt, Timothy D.},
	month = jun,
	year = {2021},
	note = {ADS Bibcode: 2021ApJS..254...42B},
	pages = {42},
}

@article{wogan_sonora_2025,
	title = {The {Sonora} {Substellar} {Atmosphere} {Models}. {V}. {A} {Correction} to the {Disequilibrium} {Abundance} of {CO2} for {Sonora} {Elf} {Owl}},
	volume = {9},
	issn = {2515-5172},
	url = {https://doi.org/10.3847/2515-5172/add407},
	doi = {10.3847/2515-5172/add407},
	language = {en},
	number = {5},
	urldate = {2025-11-06},
	journal = {Research Notes of the AAS},
	publisher = {The American Astronomical Society},
	author = {Wogan, Nicholas F. and Mang, James and Batalha, Natasha E. and Zahnle, Kevin and Mukherjee, Sagnick and Visscher, Channon and Fortney, Jonathan J. and Marley, Mark S. and Morley, Caroline V.},
	month = may,
	year = {2025},
	pages = {108},
}

@article{pepe_espresso_2021,
	title = {{ESPRESSO} at {VLT} - {On}-sky performance and first results},
	volume = {645},
	copyright = {© ESO 2021},
	issn = {0004-6361, 1432-0746},
	url = {https://www.aanda.org/articles/aa/abs/2021/01/aa38306-20/aa38306-20.html},
	doi = {10.1051/0004-6361/202038306},
	language = {en},
	urldate = {2025-11-06},
	journal = {Astronomy \& Astrophysics},
	publisher = {EDP Sciences},
	author = {Pepe, F. and Cristiani, S. and Rebolo, R. and Santos, N. C. and Dekker, H. and Cabral, A. and Marcantonio, P. Di and Figueira, P. and Curto, G. Lo and Lovis, C. and Mayor, M. and Mégevand, D. and Molaro, P. and Riva, M. and Osorio, M. R. Zapatero and Amate, M. and Manescau, A. and Pasquini, L. and Zerbi, F. M. and Adibekyan, V. and Abreu, M. and Affolter, M. and Alibert, Y. and Aliverti, M. and Allart, R. and Prieto, C. Allende and Álvarez, D. and Alves, D. and Avila, G. and Baldini, V. and Bandy, T. and Barros, S. C. C. and Benz, W. and Bianco, A. and Borsa, F. and Bourrier, V. and Bouchy, F. and Broeg, C. and Calderone, G. and Cirami, R. and Coelho, J. and Conconi, P. and Coretti, I. and Cumani, C. and Cupani, G. and D’Odorico, V. and Damasso, M. and Deiries, S. and Delabre, B. and Demangeon, O. D. S. and Dumusque, X. and Ehrenreich, D. and Faria, J. P. and Fragoso, A. and Genolet, L. and Genoni, M. and Santos, R. Génova and Hernández, J. I. González and Hughes, I. and Iwert, O. and Kerber, F. and Knudstrup, J. and Landoni, M. and Lavie, B. and Lillo-Box, J. and Lizon, J.-L. and Maire, C. and Martins, C. J. a. P. and Mehner, A. and Micela, G. and Modigliani, A. and Monteiro, M. A. and Monteiro, M. J. P. F. G. and Moschetti, M. and Murphy, M. T. and Nunes, N. and Oggioni, L. and Oliveira, A. and Oshagh, M. and Pallé, E. and Pariani, G. and Poretti, E. and Rasilla, J. L. and Rebordão, J. and Redaelli, E. M. and Tschudi, S. Santana and Santin, P. and Santos, P. and Ségransan, D. and Schmidt, T. M. and Segovia, A. and Sosnowska, D. and Sozzetti, A. and Sousa, S. G. and Spanò, P. and Mascareño, A. Suárez and Tabernero, H. and Tenegi, F. and Udry, S. and Zanutta, A.},
	month = jan,
	year = {2021},
	pages = {A96},
}

@article{foreman-mackey_emcee_2013,
	title = {emcee: {The} {MCMC} {Hammer}},
	volume = {125},
	issn = {0004-6280},
	shorttitle = {emcee},
	url = {https://ui.adsabs.harvard.edu/abs/2013PASP..125..306F},
	doi = {10.1086/670067},
	urldate = {2025-11-05},
	journal = {Publications of the Astronomical Society of the Pacific},
	publisher = {IOP},
	author = {Foreman-Mackey, Daniel and Hogg, David W. and Lang, Dustin and Goodman, Jonathan},
	month = mar,
	year = {2013},
	note = {ADS Bibcode: 2013PASP..125..306F},
	pages = {306},
}

@article{pace_chromospheric_2013,
	title = {Chromospheric activity as age indicator. {An} {L}-shaped chromospheric-activity versus age diagram},
	volume = {551},
	issn = {0004-6361},
	url = {https://ui.adsabs.harvard.edu/abs/2013A&A...551L...8P},
	doi = {10.1051/0004-6361/201220364},
	urldate = {2025-11-04},
	journal = {Astronomy and Astrophysics},
	publisher = {EDP},
	author = {Pace, G.},
	month = mar,
	year = {2013},
	note = {ADS Bibcode: 2013A\&A...551L...8P},
	pages = {L8},
}

@article{creevey_gaia_2023,
	title = {Gaia {Data} {Release} 3. {Astrophysical} parameters inference system ({Apsis}). {I}. {Methods} and content overview},
	volume = {674},
	issn = {0004-6361},
	url = {https://ui.adsabs.harvard.edu/abs/2023A&A...674A..26C},
	doi = {10.1051/0004-6361/202243688},
	urldate = {2025-11-04},
	journal = {Astronomy and Astrophysics},
	publisher = {EDP},
	author = {Creevey, O. L. and Sordo, R. and Pailler, F. and Frémat, Y. and Heiter, U. and Thévenin, F. and Andrae, R. and Fouesneau, M. and Lobel, A. and Bailer-Jones, C. A. L. and Garabato, D. and Bellas-Velidis, I. and Brugaletta, E. and Lorca, A. and Ordenovic, C. and Palicio, P. A. and Sarro, L. M. and Delchambre, L. and Drimmel, R. and Rybizki, J. and Torralba Elipe, G. and Korn, A. J. and Recio-Blanco, A. and Schultheis, M. S. and De Angeli, F. and Montegriffo, P. and Abreu Aramburu, A. and Accart, S. and Álvarez, M. A. and Bakker, J. and Brouillet, N. and Burlacu, A. and Carballo, R. and Casamiquela, L. and Chiavassa, A. and Contursi, G. and Cooper, W. J. and Dafonte, C. and Dapergolas, A. and de Laverny, P. and Dharmawardena, T. E. and Edvardsson, B. and Le Fustec, Y. and García-Lario, P. and García-Torres, M. and Gomez, A. and González-Santamaría, I. and Hatzidimitriou, D. and Jean-Antoine Piccolo, A. and Kontiza, M. and Kordopatis, G. and Lanzafame, A. C. and Lebreton, Y. and Licata, E. L. and Lindstrøm, H. E. P. and Livanou, E. and Magdaleno Romeo, A. and Manteiga, M. and Marocco, F. and Marshall, D. J. and Mary, N. and Nicolas, C. and Pallas-Quintela, L. and Panem, C. and Pichon, B. and Poggio, E. and Riclet, F. and Robin, C. and Santoveña, R. and Silvelo, A. and Slezak, I. and Smart, R. L. and Soubiran, C. and Süveges, M. and Ulla, A. and Utrilla, E. and Vallenari, A. and Zhao, H. and Zorec, J. and Barrado, D. and Bijaoui, A. and Bouret, J.-C. and Blomme, R. and Brott, I. and Cassisi, S. and Kochukhov, O. and Martayan, C. and Shulyak, D. and Silvester, J.},
	month = jun,
	year = {2023},
	note = {ADS Bibcode: 2023A\&A...674A..26C},
	pages = {A26},
}

@article{torres_search_2006,
	title = {Search for associations containing young stars ({SACY}). {I}. {Sample} and searching method},
	volume = {460},
	issn = {0004-6361},
	url = {https://ui.adsabs.harvard.edu/abs/2006A&A...460..695T},
	doi = {10.1051/0004-6361:20065602},
	urldate = {2025-11-04},
	journal = {Astronomy and Astrophysics},
	publisher = {EDP},
	author = {Torres, C. A. O. and Quast, G. R. and da Silva, L. and de La Reza, R. and Melo, C. H. F. and Sterzik, M.},
	month = dec,
	year = {2006},
	note = {ADS Bibcode: 2006A\&A...460..695T},
	pages = {695--708},
}

@article{gaia_collaboration_gaia_2023,
	title = {Gaia {Data} {Release} 3. {Summary} of the content and survey properties},
	volume = {674},
	issn = {0004-6361},
	url = {https://ui.adsabs.harvard.edu/abs/2023A&A...674A...1G},
	doi = {10.1051/0004-6361/202243940},
	urldate = {2025-11-04},
	journal = {Astronomy and Astrophysics},
	publisher = {EDP},
	author = {{Gaia Collaboration} and Vallenari, A. and Brown, A. G. A. and Prusti, T. and de Bruijne, J. H. J. and Arenou, F. and Babusiaux, C. and Biermann, M. and Creevey, O. L. and Ducourant, C. and Evans, D. W. and Eyer, L. and Guerra, R. and Hutton, A. and Jordi, C. and Klioner, S. A. and Lammers, U. L. and Lindegren, L. and Luri, X. and Mignard, F. and Panem, C. and Pourbaix, D. and Randich, S. and Sartoretti, P. and Soubiran, C. and Tanga, P. and Walton, N. A. and Bailer-Jones, C. A. L. and Bastian, U. and Drimmel, R. and Jansen, F. and Katz, D. and Lattanzi, M. G. and van Leeuwen, F. and Bakker, J. and Cacciari, C. and Castañeda, J. and De Angeli, F. and Fabricius, C. and Fouesneau, M. and Frémat, Y. and Galluccio, L. and Guerrier, A. and Heiter, U. and Masana, E. and Messineo, R. and Mowlavi, N. and Nicolas, C. and Nienartowicz, K. and Pailler, F. and Panuzzo, P. and Riclet, F. and Roux, W. and Seabroke, G. M. and Sordo, R. and Thévenin, F. and Gracia-Abril, G. and Portell, J. and Teyssier, D. and Altmann, M. and Andrae, R. and Audard, M. and Bellas-Velidis, I. and Benson, K. and Berthier, J. and Blomme, R. and Burgess, P. W. and Busonero, D. and Busso, G. and Cánovas, H. and Carry, B. and Cellino, A. and Cheek, N. and Clementini, G. and Damerdji, Y. and Davidson, M. and de Teodoro, P. and Nuñez Campos, M. and Delchambre, L. and Dell'Oro, A. and Esquej, P. and Fernández-Hernández, J. and Fraile, E. and Garabato, D. and García-Lario, P. and Gosset, E. and Haigron, R. and Halbwachs, J.-L. and Hambly, N. C. and Harrison, D. L. and Hernández, J. and Hestroffer, D. and Hodgkin, S. T. and Holl, B. and Janßen, K. and Jevardat de Fombelle, G. and Jordan, S. and Krone-Martins, A. and Lanzafame, A. C. and Löffler, W. and Marchal, O. and Marrese, P. M. and Moitinho, A. and Muinonen, K. and Osborne, P. and Pancino, E. and Pauwels, T. and Recio-Blanco, A. and Reylé, C. and Riello, M. and Rimoldini, L. and Roegiers, T. and Rybizki, J. and Sarro, L. M. and Siopis, C. and Smith, M. and Sozzetti, A. and Utrilla, E. and van Leeuwen, M. and Abbas, U. and Ábrahám, P. and Abreu Aramburu, A. and Aerts, C. and Aguado, J. J. and Ajaj, M. and Aldea-Montero, F. and Altavilla, G. and Álvarez, M. A. and Alves, J. and Anders, F. and Anderson, R. I. and Anglada Varela, E. and Antoja, T. and Baines, D. and Baker, S. G. and Balaguer-Núñez, L. and Balbinot, E. and Balog, Z. and Barache, C. and Barbato, D. and Barros, M. and Barstow, M. A. and Bartolomé, S. and Bassilana, J.-L. and Bauchet, N. and Becciani, U. and Bellazzini, M. and Berihuete, A. and Bernet, M. and Bertone, S. and Bianchi, L. and Binnenfeld, A. and Blanco-Cuaresma, S. and Blazere, A. and Boch, T. and Bombrun, A. and Bossini, D. and Bouquillon, S. and Bragaglia, A. and Bramante, L. and Breedt, E. and Bressan, A. and Brouillet, N. and Brugaletta, E. and Bucciarelli, B. and Burlacu, A. and Butkevich, A. G. and Buzzi, R. and Caffau, E. and Cancelliere, R. and Cantat-Gaudin, T. and Carballo, R. and Carlucci, T. and Carnerero, M. I. and Carrasco, J. M. and Casamiquela, L. and Castellani, M. and Castro-Ginard, A. and Chaoul, L. and Charlot, P. and Chemin, L. and Chiaramida, V. and Chiavassa, A. and Chornay, N. and Comoretto, G. and Contursi, G. and Cooper, W. J. and Cornez, T. and Cowell, S. and Crifo, F. and Cropper, M. and Crosta, M. and Crowley, C. and Dafonte, C. and Dapergolas, A. and David, M. and David, P. and de Laverny, P. and De Luise, F. and De March, R.},
	month = jun,
	year = {2023},
	note = {ADS Bibcode: 2023A\&A...674A...1G},
	pages = {A1},
}

@article{rains_precision_2020,
	title = {Precision angular diameters for 16 southern stars with {VLTI}/{PIONIER}},
	volume = {493},
	issn = {0035-8711},
	url = {https://doi.org/10.1093/mnras/staa282},
	doi = {10.1093/mnras/staa282},
	number = {2},
	urldate = {2025-11-02},
	journal = {Monthly Notices of the Royal Astronomical Society},
	author = {Rains, Adam D and Ireland, Michael J and White, Timothy R and Casagrande, Luca and Karovicova, I},
	month = apr,
	year = {2020},
	pages = {2377--2394},
}

@article{chen_precise_2022,
	title = {Precise {Dynamical} {Masses} of ε {Indi} {Ba} and {Bb}: {Evidence} of {Slowed} {Cooling} at the {L}/{T} {Transition}},
	volume = {163},
	issn = {1538-3881},
	shorttitle = {Precise {Dynamical} {Masses} of ε {Indi} {Ba} and {Bb}},
	url = {https://doi.org/10.3847/1538-3881/ac66d2},
	doi = {10.3847/1538-3881/ac66d2},
	language = {en},
	number = {6},
	urldate = {2025-11-02},
	journal = {The Astronomical Journal},
	publisher = {The American Astronomical Society},
	author = {Chen, Minghan and Li, Yiting and Brandt, Timothy D. and Dupuy, Trent J. and Cardoso, Cátia V. and McCaughrean, Mark J.},
	month = may,
	year = {2022},
	pages = {288},
}

@article{feng_detection_2019,
	title = {Detection of the nearest {Jupiter} analogue in radial velocity and astrometry data},
	volume = {490},
	issn = {0035-8711},
	url = {https://doi.org/10.1093/mnras/stz2912},
	doi = {10.1093/mnras/stz2912},
	number = {4},
	urldate = {2025-11-02},
	journal = {Monthly Notices of the Royal Astronomical Society},
	author = {Feng, Fabo and Anglada-Escudé, Guillem and Tuomi, Mikko and Jones, Hugh R A and Chanamé, Julio and Butler, Paul R and Janson, Markus},
	month = dec,
	year = {2019},
	pages = {5002--5016},
}

@article{lundkvist_low-amplitude_2024,
	title = {Low-amplitude {Solar}-like {Oscillations} in the {K5} {V} {Star} ϵ {Indi} {A}},
	volume = {964},
	issn = {0004-637X},
	url = {https://doi.org/10.3847/1538-4357/ad25f2},
	doi = {10.3847/1538-4357/ad25f2},
	language = {en},
	number = {2},
	urldate = {2025-11-02},
	journal = {The Astrophysical Journal},
	publisher = {The American Astronomical Society},
	author = {Lundkvist, Mia S. and Kjeldsen, Hans and Bedding, Timothy R. and McCaughrean, Mark J. and Butler, R. Paul and Slumstrup, Ditte and Campante, Tiago L. and Aerts, Conny and Arentoft, Torben and Bruntt, Hans and Cardoso, Cátia V. and Carrier, Fabien and Close, Laird M. and da Silva, João Gomes and Kallinger, Thomas and King, Robert R. and Li, Yaguang and Murphy, Simon J. and Rørsted, Jakob L. and Stello, Dennis},
	month = mar,
	year = {2024},
	pages = {110},
}

@article{feng_revised_2023,
	title = {Revised orbits of the two nearest {Jupiters}},
	volume = {525},
	issn = {0035-8711},
	url = {https://doi.org/10.1093/mnras/stad2297},
	doi = {10.1093/mnras/stad2297},
	number = {1},
	urldate = {2025-10-27},
	journal = {Monthly Notices of the Royal Astronomical Society},
	author = {Feng, Fabo and Butler, R Paul and Vogt, Steven S and Holden, Bradford and Rui, Yicheng},
	month = oct,
	year = {2023},
	pages = {607--619},
}

@article{feng_lessons_2025,
	title = {Lessons learned from the detection of wide companions by radial velocity and astrometry},
	volume = {539},
	issn = {0035-8711},
	url = {https://doi.org/10.1093/mnras/staf689},
	doi = {10.1093/mnras/staf689},
	number = {4},
	urldate = {2025-10-27},
	journal = {Monthly Notices of the Royal Astronomical Society},
	author = {Feng, Fabo and Xiao, Guang-Yao and Jones, Hugh R A and Jenkins, James S and Pena, Pablo and Sun, Qinghui},
	month = jun,
	year = {2025},
	pages = {3180--3200},
}

@article{oneil_improving_2019,
	title = {Improving {Orbit} {Estimates} for {Incomplete} {Orbits} with a {New} {Approach} to {Priors}: with {Applications} from {Black} {Holes} to {Planets}},
	volume = {158},
	issn = {0004-6256},
	shorttitle = {Improving {Orbit} {Estimates} for {Incomplete} {Orbits} with a {New} {Approach} to {Priors}},
	url = {https://ui.adsabs.harvard.edu/abs/2019AJ....158....4O},
	doi = {10.3847/1538-3881/ab1d66},
	urldate = {2025-10-24},
	journal = {The Astronomical Journal},
	publisher = {IOP},
	author = {O'Neil, K. Kosmo and Martinez, G. D. and Hees, A. and Ghez, A. M. and Do, T. and Witzel, G. and Konopacky, Q. and Becklin, E. E. and Chu, D. S. and Lu, J. R. and Matthews, K. and Sakai, S.},
	month = jul,
	year = {2019},
	note = {ADS Bibcode: 2019AJ....158....4O},
	pages = {4},
}

@misc{madurowicz_direct_2025,
	title = {Direct {Spectroscopy} of 51 {Eridani} b with {JWST} {NIRSpec}},
	url = {http://arxiv.org/abs/2510.08327},
	doi = {10.3847/1538-3881/ae1028},
	urldate = {2025-10-10},
	author = {Madurowicz, Alexander and Ruffio, Jean-Baptiste and Macintosh, Bruce and Perrin, Marshall and Konopacky, Quinn M. and Baburaj, Aneesh and Hoch, Kielan},
	month = oct,
	year = {2025},
	note = {arXiv:2510.08327 [astro-ph]},
}

@article{franson_jwstnircam_2024,
	title = {{JWST}/{NIRCam} 4–5 μm {Imaging} of the {Giant} {Planet} {AF} {Lep} b},
	volume = {974},
	issn = {2041-8205},
	url = {https://dx.doi.org/10.3847/2041-8213/ad736a},
	doi = {10.3847/2041-8213/ad736a},
	language = {en},
	number = {1},
	urldate = {2025-09-01},
	journal = {The Astrophysical Journal Letters},
	publisher = {The American Astronomical Society},
	author = {Franson, Kyle and Balmer, William O. and Bowler, Brendan P. and Pueyo, Laurent and Zhou, Yifan and Rickman, Emily and Zhang, Zhoujian and Mukherjee, Sagnick and Pearce, Tim D. and Bardalez Gagliuffi, Daniella C. and Biddle, Lauren I. and Brandt, Timothy D. and Bowens-Rubin, Rachel and Crepp, Justin R. and Davidson, James W. and Faherty, Jacqueline and Ginski, Christian and Horch, Elliott P. and Morgan, Marvin and Morley, Caroline V. and Perrin, Marshall D. and Sanghi, Aniket and Salama, Maïssa and Theissen, Christopher A. and Tran, Quang H. and Wolf, Trevor N.},
	month = oct,
	year = {2024},
	pages = {L11},
}

@article{sanghi_worlds_2025,
	title = {Worlds {Next} {Door}: {A} {Candidate} {Giant} {Planet} {Imaged} in the {Habitable} {Zone} of α {Centauri} {A}. {II}. {Binary} {Star} {Modeling}, {Planet} and {Exozodi} {Search}, and {Sensitivity} {Analysis}},
	volume = {989},
	issn = {2041-8205},
	shorttitle = {Worlds {Next} {Door}},
	url = {https://dx.doi.org/10.3847/2041-8213/adf53e},
	doi = {10.3847/2041-8213/adf53e},
	language = {en},
	number = {2},
	urldate = {2025-08-18},
	journal = {The Astrophysical Journal Letters},
	publisher = {The American Astronomical Society},
	author = {Sanghi, Aniket and Beichman, Charles and Mawet, Dimitri and Balmer, William O. and Godoy, Nicolas and Pueyo, Laurent and Boccaletti, Anthony and Sommer, Max and Bidot, Alexis and Choquet, Elodie and Kervella, Pierre and Lagage, Pierre-Olivier and Leisenring, Jarron and Llop-Sayson, Jorge and Ressler, Michael and Wagner, Kevin and Wyatt, Mark},
	month = aug,
	year = {2025},
	pages = {L23},
}

@article{beichman_worlds_2025,
	title = {Worlds {Next} {Door}: {A} {Candidate} {Giant} {Planet} {Imaged} in the {Habitable} {Zone} of α {Centauri} {A}. {I}. {Observations}, {Orbital} and {Physical} {Properties}, and {Exozodi} {Upper} {Limits}},
	volume = {989},
	issn = {2041-8205},
	shorttitle = {Worlds {Next} {Door}},
	url = {https://dx.doi.org/10.3847/2041-8213/adf53f},
	doi = {10.3847/2041-8213/adf53f},
	language = {en},
	number = {2},
	urldate = {2025-08-14},
	journal = {The Astrophysical Journal Letters},
	publisher = {The American Astronomical Society},
	author = {Beichman, Charles and Sanghi, Aniket and Mawet, Dimitri and Kervella, Pierre and Wagner, Kevin and Quarles, Billy and Lissauer, Jack J. and Sommer, Max and Wyatt, Mark and Godoy, Nicolas and Balmer, William O. and Pueyo, Laurent and Llop-Sayson, Jorge and Aguilar, Jonathan and Akeson, Rachel and Belikov, Ruslan and Boccaletti, Anthony and Choquet, Elodie and Fomalont, Edward and Henning, Thomas and Hines, Dean and Hu, Renyu and Lagage, Pierre-Olivier and Leisenring, Jarron and Mang, James and Ressler, Michael and Serabyn, Eugene and Tremblin, Pascal and Ygouf, Marie and Zilinskas, Mantas},
	month = aug,
	year = {2025},
	pages = {L22},
}

@article{wang_orbit_2016,
	title = {{THE} {ORBIT} {AND} {TRANSIT} {PROSPECTS} {FOR} β {PICTORIS} b {CONSTRAINED} {WITH} {ONE} {MILLIARCSECOND} {ASTROMETRY}},
	volume = {152},
	issn = {1538-3881},
	url = {https://dx.doi.org/10.3847/0004-6256/152/4/97},
	doi = {10.3847/0004-6256/152/4/97},
	language = {en},
	number = {4},
	urldate = {2025-07-04},
	journal = {The Astronomical Journal},
	publisher = {The American Astronomical Society},
	author = {Wang, Jason J. and Graham, James R. and Pueyo, Laurent and Kalas, Paul and Millar-Blanchaer, Maxwell A. and Ruffio, Jean-Baptiste and Rosa, Robert J. De and Ammons, S. Mark and Arriaga, Pauline and Bailey, Vanessa P. and Barman, Travis S. and Bulger, Joanna and Burrows, Adam S. and Cardwell, Andrew and Chen, Christine H. and Chilcote, Jeffrey K. and Cotten, Tara and Fitzgerald, Michael P. and Follette, Katherine B. and Doyon, René and Duchêne, Gaspard and Greenbaum, Alexandra Z. and Hibon, Pascale and Hung, Li-Wei and Ingraham, Patrick and Konopacky, Quinn M. and Larkin, James E. and Macintosh, Bruce and Maire, Jérôme and Marchis, Franck and Marley, Mark S. and Marois, Christian and Metchev, Stanimir and Nielsen, Eric L. and Oppenheimer, Rebecca and Palmer, David W. and Patel, Rahul and Patience, Jenny and Perrin, Marshall D. and Poyneer, Lisa A. and Rajan, Abhijith and Rameau, Julien and Rantakyrö, Fredrik T. and Savransky, Dmitry and Sivaramakrishnan, Anand and Song, Inseok and Soummer, Remi and Thomas, Sandrine and Vasisht, Gautam and Vega, David and Wallace, J. Kent and Ward-Duong, Kimberly and Wiktorowicz, Sloane J. and Wolff, Schuyler G.},
	month = oct,
	year = {2016},
	pages = {97},
}

@article{limbach_miri_2024,
	title = {The {MIRI} {Exoplanets} {Orbiting} {White} dwarfs ({MEOW}) {Survey}: {Mid}-infrared {Excess} {Reveals} a {Giant} {Planet} {Candidate} around a {Nearby} {White} {Dwarf}},
	volume = {973},
	issn = {2041-8205},
	shorttitle = {The {MIRI} {Exoplanets} {Orbiting} {White} dwarfs ({MEOW}) {Survey}},
	url = {https://dx.doi.org/10.3847/2041-8213/ad74ed},
	doi = {10.3847/2041-8213/ad74ed},
	language = {en},
	number = {1},
	urldate = {2025-07-04},
	journal = {The Astrophysical Journal Letters},
	publisher = {The American Astronomical Society},
	author = {Limbach, Mary Anne and Vanderburg, Andrew and Venner, Alexander and Blouin, Simon and Stevenson, Kevin B. and MacDonald, Ryan J. and Jenkins, Sydney and Bowens-Rubin, Rachel and Soares-Furtado, Melinda and Morley, Caroline and Janson, Markus and Debes, John and Xu, Siyi and Kleisioti, Evangelia and Kenworthy, Matthew and Butler, Paul and Crane, Jeffrey D. and Osip, Dave and Shectman, Stephen and Teske, Johanna},
	month = sep,
	year = {2024},
	pages = {L11},
}

@article{limbach_thermal_2025,
	title = {Thermal {Emission} and {Confirmation} of the {Frigid} {White} {Dwarf} {Exoplanet} {WD} 1856+534 b},
	volume = {984},
	issn = {2041-8205},
	url = {https://dx.doi.org/10.3847/2041-8213/adc9ad},
	doi = {10.3847/2041-8213/adc9ad},
	language = {en},
	number = {1},
	urldate = {2025-07-04},
	journal = {The Astrophysical Journal Letters},
	publisher = {The American Astronomical Society},
	author = {Limbach, Mary Anne and Vanderburg, Andrew and MacDonald, Ryan J. and Stevenson, Kevin B. and Jenkins, Sydney and Blouin, Simon and Rauscher, Emily and Bowens-Rubin, Rachel and Gallo, Elena and Mang, James and Morley, Caroline V. and Sing, David K. and O’Connor, Christopher and Venner, Alexander and Xu, Siyi},
	month = apr,
	year = {2025},
	pages = {L28},
}

@misc{crotts_follow-up_2025,
	title = {Follow-{Up} {Exploration} of the {TWA} 7 {Planet}-{Disk} {System} with {JWST} {NIRCam}},
	url = {http://arxiv.org/abs/2506.19932},
	doi = {10.48550/arXiv.2506.19932},
	urldate = {2025-07-04},
	publisher = {arXiv},
	author = {Crotts, Katie A. and Carter, Aarynn L. and Lawson, Kellen and Mang, James and Biller, Beth and Booth, Mark and Ferrer-Chavez, Rodrigo and Girard, Julien H. and Lagrange, Anne-Marie and Liu, Michael C. and Marino, Sebastian and Millar-Blanchaer, Maxwell A. and Skemer, Andy and Strampelli, Giovanni M. and Wang, Jason and Absil, Olivier and Balmer, William O. and Bendahan-West, Raphaël and Bogat, Ellis and Bowens-Rubin, Rachel and Chauvin, Gaël and Fontanive, Clémence and Franson, Kyle and Kammerer, Jens and Leisenring, Jarron and Morley, Caroline V. and Rebollido, Isabel and Skaf, Nour and Sutlieff, Ben J. and Bruinsma, Evelyn L. and Hinkley, Sasha and Hoch, Kielan and James, Andrew D. and Kane, Rohan and Mawet, Dimitri and Meyer, Michael R. and Palatnick, Skyler and Perrin, Marshall D. and Ray, Shrishmoy and Rickman, Emily and Sanghi, Aniket and Stephenson, Klaus Subbotina},
	month = jun,
	year = {2025},
	note = {arXiv:2506.19932 [astro-ph]
version: 1},
}

@article{lagrange_evidence_2025,
	title = {Evidence for a sub-{Jovian} planet in the young {TWA} 7 disk},
	volume = {642},
	copyright = {2025 The Author(s)},
	issn = {1476-4687},
	url = {https://www.nature.com/articles/s41586-025-09150-4},
	doi = {10.1038/s41586-025-09150-4},
	language = {en},
	number = {8069},
	urldate = {2025-07-04},
	journal = {Nature},
	publisher = {Nature Publishing Group},
	author = {Lagrange, A.-M. and Wilkinson, C. and Mâlin, M. and Boccaletti, A. and Perrot, C. and Matrà, L. and Combes, F. and Beust, H. and Rouan, D. and Chomez, A. and Milli, J. and Charnay, B. and Mazevet, S. and Flasseur, O. and Olofsson, J. and Bayo, A. and Kral, Q. and Carter, A. and Crotts, K. A. and Delorme, P. and Chauvin, G. and Thebault, P. and Rubini, P. and Kiefer, F. and Radcliffe, A. and Mazoyer, J. and Bodrito, T. and Stasevic, S. and Langlois, M.},
	month = jun,
	year = {2025},
	pages = {905--908},
}

@article{harris_array_2020,
	title = {Array programming with {NumPy}},
	volume = {585},
	url = {https://doi.org/10.1038/s41586-020-2649-2},
	doi = {10.1038/s41586-020-2649-2},
	number = {7825},
	journal = {Nature},
	publisher = {Springer Science and Business Media LLC},
	author = {Harris, Charles R. and Millman, K. Jarrod and Walt, Stéfan J. van der and Gommers, Ralf and Virtanen, Pauli and Cournapeau, David and Wieser, Eric and Taylor, Julian and Berg, Sebastian and Smith, Nathaniel J. and Kern, Robert and Picus, Matti and Hoyer, Stephan and Kerkwijk, Marten H. van and Brett, Matthew and Haldane, Allan and Río, Jaime Fernández del and Wiebe, Mark and Peterson, Pearu and Gérard-Marchant, Pierre and Sheppard, Kevin and Reddy, Tyler and Weckesser, Warren and Abbasi, Hameer and Gohlke, Christoph and Oliphant, Travis E.},
	month = sep,
	year = {2020},
	pages = {357--362},
}

@article{hunter_matplotlib_2007,
	title = {Matplotlib: {A} {2D} graphics environment},
	volume = {9},
	doi = {10.1109/MCSE.2007.55},
	number = {3},
	journal = {Computing in Science \& Engineering},
	publisher = {IEEE COMPUTER SOC},
	author = {Hunter, J. D.},
	year = {2007},
	pages = {90--95},
}

@misc{ginsburg_astropyastroquery_2024,
	title = {astropy/astroquery: v0.4.7},
	url = {https://doi.org/10.5281/zenodo.10799414},
	doi = {10.5281/zenodo.10799414},
	publisher = {Zenodo},
	author = {Ginsburg, Adam and Sipőcz, Brigitta and Brasseur, C. E. and Parikh, Madhura and {jcsegovia} and Groener, Austen and Norman, Henrik and {derdon} and Kelley, Michael and Robitaille, Thomas and Lim, P. L. and Vaher, Eero and Deil, Christoph and Mommert, Michael and Medina, Jenny V. and Tollerud, Erik and Nilsson, Ricky and Baumann, Matthieu and Craig, Matt and Val-Borro, Miguel de and Weaver, Benjamin Alan and {jespinosaar} and Davies, James and Adeleke, Tinuade and Cowboy, Nick Space and Persson, Magnus Vilhelm and Dempsey, James and {syed-gilani} and Mesh, Christian and Mirocha, Jordan},
	month = mar,
	year = {2024},
}

@misc{gommers_scipyscipy_2023,
	title = {scipy/scipy: {SciPy} 1.11.2},
	url = {https://doi.org/10.5281/zenodo.8259693},
	doi = {10.5281/zenodo.8259693},
	publisher = {Zenodo},
	author = {Gommers, Ralf and Virtanen, Pauli and Burovski, Evgeni and Haberland, Matt and Weckesser, Warren and Oliphant, Travis E. and Reddy, Tyler and Cournapeau, David and Nelson, Andrew and {alexbrc} and Roy, Pamphile and Peterson, Pearu and Wilson, Josh and Polat, Ilhan and {endolith} and Mayorov, Nikolay and Walt, Stefan van der and Brett, Matthew and Laxalde, Denis and Larson, Eric and Millman, Jarrod and Sakai, Atsushi and {Lars} and {peterbell10} and Carey, C. J. and Mulbregt, Paul van and {eric-jones} and McKibben, Nicholas and Kern, Robert and {Kai}},
	month = aug,
	year = {2023},
}

@misc{team_pandas-devpandas_2025,
	title = {pandas-dev/pandas: {Pandas}},
	url = {https://doi.org/10.5281/zenodo.15597513},
	doi = {10.5281/zenodo.15597513},
	publisher = {Zenodo},
	author = {team, The pandas development},
	month = jun,
	year = {2025},
}

@inproceedings{perrin_updated_2014,
	series = {Society of {Photo}-{Optical} {Instrumentation} {Engineers} ({SPIE}) {Conference} {Series}},
	title = {Updated point spread function simulations for {JWST} with {WebbPSF}},
	volume = {9143},
	doi = {10.1117/12.2056689},
	booktitle = {Space {Telescopes} and {Instrumentation} 2014: {Optical}, {Infrared}, and {Millimeter} {Wave}},
	author = {Perrin, Marshall D. and Sivaramakrishnan, Anand and Lajoie, Charles-Philippe and Elliott, Erin and Pueyo, Laurent and Ravindranath, Swara and Albert, Loïc.},
	editor = {Oschmann, Jacobus M., Jr. and Clampin, Mark and Fazio, Giovanni G. and MacEwen, Howard A.},
	month = aug,
	year = {2014},
	pages = {91433X},
}

@inproceedings{perrin_simulating_2012,
	series = {Society of {Photo}-{Optical} {Instrumentation} {Engineers} ({SPIE}) {Conference} {Series}},
	title = {Simulating point spread functions for the {James} {Webb} {Space} {Telescope} with {WebbPSF}},
	volume = {8442},
	doi = {10.1117/12.925230},
	booktitle = {Space {Telescopes} and {Instrumentation} 2012: {Optical}, {Infrared}, and {Millimeter} {Wave}},
	author = {Perrin, Marshall D. and Soummer, Rémi and Elliott, Erin M. and Lallo, Matthew D. and Sivaramakrishnan, Anand},
	editor = {Clampin, Mark C. and Fazio, Giovanni G. and MacEwen, Howard A. and Oschmann, Jacobus M., Jr.},
	month = sep,
	year = {2012},
	pages = {84423D},
}

@article{van_der_walt_scikit-image_2014,
	title = {scikit-image: image processing in {Python}},
	volume = {2},
	issn = {2167-8359},
	url = {https://doi.org/10.7717/peerj.453},
	doi = {10.7717/peerj.453},
	journal = {PeerJ},
	author = {van der Walt, Stéfan and Schönberger, Johannes L. and Nunez-Iglesias, Juan and Boulogne, François and Warner, Joshua D. and Yager, Neil and Gouillart, Emmanuelle and Yu, Tony and contributors, the scikit-image},
	month = jun,
	year = {2014},
	pages = {e453},
}

@article{ginsburg_astroquery_2019,
	title = {astroquery: {An} {Astronomical} {Web}-querying {Package} in {Python}},
	volume = {157},
	doi = {10.3847/1538-3881/aafc33},
	journal = {{\textbackslash}aj},
	author = {Ginsburg, A. and Sipőcz, B. M. and Brasseur, C. E. and Cowperthwaite, P. S. and Craig, M. W. and Deil, C. and Guillochon, J. and Guzman, G. and Liedtke, S. and Lian Lim, P. and Lockhart, K. E. and Mommert, M. and Morris, B. M. and Norman, H. and Parikh, M. and Persson, M. V. and Robitaille, T. P. and Segovia, J.-C. and Singer, L. P. and Tollerud, E. J. and de Val-Borro, M. and Valtchanov, I. and Woillez, J. and {The Astroquery collaboration} and {a subset of the astropy collaboration}},
	month = mar,
	year = {2019},
	note = {\_eprint: 1901.04520},
	pages = {98},
}

@book{van_rossum_python_2009,
	address = {Scotts Valley, CA},
	title = {Python 3 {Reference} {Manual}},
	isbn = {1-4414-1269-7},
	publisher = {CreateSpace},
	author = {Van Rossum, Guido and Drake, Fred L.},
	year = {2009},
}

@inproceedings{mckinney_data_2010,
	title = {Data {Structures} for {Statistical} {Computing} in {Python}},
	doi = {10.25080/Majora-92bf1922-00a},
	booktitle = {Proceedings of the 9th {Python} in {Science} {Conference}},
	author = {McKinney, Wes},
	editor = {Walt, Stéfan van der and Millman, Jarrod},
	year = {2010},
	pages = {56 -- 61},
}

@article{virtanen_scipy_2020,
	title = {{SciPy} 1.0: {Fundamental} {Algorithms} for {Scientific} {Computing} in {Python}},
	volume = {17},
	doi = {10.1038/s41592-019-0686-2},
	journal = {Nature Methods},
	author = {Virtanen, Pauli and Gommers, Ralf and Oliphant, Travis E. and Haberland, Matt and Reddy, Tyler and Cournapeau, David and Burovski, Evgeni and Peterson, Pearu and Weckesser, Warren and Bright, Jonathan and van der Walt, Stéfan J. and Brett, Matthew and Wilson, Joshua and Millman, K. Jarrod and Mayorov, Nikolay and Nelson, Andrew R. J. and Jones, Eric and Kern, Robert and Larson, Eric and Carey, C J and Polat, İlhan and Feng, Yu and Moore, Eric W. and VanderPlas, Jake and Laxalde, Denis and Perktold, Josef and Cimrman, Robert and Henriksen, Ian and Quintero, E. A. and Harris, Charles R. and Archibald, Anne M. and Ribeiro, Antônio H. and Pedregosa, Fabian and van Mulbregt, Paul and {SciPy 1.0 Contributors}},
	year = {2020},
	pages = {261--272},
}

@article{astropy_collaboration_astropy_2022,
	title = {The {Astropy} {Project}: {Sustaining} and {Growing} a {Community}-oriented {Open}-source {Project} and the {Latest} {Major} {Release} (v5.0) of the {Core} {Package}},
	volume = {935},
	doi = {10.3847/1538-4357/ac7c74},
	number = {2},
	journal = {{\textbackslash}apj},
	author = {{Astropy Collaboration} and Price-Whelan, Adrian M. and Lim, Pey Lian and Earl, Nicholas and Starkman, Nathaniel and Bradley, Larry and Shupe, David L. and Patil, Aarya A. and Corrales, Lia and Brasseur, C. E. and N"othe, Maximilian and Donath, Axel and Tollerud, Erik and Morris, Brett M. and Ginsburg, Adam and Vaher, Eero and Weaver, Benjamin A. and Tocknell, James and Jamieson, William and van Kerkwijk, Marten H. and Robitaille, Thomas P. and Merry, Bruce and Bachetti, Matteo and G"unther, H. Moritz and Aldcroft, Thomas L. and Alvarado-Montes, Jaime A. and Archibald, Anne M. and B'odi, Attila and Bapat, Shreyas and Barentsen, Geert and Baz'an, Juanjo and Biswas, Manish and Boquien, M'ed'eric and Burke, D. J. and Cara, Daria and Cara, Mihai and Conroy, Kyle E. and Conseil, Simon and Craig, Matthew W. and Cross, Robert M. and Cruz, Kelle L. and D'Eugenio, Francesco and Dencheva, Nadia and Devillepoix, Hadrien A. R. and Dietrich, J"org P. and Eigenbrot, Arthur Davis and Erben, Thomas and Ferreira, Leonardo and Foreman-Mackey, Daniel and Fox, Ryan and Freij, Nabil and Garg, Suyog and Geda, Robel and Glattly, Lauren and Gondhalekar, Yash and Gordon, Karl D. and Grant, David and Greenfield, Perry and Groener, Austen M. and Guest, Steve and Gurovich, Sebastian and Handberg, Rasmus and Hart, Akeem and Hatfield-Dodds, Zac and Homeier, Derek and Hosseinzadeh, Griffin and Jenness, Tim and Jones, Craig K. and Joseph, Prajwel and Kalmbach, J. Bryce and Karamehmetoglu, Emir and Kaluszy'nski, Mikolaj and Kelley, Michael S. P. and Kern, Nicholas and Kerzendorf, Wolfgang E. and Koch, Eric W. and Kulumani, Shankar and Lee, Antony and Ly, Chun and Ma, Zhiyuan and MacBride, Conor and Maljaars, Jakob M. and Muna, Demitri and Murphy, N. A. and Norman, Henrik and O'Steen, Richard and Oman, Kyle A. and Pacifici, Camilla and Pascual, Sergio and Pascual-Granado, J. and Patil, Rohit R. and Perren, Gabriel I. and Pickering, Timothy E. and Rastogi, Tanuj and Roulston, Benjamin R. and Ryan, Daniel F. and Rykoff, Eli S. and Sabater, Jose and Sakurikar, Parikshit and Salgado, Jes'us and Sanghi, Aniket and Saunders, Nicholas and Savchenko, Volodymyr and Schwardt, Ludwig and Seifert-Eckert, Michael and Shih, Albert Y. and Jain, Anany Shrey and Shukla, Gyanendra and Sick, Jonathan and Simpson, Chris and Singanamalla, Sudheesh and Singer, Leo P. and Singhal, Jaladh and Sinha, Manodeep and SipHocz, Brigitta M. and Spitler, Lee R. and Stansby, David and Streicher, Ole and Sumak, Jani and Swinbank, John D. and Taranu, Dan S. and Tewary, Nikita and Tremblay, Grant R. and Val-Borro, Miguel de and Van Kooten, Samuel J. and Vasovi'c, Zlatan and Verma, Shresth and de Miranda Cardoso, Jos'e Vin'icius and Williams, Peter K. G. and Wilson, Tom J. and Winkel, Benjamin and Wood-Vasey, W. M. and Xue, Rui and Yoachim, Peter and Zhang, Chen and Zonca, Andrea and {Astropy Project Contributors}},
	month = aug,
	year = {2022},
	note = {\_eprint: 2206.14220},
	pages = {167},
}

@article{astropy_collaboration_astropy_2018,
	title = {The {Astropy} {Project}: {Building} an {Open}-science {Project} and {Status} of the v2.0 {Core} {Package}},
	volume = {156},
	doi = {10.3847/1538-3881/aabc4f},
	number = {3},
	journal = {{\textbackslash}aj},
	author = {{Astropy Collaboration} and Price-Whelan, A. M. and Sip{\textbackslash}Hocz, B. M. and Günther, H. M. and Lim, P. L. and Crawford, S. M. and Conseil, S. and Shupe, D. L. and Craig, M. W. and Dencheva, N. and Ginsburg, A. and Vand erPlas, J. T. and Bradley, L. D. and Pérez-Suárez, D. and de Val-Borro, M. and Aldcroft, T. L. and Cruz, K. L. and Robitaille, T. P. and Tollerud, E. J. and Ardelean, C. and Babej, T. and Bach, Y. P. and Bachetti, M. and Bakanov, A. V. and Bamford, S. P. and Barentsen, G. and Barmby, P. and Baumbach, A. and Berry, K. L. and Biscani, F. and Boquien, M. and Bostroem, K. A. and Bouma, L. G. and Brammer, G. B. and Bray, E. M. and Breytenbach, H. and Buddelmeijer, H. and Burke, D. J. and Calderone, G. and Cano Rodríguez, J. L. and Cara, M. and Cardoso, J. V. M. and Cheedella, S. and Copin, Y. and Corrales, L. and Crichton, D. and D'Avella, D. and Deil, C. and Depagne, É. and Dietrich, J. P. and Donath, A. and Droettboom, M. and Earl, N. and Erben, T. and Fabbro, S. and Ferreira, L. A. and Finethy, T. and Fox, R. T. and Garrison, L. H. and Gibbons, S. L. J. and Goldstein, D. A. and Gommers, R. and Greco, J. P. and Greenfield, P. and Groener, A. M. and Grollier, F. and Hagen, A. and Hirst, P. and Homeier, D. and Horton, A. J. and Hosseinzadeh, G. and Hu, L. and Hunkeler, J. S. and Ivezić, Ž. and Jain, A. and Jenness, T. and Kanarek, G. and Kendrew, S. and Kern, N. S. and Kerzendorf, W. E. and Khvalko, A. and King, J. and Kirkby, D. and Kulkarni, A. M. and Kumar, A. and Lee, A. and Lenz, D. and Littlefair, S. P. and Ma, Z. and Macleod, D. M. and Mastropietro, M. and McCully, C. and Montagnac, S. and Morris, B. M. and Mueller, M. and Mumford, S. J. and Muna, D. and Murphy, N. A. and Nelson, S. and Nguyen, G. H. and Ninan, J. P. and Nöthe, M. and Ogaz, S. and Oh, S. and Parejko, J. K. and Parley, N. and Pascual, S. and Patil, R. and Patil, A. A. and Plunkett, A. L. and Prochaska, J. X. and Rastogi, T. and Reddy Janga, V. and Sabater, J. and Sakurikar, P. and Seifert, M. and Sherbert, L. E. and Sherwood-Taylor, H. and Shih, A. Y. and Sick, J. and Silbiger, M. T. and Singanamalla, S. and Singer, L. P. and Sladen, P. H. and Sooley, K. A. and Sornarajah, S. and Streicher, O. and Teuben, P. and Thomas, S. W. and Tremblay, G. R. and Turner, J. E. H. and Terrón, V. and van Kerkwijk, M. H. and de la Vega, A. and Watkins, L. L. and Weaver, B. A. and Whitmore, J. B. and Woillez, J. and Zabalza, V. and {Astropy Contributors}},
	month = sep,
	year = {2018},
	note = {\_eprint: 1801.02634},
	pages = {123},
}

@article{astropy_collaboration_astropy_2013,
	title = {Astropy: {A} community {Python} package for astronomy},
	volume = {558},
	doi = {10.1051/0004-6361/201322068},
	journal = {åp},
	author = {{Astropy Collaboration} and Robitaille, T. P. and Tollerud, E. J. and Greenfield, P. and Droettboom, M. and Bray, E. and Aldcroft, T. and Davis, M. and Ginsburg, A. and Price-Whelan, A. M. and Kerzendorf, W. E. and Conley, A. and Crighton, N. and Barbary, K. and Muna, D. and Ferguson, H. and Grollier, F. and Parikh, M. M. and Nair, P. H. and Unther, H. M. and Deil, C. and Woillez, J. and Conseil, S. and Kramer, R. and Turner, J. E. H. and Singer, L. and Fox, R. and Weaver, B. A. and Zabalza, V. and Edwards, Z. I. and Azalee Bostroem, K. and Burke, D. J. and Casey, A. R. and Crawford, S. M. and Dencheva, N. and Ely, J. and Jenness, T. and Labrie, K. and Lim, P. L. and Pierfederici, F. and Pontzen, A. and Ptak, A. and Refsdal, B. and Servillat, M. and Streicher, O.},
	month = oct,
	year = {2013},
	note = {\_eprint: 1307.6212},
	pages = {A33},
}

@article{zhou_hubble_2021,
	title = {Hubble {Space} {Telescope} {UV} and {Hα} {Measurements} of the {Accretion} {Excess} {Emission} from the {Young} {Giant} {Planet} {PDS} 70 b},
	volume = {161},
	issn = {1538-3881},
	url = {https://dx.doi.org/10.3847/1538-3881/abeb7a},
	doi = {10.3847/1538-3881/abeb7a},
	language = {en},
	number = {5},
	urldate = {2025-06-16},
	journal = {The Astronomical Journal},
	publisher = {The American Astronomical Society},
	author = {Zhou, Yifan and Bowler, Brendan P. and Wagner, Kevin R. and Schneider, Glenn and Apai, Dániel and Kraus, Adam L. and Close, Laird M. and Herczeg, Gregory J. and Fang, Min},
	month = apr,
	year = {2021},
	pages = {244},
}

@article{pueyo_detection_2016,
	title = {{DETECTION} {AND} {CHARACTERIZATION} {OF} {EXOPLANETS} {USING} {PROJECTIONS} {ON} {KARHUNEN}–{LOEVE} {EIGENIMAGES}: {FORWARD} {MODELING}},
	volume = {824},
	issn = {0004-637X},
	shorttitle = {{DETECTION} {AND} {CHARACTERIZATION} {OF} {EXOPLANETS} {USING} {PROJECTIONS} {ON} {KARHUNEN}–{LOEVE} {EIGENIMAGES}},
	url = {https://dx.doi.org/10.3847/0004-637X/824/2/117},
	doi = {10.3847/0004-637X/824/2/117},
	language = {en},
	number = {2},
	urldate = {2025-06-16},
	journal = {The Astrophysical Journal},
	publisher = {The American Astronomical Society},
	author = {Pueyo, Laurent},
	month = jun,
	year = {2016},
	pages = {117},
}

@article{sanghi_efficiently_2022,
	title = {Efficiently {Imaging} {Accreting} {Protoplanets} from {Space}: {Reference} {Star} {Differential} {Imaging} of the {PDS} 70 {Planetary} {System} {Using} the {HST}/{WFC3} {Archival} {PSF} {Library}},
	volume = {163},
	issn = {1538-3881},
	shorttitle = {Efficiently {Imaging} {Accreting} {Protoplanets} from {Space}},
	url = {https://dx.doi.org/10.3847/1538-3881/ac477e},
	doi = {10.3847/1538-3881/ac477e},
	language = {en},
	number = {3},
	urldate = {2025-06-14},
	journal = {The Astronomical Journal},
	publisher = {The American Astronomical Society},
	author = {Sanghi, Aniket and Zhou, Yifan and Bowler, Brendan P.},
	month = feb,
	year = {2022},
	pages = {119},
}

@misc{gagliuffi_jwst_2025,
	title = {{JWST} {Coronagraphic} {Images} of 14 {Her} c: a {Cold} {Giant} {Planet} in a {Dynamically} {Hot}, {Multi}-planet {System}},
	shorttitle = {{JWST} {Coronagraphic} {Images} of 14 {Her} c},
	url = {http://arxiv.org/abs/2506.09201},
	doi = {10.48550/arXiv.2506.09201},
	urldate = {2025-06-12},
	publisher = {arXiv},
	author = {Gagliuffi, Daniella C. Bardalez and Balmer, William O. and Pueyo, Laurent and Brandt, Timothy D. and Giovinazzi, Mark R. and Millholland, Sarah and Black, Brennen and Lu, Tiger and Rice, Malena and Mang, James and Morley, Caroline and Lacy, Brianna and Girard, Julien and Matthews, Elisabeth and Carter, Aarynn and Bowler, Brendan P. and Faherty, Jacqueline K. and Fontanive, Clemence and Rickman, Emily},
	month = jun,
	year = {2025},
	note = {arXiv:2506.09201 [astro-ph]},
}

@article{wang_pyklip_2015,
	title = {{pyKLIP}: {PSF} {Subtraction} for {Exoplanets} and {Disks}},
	shorttitle = {{pyKLIP}},
	url = {https://ui.adsabs.harvard.edu/abs/2015ascl.soft06001W/abstract},
	language = {en},
	urldate = {2025-06-04},
	journal = {Astrophysics Source Code Library},
	author = {Wang, Jason J. and Ruffio, Jean-Baptise and De Rosa, Robert J. and Aguilar, Jonathan and Wolff, Schuyler G. and Pueyo, Laurent},
	month = jun,
	year = {2015},
	pages = {ascl:1506.001},
}

@misc{bowens-rubin_nircam_2025,
	title = {{NIRCam} yells at cloud: {JWST} {MIRI} imaging can directly detect exoplanets of the same temperature, mass, age, and orbital separation as {Saturn} and {Jupiter}},
	shorttitle = {{NIRCam} yells at cloud},
	url = {http://arxiv.org/abs/2505.15995},
	doi = {10.48550/arXiv.2505.15995},
	urldate = {2025-05-23},
	publisher = {arXiv},
	author = {Bowens-Rubin, Rachel and Mang, James and Limbach, Mary Anne and Carter, Aarynn L. and Stevenson, Kevin B. and Wagner, Kevin and Strampelli, Giovanni and Morley, Caroline V. and Kennedy, Grant and Matthews, Elisabeth and Vanderburg, Andrew and Salama, Maïssa},
	month = may,
	year = {2025},
	note = {arXiv:2505.15995 [astro-ph]},
}

@misc{bushouse_jwst_2025,
	title = {{JWST} {Calibration} {Pipeline}},
	url = {https://zenodo.org/records/15178003},
	doi = {10.5281/zenodo.15178003},
	urldate = {2025-05-01},
	publisher = {Zenodo},
	author = {Bushouse, Howard and Eisenhamer, Jonathan and Dencheva, Nadia and Davies, James and Greenfield, Perry and Morrison, Jane and Hodge, Phil and Simon, Bernie and Grumm, David and Droettboom, Michael and Slavich, Edward and Sosey, Megan and Pauly, Tyler and Miller, Todd and Jedrzejewski, Robert and Hack, Warren and Davis, David and Crawford, Steven and Law, David and Gordon, Karl and Regan, Michael and Cara, Mihai and MacDonald, Ken and Bradley, Larry and Shanahan, Clare and Jamieson, William and Teodoro, Mairan and Williams, Thomas and Pena-Guerrero, Maria and Graham, Brett and Molter, Edward and Brandt, Timothy and Hayes, Christian and Cooper, Rachel and Clarke, Melanie and Filippazzo, Joseph},
	month = apr,
	year = {2025},
}

@article{brandt_likelihood-based_2024,
	title = {Likelihood-based {Jump} {Detection} and {Cosmic} {Ray} {Rejection} for {Detectors} {Read} {Out} {Up}-the-ramp},
	volume = {136},
	issn = {1538-3873},
	url = {https://dx.doi.org/10.1088/1538-3873/ad38da},
	doi = {10.1088/1538-3873/ad38da},
	language = {en},
	number = {4},
	urldate = {2025-05-01},
	journal = {Publications of the Astronomical Society of the Pacific},
	publisher = {The Astronomical Society of the Pacific},
	author = {Brandt, Timothy D.},
	month = may,
	year = {2024},
	pages = {045005},
}

@article{brandt_optimal_2024,
	title = {Optimal {Fitting} and {Debiasing} for {Detectors} {Read} {Out} {Up}-the-{Ramp}},
	volume = {136},
	issn = {1538-3873},
	url = {https://dx.doi.org/10.1088/1538-3873/ad38d9},
	doi = {10.1088/1538-3873/ad38d9},
	language = {en},
	number = {4},
	urldate = {2025-05-01},
	journal = {Publications of the Astronomical Society of the Pacific},
	publisher = {The Astronomical Society of the Pacific},
	author = {Brandt, Timothy D.},
	month = may,
	year = {2024},
	pages = {045004},
}

@article{carter_spaceklip_2025,
	title = {{spaceKLIP}: {JWST} coronagraphy data data reduction and analysis pipeline},
	shorttitle = {{spaceKLIP}},
	url = {https://ui.adsabs.harvard.edu/abs/2025ascl.soft02014C/abstract},
	language = {en},
	urldate = {2025-05-01},
	journal = {Astrophysics Source Code Library},
	author = {Carter, Aarynn and Kammerer, Jens and Leisenring, Jarron and Perrin, Marshall and Balmer, William O. and Glidic, Kayli and Strampelli, Giovanni Maria and Millar-Blanchaer, Max},
	month = feb,
	year = {2025},
	pages = {ascl:2502.014},
}

@inproceedings{kammerer_performance_2022,
	title = {Performance of near-infrared high-contrast imaging methods with {JWST} from commissioning},
	volume = {12180},
	url = {https://www.spiedigitallibrary.org/conference-proceedings-of-spie/12180/121803N/Performance-of-near-infrared-high-contrast-imaging-methods-with-JWST/10.1117/12.2628865.full},
	doi = {10.1117/12.2628865},
	urldate = {2025-05-01},
	booktitle = {Space {Telescopes} and {Instrumentation} 2022: {Optical}, {Infrared}, and {Millimeter} {Wave}},
	publisher = {SPIE},
	author = {Kammerer, Jens and Girard, Julien and Carter, Aarynn L. and Perrin, Marshall D. and Cooper, Rachel and Thatte, Deepashri and Vandal, Thomas and Leisenring, Jarron and Wang, Jason and Balmer, William O. and Sivaramakrishnan, Anand and Pueyo, Laurent and Ward-Duong, Kimberly and Sunnquist, Ben and Redai, Jéa Adams},
	month = aug,
	year = {2022},
	pages = {1369--1388},
}

@article{liu_substructure_2004,
	title = {Substructure in the {Circumstellar} {Disk} {Around} the {Young} {Star} {AU} {Microscopii}},
	volume = {305},
	url = {https://www.science.org/doi/10.1126/science.1102929},
	doi = {10.1126/science.1102929},
	number = {5689},
	urldate = {2025-04-22},
	journal = {Science},
	publisher = {American Association for the Advancement of Science},
	author = {Liu, Michael C.},
	month = sep,
	year = {2004},
	pages = {1442--1444},
}

@article{marois_angular_2006,
	title = {Angular {Differential} {Imaging}: {A} {Powerful} {High}-{Contrast} {Imaging} {Technique}*},
	volume = {641},
	issn = {0004-637X},
	shorttitle = {Angular {Differential} {Imaging}},
	url = {https://iopscience.iop.org/article/10.1086/500401/meta},
	doi = {10.1086/500401},
	language = {en},
	number = {1},
	urldate = {2025-04-22},
	journal = {The Astrophysical Journal},
	publisher = {IOP Publishing},
	author = {Marois, Christian and Lafrenière, David and Doyon, René and Macintosh, Bruce and Nadeau, Daniel},
	month = apr,
	year = {2006},
	pages = {556},
}

@misc{leisenring_webbpsf_2025,
	title = {{WebbPSF} {Extensions}},
	url = {https://zenodo.org/records/15086592},
	doi = {10.5281/zenodo.15086592},
	urldate = {2025-04-20},
	publisher = {Zenodo},
	author = {Leisenring, Jarron},
	month = mar,
	year = {2025},
}

@article{mang_microphysics_2022,
	title = {Microphysics of {Water} {Clouds} in the {Atmospheres} of {Y} {Dwarfs} and {Temperate} {Giant} {Planets}},
	volume = {927},
	issn = {0004-637X},
	url = {https://dx.doi.org/10.3847/1538-4357/ac51d3},
	doi = {10.3847/1538-4357/ac51d3},
	language = {en},
	number = {2},
	urldate = {2025-04-20},
	journal = {The Astrophysical Journal},
	publisher = {The American Astronomical Society},
	author = {Mang, James and Gao, Peter and Hood, Callie E. and Fortney, Jonathan J. and Batalha, Natasha and Yu, Xinting and de Pater, Imke},
	month = mar,
	year = {2022},
	pages = {184},
}

@article{madhusudhan_exoplanetary_2019,
	title = {Exoplanetary {Atmospheres}: {Key} {Insights}, {Challenges}, and {Prospects}},
	volume = {57},
	issn = {0066-4146, 1545-4282},
	shorttitle = {Exoplanetary {Atmospheres}},
	url = {https://www.annualreviews.org/content/journals/10.1146/annurev-astro-081817-051846},
	doi = {10.1146/annurev-astro-081817-051846},
	language = {en},
	number = {Volume 57, 2019},
	urldate = {2025-04-20},
	journal = {Annual Review of Astronomy and Astrophysics},
	publisher = {Annual Reviews},
	author = {Madhusudhan, Nikku},
	month = aug,
	year = {2019},
	pages = {617--663},
}

@article{leggett_first_2023,
	title = {The {First} {Y} {Dwarf} {Data} from {JWST} {Show} that {Dynamic} and {Diabatic} {Processes} {Regulate} {Cold} {Brown} {Dwarf} {Atmospheres}},
	volume = {959},
	issn = {0004-637X},
	url = {https://dx.doi.org/10.3847/1538-4357/acfdad},
	doi = {10.3847/1538-4357/acfdad},
	language = {en},
	number = {2},
	urldate = {2025-04-14},
	journal = {The Astrophysical Journal},
	publisher = {The American Astronomical Society},
	author = {Leggett, S. K. and Tremblin, Pascal},
	month = dec,
	year = {2023},
	pages = {86},
}

@article{morley_water_2014,
	title = {{WATER} {CLOUDS} {IN} {Y} {DWARFS} {AND} {EXOPLANETS}},
	volume = {787},
	issn = {0004-637X},
	url = {https://dx.doi.org/10.1088/0004-637X/787/1/78},
	doi = {10.1088/0004-637X/787/1/78},
	language = {en},
	number = {1},
	urldate = {2025-04-12},
	journal = {The Astrophysical Journal},
	publisher = {The American Astronomical Society},
	author = {Morley, Caroline V. and Marley, Mark S. and Fortney, Jonathan J. and Lupu, Roxana and Saumon, Didier and Greene, Tom and Lodders, Katharina},
	month = may,
	year = {2014},
	pages = {78},
}

@article{marley_sonora_2021,
	title = {The {Sonora} {Brown} {Dwarf} {Atmosphere} and {Evolution} {Models}. {I}. {Model} {Description} and {Application} to {Cloudless} {Atmospheres} in {Rainout} {Chemical} {Equilibrium}},
	volume = {920},
	issn = {0004-637X},
	url = {https://dx.doi.org/10.3847/1538-4357/ac141d},
	doi = {10.3847/1538-4357/ac141d},
	language = {en},
	number = {2},
	urldate = {2025-04-12},
	journal = {The Astrophysical Journal},
	publisher = {The American Astronomical Society},
	author = {Marley, Mark S. and Saumon, Didier and Visscher, Channon and Lupu, Roxana and Freedman, Richard and Morley, Caroline and Fortney, Jonathan J. and Seay, Christopher and Smith, Adam J. R. W. and Teal, D. J. and Wang, Ruoyan},
	month = oct,
	year = {2021},
	pages = {85},
}

@article{lacy_self-consistent_2023,
	title = {Self-consistent {Models} of {Y} {Dwarf} {Atmospheres} with {Water} {Clouds} and {Disequilibrium} {Chemistry}},
	volume = {950},
	issn = {0004-637X},
	url = {https://dx.doi.org/10.3847/1538-4357/acc8cb},
	doi = {10.3847/1538-4357/acc8cb},
	language = {en},
	number = {1},
	urldate = {2025-04-12},
	journal = {The Astrophysical Journal},
	publisher = {The American Astronomical Society},
	author = {Lacy, Brianna and Burrows, Adam},
	month = jun,
	year = {2023},
	pages = {8},
}

@article{leggett_measuring_2021,
	title = {Measuring and {Replicating} the 1–20 μm {Energy} {Distributions} of the {Coldest} {Brown} {Dwarfs}: {Rotating}, {Turbulent}, and {Nonadiabatic} {Atmospheres}},
	volume = {918},
	issn = {0004-637X},
	shorttitle = {Measuring and {Replicating} the 1–20 μm {Energy} {Distributions} of the {Coldest} {Brown} {Dwarfs}},
	url = {https://dx.doi.org/10.3847/1538-4357/ac0cfe},
	doi = {10.3847/1538-4357/ac0cfe},
	language = {en},
	number = {1},
	urldate = {2025-04-12},
	journal = {The Astrophysical Journal},
	publisher = {The American Astronomical Society},
	author = {Leggett, S. K. and Tremblin, Pascal and Phillips, Mark W. and Dupuy, Trent J. and Marley, Mark and Morley, Caroline and Schneider, Adam and Caselden, Dan and Guillaume, Colin and Logsdon, Sarah E.},
	month = aug,
	year = {2021},
	pages = {11},
}

@article{phillips_new_2020,
	title = {A new set of atmosphere and evolution models for cool {T}-{Y} brown dwarfs and giant exoplanets},
	volume = {637},
	issn = {0004-6361},
	url = {https://ui.adsabs.harvard.edu/abs/2020A&A...637A..38P/abstract},
	doi = {10.1051/0004-6361/201937381},
	language = {en},
	urldate = {2025-04-12},
	journal = {Astronomy and Astrophysics},
	author = {Phillips, M. W. and Tremblin, P. and Baraffe, I. and Chabrier, G. and Allard, N. F. and Spiegelman, F. and Goyal, J. M. and Drummond, B. and Hébrard, E.},
	month = may,
	year = {2020},
	pages = {A38},
}

@article{meisner_exploring_2023,
	title = {Exploring the {Extremes}: {Characterizing} a {New} {Population} of {Old} and {Cold} {Brown} {Dwarfs}},
	volume = {166},
	issn = {0004-6256},
	shorttitle = {Exploring the {Extremes}},
	url = {https://ui.adsabs.harvard.edu/abs/2023AJ....166...57M/abstract},
	doi = {10.3847/1538-3881/acdb68},
	language = {en},
	number = {2},
	urldate = {2025-04-12},
	journal = {The Astronomical Journal},
	author = {Meisner, Aaron M. and Leggett, S. K. and Logsdon, Sarah E. and Schneider, Adam C. and Tremblin, Pascal and Phillips, Mark},
	month = aug,
	year = {2023},
	pages = {57},
}

@article{balmer_jwst-tst_2025,
	title = {{JWST}-{TST} {High} {Contrast}: {Living} on the {Wedge}, or, {NIRCam} {Bar} {Coronagraphy} {Reveals} {CO}$_{\textrm{2}}$ in the {HR} 8799 and 51 {Eri} {Exoplanets}’ {Atmospheres}},
	volume = {169},
	issn = {0004-6256, 1538-3881},
	shorttitle = {{JWST}-{TST} {High} {Contrast}},
	url = {https://iopscience.iop.org/article/10.3847/1538-3881/adb1c6},
	doi = {10.3847/1538-3881/adb1c6},
	language = {en},
	number = {4},
	urldate = {2025-03-17},
	journal = {The Astronomical Journal},
	author = {Balmer, William O. and Kammerer, Jens and Pueyo, Laurent and Perrin, Marshall D. and Girard, Julien H. and Leisenring, Jarron M. and Lawson, Kellen and Dennen, Henry and Van Der Marel, Roeland P. and Beichman, Charles A. and Bryden, Geoffrey and Llop-Sayson, Jorge and Valenti, Jeff A. and Lothringer, Joshua D. and Lewis, Nikole K. and Mâlin, Mathilde and Rebollido, Isabel and Rickman, Emily and Hoch, Kielan K. W. and Soummer, Rémi and Clampin, Mark and Mountain, C. Matt},
	month = apr,
	year = {2025},
	pages = {209},
}

@article{mukherjee_sonora_2024,
	title = {The {Sonora} {Substellar} {Atmosphere} {Models}. {IV}. {Elf} {Owl}: {Atmospheric} {Mixing} and {Chemical} {Disequilibrium} with {Varying} {Metallicity} and {C}/{O} {Ratios}},
	volume = {963},
	issn = {0004-637X},
	shorttitle = {The {Sonora} {Substellar} {Atmosphere} {Models}. {IV}. {Elf} {Owl}},
	url = {https://dx.doi.org/10.3847/1538-4357/ad18c2},
	doi = {10.3847/1538-4357/ad18c2},
	language = {en},
	number = {1},
	urldate = {2025-02-11},
	journal = {The Astrophysical Journal},
	publisher = {The American Astronomical Society},
	author = {Mukherjee, Sagnick and Fortney, Jonathan J. and Morley, Caroline V. and Batalha, Natasha E. and Marley, Mark S. and Karalidi, Theodora and Visscher, Channon and Lupu, Roxana and Freedman, Richard and Gharib-Nezhad, Ehsan},
	month = feb,
	year = {2024},
	pages = {73},
}

@article{lagrange_giant_2010,
	title = {A {Giant} {Planet} {Imaged} in the {Disk} of the {Young} {Star} β {Pictoris}},
	volume = {329},
	issn = {0036-8075},
	url = {https://ui.adsabs.harvard.edu/abs/2010Sci...329...57L},
	doi = {10.1126/science.1187187},
	urldate = {2025-01-23},
	journal = {Science},
	author = {Lagrange, A. -M. and Bonnefoy, M. and Chauvin, G. and Apai, D. and Ehrenreich, D. and Boccaletti, A. and Gratadour, D. and Rouan, D. and Mouillet, D. and Lacour, S. and Kasper, M.},
	month = jul,
	year = {2010},
	note = {ADS Bibcode: 2010Sci...329...57L},
	pages = {57},
}

@article{mawet_fundamental_2014,
	title = {Fundamental {Limitations} of {High} {Contrast} {Imaging} {Set} by {Small} {Sample} {Statistics}},
	volume = {792},
	issn = {0004-637X},
	url = {https://ui.adsabs.harvard.edu/abs/2014ApJ...792...97M},
	doi = {10.1088/0004-637X/792/2/97},
	urldate = {2025-01-09},
	journal = {The Astrophysical Journal},
	publisher = {IOP},
	author = {Mawet, D. and Milli, J. and Wahhaj, Z. and Pelat, D. and Absil, O. and Delacroix, C. and Boccaletti, A. and Kasper, M. and Kenworthy, M. and Marois, C. and Mennesson, B. and Pueyo, L.},
	month = sep,
	year = {2014},
	note = {ADS Bibcode: 2014ApJ...792...97M},
	pages = {97},
}

@article{rigby_science_2023,
	title = {The {Science} {Performance} of {JWST} as {Characterized} in {Commissioning}},
	volume = {135},
	issn = {0004-6280},
	url = {https://ui.adsabs.harvard.edu/abs/2023PASP..135d8001R},
	doi = {10.1088/1538-3873/acb293},
	urldate = {2025-01-09},
	journal = {Publications of the Astronomical Society of the Pacific},
	publisher = {IOP},
	author = {Rigby, Jane and Perrin, Marshall and McElwain, Michael and Kimble, Randy and Friedman, Scott and Lallo, Matt and Doyon, René and Feinberg, Lee and Ferruit, Pierre and Glasse, Alistair and Rieke, Marcia and Rieke, George and Wright, Gillian and Willott, Chris and Colon, Knicole and Milam, Stefanie and Neff, Susan and Stark, Christopher and Valenti, Jeff and Abell, Jim and Abney, Faith and Abul-Huda, Yasin and Acton, D. Scott and Adams, Evan and Adler, David and Aguilar, Jonathan and Ahmed, Nasif and Albert, Loïc and Alberts, Stacey and Aldridge, David and Allen, Marsha and Altenburg, Martin and Álvarez-Márquez, Javier and Alves de Oliveira, Catarina and Andersen, Greg and Anderson, Harry and Anderson, Sara and Argyriou, Ioannis and Armstrong, Amber and Arribas, Santiago and Artigau, Etienne and Arvai, Amanda and Atkinson, Charles and Bacon, Gregory and Bair, Thomas and Banks, Kimberly and Barrientes, Jaclyn and Barringer, Bruce and Bartosik, Peter and Bast, William and Baudoz, Pierre and Beatty, Thomas and Bechtold, Katie and Beck, Tracy and Bergeron, Eddie and Bergkoetter, Matthew and Bhatawdekar, Rachana and Birkmann, Stephan and Blazek, Ronald and Blome, Claire and Boccaletti, Anthony and Böker, Torsten and Boia, John and Bonaventura, Nina and Bond, Nicholas and Bosley, Kari and Boucarut, Ray and Bourque, Matthew and Bouwman, Jeroen and Bower, Gary and Bowers, Charles and Boyer, Martha and Bradley, Larry and Brady, Greg and Braun, Hannah and Breda, David and Bresnahan, Pamela and Bright, Stacey and Britt, Christopher and Bromenschenkel, Asa and Brooks, Brian and Brooks, Keira and Brown, Bob and Brown, Matthew and Brown, Patricia and Bunker, Andy and Burger, Matthew and Bushouse, Howard and Cale, Steven and Cameron, Alex and Cameron, Peter and Canipe, Alicia and Caplinger, James and Caputo, Francis and Cara, Mihai and Carey, Larkin and Carniani, Stefano and Carrasquilla, Maria and Carruthers, Margaret and Case, Michael and Catherine, Riggs and Chance, Don and Chapman, George and Charlot, Stéphane and Charlow, Brian and Chayer, Pierre and Chen, Bin and Cherinka, Brian and Chichester, Sarah and Chilton, Zack and Chonis, Taylor and Clampin, Mark and Clark, Charles and Clark, Kerry and Coe, Dan and Coleman, Benee and Comber, Brian and Comeau, Tom and Connolly, Dennis and Cooper, James and Cooper, Rachel and Coppock, Eric and Correnti, Matteo and Cossou, Christophe and Coulais, Alain and Coyle, Laura and Cracraft, Misty and Curti, Mirko and Cuturic, Steven and Davis, Katherine and Davis, Michael and Dean, Bruce and DeLisa, Amy and deMeester, Wim and Dencheva, Nadia and Dencheva, Nadezhda and DePasquale, Joseph and Deschenes, Jeremy and Hunor Detre, Örs and Diaz, Rosa and Dicken, Dan and DiFelice, Audrey and Dillman, Matthew and Dixon, William and Doggett, Jesse and Donaldson, Tom and Douglas, Rob and DuPrie, Kimberly and Dupuis, Jean and Durning, John and Easmin, Nilufar and Eck, Weston and Edeani, Chinwe and Egami, Eiichi and Ehrenwinkler, Ralf and Eisenhamer, Jonathan and Eisenhower, Michael and Elie, Michelle and Elliott, James and Elliott, Kyle and Ellis, Tracy and Engesser, Michael and Espinoza, Nestor and Etienne, Odessa and Etxaluze, Mireya and Falini, Patrick and Feeney, Matthew and Ferry, Malcolm and Filippazzo, Joseph and Fincham, Brian and Fix, Mees and Flagey, Nicolas and Florian, Michael and Flynn, Jim and Fontanella, Erin and Ford, Terrance and Forshay, Peter and Fox, Ori and Franz, David and Fu, Henry and Fullerton, Alexander and Galkin, Sergey and Galyer, Anthony and García Marín, Macarena and Gardner, Jonathan P. and Gardner, Lisa and Garland, Dennis and Garrett, Bruce and Gasman, Danny and Gaspar, Andras and Gaudreau, Daniel and Gauthier, Peter and Geers, Vincent and Geithner, Paul and Gennaro, Mario and Giardino, Giovanna and Girard, Julien and Giuliano, Mark and Glassmire, Kirk and Glauser, Adrian},
	month = apr,
	year = {2023},
	note = {ADS Bibcode: 2023PASP..135d8001R},
	pages = {048001},
}

@article{christiaens_vip_2023,
	title = {{VIP}: {A} {Python} package for high-contrast imaging},
	volume = {8},
	shorttitle = {{VIP}},
	url = {https://ui.adsabs.harvard.edu/abs/2023JOSS....8.4774C},
	doi = {10.21105/joss.04774},
	urldate = {2025-01-08},
	journal = {The Journal of Open Source Software},
	author = {Christiaens, Valentin and Gonzalez, Carlos and Farkas, Ralf and Dahlqvist, Carl-Henrik and Nasedkin, Evert and Milli, Julien and Absil, Olivier and Ngo, Henry and Cantero, Carles and Rainot, Alan and Hammond, Iain and Bonse, Markus and Cantalloube, Faustine and Vigan, Arthur and Kompella, Vijay and Hancock, Paul},
	month = jan,
	year = {2023},
	note = {ADS Bibcode: 2023JOSS....8.4774C},
	pages = {4774},
}

@article{gomez_gonzalez_vip_2017,
	title = {{VIP}: {Vortex} {Image} {Processing} {Package} for {High}-contrast {Direct} {Imaging}},
	volume = {154},
	issn = {0004-6256},
	shorttitle = {{VIP}},
	url = {https://ui.adsabs.harvard.edu/abs/2017AJ....154....7G},
	doi = {10.3847/1538-3881/aa73d7},
	urldate = {2025-01-08},
	journal = {The Astronomical Journal},
	publisher = {IOP},
	author = {Gomez Gonzalez, Carlos Alberto and Wertz, Olivier and Absil, Olivier and Christiaens, Valentin and Defrère, Denis and Mawet, Dimitri and Milli, Julien and Absil, Pierre-Antoine and Van Droogenbroeck, Marc and Cantalloube, Faustine and Hinz, Philip M. and Skemer, Andrew J. and Karlsson, Mikael and Surdej, Jean},
	month = jul,
	year = {2017},
	note = {ADS Bibcode: 2017AJ....154....7G},
	pages = {7},
}

@article{soummer_detection_2012,
	title = {Detection and {Characterization} of {Exoplanets} and {Disks} {Using} {Projections} on {Karhunen}-{Loève} {Eigenimages}},
	volume = {755},
	issn = {0004-637X},
	url = {https://ui.adsabs.harvard.edu/abs/2012ApJ...755L..28S},
	doi = {10.1088/2041-8205/755/2/L28},
	urldate = {2025-01-06},
	journal = {The Astrophysical Journal},
	publisher = {IOP},
	author = {Soummer, Rémi and Pueyo, Laurent and Larkin, James},
	month = aug,
	year = {2012},
	note = {ADS Bibcode: 2012ApJ...755L..28S},
	pages = {L28},
}

@article{matthews_temperate_2024,
	title = {A temperate super-{Jupiter} imaged with {JWST} in the mid-infrared},
	volume = {633},
	copyright = {2024 The Author(s)},
	issn = {1476-4687},
	url = {https://www.nature.com/articles/s41586-024-07837-8},
	doi = {10.1038/s41586-024-07837-8},
	language = {en},
	number = {8031},
	urldate = {2024-12-15},
	journal = {Nature},
	publisher = {Nature Publishing Group},
	author = {Matthews, E. C. and Carter, A. L. and Pathak, P. and Morley, C. V. and Phillips, M. W. and P. M., S. Krishanth and Feng, F. and Bonse, M. J. and Boogaard, L. A. and Burt, J. A. and Crossfield, I. J. M. and Douglas, E. S. and Henning, Th and Hom, J. and Ko, C.-L. and Kasper, M. and Lagrange, A.-M. and Petit dit de la Roche, D. and Philipot, F.},
	month = sep,
	year = {2024},
	pages = {789--792},
}

@article{bowler_imaging_2016,
	title = {Imaging {Extrasolar} {Giant} {Planets}},
	volume = {128},
	issn = {0004-6280},
	url = {https://ui.adsabs.harvard.edu/abs/2016PASP..128j2001B},
	doi = {10.1088/1538-3873/128/968/102001},
	urldate = {2024-12-15},
	journal = {Publications of the Astronomical Society of the Pacific},
	publisher = {IOP},
	author = {Bowler, Brendan P.},
	month = oct,
	year = {2016},
	note = {ADS Bibcode: 2016PASP..128j2001B},
	pages = {102001},
}

@misc{leggett_redshifting_2024,
	title = {Redshifting the {Study} of {Cold} {Brown} {Dwarfs} and {Exoplanets}: the {Mid}-{Infrared} {Wavelength} {Region} as an {Indicator} of {Surface} {Gravity} and {Mass}},
	shorttitle = {Redshifting the {Study} of {Cold} {Brown} {Dwarfs} and {Exoplanets}},
	url = {http://arxiv.org/abs/2411.03549},
	urldate = {2024-11-07},
	publisher = {arXiv},
	author = {Leggett, S. K. and Tremblin, Pascal},
	month = nov,
	year = {2024},
	note = {arXiv:2411.03549},
}

@misc{mang_microphysical_2024,
	title = {Microphysical {Prescriptions} for {Parameterized} {Water} {Cloud} {Formation} on {Ultra}-cool {Substellar} {Objects}},
	url = {http://arxiv.org/abs/2408.08958},
	doi = {10.48550/arXiv.2408.08958},
	urldate = {2024-08-20},
	publisher = {arXiv},
	author = {Mang, James and Morley, Caroline V. and Robinson, Tyler D. and Gao, Peter},
	month = aug,
	year = {2024},
	note = {arXiv:2408.08958 [astro-ph]},
}

@article{wagner_imaging_2021,
	title = {Imaging low-mass planets within the habitable zone of α {Centauri}},
	volume = {12},
	issn = {2041-1723},
	url = {https://ui.adsabs.harvard.edu/abs/2021NatCo..12..922W},
	doi = {10.1038/s41467-021-21176-6},
	urldate = {2024-06-30},
	journal = {Nature Communications},
	author = {Wagner, K. and Boehle, A. and Pathak, P. and Kasper, M. and Arsenault, R. and Jakob, G. and Käufl, U. and Leveratto, S. and Maire, A. -L. and Pantin, E. and Siebenmorgen, R. and Zins, G. and Absil, O. and Ageorges, N. and Apai, D. and Carlotti, A. and Choquet, É. and Delacroix, C. and Dohlen, K. and Duhoux, P. and Forsberg, P. and Fuenteseca, E. and Gutruf, S. and Guyon, O. and Huby, E. and Kampf, D. and Karlsson, M. and Kervella, P. and Kirchbauer, J. -P. and Klupar, P. and Kolb, J. and Mawet, D. and N'Diaye, M. and Orban de Xivry, G. and Quanz, S. P. and Reutlinger, A. and Ruane, G. and Riquelme, M. and Soenke, C. and Sterzik, M. and Vigan, A. and de Zeeuw, T.},
	month = jan,
	year = {2021},
	note = {ADS Bibcode: 2021NatCo..12..922W},
	pages = {922},
}

@article{brandt_hipparcos-gaia_2018,
	title = {The {Hipparcos}-{Gaia} {Catalog} of {Accelerations}},
	volume = {239},
	issn = {0067-0049},
	url = {https://ui.adsabs.harvard.edu/abs/2018ApJS..239...31B},
	doi = {10.3847/1538-4365/aaec06},
	urldate = {2024-05-15},
	journal = {The Astrophysical Journal Supplement Series},
	publisher = {IOP},
	author = {Brandt, Timothy D.},
	month = dec,
	year = {2018},
	note = {ADS Bibcode: 2018ApJS..239...31B},
	pages = {31},
}

@article{carter_jwst_2023,
	title = {The {JWST} {Early} {Release} {Science} {Program} for {Direct} {Observations} of {Exoplanetary} {Systems} {I}: {High}-contrast {Imaging} of the {Exoplanet} {HIP} 65426 b from 2 to 16 μm},
	language = {en},
	journal = {The Astrophysical Journal Letters},
	author = {Carter, Aarynn L and Hinkley, Sasha and Kammerer, Jens and Skemer, Andrew and Biller, Beth A and Leisenring, Jarron M and Millar-Blanchaer, Maxwell A and Petrus, Simon and Stone, Jordan M and Ward-Duong, Kimberly and Wang, Jason J and Girard, Julien H and Hines, Dean C and Perrin, Marshall D and Pueyo, Laurent and Balmer, William O and Bonavita, Mariangela and Bonnefoy, Mickael and Chauvin, Gael and Choquet, Elodie and Christiaens, Valentin and Danielski, Camilla and Kennedy, Grant M and Matthews, Elisabeth C and Miles, Brittany E and Patapis, Polychronis and Ray, Shrishmoy and Rickman, Emily and Sallum, Steph and Stapelfeldt, Karl R and Whiteford, Niall and Zhou, Yifan and Absil, Olivier and Boccaletti, Anthony and Booth, Mark and Bowler, Brendan P and Chen, Christine H and Currie, Thayne and Fortney, Jonathan J and Grady, Carol A and Greebaum, Alexandra Z and Henning, Thomas and Hoch, Kielan K W and Janson, Markus and Kalas, Paul and Kenworthy, Matthew A and Kervella, Pierre and Kraus, Adam L and Lagage, Pierre-Olivier and Liu, Michael C and Macintosh, Bruce and Marino, Sebastian and Marley, Mark S and Marois, Christian and Matthews, Brenda C and Mawet, Dimitri and McElwain, Michael W and Metchev, Stanimir and Meyer, Michael R and Molliere, Paul and Moran, Sarah E and Morley, Caroline V and Mukherjee, Sagnick and Pantin, Eric and Quirrenbach, Andreas and Rebollido, Isabel and Ren, Bin B and Schneider, Glenn and Vasist, Malavika and Worthen, Kadin and Wyatt, Mark C and Briesemeister, Zackery W and Bryan, Marta L and Calissendorff, Per and Cantalloube, Faustine and Cugno, Gabriele and Furio, Matthew De and Dupuy, Trent J and Factor, Samuel M and Faherty, Jacqueline K and Fitzgerald, Michael P and Franson, Kyle and Gonzales, Eileen C and Hood, Callie E and Howe, Alex R and Kuzuhara, Masayuki and Lagrange, Anne-Marie and Lawson, Kellen and Lazzoni, Cecilia and Lew, Ben W P and Liu, Pengyu and Llop-Sayson, Jorge and Lloyd, James P and Martinez, Raquel A and Mazoyer, Johan and Palma-Bifani, Paulina and Quanz, Sascha P and Redai, Jea Adams and Samland, Matthias and Schlieder, Joshua E and Tamura, Motohide and Tan, Xianyu and Uyama, Taichi and Vigan, Arthur and Vos, Johanna M and Wagner, Kevin and Wolff, Schuyler G and Ygouf, Marie and Zhang, Xi and Zhang, Keming and Zhang, Zhoujian},
	year = {2023},
}

@article{nelder_simplex_1965,
	title = {A {Simplex} {Method} for {Function} {Minimization}},
	volume = {7},
	issn = {0010-4620},
	url = {https://doi.org/10.1093/comjnl/7.4.308},
	doi = {10.1093/comjnl/7.4.308},
	number = {4},
	journal = {The Computer Journal},
	author = {Nelder, J. A. and Mead, R.},
	month = jan,
	year = {1965},
	note = {\_eprint: https://academic.oup.com/comjnl/article-pdf/7/4/308/1013182/7-4-308.pdf},
	pages = {308--313},
}
\bibliographystyle{aasjournalv7}
\end{document}